\documentclass[journal]{IEEEtran}
\usepackage{amsmath,amsfonts}
\usepackage{array}
\usepackage{textcomp}
\usepackage{stfloats}
\usepackage{url}
\usepackage{verbatim}
\usepackage{graphicx}
\usepackage{xcolor}
\usepackage{mathtools} %splifrac
\usepackage{longtable}
\usepackage{adjustbox}

\usepackage{threeparttable}  
\usepackage{booktabs} 

\usepackage{float}
\usepackage{pifont}       % \ding{xx}
\usepackage{bbding}       % \Checkmark,\XSolid,... (需要和pifont宏包共同使用)
\usepackage{fontawesome}  % \faCheck,\faTimes
\usepackage{enumitem}

\usepackage{tikz}
\usetikzlibrary{trees,shadows.blur, matrix, positioning, calc, shapes.multipart}
\usepackage{forest}
\usepackage{hyperref}  %%
	
\usepackage{mdframed}

\usepackage[normalem]{ulem}
\usepackage{soul}
\definecolor{hlgrey}{rgb}{0.2,0.2,0.4}
\definecolor{hlgreen}{rgb}{0.0,0.7,0.0}
\definecolor{hlyellow}{rgb}{0.5,0.6,0.0}
\definecolor{hlcolortemp}{rgb}{1.0,0.0,0.5}
\definecolor{hlblue}{rgb}{0.0,0.0,0.99}
\definecolor{hlmagenta}{rgb}{1.0,0.0,0.6}
\definecolor{hlpurple}{rgb}{1.0,0.0,0.1}

\definecolor{boxtitlecolor}{gray}{0.7} % ??boxbackground???60%???
\definecolor{boxbgcolor}{gray}{0.96} % ??boxbackground???60%???
\definecolor{hlclcomment}{rgb}{1.0, 0.0, 0.0}
\definecolor{hlclcommentnc}{rgb}{0.9, 0.5, 0.2}
\definecolor{hlclmodify}{rgb}{0.0, 0.0,1.0}
\definecolor{hlcllink}{rgb}{0.0, 0.0,1.0}

\definecolor{mjclrevision}{rgb}{0.0, 0.0, 0.0}
\definecolor{mjclmanuscript}{rgb}{0.0,0.0,0.0}
\definecolor{mjclinternalchat}{rgb}{0.0,0.0,0.0}

\definecolor{hlunsurecolor}{rgb}{0.0, 0.7, 0.0}

\definecolor{blue-green}{rgb}{0.0, 0.87, 0.87}
\newcommand{\omark}{\ding{108}}
\newcommand{\xmark}{\ding{55}}
\newcommand{\lmark}{\ding{51}}
\newcommand{\wimark}{\ding{40}}
\newcommand{\acmark}{\ding{44}}

\newcommand{\mimark}{\ding{109}}

\newcommand{\hloitem}{\textcolor{blue-green}{\omark}}
\newcommand{\hlxitem}{\textcolor{red}{\xmark}}
\newcommand{\hllitem}{\textcolor{hlgreen}{\lmark}}
\newcommand{\hlwiitem}{\textcolor{hlblue}{\wimark}}
\newcommand{\hlacitem}{\textcolor{hlyellow}{\acmark}}

\newcommand{\hlmiitem}{\textcolor{hlpurple}{\mimark}}

\newcommand{\hlprioritem}{\textcolor{red}{\ding{88}}}

\newcommand{\hlnegmark}{\textcolor{red}{\ding{216}}}
\newcommand{\hlposmark}{\textcolor{hlgreen}{\ding{218}}}

\newcommand{\hlrevisioneq}[1]{\textcolor{mjclrevision}{#1}}

\usepackage{array} % 加载 array 包（尽管在这个例子中不是必需的，但通常用于表格定制）  

\allowdisplaybreaks[4]

\usepackage{cite}

\usepackage[linesnumbered,ruled]{algorithm2e} %hlma
\usepackage{multirow} %hlma
\usepackage{cellspace}
\usepackage{makecell}
\usepackage{amssymb}
\usepackage{cases}

\usepackage{theorem}
\usepackage[tight,footnotesize]{subfigure}

\def\myproof{\noindent{{\textbf{Proof:}}}} %hlma
\ifdefined\myproof
\else
\def\myproof{\proof}

\fi

\newtheorem{proof}{Proof}

\makeatletter

\newcommand{\Rmnum}[1]{\expandafter\@slowromancap\romannumeral #1@}
\makeatother

\hypersetup{
	colorlinks=true, % 启用彩色链接
	linkcolor=hlcllink,   % 内部链接（如章节、引用等）颜色
	filecolor=hlcllink,   % 文件链接颜色
	urlcolor=hlcllink,    % URL 链接颜色
	citecolor=hlcllink    % 引用链接颜色
}

\begin{document}

%\input{manuscript.tex}

%%%%%%%%%%%%%%%%%%%%%%%%%%%%Our customer commanc%%%%%%%%%%%%%%%%%%%%%%%%%%%%%%%%%%%%%%%%%%%%%%%%%%%%%%%
%  \hlrevision{Red underlined text (no reference citations)} %the reference citation can lead to compilation error

%   \hlrevisioneq{Red  text (reference citations allowed)}  % marked with Red  text (reference citations allowed)

%%%%%%%%%%%%%%%%%%%%%%%%%%%%%%%%%%%%%%%%%%%%%%%%%%%%%%%%%%%%%%%%%%%%%%%

\title{ 
	Through-the-Earth Magnetic Induction Communication and Networking: A Comprehensive Survey
}

\author{Honglei~Ma,
	Erwu~Liu,~\IEEEmembership{Senior Member, IEEE,}
	Wei~Ni,~\IEEEmembership{Fellow, IEEE,}
	Zhijun~Fang,
	Rui~Wang,
	Yongbin~Gao, \\
	Dusit Niyato,~\IEEEmembership{Fellow,~IEEE,}
	and Ekram Hossain,~\IEEEmembership{Fellow,~IEEE}
	%~Yu,
	%and Dongming~Zhang	
	\thanks{This work is supported in part by grants from the National Science Foundation of China (Nos. 42171404, 62271352), in part by Shanghai Engineering Research Center for Blockchain Applications And Services (No. 19DZ2255100),  in part by Seatrium New Energy Laboratory, Singapore Ministry of Education (MOE) Tier 1 (RT5/23 and RG24/24), the NTU Centre for Computational Technologies in Finance (NTU-CCTF), and the RIE2025 Industry Alignment Fund-Industry Collaboration Projects (IAF-ICP) (Award I2301E0026), administered by A*STAR, and in part by  the Fundamental Research Funds for the Central Universities under Grant 22120250094. 
	}
	\thanks{ H. Ma, Z. Fang, and Y. Gao are with the
		School of Electronic and Electrical Engineering, Shanghai University of Engineering Science, Shanghai, China (e-mail: holyma@yeah.net,   Zjfang@gmail.com, gaoyongbin@sues.edu.cn). }% <-this % stops a space
	\thanks{ E. Liu and R. Wang are with the College of Electronic and Information
		Engineering, Tongji University, Shanghai, China. R. Wang is also  with the Shanghai Institute of Intelligent Science and Technology, Tongji University, Shanghai, China (e-mail: erwu.liu@ieee.org, ruiwang@tongji.edu.cn).}% <-this % stops a space
	\thanks{W. Ni is with the School of Engineering, Edith Cowan University, Perth, WA 6027, and the School of Computer Science and Engineering, The University of New South Wales, Sydney, NSW 2033, Australia (e-mail: wei.ni@ieee.org).}
	\thanks{D. Niyato is with the College of Computing and Data Science, Nanyang Technological University, Singapore 639798 (e-mail: dniyato@ntu.edu.sg).}
	\thanks{E. Hossain is with the Department of Electrical and Computer Engineering, University of Manitoba, Winnipeg, Manitoba, Canada (e-mail: Ekram.Hossain@umanitoba.ca).}
	\thanks{ (\emph{Corresponding author: E. Liu})}
\thanks{DOI: 10.1109/COMST.2025.3623258}
	% <-this % stops a space
	\vspace{-1.2em}
} 

%\markboth{IEEE }%
%{Shell \MakeLowercase{\textit{\emph{et al.}}}:Fast-Fading Channel and Power  Optimization of the Magnetic Inductive Cellular Network}


\markboth{Accepted by IEEE Communications Surveys \& Tutorials for publication, DOI: 10.1109/COMST.2025.3623258 }
{Ma \MakeLowercase{\textit{et al.}}: 	Through-the-Earth Magnetic Induction Communication and Networking: A Comprehensive Survey}

\IEEEpubid{https://doi.org/10.1109/COMST.2025.3623258; \ \ \ \ \ \ \ \  }
%\IEEEpubid{ https://doi.org/10.1109/COMST.2025.3623258}

\maketitle

\begin{abstract}

	Magnetic induction (MI) communication (MIC) has emerged as a promising candidate for underground communication networks due to its excellent penetration capabilities. Integration with Space-Air-Ground-Underground (SAGUI) networks in next-generation mobile communication systems requires a well-defined network architecture. A recent discovery in MIC research, MI fast fading, remains in its early stages and presents unique challenges. This paper provides a comprehensive survey on through-the-earth (TTE) MIC, covering MI applications, channel modeling, point-to-point MIC design, relay techniques, network frameworks, and emerging technologies. We compare various MIC applications to highlight TTE-specific challenges and review the principles of channel modeling, addressing both MI slow fading and MI fast fading, along with its potential impact on existing MIC theories. We conduct a fine-grained decomposition of MI channel power gain into four distinct physical parameters, and propose a novel geometric model to analyze MI fast fading. We also summarize MI relay techniques, examine crosstalk effects in relay and high-density networks, and explore key research tasks within the OSI framework for a holistic MI network protocol in SAGUI. To bridge the gaps identified, we propose a MIC framework that supports TCP/IP and Linux, enabling full implementation of existing and emerging MIC solutions. This framework empowers researchers to leverage Linux resources and deep learning platforms for accelerated development of MIC in SAGUI networks. Remaining research challenges, open issues, and promising novel techniques are further identified to advance MIC research.

\end{abstract}

\begin{IEEEkeywords}
	Magnetic induction (MI), underground wireless communication, through-the-earth (TTE), fast fading,  network architecture, TCP/IP.
\end{IEEEkeywords}

\begin{table}[H]  
	\caption{Acronyms and definitions %\color{blue} \textbf{Comment:} please update the table alphabetically 
	}
	%\scriptsize \footnotesize \tiny \small
	\centering
	\vspace{-0.8em}
	\begin{threeparttable} 
		\scalebox{0.80}{
			\begin{tabular}{l |l }
				\hline 
				\textbf{Acronym} & \textbf{Full name / Definition}   \\
				\hline % \hline
				\hlprioritem / \hlprioritem\hlprioritem / \hlprioritem\hlprioritem\hlprioritem  & Low / Medium / High (priority used in tables and figures)  \\
				AF     & Amplified-and-forward   \\ %\hline
				AG     & Aboveground   \\ %\hline
				APO    & Antenna position and orientation   \\ %\hline
				AUV    & Autonomous underwater vehicle \\
				AVI    & Antenna vibration intensity \\ %\hline
				BAN    & Body area network  \\ %\hline
				BCS    & Boundary Chi-square (Boundary $\chi^2$) \\ %\hline
				BCH    & Bose, Ray-Chaudhuri, Hocquenghem \\ %\hline
				CDF    & Cumulative distribution function \\ %\hline
				CLO    & Cross-layer optimization  \\
				CLT    & Central limit theorem \\ %\hline
				CMG    & CMIC achievable rate gain \\ %\hline
				CMI    & Cooperative magnetic induction \\ %\hline   
				CMIC   & Cooperative magnetic induction communication\\ %\hline 
				CMIC-1NR & CMIC with one non-aligned relay  \\ %\hline 
				CMIC-$n$AR  & CMIC with multiple aligned relays  \\ %\hline     
				CSI    & Channel state information \\ %\hline
				CSMA   & Carrier sense multiple access \\ %\hline
				DF     & Decode-and-forward \\ %\hline
				DMI    & Direct magnetic induction \\ %\hline   
				DWE    & dynamic weighted evolution/learning \\
				EMW    & Electromagnetic wave \\ %\hline
				EMWC   & Electromagnetic wave communication\\ %\hline
				EPR    & Effective payload ratio \\ %\hline
				FEC    & Forward error correction \\ %\hline
				FEM    & Finite element method \\
				FF     & Filter-and-forward \\ %\hline
				FSOC   & Free-space optical communication \\ %\hline
				GAN    & Generative adversarial networks                   \\
				GTS    & Guaranteed time slot \\ %\hline                       \\ 
				IP-HC  & Internet protocol header compression \\
				IoV    & Internet of Vehicles \\
				JSCC   & Joint source-channel coding \\ %\hline     
				KVL    & Kirchhoff's voltage law \\ %\hline
				LLM    & Large language model \\
				LLC    & Logical link control  \\
				Ly     & Layer \\ %\hline
				MAC    & Medium access control \\ %\hline
				M$^2$I & Metamaterial-enhanced magnetic induction \\ %\hline
				MCD    & MI Linux character device\\
				MCNSI   & MI communication-navigation-sensing integrated \\ %\hline 
				MI     & Magnetic induction \\ %\hline
				MIC    & Magnetic induction communication\\ %\hline
				MIMO   & Multiple-input multiple-output \\ %\hline
				MND &  MI Linux network device \\
				MPRlA  & MI passive relay array \\ %\hline
				NFC    & near field communication \\
				OSI    & Open Systems Interconnection \\ %\hline
				P2P    & Point-to-point \\ %\hline
				PDF    & Probability density function \\ %\hline
				PSD    & Power spectral density \\ %\hline
				RL     & Reinforcement learning \\ %\hline
				RPMA   & Rotating permanent magnet antenna \\ %\hline
				RTT    & Round-trip time \\
				Rx     & Receive \\ %\hline
				SAGUMI & Space-Air-Ground-Underground multi-network integration\\
				SISO   & Single-in single-out \\ %\hline
				SNR    & Signal-to-noise ratio \\ %\hline
				TCP / IP & Transmission control protocol / Internet protocol \\
				TMR    & Tunneling Magnetoresistance \\
				ToA    & Time of arrival \\ %\hline 
				TTE    & Through-the-earth \\ %\hline
				Tx     & Transmit \\ %\hline
				UDP    & User datagram protocol \\ %\hline
				UG     & Underground   \\
				UG-WSN  & Underground wireless underground sensor network \\ %\hline
				UW-WSN  & Underwater wireless  sensor network \\ %\hline
				VMI    & Vehicle magnetic induction \\ %\hline   
				VMIC   & Vehicle magnetic induction communication\\ %\hline  
				VLF-LA & Very low frequency and a larger antenna \\ %\hline
				
				WSN    & Wireless sensor network \\
				
				\hline
			\end{tabular}
		}
	\end{threeparttable}
	\vspace{-2.0em}
	\label{tbl_acronyms}
\end{table}

%\vspace{-20em}

\section{Introduction} \label{sect_introduction}
\subsection{Underground Communication and TTE Communication}\label{sect_introduction_uctc}

\IEEEPARstart{A}{s} human activities increasingly extend into underground environments, and reliable underground communication has become essential for applications, such as resource exploration, disaster rescue, underground blasting, and robotic operations in hazardous scenarios~\cite{Mahrokh2021TVT}. DARPA launched the Subterranean Challenge in 2017 to drive innovation in mapping, navigation, and search solutions for complex underground spaces, including tunnel systems, urban underground areas, and natural caves~\cite{Orekhov2022Darpa}. One of the primary challenges in this initiative was ensuring robust underground communication. As next-generation mobile communication technologies, such as 6G~\cite{CuiQimei2025Survey}, evolve, underground communication is expected to play a critical role in integrated networks, where protocols, such as TCP/IP, will be indispensable for enabling essential functionalities, such as dynamic passwords, node registration, and licensing approval by regional authorities (e.g., national security centers). However, without standardized protocols, these advanced features are nearly impossible to achieve.

\begin{table}[H]  
	\caption{Notation and definition }
	%\begin{minipage}{\textwidth} 
	%\tiny
	\footnotesize
	\centering
	\scalebox{0.93}{
		\begin{threeparttable} 
			%\begin{tabular}{ p{2.3cm} p{4.5cm}}
			\begin{tabular}{l |l }
				\hline 
				\textbf{Definition$^\dagger$ } &\textbf{Notation}   \\
				\hline \hline
				%\raggedleft
				\centering
				Basic parameters & See Table \ref{tbl_sim} \\ %\hline
				Probability     & $\mathbb{P}(\cdot)$ \\ %\hline
				Expectation    & $\mathbb{E}(\cdot)$ \\ %\hline
				Imaginary unit$\sqrt{-1}$ &$j$ \\ %\hline
				Source/Tx          &$\mathrm{S}$   \\ %\hline
				Destination/Rx     &$\mathrm{D}$  \\ %\hline
				Relay              & $\mathrm{R}$   \\ %\hline
				$(\cdot)$ of source/Tx antenna & $(.)_{\mathrm{S}}$  \\ %\hline
				$(\cdot)$ of destination/Rx antenna  &	$(.)_{\mathrm{D}}$ \\ %\hline
				$(\cdot)$  of relay  antenna  &  $(.)_{\mathrm{R}}$\\ %\hline
				$(\cdot)$ of link $\mathrm{S}$$\rightarrow$$\mathrm{D}$ &  $(.)_{\mathrm{SD}}$ \\ %\hline
				Magnetic moment by antenna $\mathrm{S}$ & $\mathbf{m}_{\mathrm{S}}$ \\ %\hline
				Channel power gain [of link $\mathrm{S}$$\rightarrow$$\mathrm{D}$]  & $G_{\mathrm{SD}}$\\ %\hline
				Circuit  gain [of link $\mathrm{S}$$\rightarrow$$\mathrm{D}$] & $\mathcal{C}_{\mathrm{SD}}$ \\ %\hline
				Space gain [of link $\mathrm{S}$$\rightarrow$$\mathrm{D}$] & $\mathcal{S}_{\mathrm{SD}}$ \\ %\hline
				Eddy gain [of link $\mathrm{S}$$\rightarrow$$\mathrm{D}$] &$\mathcal{E}_{\mathrm{SD}}$ \\ %\hline
				Polarization gain or MI fast fading [of link $\mathrm{S}$$\rightarrow$$\mathrm{D}$] &$J_{\mathrm{SD}}$ \\ %\hline
				Mutual inductance [of link $k$$\rightarrow$$l$] & $M_{kl}$ \\ %\hline
				Compensated in-device gain/loss [for link $\mathrm{S}$$\rightarrow$$\mathrm{D}$]  & $\aleph_{\mathrm{SD}}$ \\ %\hline
				Wavenumber in the medium    & $k_0$ \\ %\hline
				Horizontal vibration angle [of the antenna $\mathrm{S}$]  & $\phi_{\mathrm{S}}$    \\ %\hline
				Vertical vibration angle [of the antenna $\mathrm{D}$]  & $\theta'_{\mathrm{D}}$    \\ %\hline 
				Average AVI  [of antenna $\mathrm{S}$]& $\sigma_{\mathrm{S}}$ \\ %\hline
				Boundary of the antenna vibration & $\varsigma$ \\ %\hline
				Dirac's function   &$\delta_{\rm pu}(\cdot)$  \\ %\hline
				Overall circuit impedance [of coil $\mathrm{S}$]  &$Z_{\mathrm{S}}$ \\ %\hline
				%   Equivalent impedance by mutual inductance $M_{kl}$ & $Z_{kl}$ \\ %\hline 
				Current or divisor [at coil/antenna $\mathrm{S}$] &$I_{\mathrm{S}}$ \\ %\hline 
				PSD    [of Tx $\mathrm{S}$] &$P_{\mathrm{S}f}$ \\ %\hline 
				Noise power       & $N_{o}$ \\ %\hline
				Capacitance of match capacitor for $f_0$  [at coil $\mathrm{S}$]  &  $C_{\mathrm{cS}}$ \\ %\hline
				Resistance of the load      & $R_L$ \\ %\hline
				Resistance [of coil $\mathrm{S}$] & $R_{\mathrm{cS}}$ \\ %\hline
				Time or time slot [of Tx $\mathrm{S}$] & $t_{\mathrm{S}}$           \\
				Inductance   [of coil  $\mathrm{S}$]  &  $L_{\mathrm{cS}}$ \\ %\hline
				Bandwidth     &   $B_{\rm w}$ \\ %\hline
				Function of 3-dB bandwidth&  $B_{\rm w}(\cdot)$  \\ %\hline 
				3-dB bandwidth  for CMIC-1NR using AF & $B_{\rm AF}$  \\ %\hline 
				
				Achievable rate [of link/system $\mathrm{S}$$\rightarrow$$\mathrm{D}$] & $\mathfrak{C}_{\mathrm{SD}}$ \\ \hline
			\end{tabular}
			\begin{tablenotes}  
				\footnotesize  
				\item[$\dagger$] The brackets [$\cdot$] represents the optional part. For example,    the average AVI  [of antenna $\mathrm{S}$] is denoted by $\sigma_{\mathrm{S}}$. Optionally,  the average AVI  [of antenna $\mathrm{D}$] is denoted by $\sigma_{\mathrm{D}}$.
			\end{tablenotes}  
		\end{threeparttable}
	}
	\label{tbl_symbol}
	\vspace{-1.31em}
\end{table}

\begin{figure}[t]
	\centering
	%\includegraphics[width=3.2in]{poutpwrbi.eps}\\
	\includegraphics[width=3.2in]{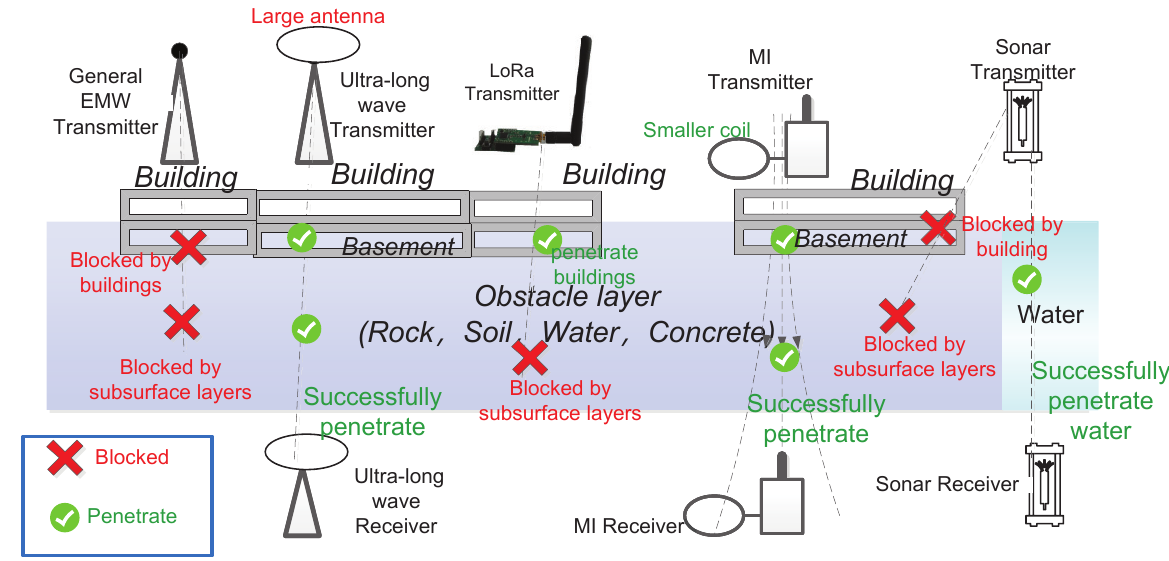}\\
	\vspace{-0.4em}
	\caption{ Comparison of the penetration abilities of communication approaches. Green  and red texts indicate the advantages and disadvantages, respectively. Performance and deployment comparisons are presented in Table \ref{tbl_cmpcommperf}.}\label{fig_commopt}
	\vspace{-1.8em}
\end{figure}

Despite its importance, underground communication has received significantly less research attention compared to aboveground technologies such as electromagnetic wave communication (EMWC). It remains a bottleneck in applications ranging from general underground operations to DARPA's robotic systems and next-generation multi-network integration, particularly when supporting TCP/IP protocols. Known as through-the-earth (TTE) communication due to its ability to penetrate over 10 m of soil and rock, this technology is especially critical for scenarios such as mining and drilling, where depths often exceed 1,000 m. Notably, the Kola Superdeep Borehole even extends beyond 12,000 m~\cite{Wrigley2023Going}. During disasters, when EMWC and wired infrastructures are often rendered inoperative, TTE communication's flexible deployment capabilities can expedite rescue operations, potentially saving countless lives.

{}\label{resppage_illustratesvariousTTE} %for response letter page reffering

{
	Fig. \ref{fig_commopt} illustrates various TTE communication techniques with their performance and deployments  compared in Table \ref{tbl_cmpcommperf}. General EMW signals are limited in penetration and may fail to reach basement levels. Acoustic communication systems are unsuitable for TTE communication due to severe multipath effects in complex underground materials~\cite{Shovon2022Survey}. Ultra-long wave (ULW) radio-based communication, though capable of penetrating deeper, requires kilometer-scale antennas, which are impractical in confined underground spaces~\cite{zhang2014cooperative}. In contrast, MIC has proven effective for a wide range of underground applications. Time-varying magnetic fields experience less attenuation than EMWs~\cite{Ma2019Antenna}, and MIC antennas (coils) are compact compared to ULW devices. For instance, an 8-meter diameter MIC antenna can achieve a TTE communication range of up to 310 m~\cite{Zhang2017Connectivity}, while smaller antennas (less than 1 m in diameter) are sufficient for shorter distances, making them suitable for vehicle-mounted applications. This flexibility enables mobile MI networks, ideal for emergency rescue operations~\cite{Ma2020Channel, Ma2019Effect}.}

\IEEEpubidadjcol

\begin{table*}[t]  
	\caption{Key  parameters, symbols, and their default values for simulations in this paper}
	%\begin{minipage}{\textwidth} 
	%\tiny
	\centering
	\vspace{-0.8em}
	\scalebox{0.90}{
		\begin{threeparttable} 
			\begin{tabular}{ p{5.9cm}<{\centering} p{0.5cm}<{\centering} p{2.0cm}<{\centering} p{2.2cm}<{\centering} p{1.5cm}<{\centering} p{5.3cm}<{\centering}}
				\hline
				\textbf{Parameters}  & \hspace{-2.0em}\textbf{Symbol\tnote{$\dagger$}} &\textbf{Default values} & \textbf{Unit} & \textbf{\emph{Refs.}}  & \textbf{Related simulations in this survey} \\
				\hline 
				%\raggedleft
				\centering
				
				Tx or CMI coil radius & $a_{\mathrm{cS}}$ &0.6  & meter& \cite{Ma2024Fast}   & Figs. \ref{fig_sec2fem}, \ref{fig_chp2capadsd}, \ref{fig_chp2snrdsd}, \ref{fig_sec3crosstalk}, \ref{fig_sec3bafb1}, \ref{fig_sec3cmgxyth}\\ 
				
				Rx coil radius &$a_{\mathrm{cD}}$ &0.4  &  meter & \cite{Ma2024Fast}   & Fig. \ref{fig_chp2capadsd}, \ref{fig_chp2snrdsd}\\ 
				
				Number of turns of Tx or CMI coils & $N_{\mathrm{S}}$ &15  & -  & \cite{Ma2024Fast}  & Figs. \ref{fig_chp2capadsd}, \ref{fig_chp2snrdsd}, \ref{fig_sec3crosstalk}, \ref{fig_sec3bafb1}, \ref{fig_sec3cmgxyth}\\ 
				
				Number of turns of Rx coil & $N_{\mathrm{D}}$ &30  & -  & \cite{Ma2024Fast}   & Fig. \ref{fig_chp2capadsd}, \ref{fig_chp2snrdsd}\\ 
				
				Coil resistance&$\rho_\mathrm{w}$ &0.166  & $\Omega$ / m & \cite{Zhang2017Connectivity,Ma2024Fast}    & Figs. \ref{fig_chp2capadsd}, \ref{fig_chp2snrdsd}, \ref{fig_sec3crosstalk}, \ref{fig_sec3bafb1}, \ref{fig_sec3cmgxyth}\\ 
				
				Tx Power &$P_s, P_S$ &5  &   Watt & \cite{Ma2019Antenna,Ma2024Fast}    & Figs. \ref{fig_chp2capadsd}, \ref{fig_chp2snrdsd}, \ref{fig_sec3crosstalk}, \ref{fig_sec3bafb1}, \ref{fig_sec3cmgxyth}\\ 
				
				Resonance (center) frequency &$f_0$ &10000  &  Hz & \cite{Ma2019Antenna,Ma2024Fast}    & Figs. \ref{fig_sec3crosstalk}, \ref{fig_sec3cmgxyth}\\

				Carrier frequency &$f$ &10000  &  Hz & \cite{Ma2019Antenna,Ma2024Fast}    & Figs. \ref{fig_sec3crosstalk}, \ref{fig_sec3cmgxyth}\\ 
				
				%\makecell{\vspace{-0.6em}\\\hspace{-0.6em} Rx BER threshold} &BER &\makecell{\vspace{-0.6em}\\10$^{-3} $}  &  \makecell{\vspace{-0.9em}\\\hspace{-1.9em} bit / s} & \makecell{\vspace{-0.9em}\\\hspace{-0.6em}\cite{Sun2010Magnetic, Kisseleff2014Modulation} }   & \makecell{\vspace{-0.9em}\\\hspace{-0.6em}Device parameter}\\ 
				
				Tx coil orientation& $\theta_s$ &0 & - & \cite{Sun2010Magnetic}    & Figs. \ref{fig_sec2Ejsd}, \ref{fig_chp2performance}, \ref{fig_chp2capadsd}, \ref{fig_chp2snrdsd},  \ref{fig_sec3crosstalk}, \ref{fig_sec3bafb1}, \ref{fig_sec3cmgxyth}\\

				DMI Rx coil orientation & $\theta_{\mathrm{D}}$ &$\pi$  &  -  & \cite{Sun2010Magnetic}   & Figs. \ref{fig_chp2capadsd}, \ref{fig_chp2snrdsd},  \ref{fig_sec3crosstalk}\\ 
				
				VMI Rx coil orientation & $\theta_{\mathrm{D}}$ &$-\frac{\pi}{2}$  &  - & \cite{Ma2020Channel}   & Fig. \ref{fig_sec2Ejsd}\\

				Ambient noise PSD & $N_{\mathrm{o}f}$ &-103 &  dBm / 2 kHz & \cite{Sun2010Magnetic}   & Figs. \ref{fig_chp2capadsd}, \ref{fig_chp2snrdsd}, \ref{fig_sec3crosstalk}, \ref{fig_sec3bafb1}, \ref{fig_sec3cmgxyth}\\ 
				
				Permeability of medium & $\mu_\mathrm{u}$ &4$\pi$ $\times$ 10$^{-7}$  &  H / m & \cite{kisseleff2013channel,Ma2019Antenna}    & Figs. \ref{fig_chp2capadsd}, \ref{fig_chp2snrdsd},  \ref{fig_sec3crosstalk}, \ref{fig_sec3bafb1}, \ref{fig_sec3cmgxyth}, \ref{fig_sec5nearfieldrange}\\

				Conductivity of medium & $\sigma_\mathrm{u}$ &10$^{-2}$  &  S / m & \cite{kisseleff2013channel,Ma2019Antenna}   & Figs. \ref{fig_chp2capadsd}, \ref{fig_chp2snrdsd}, \ref{fig_sec3crosstalk}, \ref{fig_sec3bafb1}, \ref{fig_sec3cmgxyth}\\ 
				
				Permittivity of medium & $\epsilon_\mathrm{u}$ & 6.978$\times$10$^{-11}$  &  F / m & \cite{kisseleff2013channel}   & Figs. \ref{fig_chp2capadsd}, \ref{fig_chp2snrdsd}, \ref{fig_sec3crosstalk}, \ref{fig_sec3bafb1}, \ref{fig_sec3cmgxyth}, \ref{fig_sec5nearfieldrange}\\

				Distance between $\mathrm{S}$ and  $\mathrm{D}$  & $d_{\mathrm{SD}}$ & 60  & meter & \cite{Ma2019Antenna}    & Figs.  \ref{fig_sec3crosstalk}, \ref{fig_sec3bafb1}, \ref{fig_sec3cmgxyth}\\ 
				
				\hline
				
			\end{tabular}
			%	\item[a]{}
			%\end{minipage}
			\begin{tablenotes}  
				\footnotesize  
				\item[$\dagger$] In this paper,  Italic subscript `$f$' indicates ``$(\cdot)$ per Hz". All the normal (upright) subscripts indicate the description which does not represent any number or node ID.
			\end{tablenotes}
			
		\end{threeparttable}
	}
	\label{tbl_sim}
	\vspace{-1.5em}
\end{table*}

\begin{table*}[h!]  
	\caption{Comparison of MIC and other communications  in performance and deployment dimensions for underground environments}
	%\begin{minipage}{\textwidth} 
	%\tiny \scriptsize \footnotesize \tiny \small
	
	\vspace{-0.8em}
	\scalebox{0.89}{
		\footnotesize
		\centering
		\begin{threeparttable} 
			\begin{tabular}{  m{2.5cm}||m{2.5cm} |m{2.9cm}|m{2.5cm}|m{3.5cm}|m{3.1cm}}
				\hline
				\textbf{Performance$\dagger$}  & \textbf{MIC} &\textbf{EMWC} &\textbf{Acoustic} &\textbf{Wired} &\textbf{Hybrid}\\ 
				\hline \hline
				%%%%%%%%%%%%%%%%%%%%%%%%%%%%%%%%%%%%%%%%%%%%%%%%%%%%%%%%%%%%%%%%%%%%%%%%%%%%%%%%%%%%%%%%%%%%%%%%%%%%%%%
				Comm. ranges (m)&\hspace{-0.6em} 0.1$\sim$1,700   & \hspace{-0.6em}$<$10  ($f$$=$300MHz )\cite{Sun2010Magnetic}   &\hspace{-0.6em}$<$50 m (Position)\cite{Zeeshan2023Review}& $>$ 10,000&-  \\
				\hline
				Data rates (kbps)	& $<10$  ($d_{\mathrm{SD}}>$50 m) & $<100$  ($d_{\mathrm{SD}}>$5 m)\cite{Kisseleff2018Survey} &5$\sim$17.8 \cite{Muzzammil2020Fundamentals}& $>$1,000 & - \\
				\hline 
				%%%%%%%%%%%%%%%%%%%%%%%%%%%%%%%%%%%%%%%%%%%%%%%%%%%%%%%%%%%%%%%%%%%%%%%%%%%%%%%%%%%%%%%%%%%%%%%%%%%%%%%
				\vspace{0.5em}Channel dependency & Conductivity, antenna vibration &Multipath, conductivity, permittivity, & Multipath, Doppler effect, sound noise & Cable properties and length, shielding, connection quality 
				&Interference, protocols, power control \\
				\hline
				%%%%%%%%%%%%%%%%%%%%%%%%%%%%%%%%%%%%%%%%%%%%%%%%%%%%%%%%%%%%%%%%%%%%%%%%%%%%%%%%%%%%%%%%%%%%%%%%%%%%%%%
				Antenna/device size (m) &0.1$\sim$4 radius coil&$>10,000$ (VLF antenna)\cite{zhang2014cooperative} &$<$1 &$>$10,000 (including cables) & Smaller antenna for  subsystems \\ \hline
				Deployment costs & Low& High& Medium &Extremely High& High \\ \hline
				Maintenance costs & Low & High & High & High & High \\ \hline
				System complexity&Medium &High & High & Low & High \\ \hline
				Disaster resilience &Strong&Weak&Medium &Medium &Strong \\ \hline
				Maturity levels&Low &High &Medium &High &Medium \\ \hline
			\end{tabular}
			\begin{tablenotes}  
				\footnotesize  
				\item[$\dagger$] Optical communication is omitted due to its zero communication range underground.
			\end{tablenotes}
		\end{threeparttable}
	}
	\label{tbl_cmpcommperf}
	\vspace{-1.5em}
\end{table*}

The MIC uses a modulated magnetic field generated by a transmitter antenna. This field is received by a receiver coil or magnetic sensor, which demodulates it into symbols. The Wrathall team accomplished an underwater MIC in 1999 using a 3 kHz field \cite{Wrathall1999Magneto}. Since then, MIC has been explored as an alternative for underground (UG) WSNs. In 2006, Akyildiz \emph{et al.}~\cite{akyildiz2006wireless, akyildiz2009signal} summarized various UG-WSN scenarios, including soil monitoring, underground water monitoring, pipe leak detection, intruder detection, rescue personnel localization, and building load-bearing pressure detection. Sun \emph{et al.} \cite{Sun2010Magnetic} introduced MIC into UG-WSNs with a small device featuring a 10 cm coil in 2010.  Recently, researchers have expanded on applications of the non-coil-based MIC and large-scale MI networks. For example, Zhang \emph{et al.} \cite{Zhang2023Rotation} developed a  rotating permanent magnet antenna (RPMA) array device for the through-the-sea MIC application in 2023, and Ma \emph{et al.} \cite{Ma2024Fast} focused on MI multi-cellular networks in 2024.

\subsection{Related Surveys  and Motivations of This Survey}\label{sectsub_rsmts}
\vspace{-0.2em}
\begin{table*}[t!]  
	\caption{Related Surveys and their differences from this survey }
	%\begin{minipage}{\textwidth} 
	%\tiny \scriptsize \footnotesize \tiny \small
	\vspace{-0.9em}
	\scalebox{0.89}{
		\footnotesize
		\centering
		\begin{threeparttable} 
			\begin{tabular}{  p{2.3cm}|p{1.2cm} |p{7.9cm}|p{6.9cm}}
				\hline
				\makecell{\vspace{-0.6em}\\\hspace{-0.6em}\textbf{Aspects} } & \textbf{\emph{Refs.}} &\textbf{Most important contribution} &\textbf{Differences from this survey}\\ 
				\hline

				%	\makecell{\vspace{-0.6em}\\\hspace{-0.6em}MICs}&\cite{Liu2024Magnetic} & \makecell{\vspace{-0.6em}\\Reference book for explaining MIC theory \\step by step with examples  } &  \makecell{\vspace{-0.6em}\\ A reference book focusing on the basic concepts \\proposed before 2020 and lack of a complete \\MI network architecture (\emph{e.g.} routing)}\\
				\hline
				%%%%%%%%%%%%%%%%%%%%%%%%%%%%%%%%%%%%%%%%%%%%%%%%%%%%%%%%%%%%%%%%%%%%%%%%%%%%%%%%%%%%%%%%%%%%%%%%%%%%%%%
				\makecell{\vspace{-0.6em}\\\hspace{-0.6em}UG communications}&\cite{Yarkan2009Underground, Forooshani2013Survey, Sheikhpour2017Agricultural, Saeed2019Toward,Hancke2021Wireless, Wohwe2023Wireless, Raza2020Survey, Hrovat2014Survey, Riurean2021Conventional} & Issues of acoustic wave communication, EMWC, wired communication, mud pulse telemetry communication and MIC for UG-WSNs &  Diluting  comprehensiveness due to non-exclusive to MICs, and no exploration of issues of  the MI fast fading, RPMA\\
				\hline
				
				%%%%%%%%%%%%%%%%%%%%%%%%%%%%%%%%%%%%%%%%%%%%%%%%%%%%%%%%%%%%%%%%%%%%%%%%%%%%%%%%%%%%%%%%%%%%%%%%%%%%%%%		
				\makecell{\vspace{-0.6em}\\\hspace{-0.6em} MICs} &\cite{Liu2024Magnetic}&Reference book covering  antennas, channels, performance, and protocols related to  MICs &No exploration of issues of MI fast fading, RPMA and routing algorithm\\ 
				\hline
				%%%%%%%%%%%%%%%%%%%%%%%%%%%%%%%%%%%%%%%%%%%%%%%%%%%%%%%%%%%%%%%%%%%%%%%%%%%%%%%%%%%%%%%%%%%%%%%%%%%%%%%
				\makecell{\vspace{-0.6em}\\\hspace{-0.6em} Underground MICs} &\cite{Sharma2017Magnetic, Kisseleff2018Survey}&Issues for general MI UG-WSNs, primarily  concerning short-to-mid range MICs &No exploration of issues of MI fast fading, RPMA, a complete MI network architecture\\ 
				\hline
				
				\makecell{\vspace{-0.6em}\\\hspace{-0.6em}Underwater MICs}& \cite{Li2019Survey, Muzzammil2020Fundamentals}&  Fundamental issues and advances in  underwater MICs & No exploration of  issues of MI fast fading, RPMA, and a complete MI network architecture\\ 
				\hline
				%%%%%%%%%%%%%%%%%%%%%%%%%%%%%%%%%%%%%%%%%%%%%%%%%%%%%%%%%%%%%%%%%%%%%%%%%%%%%%%%%%%%%%%%%%%%%%%%%%%%%%%
				\makecell{\vspace{-0.6em}\\\hspace{-0.6em}TTE MICs}& This survey&  Fundamental issues and advances in underground MIC, primarily concerning long-range MICs, MI fast fading,  and a complete MI network architecture& -\\ 
				\hline

			\end{tabular}
			
		\end{threeparttable}
	}
	\label{tbl_relatedsurvey}
	\vspace{-1.0em}
\end{table*}

While many surveys on underground communication have been published as listed in Table \ref{tbl_relatedsurvey}, only a handful of the surveys have focused on underground MICs. For example, Sharma \emph{et al.} \cite{Sharma2017Magnetic} reviewed  MIC research until 2017 for non-conventional media applications. They introduced the applications and advantages of MICs and briefly introduced the channel modeling of the P2P MIC and a hardware testbed for MIC research.    Kisseleff \emph{et al.} \cite{Kisseleff2018Survey} conducted a comprehensive review of underground MIC studies up to that point. Although their work primarily focused on P2P MIC and MI waveguide issues, they also discussed physical protocols such as channel estimation and node deployments.  Recently, Liu \emph{et al.} \cite{Liu2024Magnetic} offered a thorough introduction to general MICs in a monograph. This monograph covers the basic concepts and theories of background, developments, antennas, channels, performance, and protocols related to  MICs proposed before 2020. 

Compared to this current survey, the existing review~\cite{Sharma2017Magnetic} has overlooked multi-node MICs. The survey~\cite{Kisseleff2018Survey} does not comprehensively review MI network architectures and the issues of mobile MIC, including the data link layer, network layers, RPMAs, and MI fast fading. The monograph~\cite{Liu2024Magnetic} does not review MI network architectures and MI fast fading, especially the recent routing algorithms and RPMAs developed in the past five years.

However, the surveys conducted since 2020 have not yet provided comprehensive reviews on expanded MIC techniques, such as MAC and routing protocols for a large-scale MI network and a novel mechanical antenna. This is due to the great surge in mobile MICs, mechanical antennas, and upper-layer MI research since 2020. %This identified gap serves as the first motivation to conduct this survey.
Moreover, almost all existing articles on MICs presented the common conclusion that the MI channel is a quasi-static and predictable channel without a small-scale fading. These articles also include the surveys (e.g.,\cite{Li2019Survey,Kisseleff2018Survey}) and reviews (e.g.,\cite{Sharma2017Magnetic}). However, several studies have described the MI fast fading channel recently. The common conclusion may no longer hold.

Although most related articles claim that their research on MIC is compatible with TTE communication, few surveys and reviews highlight the potential or specific issues and methodologies when existing MIC techniques are applied in TTE or long-distance MIC environments. % Revealing these issues serves as the third motivation to conduct this survey.
%Although research articles on upper-layer solutions are increasing, they are still insufficient to form a functional MI-based network protocol stack. By referring to the OSI-originated framework (e.g., TCP/IP),  researchers can identify and address remaining tasks that aid in reusing interdisciplinary resources. This serves as the third motivation.  %which encompasses the TCP/IP and IEEE 802 standards,  
Currently, there is no agreed protocol stack for larger-scale MI networks. By organizing the existing MIC research with reference to the OSI-originated framework, we can identify the remaining issues related to MIC to establish a standard MIC protocol stack. This is crucial for  Space-Air-Ground-Underground multi-network integration (SAGUIMI) for the next generation of mobile communication.

\vspace{-0.0em}

\subsection{Contributions and Organization of This  Survey}\label{sectsub_cots}

% \subsubsection{Contributions} 
This survey reviews research on underground MICs, particularly TTE applications, covering point-to-point TTE MIC techniques and the impact of MI fast fading on existing MIC theories. To guide optimization efforts, we  decompose the MI channel power gain into four components with low inter-coupling and distinct physically interpretable meanings. The survey also covers MI relay techniques, analyzing crosstalk effects in both relay and high-density networks. Moreover, we identify the remaining research tasks for a comprehensive MI network protocol in SAGUI. Based on the surveyed literature, we propose an advanced MIC framework that supports TCP/IP and Linux, addressing both the current state and future challenges of MIC. This framework enables researchers to utilize extensive Linux resources and deep learning platforms, accelerating research and implementation in SAGUI applications. The key research challenges, open issues in MICs, and promising novel techniques to address them are highlighted.

The key contributions of this survey include: 

\begin{itemize}

	\item {\textbf{First Survey on MI Fast-Fading Channels:} This survey identifies that MI fast fading challenges the prevailing notion of quasi-static MI channels. We highlight that research on MI fast fading remains in its early stages due to the lack of a universal statistical model. To address this, we introduce an antenna vibration model and corresponding simulations. We also analyze the potential impacts on existing MIC theorems, a topic not yet covered in any previous surveys.}
	
	\item {\textbf{Comprehensive Review of MI Network Architecture:} We present a complete review of the MI network architecture across the OSI framework layers, identifying remaining issues and possible solutions. A significant finding is the absence of standardized MI protocol stacks, which represents a major barrier to achieving SAGUIMI integration in next-generation mobile communications. Existing surveys often focus on specific applications without providing a holistic and runnable framework.}
	
	\item {\textbf{Fine-grained Decomposition of Channel Power Gain:} We introduce a detailed conceptual modeling approach for channel power gain and provide optimization directions for MIC systems, including antenna designs, bandwidth, and MIC range improvements. This contributes to simplifying MIC optimization for future research by narrowing the scope of MI parameters to focus on relevant variables while fixing others.}\label{resppage_fixingirrelevantvariables}
	
	\item {\textbf{Identification of Positive MI Crosstalk Effects:} This study uncovers the positive MI crosstalk effects, which are crucial for addressing challenges in MI waveguides and massive MIMO systems. Previous literature primarily focused on negative crosstalk effects, while the positive crosstalk aspect has been largely overlooked.}

\end{itemize}

\begin{figure*}[htb]
	\scriptsize
	\centering
	\tikzset{
		my node/.style={
			draw=gray,
			thick,
			font=\sffamily,
			drop shadow,
			minimum height=0.1cm,
		},
		my node level 1/.style={
			my node,
			inner color=blue!5,
			outer color=blue!10,
			minimum width=1cm,
			rounded corners=2,
		},
		my node level 2/.style={
			my node,
			inner color=green!5,
			outer color=green!10,
			minimum width=2cm,
			rounded corners=4,
		},
		my node level 3/.style={
			my node,
			inner color=orange!5,
			outer color=orange!10,
			minimum width=3cm,
		},
		% 更多层级样式可以在这里定义
	}
	\scalebox{0.71}{
		\begin{forest}
			for tree={%
				my node,
				l sep+=8pt,
				%s sep+=80pt
				grow'=east,
				edge={gray, thick},
				parent anchor=east,
				child anchor=west,
				tier/.option=level, 
				edge path={
					\noexpand\path [draw, \forestoption{edge}] (!u.parent anchor) -- +(10pt,0) |- (.child anchor)\forestoption{edge label};
				},
				if level=1{ % 这里尝试识别根节点，但可能需要根据上下文调整条件
					edge path={
						\noexpand\path [draw, \forestoption{edge}] (!u.south) -- +(10pt,0) |- (.child anchor)\forestoption{edge label};
					},
				}{},
				if level=1{my node level 1}{},
				if level=2{my node level 2}{},
				if level=3{my node level 3}{},
				% if={isodd(n_children())}{
					%     for children={
						%         if={equal(n,(n_children("!u")+1)/2)}{calign with current}{}
						%     }
					% }{},
				align=center, % 居中对齐文本
				baseline, % 设置基线对齐方式
				% if level=3{ % 这里尝试识别根节点，但可能需要根据上下文调整条件
					%     align=left,
					% },
			}
			[Through-the-Earth Magnetic Induction Communication and Networking, rotate=90
			[Introduction, minimum width=4.0cm,
			[Underground Communication and TTE Communication, minimum width=4.0cm, 
			[{DARPA   Subterranean Challenge \cite{Orekhov2022Darpa}; SAGUMI; etc.
			}, minimum width=7.0cm, minimum height=0.18cm, text height=0.22cm, base=midpoint
			]
			]
			[Related Surveys  and Motivations of  This Survey,  minimum width=4.0cm
			[{UG-WSN\cite{Wohwe2023Wireless}; MIC\cite{Liu2024Magnetic}; MI UG-WSN\cite{Kisseleff2018Survey}; MI UW-WSN\cite{Li2019Survey}; TTE WSN [this survey]}, minimum width=7.0cm, text height=0.23cm, base=midpoint]
			]
			[Contributions and Organization of This  Survey, minimum width=4.0cm, minimum height=0.13cm]
			[{Research Gap Summary}, minimum width=4.0cm, minimum height=0.13cm]
			]
			[Applications of MICs: General Scenarios\\ and TTE-Specific Features, minimum width=4.0cm, base=midpoint,
			[\textit{No subsections}, minimum width=4.0cm, minimum height=0.18cm, text height=0.22cm, base=midpoint
			[{Agriculture~\cite{Silva2015Strategic}; Industry~\cite{Sun2011MISE}; BAN~\cite{Jenkins2023Wearable}; TTE~\cite{zhang2014cooperative}; etc.
			}, minimum width=7.0cm, text height=0.23cm, base=midpoint
			]
			]
			]
			% [Data Reconstruction \\and Augmentation, minimum width=4.0cm, base=midpoint,
			% [Data Reconstruction, minimum width=4.0cm, [{SARGAN \cite{tran2018generative}, VAE-GAN \cite{feng2022waveform}, CDDM \cite{wu2023cddm}, etc.}, minimum width=7.0cm]][Data Augmentation, minimum width=4.0cm,  [{SAGA \cite{davaslioglu2018generative}, ACGAN \cite{tang2018digital}, MSGAN \cite{lu2021gan}, etc.}, minimum width=7.0cm]]
			% ]
			[Channel Models for TTE MIC, minimum width=4.0cm, base=midpoint,
			[System and Channel Modeling, minimum width=4.0cm, minimum height=0.18cm, text height=0.22cm, base=midpoint
			[{General MI link \cite{Sun2010Magnetic}; MIMO MI link \cite{Li2015Capacity}; M$^2$I link\cite{Guo2015M2I}; RPMA link \cite{Zhang2023Rotation}; MI fast fading\cite{Ma2024Fast} }, minimum width=7.0cm, base=midpoint, text height=0.22cm]
			][Decomposition of  Channel Power Gain, minimum width=4.0cm, minimum   height=0.18cm, text height=0.22cm, base=midpoint
			[{Circuit gain \cite{Wang2024Multi};Space gain \cite{Guo2015M2I}; Eddy gain\cite{sun2012capacity}; Polarization gain \cite{zhang2014cooperative} }, minimum width=7.0cm, base=midpoint, text height=0.22cm]
			]
			%		    [Circuit Gain, minimum width=4.0cm, minimum height=0.18cm, text height=0.22cm, base=midpoint
			%		        [{Loss in electrical circuit \cite{Sun2010Magnetic, Fawole2017Electromechanically,Wang2024Multi,Dionigi2012Multiband, Agbinya2019Principles, Sun2013Increasing}, etc.}, minimum width=7.0cm, base=midpoint, text height=0.22cm]
			%			] [Space Gain, minimum width=4.0cm, minimum height=0.18cm, text height=0.22cm, base=midpoint
			%			[{Due to space expansion\cite{Guo2015M2I,lin2015distributed,Wang2019Efficient,Alsalman2021Balanced}, etc.}, minimum width=7.0cm, base=midpoint, text height=0.22cm]
			%			][Eddy Gain, minimum width=4.0cm, minimum height=0.18cm, text height=0.22cm, base=midpoint
			%			[{ Eddy current loss of underground materials\cite{Sun2013Increasing, Sogade2004Electromagnetic}, etc.}, minimum width=7.0cm, base=midpoint, text height=0.22cm]
			%			][Polarization  Gain, minimum width=4.0cm, minimum height=0.18cm, text height=0.22cm, base=midpoint
			%			[{ Due to antenna misalignment \cite{Dumphart2016Stochastic, Li2015Capacity,zhang2014cooperative},  etc.}, minimum width=7.0cm, base=midpoint, text height=0.22cm]
			%			]
			[MI Fast-Fading Channel, minimum width=4.0cm, minimum height=0.18cm, text height=0.22cm, base=midpoint
			[{Antenna-vibration-oriented fading \cite{Ma2024Fast};  Our proposed geometric model and simulation.}, minimum width=7.0cm, base=midpoint, text height=0.22cm]
			][Summary and Lessons Learned, minimum width=4.0cm, minimum height=0.18cm, text height=0.22cm, base=midpoint
			]
			]
			[Design of Point-to-point TTE MIC, minimum width=4.0cm, base=midpoint,
			[Antenna Design, minimum width=4.0cm, minimum height=0.18cm, text height=0.22cm, base=midpoint
			[{Coil \cite{Sun2010Magnetic}; M2I \cite{Guo2015M2I}; RPMA \cite{Zhang2023Rotation};    etc.}, minimum width=7.0cm, base=midpoint, text height=0.22cm]
			][Channel Capacity and Bandwidth, minimum width=4.0cm, minimum height=0.18cm, text height=0.22cm, base=midpoint
			[{Frequency  bandwidth calculations \cite{Ma2019Effect,Azad2012Link,Jiang2015Capacity}, etc.}, minimum width=7.0cm, base=midpoint, text height=0.22cm]
			][Communication Range , minimum width=4.0cm, minimum height=0.18cm, text height=0.22cm, base=midpoint
			[{MIC range calculation \cite{Zhou2017Maximum};  Influence of underground materials;  etc.}, minimum width=7.0cm, base=midpoint, text height=0.22cm]
			][{Upper-Layer P2P techniques}, minimum width=4.0cm, minimum height=0.18cm, text height=0.22cm, base=midpoint [{{Channel estimation} \cite{Guo2019Inter}; {Modulation}\cite{Liu2023Chirp,Kisseleff2014Modulation}; {channel coding} \cite{Chen2023Novel};     etc.}, minimum width=7.0cm, base=midpoint, text height=0.22cm]]
			[Summary and Lessons Learned, minimum width=4.0cm, minimum height=0.18cm, text height=0.22cm, base=midpoint
			]
			]
			[MI Relay and Cooperative MIC, minimum width=4.0cm, base=midpoint,
			[{Use Cases and Scenarios}, minimum width=4.0cm, minimum height=0.18cm, text height=0.22cm,   base=midpoint
			[{ {Pipeline case}\cite{Alshehri2017Optimal,guo2014channel}; {Mobile MIC case} \cite{Ma2019Antenna}}, minimum width=7.0cm, base=midpoint, text height=0.22cm]
			]
			[MI Waveguide and Passive Relay Techniques, minimum width=4.0cm, minimum height=0.18cm, text height=0.22cm, base=midpoint
			[{ Waveguide\cite{Sun2010Magnetic}; MPRlA\cite{Ma2015Topology}; Crosstalk effect;    etc.}, minimum width=7.0cm, base=midpoint, text height=0.22cm]
			][Cooperative MIC, minimum width=4.0cm, minimum height=0.18cm, text height=0.22cm, base=midpoint
			[{CMIC-$n$AR  \cite{Khalil2021Optimal}; CMIC-1NR\cite{Ma2019Effect}; Mobile CMIC\cite{Zhang2024Cooperative},    etc.}, minimum width=7.0cm, base=midpoint, text height=0.22cm]
			][Summary and Lessons Learned, minimum width=4.0cm, minimum height=0.18cm, text height=0.22cm, base=midpoint
			]
			]
			[MI Network and Its Architecture, minimum width=4.0cm, base=midpoint,
			[Physical Layer (Ly1), minimum width=4.0cm, minimum height=0.18cm, text height=0.22cm, base=midpoint
			[{{Power allocation} \cite{Ma2024Fast, lin2015distributed}; {Bandwidth allocation}\cite{Li2024Resource};   etc.}, minimum width=7.0cm, base=midpoint, text height=0.22cm]
			][Data Link Layer (Ly2), minimum width=4.0cm, minimum height=0.18cm, text height=0.22cm, base=midpoint
			[{MI MAC solutions\cite{Ahmed2016Multi,Ahmed2024Design}, etc.}, minimum width=7.0cm, base=midpoint, text height=0.22cm]
			][Network Layer (Ly3), minimum width=4.0cm, minimum height=0.18cm, text height=0.22cm, base=midpoint
			[{Connectivity\cite{Zhang2017Connectivity}; deployment\cite{Wang2018Data};  routing \cite{Liu2024Frequency}; etc.}, minimum width=7.0cm, base=midpoint, text height=0.22cm]
			][Transport Layer (Ly4) and TCP, minimum width=4.0cm, minimum height=0.18cm, text height=0.22cm, base=midpoint
			[{Lacking MI-specific solutions}, minimum width=7.0cm, base=midpoint, text height=0.22cm]
			][Cross-layer Optimization (CLO), minimum width=4.0cm, minimum height=0.18cm, text height=0.22cm, base=midpoint
			[{Distributed algorithms with Nash games \cite{lin2015distributed, Singh2021Optimal,Ma2024Fast},    etc.}, minimum width=7.0cm, base=midpoint, text height=0.22cm]
			][Summary and Lessons Learned, minimum width=4.0cm, minimum height=0.18cm, text height=0.22cm, base=midpoint
			]
			]
			[Promising MI Network Framework with \\TCP/IP and Linux Support, minimum width=4.0cm, base=midpoint,
			[{Significance and Architecture Overview}, minimum width=4.0cm, minimum height=0.18cm, text height=0.22cm,
			[{Also with deep learning platforms (e.g., TensowFlow) support }, minimum width=7.0cm, base=midpoint, text height=0.22cm]
			]
			[{System  Architecture and Implementation}, minimum width=4.0cm, minimum height=0.18cm, text height=0.22cm,
			[{{Fig. \ref{fig_sec5tcp}, Algorithm \ref{alg_tcp} }}, minimum width=7.0cm, base=midpoint, text height=0.22cm]
			]
			[{Summary}, minimum width=4.0cm, minimum height=0.18cm, text height=0.22cm]
			]
			[Research Challenges and Future Directions, minimum width=4.0cm, base=midpoint,
			[MI Fast Fading in Mobile MIC Systems, minimum width=4.0cm, minimum height=0.18cm, text height=0.22cm, base=midpoint
			[{Lacking a universal model, {A Transformer-based framework (Fig. \ref{fig_sec5avipre})}}, minimum width=7.0cm, base=midpoint, text height=0.22cm]
			][Antenna Design, minimum width=4.0cm, minimum height=0.18cm, text height=0.22cm, base=midpoint
			[{Inertia issue of RPMA; massive MI MIMO application}, minimum width=7.0cm, base=midpoint, text height=0.22cm]
			][{MI Crosstalk Effect Prediction Strateges}, minimum width=4.0cm, minimum height=0.18cm, text height=0.22cm, base=midpoint
			[{{A Transformer-based framework (Fig. \ref{fig_sec5crossmig})}}, minimum width=7.0cm, base=midpoint, text height=0.22cm]
			] [MI Communication-Navigation- Sensing Integrated System, minimum width=4.0cm, minimum height=0.18cm, text height=0.22cm, base=midpoint  [{Balancing performance metrics among communication, navigation and sensing}, minimum width=7.0cm, base=midpoint, text height=0.22cm]  
			] [Mixed-Field MI Channel Model, minimum width=4.0cm, minimum height=0.18cm, text height=0.22cm, base=midpoint
			[{Challenging modeling for upper-layer protocols}, minimum width=7.0cm, base=midpoint, text height=0.22cm]
			][Multi-Layer and Inhomogeneous Media, minimum width=4.0cm, minimum height=0.18cm, text height=0.22cm, base=midpoint
			[{Issue of irregular shapes of geological strata}, minimum width=7.0cm, base=midpoint, text height=0.22cm]
			] [Image Transmission and  Deep JSCC, minimum width=4.0cm, minimum height=0.18cm, text height=0.22cm, base=midpoint
			[{Multimodal/content-based semantic communication for surpassing Shannon's limit}, minimum width=7.0cm, base=midpoint, text height=0.22cm]
			] [Cooperative MIC, minimum width=4.0cm, minimum height=0.18cm, text height=0.22cm, base=midpoint
			[{ Multiple active relays with misaligned antenna}, minimum width=7.0cm, base=midpoint, text height=0.22cm]
			] [MI Network and Architecture, minimum width=4.0cm, minimum height=0.18cm, text height=0.22cm, base=midpoint
			[{Heterogeneous network}, minimum width=7.0cm, base=midpoint, text height=0.22cm]
			] [TCP/IP Support, minimum width=4.0cm, minimum height=0.18cm, text height=0.22cm, base=midpoint
			[{IP-HC, RTT suppression, Intelligent retransmission strategy, etc}, minimum width=7.0cm, base=midpoint, text height=0.22cm]
			][{Experiments and Testing in TTE MIC Systems}, minimum width=4.0cm, minimum height=0.18cm, text height=0.22cm, base=midpoint
			[{Unrepeatable UG environments, antenna deployment challenge}, minimum width=7.0cm, base=midpoint, text height=0.22cm]
			] [Summary of Challenges and Opportunities, minimum width=4.0cm, minimum height=0.18cm, text height=0.22cm, base=midpoint
			]
			]	   
			[Conclusion, minimum width=4.0cm, ]
			]
		\end{forest}
	}
	\caption{The structure of this survey, where we perform  fine-grained decomposition of the MI power channel gain and  novel antenna vibration model for MI fast fading in Section \ref{sect_sub2channel},  MI crosstalk effect in Section \ref{sect_cmi}, and MI network framework with TCP/IP \& Linux support in Section \ref{sect_linux}.
	}
	\label{fig_sec1org}
	\vspace{-1.91em}
\end{figure*}
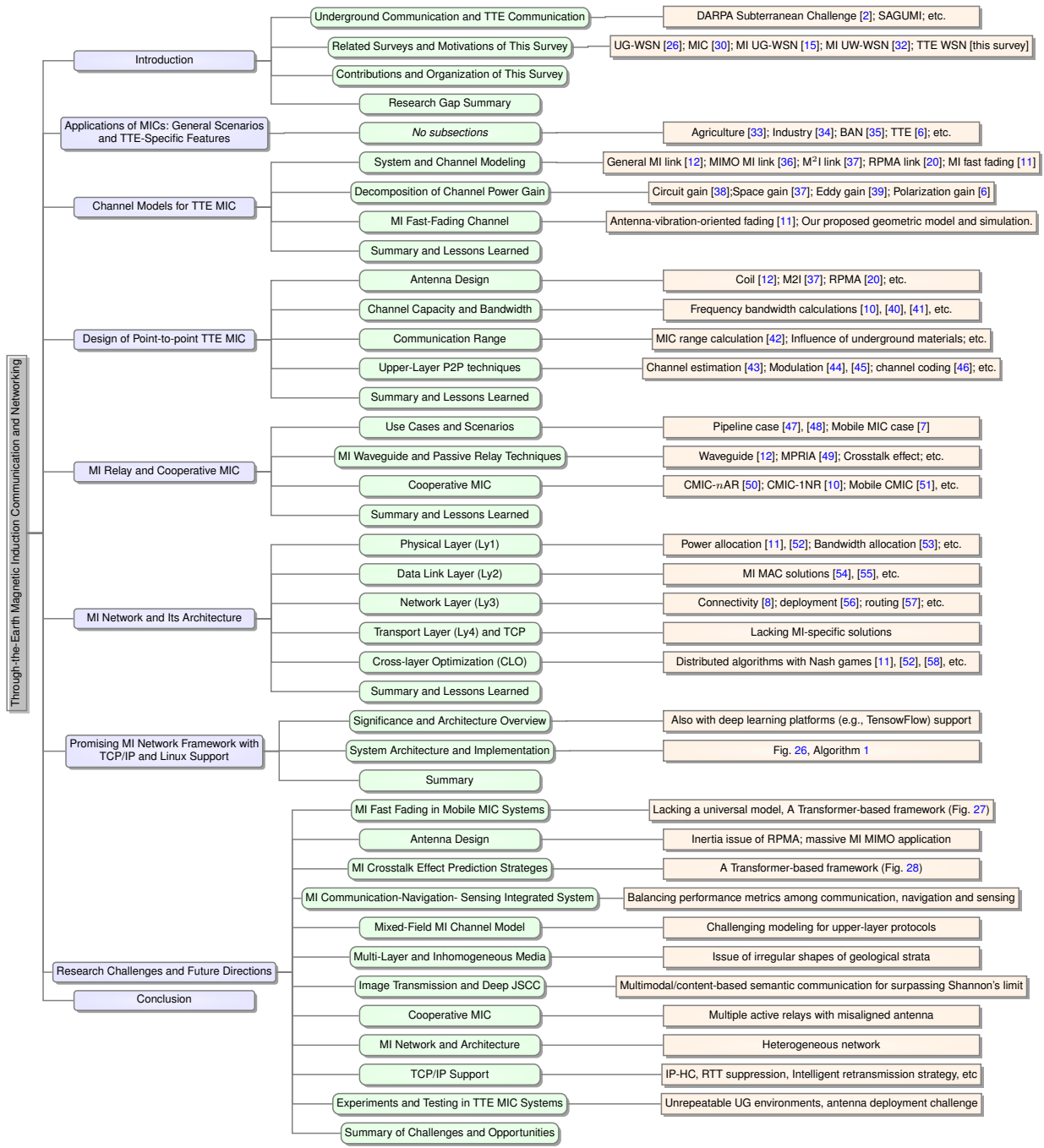

The remainder of this survey is structured as follows (more details in Fig.~\ref{fig_sec1org}): 
Section \ref{sect_sub2channel} covers MI channel modeling, including MI channel power gain and MI fast fading. 
%, the discovery of MI fast fading since 2016, and our proposed antenna vibration model and simulations to tackle the MI fast fading challenge.
Section \ref{sect_p2p} summarizes P2P MIC designs, focusing on MI antenna design, bandwidth, and MIC range. %derivation challenges for P2P links.
Section \ref{sect_cmi} surveys MI relay techniques, such as MI waveguides, MI passive relay array (MPRlA), Cooperative magnetic induction communication (CMIC), and the MI crosswalk effect.
Section \ref{sect_network} reviews multi-node MIC from the perspective of 
%MI network architecture, 
%covering studies across the physical, data link, network, and application layers, based on 
the OSI framework.
Section \ref{sect_linux} introduces a promising MI network framework with TCP/IP and Linux support in an attempt to address the challenges in current and future MIC studies.
Section \ref{sect_future} explores unresolved challenges. 
%including the universal MI fast fading model, its impact on MIC theorems, and the MI network architecture.
It also discusses promising methodologies including  novel MI antennas,  MI communication-navigation-sensing integrated (MCNSI), massive MI MIMO, deep JSCC for MIC, heterogeneous MI network techniques,  TCP/IP framework support, and Transformer-based prediction frameworks.
Section \ref{sect conclusions} concludes this survey.

% \subsubsection{Notation and Simulation Conventions}
The acronyms that appear across subsections of this survey are listed in Table~\ref{tbl_acronyms}. The physical representations of mathematical symbols are listed in Tables~\ref{tbl_sim} and \ref{tbl_symbol}.  The default values for the simulations and numerical evaluations, that we conducted in this survey, are listed in Table \ref{tbl_sim}.

\begin{comment}
	
	\begin{figure*}[htp]
		\centering
		%\includegraphics[width=3.2in]{poutpwrbi.eps}\\
		\includegraphics[width=6.0in]{sec1org.eps}\\
		\vspace{-0.4em}
		\caption{ The structure of the survey paper, where we introduce GAI methods for physical layer security through Communication Confidentiality and Authentication (Section \ref{CCA}), Communication Availability and Resilience (Section \ref{CAR}), and Communication Integrity (Section \ref{CI}).}\label{fig_commopt}
		\vspace{-0.8em}
	\end{figure*}
\end{comment}

\vspace{-0.0em}
{
	\subsection {Research Gap Summary}\label{sect_chp1rgs}

	Open  issues on MIC  are summarized in  Table \ref{tbl_sec1resgapsumm}, highlighting existing research gaps in several pivotal domains.  The high-priority domains are described in detail as follows.
	\subsubsection{MI fast fading for mobile MIC}\label{sectsubsub_mffmm}
	Research on MI fast fading still has significant gaps despite  existing efforts. The primary gap is the absence of a universal statistical model, which severely hinders the development of upper-layer protocols for mobile UG-WSNs. To address this gap, we propose a geometric antenna vibration model that can potentially tackle this challenge using Monte Carlo methods.
	
	\subsubsection{Significant gap in TCP/IP framework}  \label{sectsubsub_sgtf}
	The TCP/IP is crucial for SAGUI networks in next-generation communications.  However, when mapping the surveyed research to each OSI layer, significant gaps emerge:  certain layers and key topics remain underexplored. Specifically, no literature addresses IP applicability in MI networks, and the entire Transport Layer (Ly4) lacks dedicated MI solutions. Meanwhile, the extremely low bandwidth and channel capacity of MIC may not be compatible with existing IP and Ly4 protocols, as this low performance can lead to congestion mechanism failures and retransmission storms. Additionally, excessively large headers waste scarce MI bandwidth. To address these gaps in MI TCP/IP solutions, we propose a promising MIC framework supporting TCP/IP and Linux, which systematically incorporates MI algorithms and protocols drawn from the literature and  future potential solutions.
	
	\subsubsection{Underdeveloped  channel models and protocols for TTE-specific MIC}\label{sectsubsub_ucmptm}
	Current studies on MIC have largely overlooked the unique challenges of the TTE scenario, including mobility, heterogeneous geological materials, and constrained antenna position and orientation (APO). Key research gaps in TTE-specific MICs are as follows: (i) The eddy gain model for heterogeneous UG materials may require correction or numerical validation; (ii) shadow fading  should be considered for mobile nodes in Case (i); (iii) the existing upper-layer MI protocols lack TTE-specific design adaptations. For these gaps, the FEMs as we conducted in this survey are promising.

	\begin{table*}[htp] \scriptsize%\footnotesize
		\centering
		%	\tiny
		\caption{Research gaps and potential solutions, and their mappings}
		\label{tbl_sec1resgapsumm}
		\vspace{-0.8em}
		\scalebox{0.95}{
			\begin{threeparttable}
				%	\adjustbox{max width=\textwidth, scale=0.85}{}
				\begin{tabular}{m{0.04\textwidth}<{\centering}|m{0.12\textwidth}<{\centering}|m{0.37\textwidth}<{\centering}|m{0.36\textwidth}<{\centering} |m{0.03\textwidth}<{\centering}}
					\hline
					\hspace{-1.6em}\textbf{OSI Layers$^\dagger$}  &  \textbf{Research gaps} & \textbf{Identified open issues} & \textbf{Potential  solutions$^\ddagger$} &\textbf{Priority} \\
					\hline \hline
					%% %% \multirow{NumberOfRows}{CellWidth\textwidth}[-Fromtop]{\centering Passive Relay}
					%	\multirow{4}{0.03\textwidth}[-10pt]{\centering Ly1} &MI fast fading& \ding{72} (2) & A universal statistical modeling \\ \cline{2-5}
					%%%%%%%%%%%%%%%%%%%%%%%%%%%%%%%%%%%%%%%%%%%%%%%%%%%%%%%%%%%%%%%%%%%%%%		
					\multirow{2}{0.03\textwidth}[-3pt]{\centering Ly1}& MI fast fading &
					\begin{itemize}[leftmargin=1.9mm]	
						\item Universal statistical models
						\item Related upper-layer  solutions
						\vspace{-1.0em}
					\end{itemize}
					.	&\begin{itemize}[leftmargin=1.9mm]	
						\item Geometric antenna vibration model (cf. Fig. \ref{fig_fastfadingmodel});
						\item Monte Carlo methods (cf. Fig. \ref{fig_sec2Ejsd})
						\item Expressions of ergodic performances (cf. \eqref{eqn_chp2ec0}, \eqref{eqn_chp2ber0})
						\item Comparison simulations for modulations and FECs  (cf. Fig. \ref{fig_chp2bersnr})
						\vspace{-2.3em} 
					\end{itemize}
					& \hlprioritem\hlprioritem\hlprioritem \\ \cline{2-5}
					%%%%%%%%%%%%%%%%%%%%%%%%%%%%%%%%%%%%%%%%%%%%%%%%%%%%%%%%%%%%%%%%%%%%%%	
					&MI crosstalk  & \begin{itemize}[leftmargin=1.9mm]
						\item	Positive effects; spatial distribution  
						\vspace{-1.0em} 
					\end{itemize}
					& \begin{itemize}[leftmargin=1.9mm]
						\item 	Transformer-based prediction (cf. Fig. \ref{fig_sec5crossmig})
						\vspace{-1.0em} 
					\end{itemize}
					&\hlprioritem\hlprioritem \\ \cline{2-5}
					%%%%%%%%%%%%%%%%%%%%%%%%%%%%%%%%%%%%%%%%%%%%%%%%%%%%%%%%%%%%%%%%%%%%%%	
					&TTE-specific MIC  & \begin{itemize}[leftmargin=1.9mm]
						\item	Closed-form expressions of MIC range; optimal carrier frequency
						\item   Eddy current effects \& shadow fading from heterogeneous media 
						\vspace{-2.0em} 
					\end{itemize}
					& \begin{itemize}[leftmargin=1.9mm]
						\item 	Derivation based on Lambert-W function properties for MIC range (cf. \eqref{eqn_chp2eqrsd}) or FEM, as shown in Fig. \ref{fig_sec2fem}
						\vspace{-1.0em} 
					\end{itemize}
					&\hlprioritem\hlprioritem\hlprioritem \\ \cline{2-5}
					%%%%%%%%%%%%%%%%%%%%%%%%%%%%%%%%%%%%%%%%%%%%%%%%%%%%%%%%%%%%%%%%%%%%%%	
					& (Deep) JSCC  & \begin{itemize}[leftmargin=1.9mm]
						\item No relevant solutions (key: prospect of exceeding Shannon limits)
						\vspace{-2.3em} 
					\end{itemize}
					&  -
					&\hlprioritem\hlprioritem \\ \hline
					%%%%%%%%%%%%%%%%%%%%%%%%%%%%%%%%%%%%%%%%%%%%%%%%%%%%%%%%%%%%%%%%%%%%%%			
					Ly2& LLC &
					\begin{itemize}[leftmargin=1.9mm]	
						\item No  references (optional sublayer)
						\vspace{-1.0em}
					\end{itemize}
					&  -
					& \hlprioritem\\ \hline
					%%%%%%%%%%%%%%%%%%%%%%%%%%%%%%%%%%%%%%%%%%%%%%%%%%%%%%%%%%%%%%%%%%%%%%	    	
					Ly3& IP &
					\begin{itemize}[leftmargin=1.9mm]	
						\item  No  references (key: VLF adaptions and large packet headers)
						\vspace{-1.0em}
					\end{itemize}
					& \begin{itemize}[leftmargin=1.9mm]	
						\item MIC framework for TCP/IP \& Linux support (cf. Fig. \ref{fig_sec5tcp}, Algorithm \ref{alg_tcp})
						\vspace{-2.0em}
					\end{itemize}
					& \hlprioritem\hlprioritem\hlprioritem \\ \cline{1-3}
					%%%%%%%%%%%%%%%%%%%%%%%%%%%%%%%%%%%%%%%%%%%%%%%%%%%%%%%%%%%%%%%%%%%%%%	 	
					Ly4& TCP &
					\begin{itemize}[leftmargin=1.9mm]	
						\item No  references (key: Unstable connection and large headers)
						\vspace{-1.0em}
					\end{itemize}
					& 
					& \\ \hline
					%%%%%%%%%%%%%%%%%%%%%%%%%%%%%%%%%%%%%%%%%%%%%%%%%%%%%%%%%%%%%%%%%%%%%%			    	
					%				  	Ly1-7& Most researches &
					%				       \begin{itemize}[leftmargin=1.9mm]	
						%				      	\item Experimental validations with physical prototype/hardware
						%				     	\vspace{-1.0em}
						%				       \end{itemize}
					%				       & -
					%				       & \hlprioritem \\ 
					\hline
				\end{tabular}
				\begin{tablenotes}  
					\footnotesize  
					\item[$\dagger$] OSI Layers: Ly1: physical; Ly2: data link; Ly3: network; Ly4: transport; Ly5–7: combined application layers (i.e., session, presentation, application).
					\item [$\ddagger$]  Potential  solutions: systematically developed approaches with specific formulations (key expressions, simulations or frameworks),  distinct from generic proposals.
				\end{tablenotes}
			\end{threeparttable}
		}
		\vspace{-1.0em}
	\end{table*}

}

\section{Applications of MICs: General Scenarios and TTE-Specific Features} \label{sect_app}

In this section,  we survey the MICs for various potential applications, as summarized in Table \ref{tbl_app}.  Subsequently,  we discuss the specific challenges and considerations of applying MIC techniques to TTE scenarios.

\begin{table*}[htp!]  
	\caption{Applications using MI-based techniques} 
	%\begin{minipage}{\textwidth} 
	\vspace{-0.8em}
	\scalebox{0.88}{
		\centering
		\begin{threeparttable}
			\begin{tabular}{m{1.5cm}<{\centering}|m{1.2cm}<{\centering}|m{1.0cm}<{\centering}|m{1.6cm}<{\centering}|m{4.05cm}<{\centering}|m{5.9cm}<{\centering} |m{2.1cm}<{\centering}}
				\hline
				\textbf{Applications}&  \textbf{Predictable \& stable channels \tnote{$\dagger$}}  & \textbf{Waveguide  compatibility \tnote{$\ddag$}}& \textbf{MIC distance} & \textbf{Other characteristics}&\textbf{Impact on practical MIC systems}&\textbf{Involved \emph{refs.}}\\	             
				\hline \hline
				%\raggedleft
				\centering
				%%%%%%%%%%%%%%%%%%%%%%%%%%%%%%%%%%%%%%%%%%%%%%%%%%%%%%%%%%%%%%%%%%%%%%%%%%%%%%%%%%%%%%%%%%%%%%%%%%%%%%%
				Agriculture& Yes &Yes& Short to mid ($<$30m)&\begin{enumerate}[leftmargin=3mm]
					\item Limited  free spaces		
					\item Medium with high and time-variable VWC
					\item Shallow burial
					\vspace{-1.0em}
				\end{enumerate} &\begin{enumerate}[leftmargin=3mm]
					\item Restrict antenna size and deployments	
					\item  Variable performance due to time-variable VWC
					\item  Simple hardware, complex topologies and protocol stacks designs
					\vspace{-1.0em}
				\end{enumerate}   & \cite{Parameswaran2012Irrigation, Silva2015Strategic, Li2019Large, Sugumar2021Design, Cariou2023Internet}   \\ %\cline{2-5}
				\hline		
				%%%%%%%%%%%%%%%%%%%%%%%%%%%%%%%%%%%%%%%%%%%%%%%%%%%%%%%%%%%%%%%%%%%%%%%%%%%%%%%%%%%%%%%%%%%%%%%%%%%%%%%		
				\multirow{1}{*}{Industry} &Yes & Yes & Short to mid ($<$36 m)  & \begin{enumerate}[leftmargin=3mm]
					\item {Reliant on specific applications}	
					\vspace{-1.0em}
				\end{enumerate}  &{-} &\cite{Sun2011MISE, Tan2013An, Tan2015OnLocalization, Dong2023Novel, Li2024Rotating,Park2005DuraNode, Kisseleff2018Survey, Singh2022Magnetic, Markham2012Magnetio, Meybodi2013Feasibility}   \\ %\cline{2-5}	
				\hline	
				%%%%%%%%%%%%%%%%%%%%%%%%%%%%%%%%%%%%%%%%%%%%%%%%%%%%%%%%%%%%%%%%%%%%%%%%%%%%%%%%%%%%%%%%%%%%%%%%%%%%%%%	
				Underwater & No & Yes & Mid to long (1$\sim$100 m) &\begin{enumerate}[leftmargin=3mm]
					\item Sufficient free space	
					\item Randomly  misalignment coils
					\item  Remarkable  eddy  loss
					\vspace{-1.0em}
				\end{enumerate}&\begin{enumerate}[leftmargin=3mm]
					\item {Diverse antenna designs and configurations}
					\item {Fast fading-like phenomenon}
					\item  {Limited P2P MIC range}
					\vspace{-1.0em}
				\end{enumerate}   &\hspace{-0.4em} \cite{Li2019Survey,Domingo2012Magnetic, Gulbahar2012Communication,Erdogan2014Dynamic, Akyildiz2015Realizing, Guo2015Channel, Guo2017Multiple, Wei2018ROV,Liu2021Mechanical, Guo2021Joint, Wei2022Power, Zhang2023Rotation,Wang2023Novel, Wang2024Radiation, He2024Rotating, Zhang2023Design, Wang2019Efficient, Alsalman2021Balanced, Shovon2022Survey, Zhang2022Performance, Zhang2024Cooperative, Zhilin2023Universal, Dumphart2016Stochastic, Muzzammil2020Fundamentals} \\ %\cline{2-5} 	
				\hline	
				%%%%%%%%%%%%%%%%%%%%%%%%%%%%%%%%%%%%%%%%%%%%%%%%%%%%%%%%%%%%%%%%%%%%%%%%%%%%%%%%%%%%%%%%%%%%%%%%%%%%%%%
				\multirow{1}{*}{BAN} & No & No & Short ($<$5 m)&\begin{enumerate}[leftmargin=3mm]
					\item Low power requirements
					\item Non-magnetic dipole models
					\vspace{-1.0em}
				\end{enumerate} & \begin{enumerate}[leftmargin=3mm]
					\item {Small device size}
					\item {Strong coupling \& higher carrier frequency due to short MIC range }
					\vspace{-1.0em}
				\end{enumerate}   & \cite{Farhad2016DMac,Agbinya2010Power, Thilak2012Near,  Masihpour2013Multihop, Kisseleff2016Distributed, Dumphart2019Megneto, Golestani2018Theoretical,Banou2020MAGIC, Golestani2020Human,Mishra2021Wearable,Jenkins2023Wearable, Huang2024Field, Xu2024High} \\ %\cline{2-5}
				\hline	
				%%%%%%%%%%%%%%%%%%%%%%%%%%%%%%%%%%%%%%%%%%%%%%%%%%%%%%%%%%%%%%%%%%%%%%%%%%%%%%%%%%%%%%%%%%%%%%%%%%%%%%%	
				Military &	-&	-&	- & \begin{enumerate}[leftmargin=3mm]
					\item BorderSense architecture		
					\vspace{-1.0em}
				\end{enumerate} & \begin{enumerate}[leftmargin=3mm]
					\item {Access to EMWCs}		
					\vspace{-1.0em}
				\end{enumerate}&\cite{akyildiz2006wireless, Kisseleff2018Survey, Sun2011Border}   \\ %\cline{2-5}
				\hline	
				%%%%%%%%%%%%%%%%%%%%%%%%%%%%%%%%%%%%%%%%%%%%%%%%%%%%%%%%%%%%%%%%%%%%%%%%%%%%%%%%%%%%%%%%%%%%%%%%%%%%%%%		
				
				Environment \& disaster monitoring &	Yes&	Yes&	Short to long ($<$1,000 m)& \begin{enumerate}[leftmargin=3mm]
					\item  Lack of CSI due to disaster
					\vspace{-1.0em}
				\end{enumerate} &\begin{enumerate}[leftmargin=3mm]\item {Transmitter-side channel estimation}	\vspace{-1.0em}
				\end{enumerate}&\cite{Sharma2017Magnetic, Kisseleff2014Disaster}   \\ %\cline{2-5}
				\hline	
				%%%%%%%%%%%%%%%%%%%%%%%%%%%%%%%%%%%%%%%%%%%%%%%%%%%%%%%%%%%%%%%%%%%%%%%%%%%%%%%%%%%%%%%%%%%%%%%%%%%%%%%		
				
				\multirow{2}{*}{Localization} & Yes& No& Mid to Long  (1 $\sim$1,700 m) &\begin{enumerate}[leftmargin=3mm]
					\item Expecting  signal difference  across different positions
					\vspace{-1.0em}
				\end{enumerate}& \begin{enumerate}[leftmargin=3mm]
					\item Require specific localization algorithm (e.g., ToA)
					\vspace{-1.0em}
				\end{enumerate} & \cite{Abrudan2015Distortion, Tan2015OnLocalization, Kypris2016Magnetic,Lin2017Magnetic, Saeed20193D, Sheinker2019Localization, Wei2022iMag, Li2022Inertial,Pal2022MagLoc, Zeeshan2023Review,Chavda2023Magnetic, Chen20233D, Dong2023Novel, Abraha2024Survey, Su2016Design, Saeed2019Toward}. \\ %\cline{2-5}
				\hline		
				%%%%%%%%%%%%%%%%%%%%%%%%%%%%%%%%%%%%%%%%%%%%%%%%%%%%%%%%%%%%%%%%%%%%%%%%%%%%%%%%%%%%%%%%%%%%%%%%%%%%%%%		
				\multirow{2}{*}{TTE} & No & No & Long \quad \quad  \quad(60 $\sim$1,000 m)  &\begin{enumerate}[leftmargin=3mm]
					\item With non-uniform  eddy loss
					\item  Relatively limited free spaces
					\item  Using VLF-LA
					\item  Remarkable fast fading
					\vspace{-1.0em}
				\end{enumerate}&\begin{enumerate}[leftmargin=3mm]
					\item {Shadow fading-like phenomenon   }
					\item  {High deployment costs}
					\item  {Extremely low capacity}
					\item  {Complex hardware and upper-layer protocols   even for a simple network}
					\vspace{-1.0em}
				\end{enumerate}
				&\cite{zhang2014cooperative, Zhang2015Effective, Zhang2017Connectivity,Ma2019Effect,Ma2019Antenna, Ma2020Channel, Ma2024Fast}  \\ %\cline{2-5}
				\hline
				
			\end{tabular}
			\begin{tablenotes}  
				\footnotesize  
				\item[$\dagger$] Predictable and stable channels in most scenarios.
				\item[$\ddag$] Waveguide system compatibility in most scenarios. 
			\end{tablenotes}	
			
		\end{threeparttable}
	}
	%\end{minipage}
	\label{tbl_app}
	
	\vspace{-2.0em}
\end{table*}

\textbf{Applications in agriculture:} Soil conditions are crucial for crops, making it essential to build an MI UG-WSN  for agricultural automation. This attracts some researchers. 
%Parameswaran \emph{et al.}\cite{Parameswaran2012Irrigation} explored the  MIC system to establish the UG-WSNs for accurate soil moisture reports.  
Li \emph{et al.} \cite{Li2019Large} derived the conductivity and permittivity distribution using the Simultaneous Iterative Reconstruction Technique (SIRT) algorithm for soil moisture sensing, and obtained moisture sensing results based on an empirical model, i.e., a soil’s relative permittivity function w.r.t the volumetric water content (VWC). Sensors are spaced 5 to 10 m apart. The COMSOL simulations showed that the sensing
accuracy can achieve a root mean square error of 6\% in VWC~\cite{Li2019Large}. Agnelgo \emph{et al.} \cite{Silva2015Strategic} studied mid-range MI-based UG-WSNs for real-time soil moisture sensing, with a communication distance of 15-30 m.  A common feature of these MI agriculture applications is that the sensors are buried near the ground surface, and the communication distance among MI nodes is shorter than 30 m. In addition, the MI waveguide has been widely applied to enhance the MIC distance.

\textbf{Applications in industry:} The MI-based application has broader prospects in the industrial field, including pipeline leakage detection, infrastructure monitoring, and communication within mines.  For the pipeline leakage detection, many researchers focus on this area \cite{Sun2011MISE, Tan2013An, Tan2015OnLocalization, Dong2023Novel, Li2024Rotating}.  Sun \emph{et al.} \cite{Sun2011MISE} discussed the leakage detection and localization for the pipeline with a length of 36 m and 27 m, using the MI waveguide model. Tan's team studied the position of pipeline breakage by the MI-based method\cite{Tan2015OnLocalization}. Recently, Li's team \cite{Li2024Rotating} developed an MI-based position system for underground pipelines by applying rotating permanent magnets. For infrastructures, monitoring their health such as residential or commercial buildings, bridges, and dams is crucial to avoiding possible disasters\cite{Park2005DuraNode, Kisseleff2018Survey, Singh2022Magnetic}.  Mines provide an important application scenario. Compared to the pipeline and infrastructure scenarios,  the distance between two nodes can be much larger.  Among these studies on industry applications, the MI waveguide method was widely applied to extend the distance.
%For instance, Markham's team discussed the MIC  in the disaster scenario~\cite{Markham2012Magnetio}. They improved the data rate performance by the magnetic vector modulation. 
%In addition, Meybodi~\emph{et al.}~\cite{Meybodi2013Feasibility} proposed the method of using the metal pipeline as an ultra-long ferromagnetic core to improve the MIC performance.

\textbf{Underwater Applications:} 
Water covers over 70\% of Earth's surface, offering abundant resources. MIC technology shows promise for underwater exploration.  
Underwater applications can be categorized into under-freshwater and under-seawater applications.  As seawater has higher conductivity, the path loss in MI channels under the sea is significantly greater than that in freshwater. Li \emph{et al.}  \cite{Li2019Survey} offered a comprehensive overview of existing studies on underwater applications conducted prior to 2018, including those developed in \cite{Domingo2012Magnetic, Gulbahar2012Communication,Erdogan2014Dynamic, Akyildiz2015Realizing, Guo2015Channel, Guo2017Multiple,Wei2018ROV}. After 2018, MIC for UW-WSNs was increasingly attracting the attention of researchers. Firstly, several researchers proposed RPMAs to generate modulated magnetic fields with 30 Hz$\sim$1 kHz frequencies for MIC energy saving~\cite{Liu2021Mechanical, Zhang2023Rotation,Wang2023Novel, Wang2024Radiation, He2024Rotating, Zhang2023Design}.  
%Such MI devices can be used in AUVs  for long-range communication.
Secondly, the upper-layer protocols for MI-based UW-WSNs were investigated, including MAC protocols~\cite{Ahmed2024Design}, routing solutions~\cite{Wang2019Efficient, Alsalman2021Balanced}. Thirdly,  contrasting with UG-WSNs, the multi-antenna techniques can be applied in water thanks to ample free spaces for antenna deployment. These techniques include MI MIMO~\cite{Zhang2022Performance}, antenna array~\cite{Wang2024Radiation}, CMIC~\cite{Zhang2022Performance, Zhang2024Cooperative}, and MI waveguide. Finally, due to the sufficient free spaces,  the random misalignment between the coils would generate an unpredictable and unstable channel~\cite{Dumphart2016Stochastic}. Notably, compared to the UG-WSNs, the acoustic (sonar) and optical techniques are optional communication methods for UW-WSNs~\cite{Zhilin2023Universal}.

\textbf{Body Area Network (BAN):} The BAN is a sensor technology used to connect small nodes with sensing abilities to collect necessary information \cite{Farhad2016DMac}. In the cells-filled body,  EMW-based in the microwave range faces high attenuation, interference, and multipath effects. MI techniques can address these issues \cite{Agbinya2010Power, Thilak2012Near,  Masihpour2013Multihop, Kisseleff2016Distributed, Dumphart2019Megneto, Golestani2018Theoretical,Banou2020MAGIC, Golestani2020Human,Mishra2021Wearable,Jenkins2023Wearable, Huang2024Field, Xu2024High}. Specifically, the principle, power equations, and capacity of MI-based  BANs are demonstrated  in~\cite{Agbinya2010Power, Thilak2012Near}.  
Kisseleff \emph{et al.} \cite{Kisseleff2016Distributed} studied a distributed beamforming. Their simulation showed that a data rate of 3200 bits/s, at a distance of 5 m and a frequency of 13.56 MHz. Golestani \emph{et al.} \cite{Golestani2018Theoretical} derived the theoretical model of MI communication in wireless BANs, and analyzed its working frequencies of 0$\sim$50 MHz, based on the experimental MIC device at an MIC distance of 40 cm.
Mishra \emph{et al.} \cite{Mishra2021Wearable} proposed the MI waveguide for the BAN link.  Compared to UG-WSNs and UW-WSNs, MI-based nodes are closer, allowing a lower power requirement and a higher working frequency, e.g., 30 MHz, as reported in  \cite{Golestani2018Theoretical}. Due to the coils' size being comparable to MIC distances for the wearable requirements, these designs are similar to the standard Near-Field Communication (NFC) with simulation parameters aligned with those in ISO 14443/15693.
% some researchers did not consider the MI transmitter as a magnetic dipole, making the channel model more intricate. 

% \textbf{Military Applications:} As mentioned in  \cite{akyildiz2006wireless, Kisseleff2018Survey}, subsurface sensors are less likely to be detected by intruders.   Also,  MI-based techniques facilitate communication between submarines and AUVs in the sea. In \cite{Sun2011Border}, Sun \emph{et al.} proposed the concept of BorderSense, a hybrid wireless sensor network architecture for border patrol systems. In BorderSense, the UG-WSNs are the  MI sub-networks.

\textbf{Environment Monitoring:}   Environment monitoring covers public services, including pollution monitoring and disaster detection. MICs can perform chemical, biological, and pollution monitoring for river, lake, and water reservoirs~\cite{Sharma2017Magnetic}. Also,  MIC systems can monitor transport tunnels and mining conditions, and issue early warnings from underground to relevant departments for imminent accidents. In disaster scenarios,  obtaining CSI may be challenging, so Kisseleff \emph{et al.} \cite{Kisseleff2014Disaster} proposed a specific channel estimation within the MI transmitter circuit without explicit feedback CSI to address this issue.

\textbf{MI Localization:} MI techniques are widely employed for localization in environments where EMW propagation is challenging~\cite{NiCWLKHA15, Abrudan2015Distortion, Tan2015OnLocalization, Kypris2016Magnetic, Lin2017Magnetic, Saeed20193D, Sheinker2019Localization, Wei2022iMag, Li2022Inertial, Pal2022MagLoc, Zeeshan2023Review, Chavda2023Magnetic, Chen20233D, Dong2023Novel, Abraha2024Survey}. While the signal propagation model in MI localization shares similarities with that of MIC, the underlying methods and design principles differ significantly. MIC systems prioritize uniform signal strength around the transmitter to prevent local network congestion and maintain consistent data rates. Conversely, MI localization systems benefit from amplifying signal differences to improve positional distinction and accuracy. Additionally, various advanced techniques, such as time of arrival (ToA) \cite{Zeeshan2023Review}, time difference of arrival (TDoA) \cite{Zeeshan2023Review}, data fusion \cite{Su2016Design}, and inertial navigation \cite{Abrudan2015Distortion}, can be employed to further enhance localization accuracy.

\textbf{TTE Applications:}   
TTE applications typically have a vertical communication range of tens to hundreds of meters.  However, due to the extremely high cost of MI-based TTE devices, most studies focused on simulations and experiments within a limited distance of several meters.  Despite this challenge, several companies and research institutions have developed TTE devices. In 2011, Lockheed Martin developed the MagneLink$^{\rm TM}$ MIC system, achieving a two-way communication range of 472.5 meters using an 8-meter transmit antenna \cite{zhang2014cooperative, CDC2012Through}.   Vital Alert also developed MI-based TTE devices with a claimed MIC range of over 450 m~\cite{VitalCanaryComm}. Liu's team (our team) at Tongji University developed the TTE experiment device achieving a vertical distance of 310 m in Datong Coalmine using an 8 m transmit coil \cite{zhang2014cooperative,Zhang2017Connectivity}. Based on this device, we investigated the CMIC~\cite{zhang2014cooperative, Ma2019Effect,Ma2019Antenna} under the TTE environment using very low frequency (VLF) and larger antenna (VLF-LA) methods.  Recently, the MI fast fading channel of TTE MIC is studied in \cite{ Ma2024Fast}. With such fading, the MI channel becomes unpredictable and unstable. A comparison between TTE and general MIC settings is outlined in Table~\ref{tbl_ttesetting}.

While many studies (e.g., \cite{zhang2014cooperative, Zhang2015Effective, Zhang2017Connectivity, Ma2019Effect, Ma2019Antenna, Ma2020Channel, Ma2024Fast, CDC2012Through, VitalCanaryComm, Sun2010Magnetic, Kisseleff2018Survey}) have adapted MIC technologies for underwater and general underground applications to TTE communications, they have not fully addressed the specific challenges and considerations of applying MIC techniques to TTE scenarios, including those outlined in Table~\ref{tbl_app} and Table~\ref{tbl_ttesetting}.
\begin{itemize}  \label{resppage_ofchannelpower}
	\item {\textbf{Instability of channel power gain}: The uncertainty of the medium over a large space and the vibration of mobile MIC antenna cause random variations in channel power gain, ultimately leading to remarkable fast fading. }
	
	\item{\textbf{Extremely low bandwidth:} To extend the MIC range as much as possible, the throughput of the MI link must be greatly sacrificed, which has a significant impact on existing upper-layer solutions (e.g., centralized Q-learning-based routing \cite{Liu2022Qlearning}, and TCP/IP support).
		
		\item{\textbf{Eddy current in medium:} As the MIC distance increases, the eddy current is increasingly significant and complicated. Some relatively simple techniques, e.g., the MIC range/coverage \cite{Sun2010Magnetic,zhang2014cooperative, Zhang2017Connectivity}, and the frequency switchable routing protocol~\cite{Liu2022Qlearning}, become complicated. %such as the expression of MIC range/coverage and its optimization. 
		}
		
		\item{\textbf{Challenging deployment}: Limited space in deep underground environments poses deployment challenges for some MIC methods, such as MI waveguide and MIMO. Moreover, we must carefully consider the energy consumption of MIC devices due to the extremely high cost of replacing their batteries.}
	}
\end{itemize}
%These features are part of our motivations for this survey.

\begin{table}[t]  
	\caption{Comparison between TTE and general MIC settings}
	%\begin{minipage}{\textwidth} 
	%\tiny
	\vspace{-0.8em}
	\scalebox{0.88}{
		\centering
		\begin{threeparttable} 
			%\begin{tabular}{ p{2.3cm} p{4.5cm}}
			\begin{tabular}{m{0.160\textwidth}<{\centering}  |m{0.165\textwidth}<{\centering}| m{0.15\textwidth}<{\centering}}
				\hline 
				\textbf{Settings}  & \textbf{TTE}   & \textbf{General}   \\
				\hline \hline
				%\raggedleft
				\centering
				MIC Range (m) & 60$\sim$1,000 (Table \ref{tbl_app}) &  $<$36  (Table \ref{tbl_app}) \\\hline  
				Capacity (bps) &  $<$ 10 k  & $\sim$ 1 M \cite{Kisseleff2018Survey} \\ \hline  
				3-dB Bandwidth (Hz)  & 300$\sim$500\cite{Ma2024Fast} & 2 k$\sim$10 k\cite{Sun2010Magnetic, Kisseleff2015On} \\ \hline
				Frequency (Hz)      &   1 k$\sim$10 k\cite{Ma2024Fast} & 1 M$\sim$50 M \cite{Kisseleff2018Survey, Golestani2018Theoretical}\\ \hline
				Coil radius (m) &  0.5$\sim$4\cite{Ma2019Antenna, Zhang2017Connectivity}  & $<$ 0.2 \cite{Sun2010Magnetic}\\ \hline
				Tx Power (W)  &  5$\sim$126$^\dagger$\cite{Ma2019Effect} & $<$ 1\cite{Sun2010Magnetic} \\ \hline
				MI fast fading &  Yes\cite{Ma2024Fast} & No\cite{Li2019Survey}  \\ \hline
				Propagation medium      & Inhomogeneous & homogeneous \\ \hline
				Eddy effect & Significant & Mild \\ \hline
				%	Amplifier and filter  & Yes & No \\  \hline
				Deployment cost &  Extremely high & Low \\ \hline
				Existing TCP/IP support & Challenging & Potential \\ \hline
				Typical scenarios & Tunnel, mine, mountain & Farmland, pipeline  \\
				\hline
			\end{tabular}
			\begin{tablenotes}  
				\footnotesize  
				\item[$\dagger$] The Tx power of 126 W stems from our Datong Coalmine TTE experiment.
			\end{tablenotes}
		\end{threeparttable}
	}
	\label{tbl_ttesetting}
	\vspace{-2.0em}
\end{table}

\section {Channel Models for TTE MIC}\label{sect_sub2channel}
In this section, we introduce the MI channel power gain and MI fast fading discovered recently. Specifically, we categorize the channel power gain of the P2P MI link into four factors with distinct physically interpretable meanings and introduce their optimization strategies. Then, we discuss the MI fast fading, including its current statistical models, limitations, derivation challenges, and our proposed antenna vibration model with a simulation for the challenges.

\subsection{System and Channel Modeling} \label{sectsub3_scm}
\begin{figure}[htp]
	\centering
	\includegraphics[width=2.7in]{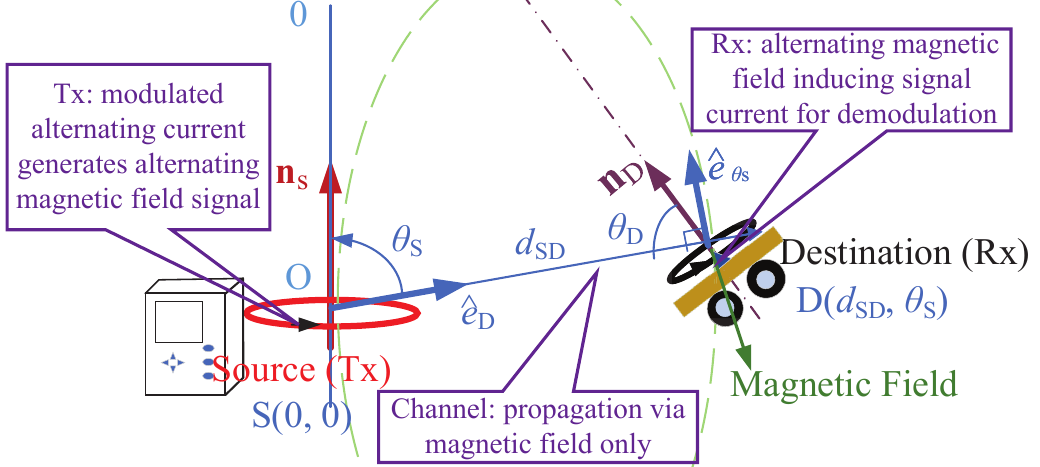}\\
	\vspace{-0.5em}
	\caption{ P2P TTE MI communication model in polar coordinates.
			%	Here, $\mathrm{S}$ and $\mathrm{D}$ represent the source magnetic dipole and destination directional magnetic sensor, respectively.
			Here,	 $\hat{e}_{\mathrm{D}}$ and $\hat{e}_{\theta_{\mathrm{S}}}$ are the radial and angular unit axial vectors, respectively;  $\mathbf{n}_{\mathrm{S}}$  and $\mathbf{n}_{\mathrm{D}}$ are the normal vectors of the dipole and sensor, respectively.
	}\label{fig_chp2HField}
	\vspace{-0.01em}
\end{figure}

The block diagram of P2P MIC link $\mathrm{S}$$\rightarrow$$\mathrm{D}$ is shown in Fig.~\ref{fig_chp2HField}, where the channel model depends on the antenna type, medium, and orientation. Specifically, for the coil-based MIC, Sun \emph{et al.} \cite{Sun2010Magnetic} derived the path loss expressions for both coil-based MI link and MI waveguide link, indicating that the path loss of P2P MI link scales with the sixth power of distance $d_{\mathrm{SD}}$. Their findings were experimentally validated in both wet and dry soil conditions. Further, Sun \emph{et al.} \cite{sun2012capacity} examined the impact of underground materials on MI communication. Li \emph{et al.} \cite{Li2015Capacity} highlighted the significant orientation sensitivity of MIC and proposed an orthogonal MIMO coil model to mitigate this effect. 
For the M$^2$I link, Guo \emph{et al.} \cite{Guo2015M2I} formulated the channel model using Maxwell’s equations, validated through COMSOL simulations \cite{Guo2015M2I}, and experimentally tested using a metamaterial shell composed of multiple small coils \cite{Guo2017Practical, Li2022Optimal}. In RPMA-based MIC, Rezaei \emph{et al.} \cite{Rezaei2020Mechanical} analyzed the magnetic moment of a single permanent magnet, while Zhang \emph{et al.} \cite{Zhang2023Rotation} extended this study to magnet arrays. Ma \emph{et al.} \cite{Ma2024Fast} investigated fast fading in downlink vehicular MI (VMI) channels.

{}\label{resppage_amongtheseresearch} %for response letter page reffering

Among these research articles,  the P2P MIC modeling start with the Maxwell's equations, the magnetic intensity $\mathbf{H}_{\mathrm{D}}$, induction intensity $\mathbf{B}_{\mathrm{D}} $, and the flux $\Psi_{\mathrm{D}}$ received by the Rx antenna D with the polar coordinates $(d_{\mathrm{SD}}, \theta_{\mathrm{S}})$ can be derived as	$\mathbf{H}_{\mathrm{D}}$$=$$\mathbf{m}_{\mathrm{S}} \vec{h}_{\mathrm{D}}$, $\mathbf{B}_{\mathrm{D}}$$=$$\mu_{\mathrm{u}} \mathbf{H}_{\mathrm{D}} $ and %$\Psi_{\mathrm{D}}$$=$$\mathbf{B}_{\mathrm{D}} \cdot \mathbf{n}_{\mathrm{D}}$, i.e.,
\begin{equation}\label{eqn_chp2HField}
	\begin{aligned}
		\Psi_{\mathrm{D}} = \mathbf{B}_{\mathrm{D}} \cdot \mathbf{n}_{\mathrm{D}} = \mu_{\mathrm{u}}\mathbf{m}_{\mathrm{S}} (\vec{h}_{\mathrm{D}} \cdot \mathbf{n}_{\mathrm{D}}) , 
	\end{aligned}
\end{equation}
respectively, where  $\vec{h}_{\mathrm{D}}$ is given by~\cite[Chapter 5]{balanis2005antenna}
% \begin{equation}\label{eqn_chp2vech}
	\begin{align}
		\vec{h}_{\mathrm{D}} &\!=\!\frac{1}{4\pi} \frac{ k_0 }{ 2d_{\mathrm{SD}}}e^{-jk_0d_{\mathrm{SD}}}\bigg[\cos\theta_{\mathrm{S}} \left(\tfrac{1}{d_{\mathrm{SD}}} (1\!+\! \tfrac{1}{jk_0d_{\mathrm{SD}}})\right)\hat{e}_{\mathrm{D}}  \nonumber\\ 
		&+ \sin\theta_{\mathrm{S}}\left( \frac{k_0}{2}\left(1\!+\!\tfrac{1}{jk_0d_{\mathrm{SD}}}\!-\!\tfrac{1}{(k_0d_{\mathrm{SD}})^2}\right)\right)\hat{e}_{\theta_{\mathrm{S}}}\bigg], \label{eqn_chp2vech}
	\end{align}
	% \end{equation}
where  $k_0 = 2\pi f \sqrt{\mu_{\mathrm{u}}   (\epsilon_{\mathrm{u}} + j (\sigma_{\mathrm{u}}/(2\pi f))) }$ \cite{Guo2021Jointchannel}.  Note that~\eqref{eqn_chp2vech} holds only in the weak coupling and linear regime, where the distance is much greater than the antenna size. It may break down in near-resonant or highly reactive systems.

In the receive coil or magnetic sensor,  the magnetic flux  $\Psi_{\mathrm{D}}$ induces a current $I_{\mathrm{D}}\propto\Psi_{\mathrm{D}}$.    
The channel power gain between link S$\rightarrow$D is given by
\begin{equation}\label{eqn_chp2Gsd}
	\begin{aligned}
		G_{\mathrm{SD}} &\!=\! \left|\tfrac{P_{\mathrm{D}f}}{P_{\mathrm{S}f}}\right| \!=\!\tfrac{1}{I_{\mathrm{S}}^2}\aleph_{\mathrm{SD}} \Psi^2_{\mathrm{D}}
		&\!=\! \tfrac{ |\mathbf{m}_{\mathrm{S}}|^2}{I_{\mathrm{S}}^2}\aleph_{\mathrm{SD}} \left(\mu_{\mathrm{u}} (\vec{h}_{\mathrm{D}}\cdot \mathbf{n}_{\mathrm{D}})\right)^2,
	\end{aligned}
\end{equation}
where the compensated variable  $\aleph_{\mathrm{SD}}$  significantly depends on the specific circuit and antenna type.

\begin{table*}[t!]  
	\caption{Partitions of channel power gains and their issues \\The marker `\hloitem'  describes the methods; The markers `\hllitem' and `\hlxitem' describe addressed and remaining issues,  respectively. }
	%\begin{minipage}{\textwidth} 
	\centering	
	\vspace{-0.8em}
	\scalebox{0.89}{	
		\begin{threeparttable}
			\begin{tabular}{m{0.05\textwidth}<{\centering}|m{0.09\textwidth}<{\centering}|m{0.075\textwidth}<{\centering}|m{0.40\textwidth}<{\centering}|m{0.32\textwidth}<{\centering}|m{0.03\textwidth}<{\centering}}	
				\hline
				\textbf{Factors} & \textbf{Physical meanings} & \textbf{Involved \emph{Refs.}} &\textbf{Addressed issues and methods}&\textbf{Remaining issues and optimization directions}&\textbf{Priority$^\dagger$}  \\  
				\hline
				\hline
				%%%%%%%%%%%%%%%%%%%%%%%%%%%%%%%%%%%%%%%%%%%%%%%%%%%%%%%%%%%%%%%%%%%%%%%%%%%%%%%%%%%%%
				Circuit gain ($\mathcal{C}_{\mathrm{SD}}$) & Local energy consumption &\cite{Sun2010Magnetic, Fawole2017Electromechanically,Wang2024Multi, Dionigi2012Multiband, Agbinya2019Principles, Sun2013Increasing}& \begin{itemize}[leftmargin=*]
					\item[\hllitem]Increase the magnetic moment with minimal energy
					\item[\hloitem]Use RPMA to reduce $\aleph_{\mathrm{SD}}$ under VLF
					\item[\hloitem]\textbf{Impedance matching} technique to maximize the receive power
					\item[\hloitem]\textbf{Multi-band MI technique} to achieve several 3dB bandwidth
					\vspace{-2.0em}
				\end{itemize} & \begin{itemize}[leftmargin=*]
					\item[\hlxitem]Challenge to derive the \textbf{capacity and energy loss} of \textbf{RPMA-based} link due to the inertia
					\item[\hlxitem] Deviation  of \textbf{3-dB bandwidth} in VLF below 100 Hz  %from existing theoretical expressions} 
				\vspace{-2.0em}
			\end{itemize}&\hlprioritem\hlprioritem\hlprioritem
			\\ \hline
			%%%%%%%%%%%%%%%%%%%%%%%%%%%%%%%%%%%%%%%%%%%%%%%%%%%%%%%%%%%%%%%%%%%%%%%%%%%%%%%%%%%%%%%%%%%%%%%%%%
			Space gain ($\mathcal{S}_{\mathrm{SD}}$) & Field attenuation  with spatial expansion &\cite{Guo2015M2I}& \begin{itemize}[leftmargin=*]			
				\item[\hllitem]Mitigate the rapidly signal attenuation with spatial expansion 
				\item[\hloitem] \textbf{M$^2$I enhanced techniques}  to enhance the overall permeability between $\mathrm{S}$ and $\mathrm{D}$
				\vspace{-1.0em}
			\end{itemize} &
			\begin{itemize}[leftmargin=*]			
				\item[\hlxitem] \textbf{Too large M$^2$I shell} occupying several times of coil volume% and  difficult to deploy in mobile MICs
				\vspace{-2.0em}
			\end{itemize} &\hlprioritem\hlprioritem \\ \hline
			%%%%%%%%%%%%%%%%%%%%%%%%%%%%%%%%%%%%%%%%%%%%%%%%%%%%%%%%%%%%%%%%%%%%%%%%%%%%%%%%%%%%%%%%%%%%%%%
			Eddy gain ($\mathcal{E}_{\mathrm{SD}}$) & Loss due to eddy current  of medium &\cite{Sun2013Increasing, Zhang2017Connectivity}& \begin{itemize}[leftmargin=*]			
				\item[\hllitem]Determining the non-negligible eddy current for TTE MIC
				\item[\hloitem]Derivation using magnetic potential
				\item[\hloitem] Using \textbf{Lambert-W function} to express the MI coverage
				\vspace{-1.0em}
			\end{itemize} & \begin{itemize}[leftmargin=*]			
				\item[\hlxitem] Unpredictable medium and \textbf{shadow fading} in the fluid  scenarios or mobile MIC
				\item[\hlxitem] Eddy gain in \textbf{inhomogeneous media} (by FEM)
				\vspace{-2.0em}
			\end{itemize} &\hlprioritem\hlprioritem\hlprioritem
			\\ \hline
			%%%%%%%%%%%%%%%%%%%%%%%%%%%%%%%%%%%%%%%%%%%%%%%%%%%%%%%%%%%%%%%%%%%%%%%%%%%%%%%%%%%%
			\hspace{-1em}Polarization gain($J_{\mathrm{SD}}$) & Loss due to antenna orientation &\cite{Zhang2017Connectivity, Li2015Capacity, Dumphart2016Stochastic, Ma2020Channel, Zhang2022Performance, Ma2024Fast}& \begin{itemize}[leftmargin=*]			
				\item[\hllitem] \textbf{Stochastic  misalignment} coils in underwater scenarios
				\item[\hllitem]\textbf{Unpredictable MI fast fading channel} in the underground/aboveground MIC
				\item[\hloitem] \textbf{Orthogonal MIMO coils} to mitigate  orientation sensitivity
				\item[\hloitem] \textbf{Boundary $p(x)$ distribution} to model MI fast fading
				%derive the CDF, PDF and expectation of MI fast fading} 
			\vspace{-1.0em}
		\end{itemize} &\begin{itemize}[leftmargin=*]			
			\item[\hlxitem] Challenge to deploy the MIMO antenna in the TTE scenarios
			\item[\hlxitem] Challenge to derive a \textbf{universal statistical model} for MI fast fading
			\vspace{-1.0em}
		\end{itemize} &\hlprioritem\hlprioritem\hlprioritem
		\\ \hline
		%%%%%%%%%%%%%%%%%%%%%%%%%%%%%%%%%%%%%%%%%%%%%%%%%%%%%%%%%%%%%%%%%%%%%%%%%%%%%%%%%%%%%%%%%%%%%%%%%
		Mixed field case & Non-near-field case (not $k_0d_{\mathrm{SD}} \ll 1$) &\cite{Liu2021Mechanical,  Guo2021Jointchannel, Zhou2017Maximum, Guo2021Joint}& \begin{itemize}[leftmargin=*]			
			\item[\hllitem] Channel modeling for P2P MI link
			\item[\hloitem]  Derivation based on Maxwell's equations
			\vspace{-1.0em}
		\end{itemize}& \begin{itemize}[leftmargin=*]			
			\item[\hlxitem]Challenge  in partitioning  the channel power gain
			\item[\hlxitem] Challenge for antenna design  and  optimization  due to excessive dependence on variables
			\item[\hlxitem]  CMIC  channel modeling
			\item[\hlxitem]  Upper-layer solutions, such as MAC and routing
			\item[\hlxitem]  Network deployment
			\item[\hlxitem] Frequency optimization and 3-dB bandwidth derivation 
			\vspace{-1.0em}
		\end{itemize} &\hlprioritem\hlprioritem
		\\ \hline
		%%%%%%%%%%%%%%%%%%%%%%%%%%%%%%%%%%%%%%%%%%%%%%%%%%%%%%%%%%%%%%%%%%%%%%%%%%%%%%%%%%%%%%%%%%%%%
		\hspace{-0.6em}	Strong coupling &Coupling coefficient$\approx$1 &\cite{Jiang2015Capacity}& \begin{itemize}[leftmargin=*]			
			\item[\hllitem]  Communication capacity in the strong coupling case  (MIC distance much smaller than coils size)
			\item[\hloitem]  Non-magnetic-dipole modeling
			\vspace{-1.0em}
		\end{itemize}& \begin{itemize}[leftmargin=*]			
			\item[\hlxitem]  Inapplicable channel power gain expressions ~\eqref{eqn_chp2Gsd} and~\eqref{eqn_chp2GSDNear} 
			\vspace{-1.0em} \end{itemize}
		&\hlprioritem
		\\ \hline
		
		\hline	
	\end{tabular}
	\begin{tablenotes}  
		\footnotesize  
		\item[$\dagger$] Priority: Priority level of remaining issues  for exploration.
	\end{tablenotes}  
\end{threeparttable}
}
%\end{minipage}
\vspace{-1.5em}
\label{tbl_gain}
\end{table*}

\subsection{Decomposition of  Channel Power Gain} \label{sectsub3_odcpg}
\begin{figure}[t]
\centering
\includegraphics[width=3.4in]{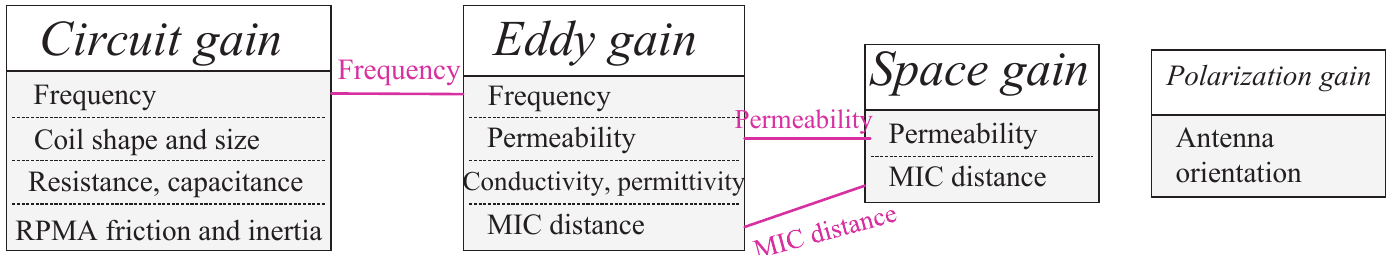}\\
\vspace{-0.5em}
\caption{Relationship among circuit, eddy, space, and polarization gains. The normal fonts with gray background represent the primary parameters.   %SU - SVU, PU->PVU P1->P0 %SU - SVU, PU->PVU P1->P0
}\label{fig_sec2channdecomp}
\vspace{-0.01em}
\end{figure}

Eq.~\eqref{eqn_chp2Gsd} is a universal expression of the channel power gain both in the near field and radiation field ranges. However, this expression is challenging to analyze and formulate into a further optimization model in MIC research. Thus,
in most MI UG-WSN studies, insufficient communication distance~\cite{Sun2010Magnetic} or VLF-LA methods~\cite{zhang2014cooperative} lead to the assumption that $k_0 d_{\mathrm{SD}} \ll 1$ holds.  
\begin{comment}
Thus, \eqref{eqn_chp2vech} is often approximated as 
\begin{equation}\label{eqn_chp2HNear}
\begin{aligned}
	\vec{h}_{\mathrm{D}} \simeq \tfrac{1}{4\pi}\tfrac{1}{4 d_{\mathrm{SD}}^3} \left(2\cos\theta_{\mathrm{S}}\hat{e}_{\mathrm{D}} + \sin\theta_{\mathrm{S}}\hat{e}_{\theta_{\mathrm{S}}}\right) \cdot e^{-jk_0d_{\mathrm{SD}}}.
\end{aligned}
\end{equation}
\end{comment}
Guided by the ``high cohesion and low coupling" principles in software engineering  and  by substituting  $k_0 d_{\mathrm{SD}}\ll1$ into~\eqref{eqn_chp2Gsd}, we propose  a decomposition of the channel power gain  $G_{\mathrm{SD}}$  into four  physically meaningful factors 
\begin{equation}\label{eqn_chp2GSDNear}
\begin{aligned}
G_{\mathrm{SD}} &\!= \! \underbrace {\tfrac{|\mathbf{m}_{\mathrm{S}}|^2\aleph_{\mathrm{SD}}}{16 \pi^2|I_{\mathrm{S}}|^2} }_{{\rm Circuit \ gain} \ \mathcal{C}_{\mathrm{SD}}} \times \underbrace{\tfrac{ |\mu_{\mathrm{u}}|^2}{( d_{\mathrm{SD}}^3 )^2}}_{{\rm Space \ gain}\ \mathcal{S}_{\mathrm{SD}}} \times    \underbrace{e^{-2jk_0d_{\mathrm{SD}}}}_{ {\rm Eddy \ gain} \ \ \mathcal{E}_{\mathrm{SD}}}
\\
& \times \underbrace{(2\cos\theta_{\mathrm{S}}\cos\theta_{\mathrm{D}}\!+\! \sin\theta_{\mathrm{S}}\sin\theta_{\mathrm{D}})^2}_{{\rm Polarization \ gain} \ J_{\mathrm{SD}}=\mathcal{J}^2_{\mathrm{SD}}},
\end{aligned}
\end{equation}
%which explicitly indicates that the P2P MI channel power gain $G_{\mathrm{SD}}$ can be divided into four factors:
i.e., circuit gain $\mathcal{C}_{\mathrm{SD}}$, space gain $\mathcal{S}_{\mathrm{SD}}$, eddy gain $\mathcal{E}_{\mathrm{SD}}$, and polarization gain $J_{\mathrm{SD}}$. As shown in Fig. \ref{fig_sec2channdecomp}, the relationships among these decomposed gains involve only three logical coupling links, conforming to the low-coupling principle. Their physically interpretable meanings are provided in the rest of this subsection.
We summarize addressed issues, methods,  remaining issues, and optimization directions in existing research in Table~\ref{tbl_gain},  as described in what follows.

\subsubsection{Circuit gain $\mathcal{C}_{\mathrm{SD}}$} \label{sectsub3_circuitgain}

The circuit gain, with two optimizable factors $\mathbf{m}_{\mathrm{S}}$ and $\aleph_{\mathrm{SD}}$,  reflects energy consumption on the circuit. We observe that the studies on antenna and hardware designs, including resonance characteristics of the coil antenna~\cite{Sun2010Magnetic},  magnetic pole strengths of the bias and rotor magnet of RPMA~\cite{Fawole2017Electromechanically}, can be categorized as the independent considerations of the optimization of the circuit gain $\mathcal{C}_{\mathrm{SD}}$.
This is crucial for improving the bandwidth through the factor $\aleph_{\mathrm{SD}}$ of $\mathcal{C}_{\mathrm{SD}}$, and MIC range through the factor $\mathbf{m}_{\mathrm{S}}$ of $\mathcal{C}_{\mathrm{SD}}$.  For example, RPMA without coils can eliminate the resonance coupling (expressed in $\aleph_{\mathrm{SD}}$)  ~\cite{Fawole2017Electromechanically}. However, RPMA may face an additional equivalent circuit loss due to mechanical inertia and friction  (also expressed in $\aleph_{\mathrm{SD}}$). Wang \emph{et al.} \cite{Wang2024Multi} used impedance matching $Z_L=R + \frac{(2 \pi f M)^2}{R}$ in $\aleph_{\mathrm{SD}}(Z_L(f))$ to maximize the receive power, where $Z_L$, $R$, and $M$ are the impedance and resistance of the coil, and mutual inductance, respectively. Their simulations showed that their approach produced two frequency resonant points, at 1 MHz and 5 MHz, respectively.  In \cite{ Dionigi2012Multiband, Agbinya2019Principles, Sun2013Increasing}, multi-band MI techniques achieve several discontinuous 3-dB bandwidths in the frequency domain through a coil array, based on  $\mathcal{\aleph}_{\mathrm{SD}}(Z_L(f))$. Unfortunately, additional loss arises due to crosstalk among the coil array. The circuit gain exhibits optimal operability, as it minimizes the need to consider the MIC environment.

%\vspace{-1.0em}

\subsubsection{Space gain $\mathcal S_{\mathrm{SD}}$}  \label{sectsub3_spacegain}

This gain represents the ideal spatial path loss due to space expansion. This gain serves as a primitive model for MIC.  It is widely used in MI network studies to analyze and optimize signal transmission.  When considering short to mid-range MIC systems, such as  MI waveguides \cite{kisseleff2013channel},  M$^2$I communications\cite{Guo2015M2I}, and upper-layer protocol for MIC\cite{lin2015distributed,Wang2019Efficient,Alsalman2021Balanced}, researchers often choose to simplify theoretical communication models by ignoring the eddy gain $\mathcal{E}_{\mathrm{SD}}$ and polarization gain $\mathcal{J}_{\mathrm{SD}}$. While $\mathcal{E}_{\mathrm{SD}}$ and  $\mathcal{J}_{\mathrm{SD}}$ can have significant effects on signal propagation and reception, ignoring them allows for more straightforward analyses and calculations, especially during the initial stages of network design and optimization.
The space gain expression $\mathcal{S}_{\mathrm{SD}} = \frac{ |\mu_{\mathrm{u}}|^2}{(16\pi d_{\mathrm{SD}}^3 )^2}$ suggests that optimizing $\mathcal{S}_{\mathrm{SD}}$  is challenging due to the fixed permeability  $\mu_{\rm u}$  determined by underground material.   Guo \emph{et al.} optimized the equivalent permeability by designing a metamaterial-enhanced antenna enclosed by a metamaterial shell. This shell has a permeability of $\mu_2 = -\mu_0 $ where $\mu_0$ is the vacuum permeability\cite{Guo2015M2I}. This enhances the overall permeability between $\mathrm{S}$ and $\mathrm{D}$, resulting in an improved space gain. In practice, the TTE environment is hardly modifiable, leading to limited operability of space gain.

\vspace{-0.0em}

\subsubsection{Eddy gain $\mathcal{E}_{\mathrm{SD}}$} \label{sectsub3_eddygain}
The eddy gain is known as the Skin Effect. It is generated by induction current from the time-varying magnetic field within underground materials.  In TTE communication, heterogeneous materials exhibit dynamic characteristics, especially for mobile MI networks. This makes MIC channel coefficients unpredictable. It also poses significant challenges to MIC research. 
For short-range MICs and early MICs research, the eddy gain can be ignored to simplify primary problems.   However, for TTE MICs, the cumulative eddy current is highly remarkable. In an infinite and uniform medium, the expression of the eddy gain $\mathcal{E}_{\mathrm{SD}}$ is given in  \cite{Sun2013Increasing, Sogade2004Electromagnetic} and can be approximately matched as \cite{Ma2019Effect}
\begin{equation}\label{eqn_chp2GeTTE}
\begin{aligned}
\sqrt{\mathcal{E}_{\mathrm{SD}}}&\!\simeq\!\exp(\tfrac{-d_{\mathrm{SD}}}{\delta_{\rm u}}) = \exp \left(- d_{\mathrm{SD}} \delta_{\rm u}^{-1} \right),  \\
\end{aligned}
\end{equation}
%%\begin{equation}\label{eqn_chp2Ge0}
%%\begin{aligned}
%%\sqrt{\mathcal{E}_{\mathrm{SD}}}&\!=\!\left|\int_{0}^{\infty} \tfrac{x^{3}\exp %%\left[-x^{2}\!-\!j\left(\tfrac{\sqrt{2}d_{\mathrm{SD}}}{\delta}\right)^{2}\right]^{\frac{1}{2}} %%}{x\!+\!\left[x^{2}+j\left(\tfrac{\sqrt{2} d_{\mathrm{SD}}}{\delta_{\mathrm{u}}}\right)^{2}\right]^{\tfrac{1}{2}}}   \mathrm{d} x\right| %%,  \\
%% 	\end{aligned}
%%  \end{equation}
with 
\begin{equation}\label{eqn_chp2Skin0}
\begin{aligned}
\delta_\mathrm{u} &= \tfrac{1}{2 \pi f \sqrt{\frac{\mu_{\rm u} \epsilon_\mathrm{u}}{2}\left(\sqrt{1+\frac{\sigma_{\mathrm{u}}^{2}}{(2 \pi f)^{2} \epsilon_\mathrm{u}^{2}}}-1 \right)}},
\end{aligned}
\end{equation}
where the permittivity $\epsilon_\mathrm{u}$ is typically very small in most underground environments. For instance, the permittivities of dry and wet soils are $7 \times 8.854 \times 10^{-12}$ F/m and $29 \times 8.854 \times 10^{-12}$ F/m, respectively.  Their conductivities $\sigma_\mathrm{u}$  are 0.01 S/m and 0.077 S/m, respectively\cite{sun2012capacity}. Hence, for the TTE MIC with VLF-LA, the condition $\sigma_\mathrm{u} \gg 2\pi f \epsilon_\mathrm{u}$ typically  holds, allowing~\eqref{eqn_chp2Skin0} to be further approximated as 
\begin{equation}\label{eqn_chp2SkinVLF}
\begin{aligned}
\delta_\mathrm{u} &\simeq \sqrt{\tfrac{1}{\pi f \mu_{\rm u} \sigma_{\rm u}}}.
\end{aligned}
\end{equation}
%% By analyzing the numerical results of~\eqref{eqn_chp2Ge0}, the eddy gain can be approximately matched as 
%%  \begin{equation}\label{eqn_chp2GeTTE}
%%	\begin{aligned}
%%		\sqrt{\mathcal{E}_{\mathrm{SD}}}&\!\simeq\!\exp(\tfrac{-d_{\mathrm{SD}}}{\delta_{\rm u}}) \simeq \exp \left(- d_{\mathrm{SD}} \sqrt{ \pi f %%\mu_{\rm u} \sigma_{\rm u}  } \right).  \\
%%	\end{aligned}
%%\end{equation}
From~\eqref{eqn_chp2GSDNear} and \eqref{eqn_chp2GeTTE}, we observe that only the circuit gain $\mathcal{C}_{\mathrm{SD}}$ and eddy gain $\mathcal{E}_{\mathrm{SD}}$ depend on the carrier frequency $f$. This suggests a potential to enhance MIC performance through frequency optimization. For the short-range and mid-range MICs, the effect of the eddy gain is not significant. Hence, there is limited literature specifically studying the eddy gain at this moment.

{}\label{resppage_thattheeddy} % for response letter page reffering

\begin{figure}[t]
\centering
\includegraphics[width=2.5in, height=1.5in]{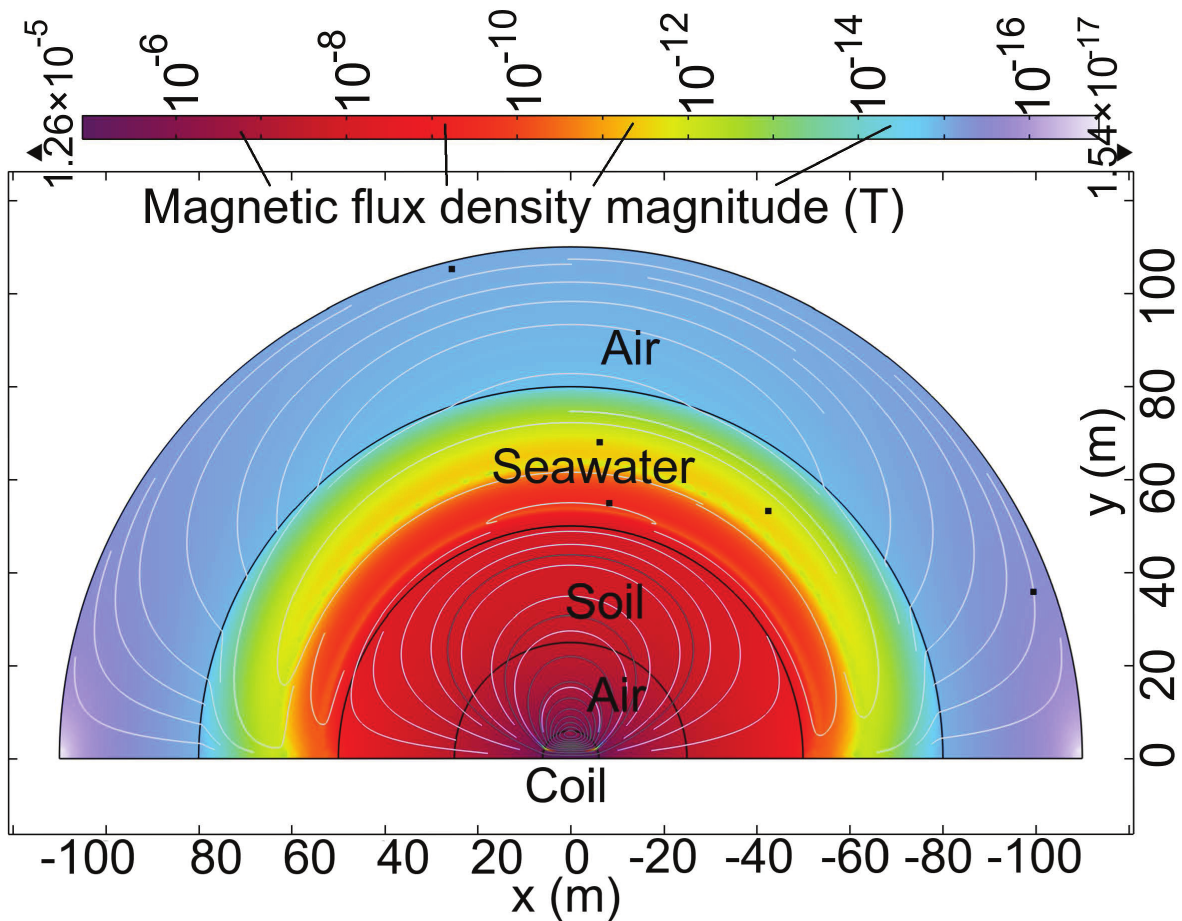}\\
\vspace{-0.5em}
\caption{FEM simulation of magnetic flux density in multilayer materials with different conductivities. Here, the conductivities of air, soil and seawater are 0, 0.01, and 4.8 S/m, respectively. The current of coil $I_{\mathrm{S}}$$=$$15\cos(2\pi\times 10^4 t)$ (A). The coil radius is 6 m.  %SU - SVU, PU->PVU P1->P0 %SU - SVU, PU->PVU P1->P0
}\label{fig_sec2fem}
\vspace{-0.01em}
\end{figure}

Note that the eddy gain expressions in \eqref{eqn_chp2GSDNear} and \eqref{eqn_chp2GeTTE} apply only to near-field homogeneous media. Fig. \ref{fig_sec2fem} presents COMSOL simulations of magnetic flux density distributions in multilayer materials. These simulations show significant distortion in both magnetic field direction and magnitude across different media due to complex eddy effects, particularly near medium boundaries. This indicates that the eddy gain model requires correction for multilayer or inhomogeneous media.

\subsubsection{Polarization gain $J_{\mathrm{SD}}$}   \label{sectsub3_pogain}
This gain $J_{\mathrm{SD}}=\mathcal{J}^2_{\mathrm{SD}}$ represents the effects of antenna orientations.  From~\eqref{eqn_chp2GSDNear},  $J_{\mathrm{SD}}$ is within $[0^2, 2^2] $. Specifically, when $\theta_{\mathrm{S}} = \frac{\pi}{2}$ and  $\theta_{\mathrm{D}} = 0$,  $J_{\mathrm{SD}}$ falls into 0. This poses several challenges as follows. 1) Aligning the Tx and Rx antennas may be difficult in practical scenarios, such as the  autonomous underwater vehicle (AUV)  scenario. To address this, the orthogonal MIMO has been introduced into the MICs. In \cite{Li2015Capacity}, Li \emph{et al.} deployed three orthogonal coils in each device, increasing the minimal channel power gain to $(\frac{1}{3})^2$ of the maximal one. 2) In MI UW-WSNs, antenna vibrations cause random misalignment, leading to stochastic polarization gain and channel coefficients \cite{Dumphart2016Stochastic}.  %In \cite{Dumphart2016Stochastic}, Dumphart \emph{et al.} derive the probability density function (PDF) of $\mathcal{J}_{\mathrm{SD}}$ assuming uniform distributed antenna orientation.  
3) Fast fading in mobile MI UG-WSNs, such as vehicle MICs (VMICs) and backpack MICs, also stems from polarization gain \cite{Ma2020Channel, Ma2024Fast}. Most challengingly, in most mobile MICs, \ $J_{\mathrm{SD}}$ is a random variable with uncharacterized distribution. %Researchers have derived its statistical characteristics. However, they have found new issues where statistical parameters are also unpredictable since the expectation and variance of $J_{\mathrm{SD}}$ deeply depend on the driver's mind. Fortunately, they propose a multiagent RL based algorithm to solve this problem.

In summary, the fine-grained decomposition of channel power gain into four low logical coupling gains $G_{\mathrm{SD}}=\mathcal{C}_{\mathrm{SD}}\mathcal{S}_{\mathrm{SD}}\mathcal{E}_{\mathrm{SD}}J_{\mathrm{SD}}$ allows us to simplify the optimization problem formulation by focusing on one or several specific factors, while fixing the rest of parameters. For instance, the MI antenna optimization prioritizes the circuit gain $\mathcal{C}_{\mathrm{SD}}$, while MI MIMO optimization focuses on enhancing polarization gain $J_{\mathrm{SD}}$. For MIC in multi-layer medium, attention can be given to the eddy gain $\mathcal{E}_{\mathrm{SD}}$. For the long-range MIC  in ultra-low-conductivity medium or the short-range MIC, focusing on space gain $\mathcal{S}_{\mathrm{SD}}$ suffices.

%\subsubsection{Coil Tx/Rx Antenna}

%To concisely clarify the channel power gain, its issues, methodologies, unresolved issues, and related references (\emph{refs.}), we summarize them in Table \ref{tbl_gain}. 

\begin{table*}[t!]  
\caption{Comparison among different models of the fast fading channel in related references }
%\begin{minipage}{\textwidth} 
\vspace{-0.9em}
\scalebox{0.94}{
\centering
\begin{tabular}{m{1.2cm}<{\centering}|m{1.5cm}<{\centering}|m{6.4cm}<{\centering}|m{6.4cm}<{\centering}|m{1.5cm}<{\centering}}
\hline
\textbf{Types} & \textbf{Models} & \textbf{Compared points}  & \textbf{Different challenges of  MI fast fading}   & \textbf{\emph{Refs.}}\\
\hline \hline
%\raggedleft
\centering
%%%%%%%%%%%%%%%%%%%%%%%%%%%%%%%%%%%%%%%%%%%%%%%%%%%%%%%%%%%%%%%%%%%%%%%%%%%%%%%%%%%%%%%%%%%%%%%%%%%%%%%		
EMWC & Rayleigh, Rician, Nakagami &\begin{enumerate}[leftmargin=3.2mm]
	\item  Relatively predictable large-scale fading
	\item  Multi-path effect 
	\item  Using CLT-based simplifications
	\item  No boundary limits
	\vspace{-1.0em}
\end{enumerate}
& \begin{enumerate}[leftmargin=4.7mm]
	\item [\maltese 1)] Unpredictable large-scale fading
	\item [\maltese 2)] Vibrating-oriented  fading without multi-path effect
	\item [\maltese 3)]  Hard to use CLT-based methods
	\item [\maltese 4)]With boundary limits
	\vspace{-1.0em}
\end{enumerate}
&\cite{Simon2005Digital,Tse2005Fundamentals, Ma2025Novel}\\ \cline{2-5}
\hline
%%%%%%%%%%%%%%%%%%%%%%%%%%%%%%%%%%%%%%%%%%%%%%%%%%%%%%%%%%%%%%%%%%%%%%%%%%%%%%%%%%%%%%%%%%%%%%%%%%%%%%%
\multirow{2}{0.06\textwidth}{FSOC} & M\'{a}laga  &\begin{enumerate}[leftmargin=3.2mm]
	\item  Simplified to Rayleigh, Rician, and Nakagami models
	\item  No boundary limits
	\vspace{-1.0em}
\end{enumerate}
& Same as \maltese1), \maltese2), \maltese3), \maltese4) & \cite{Farid2012Diversity,jurado2011unifying,Wang2021Hovering}\\ \cline{2-5}
&\hspace{-0.8em}Geometric and misalignment loss& \begin{enumerate}[leftmargin=3.2mm]
	\item  Linear propagation signals
	\item  No boundary limits
	\vspace{-1.0em}
\end{enumerate}
&  \begin{enumerate}[leftmargin=3.2mm]
	\item  Linear propagation signals
	\item   Orientation sensitivity
	\item    With boundary limits
	\vspace{-1.0em}
\end{enumerate} & \cite{Najafi2020Statistical} \\   \cline{2-5}		
\hline
%%%%%%%%%%%%%%%%%%%%%%%%%%%%%%%%%%%%%%%%%%%%%%%%%%%%%%%%%%%%%%%%%%%%%%%%%%%%%%%%%%%%%%%%%%%%%%%%%%%%%%%
\multirow{2}{0.06\textwidth}[-3pt]{Acoustic communi-cation} & Rician &\begin{enumerate}[leftmargin=3.2mm]
	\item   Be simplified to Rician 
	\item    Strong LOS
	\item    No boundary limits
	\vspace{-1.0em}
\end{enumerate} & Same as \maltese1), \maltese2), \maltese3), \maltese4) & \cite{Aik2006Characterization}\\ \cline{2-5}
%%%%%%%%%%%%%%%%%%%%%%%%%%%%%%%%%%%%%%%%%%%%%%%%%%%%%%%%%%%%%%%%%%%%%%%%%%%%%%%%%%%%%%%%%%%%%%%%%%%%%%%		
&MIMO SWA  &\begin{enumerate}[leftmargin=3.2mm]
	\item  With macro scattering and micro scattering
	\item    No boundary limits
	\vspace{-1.0em}
\end{enumerate} &  Same as \maltese1), \maltese2), \maltese3), \maltese4)  & \cite{Zajic2011Statistical} \\  \cline{2-5}
\hline		
%	\multirow{2}{*}{MIC} &\makecell[l]{\vspace{-0.6em}\\ Traditional \\MIC} & No fast fading & With fast fading & \cite{Sun2010Magnetic,sun2012capacity, kisseleff2013channel,Kisseleff2014Modulation,zhang2014cooperative,Kisseleff2014Transmitter,lin2015distributed,Tan2017environment, Zhang2017Connectivity, Kisseleff2018Survey, Li2019Survey,Ma2019Effect, Ma2019Antenna,Guo2017Practical,Guo2021Joint, Wang2022Multi,Chen2023Novel, Wang2023Backscatter}\\\cline{2-5}
%%%%%%%%%%%%%%%%%%%%%%%%%%%%%%%%%%%%%%%%%%%%%%%%%%%%%%%%%%%%%%%%%%%%%%%%%%%%%%%%%%%%%%%%%%%%%%%%%%%%%%%		
\multirow{2}{0.06\textwidth}[-23pt]{MIC} &Traditional MIC & No fast fading & With fast fading & \cite{ Kisseleff2018Survey, Li2019Survey, Wang2022Multi,Chen2023Novel}\\\cline{2-5}

%\multirow{2}{*}{MIC} &\makecell[l]{\vspace{-0.6em}\\ Traditional \\MIC} & No fast fading & With fast fading & \cite{Sun2010Magnetic, Kisseleff2018Survey, Li2019Survey}\\\cline{2-5}		
& Underwater &\begin{enumerate}[leftmargin=3.2mm]
	\item   Antenna orientation with a uniform distribution
	\item    No boundary limits
	\vspace{-1.0em}
\end{enumerate} 
&\begin{enumerate}[leftmargin=3.2mm]
	\item  Antenna angle with complex probability  distributions
	\item    With boundary limits
	\vspace{-1.0em}
\end{enumerate} 
&\cite{Dumphart2016Stochastic} \\  \cline{2-5}
& Vehicle MI& \begin{enumerate}[leftmargin=3.2mm]
	\item  Antenna angle with a BCS distribution
	\item  Antenna angle with a boundary $p(x)$ distribution
	\vspace{-1.0em}
\end{enumerate} 
&\begin{enumerate}[leftmargin=3.2mm]
	\item   Lacking universal solutions
	\vspace{-1.0em}
\end{enumerate} &\cite{Ma2020Channel, Ma2024Fast, Chen2025Statistical} \\ 
%%%%%%%%%%%%%%%%%%%%%%%%%%%%%%%%%%%%%%%%%%%%%%%%%%%%%%%%%%%%%%%%%%%%%%%%%%%%%%%%%%%%%%%%%%%%%%%%%%%%%%%		
\hline
\end{tabular}
}
\vspace{-1.5em}
%\end{minipage}
\label{tbl_fastfadingdiff}
\end{table*}

\subsection{MI Fast Fading Channel}\label{sect_sub2fastfading}

In this subsection, we introduce the recently discovered MI fast fading, including its causes, distinctions from other communications, and closed-form expressions for PDF and expectation. We highlight the derivation challenge for a more universal statistical model and propose a universal geometric model. Additionally, we perform simulations based on this geometric model. These geometric models and simulations may aid further study of this derivation challenge.

\subsubsection{MI fast fading modeling} \label{sectsub3_mffm}

The concept of MI fast fading originated in 2016 \cite{Dumphart2016Stochastic}, and was formally defined in 2020\cite{Ma2020Channel}.  Such fading is primarily due to the time-varying polarization gain $J_{\mathrm{SD}}$\cite{Dumphart2016Stochastic}. The fast fading channel is a crucial concept in the field of wireless communication. Specifically, in wireless communication, a fast fading channel significantly varies over the communication time scale. In fast fading channels, the symbol time exceeds the coherence time \cite[Chapter 1]{Tse2005Fundamentals}.
In \cite{Ma2024Fast}, Ma \emph{et al.}  provided a detailed explanation of the different issues between   MICs and other communications such as EMWC,  free-space optical communication (FSOC), and acoustic communication methods, as depicted in Table~\ref{tbl_fastfadingdiff}. Here, we summarize these issues as follows.
\begin{itemize}\label{resppage_centrallimittheorem}
\item {The central limit theorem (CLT) cannot be applied to simplify the statistical channel model in MIC, making it difficult to  derive a universal theoretical model}
\item {The expectation and variance of MI fast fading are unpredictable due to the AVI. Such AVI depends on the driver's mindset. }
\item{When vibrating, antennas may encounter boundaries, causing the PDF of MI fast fading to be discontinuous. }
\end{itemize}

In traditional MIC studies, the MI channel has been assumed to be quasi-static due to reliance on near-field signals without multi-path fading. However, this assumption has been challenged in mobile MIC research \cite{Dumphart2016Stochastic, Ma2020Channel, Ma2024Fast, Zhang2022Performance} since 2016. In 2016, Dumphart \emph{et al.} \cite{Dumphart2016Stochastic} observed the phenomenon that the polarization gain often suffers from coil misalignment due to mobility and deployment in the underwater MIC system. They simply assumed a uniform distribution for  $\theta_{\mathrm{S}}$ and $\theta_{\mathrm{D}}$, and derived the PDF of $\mathcal{J}_{\mathrm{SD}} $. For example, the SISO links, the PDF of  $\mathcal{J}_{\mathrm{SD}}$ is given by 
\begin{equation}\label{eqn_fjuniform}
\begin{aligned}
f_{\mathcal{J}_{\mathrm{SD}}}(x)   = \!\begin{cases}
\tfrac{\mathrm{arsinh}\sqrt{3}  }{\sqrt{3}},      &\text{if }|x| \leq \frac{1}{2}; \\
\tfrac{\mathrm{arsinh}\sqrt{3} - \mathrm{arsinh}\sqrt{4x^2-1} }{\sqrt{3}},     & \text{if }\frac{1}{2}< |x| \leq 1; \\
0,  & \text{if } |x|>1,
\end{cases}
\end{aligned}
\end{equation}
with moments $\mathbb{E}(\mathcal{J}^2_{\mathrm{SD}}) = \frac{1}{6}$ and $\mathbb{E}(\mathcal{J}_{\mathrm{SD}}^4) = \frac{3}{50}$.
This study \cite{Dumphart2016Stochastic} implies that the MI channel may not be quasi-static. The PDF in~\eqref{eqn_fjuniform} is suitable for the underwater environment.
% Since the antenna orientation changes throughout several to tens of seconds in the sea, such fluctuations are not sufficient to cause fast fading,
The antenna orientation changes over time, making the MI channel unstable. Since there are many more weak vibrations than strong ones, the PDF in~\eqref{eqn_fjuniform} cannot represent terrestrial/subterrestrial MI fast fading caused by the antenna vibration.   

\begin{table}[t]  
\caption{Comparison of two PDF expressions for MI fast fading }
%\begin{minipage}{\textwidth} 
%\tiny
\vspace{-0.9em}
\scalebox{0.89}{
\centering
\begin{threeparttable} 
%\begin{tabular}{ p{2.3cm} p{4.5cm}}
\begin{tabular}{m{0.030\textwidth}<{\centering}  |m{0.42\textwidth}<{\centering}| m{0.02\textwidth}<{\centering}}
	\hline 
	\textbf{\emph{Refs.}}  & \textbf{Methods, Pros \& Cons}   & \textbf{Eqs.}   \\
	\hline \hline
	%\raggedleft
	\centering
	%%%%%%%%%%%%%%%%%%%%%%%%%%%%%%%%%%%%%%%%%%%%%%%%%%%%%%%%%%%%%%%%%%%%%%%%%%%%%%%%%%%%%%%%%%%%%%%%%%%%%%%
	\cite{Dumphart2016Stochastic}	&
	\begin{itemize}[leftmargin=*]
		\item[\hllitem]  Simplified underwater model derived from uniform distribution 
		\item[\hlxitem] Incapacity to represent antenna vibration
		\item[\hlxitem] {Lacks universality; underwater-specific}
		\item [\hloitem] {Derivation based on probability theory}
		\vspace{-1.0em}
	\end{itemize}
	%%%%%%%%%%%%%%%%%%%%%%%%%%%%%%%%%%%%%%%%%%%%%%%%%%%%%%%%%%%%%%%%%%%%%%%%%%%%%%%%%%%%%%%%%%%%%%%%%%%%%%%
	& \eqref{eqn_fjuniform} \\ \hline
	\cite{Ma2024Fast, Chen2025Statistical} &
	\begin{itemize}[leftmargin=*]
		\item[\hllitem]  Two antenna-vibration models derived from  BCS distribution and boundary $p(x)$ distribution, respectively
		\item[\hlxitem]  Lacks universality; base station–mobile terminal MIC link only
		\item [\hloitem] Derivation and validated by Monte Carlo simulations
		\vspace{-1.0em}
	\end{itemize}
	& \eqref{eqn_chp2JPdf}\\ \hline
\end{tabular}
\begin{tablenotes}  
	\footnotesize  
	\item[] \hspace{-1.6em} The markers `\hllitem', `\hlxitem' and  `\hloitem' represent pros, cons and methods, respectively. 
\end{tablenotes}  
\end{threeparttable}
\vspace{-2.0em}
}
\label{tbl_pdfexpress}
\end{table}

For the mobile TTE MICs using VLF-LA, previous studies have reported a Shannon's capacity range of 0.5$\sim$10 kbps at a 50-meter MIC distance \cite{zhang2014cooperative, Ma2019Effect, Ma2019Antenna}.  The transmit rate achieved by TTE MIC experiment devices is significantly lower than this capacity, as mentioned in \cite{Ma2024Fast}.  However, the frequency spectrum range of road disturbance input, causing antenna vibrations, is approximately 10 Hz$\sim$1000 Hz, as stated by ISO/TC108 and \cite{Sheng2012Vehicle}. Since $J_{\mathrm{SD}}$ in channel power gain $G_{\mathrm{SD}}$ suffers rapid changes during a symbol time, MI fast fading was formally defined in \cite{Ma2020Channel, Ma2024Fast}. Due to the complexity of the antenna vibration statistical model compared to that in \cite{Dumphart2016Stochastic}, novel mathematical concepts such as Boundary Chi-square (BCS) distribution for VMIC systems, boundary $p(x)$ distribution, and conjugate pseudo-piecewise function were proposed to derive statistical expressions\cite{Ma2024Fast}. 

However, these expressions have only been derived for specific scenarios, 
%such as 2D and 3D models of the downlink in a cellular network. 
Specifically, for the downlink VMI channel with the same height, the PDF of  $J_{\mathrm{SD}}$ can be simplified as
\begin{equation}\label{eqn_chp2JPdf}
\begin{aligned}
f_{J_{\mathrm{SD}}}(x)=\begin{cases}
% \begin{aligned}
	\left[1 - \mathrm{erf}\left(\sqrt{\tfrac{ \varsigma}{2\sigma^2_{\mathrm{D}}}}\right)\right]  \delta_{\rm pu}(0) ,   &\text{if } x = 1-\varsigma;  \\  %J=cos theta=1-avi; avi meets boundary
	\tfrac{\exp\left(-\frac{1 -x}{2\sigma_{\mathrm{D}}^2}\right)}{\sqrt{2\pi\sigma_{\mathrm{D}}^2 (1 - x)}}   ,  &\text{if } 1-\varsigma < x \leq 1; \\   % J=cos(theta)=1-avi
	0     ,      &\text{\rm otherwise}, 
	% \end{aligned}
	\end{cases}
\end{aligned} 
\end{equation}
where the instantaneous AVI  follows the BCS distribution as proposed in \cite{Ma2020Channel}. From~\eqref{eqn_chp2JPdf}, the expectation of MI fast fading can be deduced as
\begin{equation}\label{eqn_chp2JExpect}
\begin{aligned}
\hspace{-0.1em}	\mathbb{E}({J_{\mathrm{SD}}}) 
%	&= 		\int^1_{1-\varsigma}x\cdot f_{J}(x)dx  + \sum\limits_{x\in\{0\}} x \cdot \mathbb{P}[x=0] \\
&=(1   - {\sigma_{\mathrm{D}}^2})\mathrm{erf}\left(\sqrt{\tfrac{\varsigma}{2\sigma^2_{\mathrm{D}}}}\right) + \tfrac{\sqrt{2\varsigma\sigma^2_{\mathrm{D}}}\exp(\tfrac{-\varsigma}{2\sigma^2_{\mathrm{D}}})}{\sqrt{\pi}}.
\end{aligned}
\end{equation}
Here,~\eqref{eqn_chp2JPdf} and~\eqref{eqn_chp2JExpect} represent a 2D statistical model of MI fast fading for a downlink MI link. For more universal models, it is a challenge without the support of the CLT. To address this challenge, we come up with an antenna vibration model, as depicted in Fig. \ref{fig_fastfadingmodel}, where each coil angle is divided into horizontal and vertical components. Since the model is in the geodetic reference system,  the horizontal and vertical components are independent, i.e., all random variables in this model are independent.

Since the theoretical statistical models for the universal MIC scenarios remain unexplored, there are no calculation lines in Row 2 (from top) in Fig. \ref{fig_sec2Ejsd}(b). Here,   we estimate  MI fast fading through Monte Carlo simulations, as shown in Fig.~\ref{fig_sec2Ejsd}(a) and Row 2 in Fig. \ref{fig_sec2Ejsd}(b). This may facilitate future theoretical derivations on a more universal expectation of MI fast fading.

\subsubsection{Impact on channel power gain}\label{sect_subsubiacpg}
\begin{table*}[t]  
\caption{Impact of typical MI fast fading on MIC performance}
%\begin{minipage}{\textwidth} 
%\tiny
\vspace{-0.8em}
\scalebox{0.93}{
\centering
\begin{threeparttable} 
%\begin{tabular}{ p{2.3cm} p{4.5cm}}
\begin{tabular}{m{0.145\textwidth}<{\centering}  |m{0.114\textwidth}<{\centering} |m{0.06\textwidth}<{\centering}| m{0.65\textwidth}<{\centering}}
	\hline 
	\textbf{Performance Metrics} & \textbf{\emph{Refs.}} &\textbf{Vibration input} & \textbf{Effect points (Compare to the MIC without fast fading)}    \\
	\hline \hline
	%\raggedleft
	\centering
	%%%%%%%%%%%%%%%%%%%%%%%%%%%%%%%%%%%%%%%%%%%%%%%%%%%%%%%%%%%%%%%%%%%%%%%%%%%%%%%%%%%%%%%%%%%%%%%%%%%%%%%		
	\multirow{3}{0.08\textwidth}[-10pt]{Average channel power gain}&\cite{Dumphart2016Stochastic} &Uniform
	&\begin{itemize}[leftmargin=*]
		\item[\hlnegmark] Decrease to $\frac{1}{6}$ in a SISO system
		\vspace{-1.0em}
	\end{itemize}\\ \cline{2-4}
	&\cite{Ma2024Fast}& BCS
	&\begin{itemize}[leftmargin=*]
		\item[\hlnegmark] Decrease to about 80\% at an Rx average AVI of $30^\circ$
		\item[\hlposmark] More uniform  around  transmitter
		\item[\hlnegmark]  Severe  variation in  a poor and congested road condition 
		\vspace{-1.0em}
	\end{itemize}\\ \cline{2-4}
	%%%%%%%%%%%%%%%%%%%%%%%%%%%%%%%%%%%%%%%%%%%%%%%%%%%%%%%%%%%%%%%%%%%%%%%%%%%%%%%%%%%%%%%%%%%%%%%%%%%%%%%
	&Fig. \ref{fig_sec2Ejsd}(a)& BCS
	&\begin{itemize}[leftmargin=*]
		\item[\hlposmark] Increase over 5\%   when Tx average AVI angle $=90^\circ$ and Rx average AVI angle $= 90^\circ$
		\vspace{-1.0em}
	\end{itemize}\\ \hline
	%%%%%%%%%%%%%%%%%%%%%%%%%%%%%%%%%%%%%%%%%%%%%%%%%%%%%%%%%%%%%%%%%%%%%%%%%%%%%%%%%%%%%%%%%%%%%
	Outage probability &\cite{Ma2024Fast}, Fig. \ref{fig_chp2performance:pout} &BCS
	& 	\begin{itemize}[leftmargin=*]
		\item[\hlnegmark] Increase over 10\%  when average AVI $\sigma_{\mathrm{D}} > 0.6$, even if Tx Power $P_{\mathrm{S}}\rightarrow\infty$
		\vspace{-1.0em}
	\end{itemize}\\ \hline
	%%%%%%%%%%%%%%%%%%%%%%%%%%%%%%%%%%%%%%%%%%%%%%%%%%%%%%%%%%%%%%%%%%%%%%%%%%%%%%%%%%%%%%%%%%%%%%%%%%%%%%%
	Ergodic capacity & Fig. \ref{fig_chp2performance:er} &BCS
	& \begin{itemize}[leftmargin=*]
		\item[\hlnegmark] A  decrease  from $3.4$ to $2.7$ bps/Hz when $\sigma_{\mathrm{D}} > 0.6$ (i.e., average AVI angle$=37^\circ$)
		\vspace{-1.0em}
	\end{itemize}\\ \hline
	%%%%%%%%%%%%%%%%%%%%%%%%%%%%%%%%%%%%%%%%%%%%%%%%%%%%%%%%%%%%%%%%%%%%%%%%%%%%%%%%%%%%%%%%%%%%%%%%%%%%%%%
	BER &\cite{Chen2025Statistical}, Fig. \ref{fig_chp2performance:ber} &BCS
	& \begin{itemize}[leftmargin=*]
		\item[\hlnegmark] A sharp increase  from $10^{-3}$ to $10^{-1}$ when  $\sigma_{\mathrm{D}} > 0.6$
		\vspace{-1.0em}
	\end{itemize}\\ \hline
	%%%%%%%%%%%%%%%%%%%%%%%%%%%%%%%%%%%%%%%%%%%%%%%%%%%%%%%%%%%%%%%%%%%%%%%%%%%%%%%%%%%%%%%%%%%%%%%%%%%%%%%
\end{tabular}
\begin{tablenotes}  
	\footnotesize  
	\item[] The markers `\hlposmark' and `\hlnegmark'  represent the positive and negative effects, respectively.
\end{tablenotes}  
\end{threeparttable}
}
\vspace{-1.3em}
\label{tbl_fastfadingperformance}
\end{table*}

Fig. \ref{fig_sec2Ejsd} exhibits the averages/expectations of MI fast fading gain. It is observed from Fig. \ref{fig_sec2Ejsd}(a) that most averages are below 0.9, implying a power loss of over 10\% due to vibration.  However, an interesting phenomenon arises: When the average AVIs of Tx and Rx reach a certain threshold, the channel power gain $G_{\mathrm{SD}}$ increases. This is due to the fact that $0$$\leq$$J_{\mathrm{SD}}$$\leq 4$.  Thus,  MI fast fading significantly influences the MIC channel. %Even more excitingly, \eqref{eqn_chp2JExpect} implies that the MI fast fading sometimes is beneficial to the MIC, as such fading makes the signal strength around the Tx coil more uniform.

\subsubsection{Impact on outage probability} \label{sect_subsubiop}
Outage probability is defined as the probability of instantaneous SNR below a threshold $\Upsilon_{\rm th}$\cite{Ma2024Fast}. In earlier MIC studies, as the MIC channel is regarded as a quasi-static channel\cite{Li2019Survey}, the outage probability losses its physical significance. In channels with fast fading, the outage probability is  non-negligible. The closed-form expression of the outage probability $p_{\rm out}^{\mathrm{SD}}$ for a P2P MI fast fading channel is given by\cite{Ma2020Channel}
\begin{equation}\label{eqn_chp2pout}
p_{\rm out}^{\mathrm{SD}}\!=\!\mathbb{P}\left[J_{\mathrm{SD}}\!<\!\tfrac{\Upsilon_{\rm th}}{\tfrac{P_{\mathrm{S}}}{N_o}\mathcal{C}_{\mathrm{SD}}\mathcal{S}_{\mathrm{SD}}\mathcal{E}_{\mathrm{SD}}}\right]
=\int_{-\infty}^{\frac{\Upsilon_{\rm th}}{\frac{P_S}{N_o}\mathcal{C}_{\mathrm{SD}}\mathcal{S}_{\mathrm{SD}}\mathcal{E}_{\mathrm{SD}}}}f_{J_{\mathrm{SD}}}(x)dx,
\end{equation}
In Fig. \ref{fig_chp2performance:pout}, the simulation results  based on~\eqref{eqn_chp2pout} indicate that even when the Tx power approaches infinity, the outage probability is over 0.1 at a relatively small  AVI angle of $37^\circ$ (i.e., $\sigma_{\mathrm{D}}=0.6$), thereby  exerting   a notable adverse effect.

\subsubsection{Impact on ergodic capacity} \label{sect_subsubiec}
Under fast fading, the ergodic capacity EC=$\mathbb{E}[\log_2(1 + {\rm SNR})]$ (bits/s/Hz) is widely used  for  a long term average achievable rate, i.e.,
\begin{equation}\label{eqn_chp2ec0}
\text{EC} \!=\! \int_{0}^{\infty} \left(\log_2(1 + \tfrac{P_{\mathrm{S}}}{N_o} \mathcal{C}_{\mathrm{SD}}\mathcal{S}_{\mathrm{SD}}\mathcal{E}_{\mathrm{SD}} x  )\right) f_{J_{\mathrm{SD}}}(x) \, dx.
\end{equation}
In Fig. \ref{fig_chp2performance:er}, the simulation results based on~\eqref{eqn_chp2ec0} show that  the ergodic capacity declines by nearly one-third  at an  AVI angle of $37^\circ$.

\subsubsection{Impact on BER}\label{sect_subsubiber}
\begin{figure}[t]
\centering
\hspace{-1.8em}
\subfigure[Outage probability]{
\label{fig_chp2performance:pout} %% label for second subfigure
\includegraphics[width=1.7in]{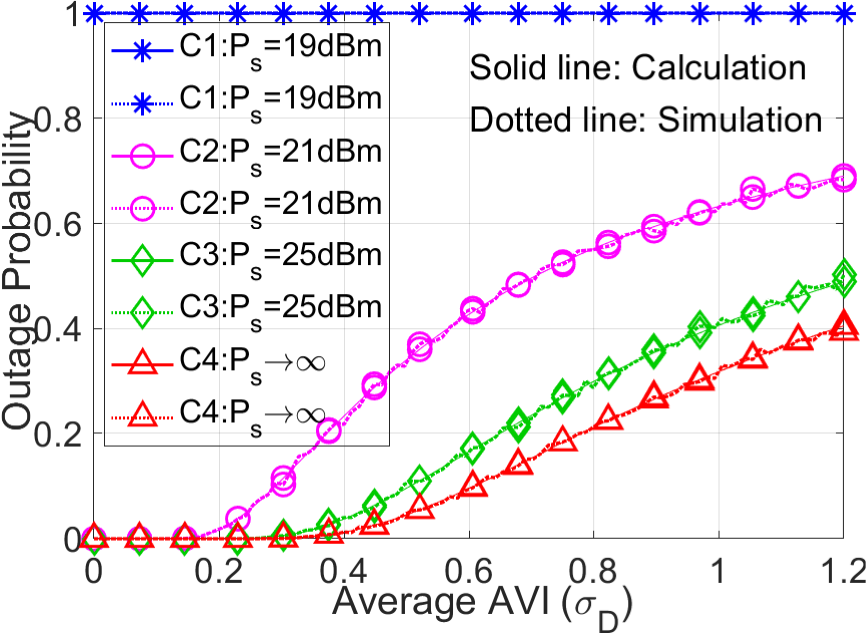}}  
\hspace{-0.8em} 
\\
\hspace{-1.8em}
\subfigure[Ergodic capacity]{
\label{fig_chp2performance:er} %% label for second subfigure
\includegraphics[width=1.7in]{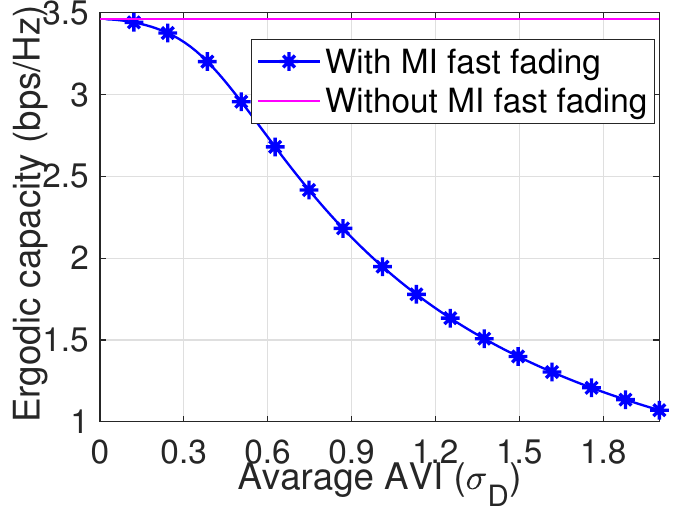}}  
\hspace{-0.7em}
\subfigure[Ergodic BER]{
\label{fig_chp2performance:ber} %% label for second subfigure
\includegraphics[width=1.80in]{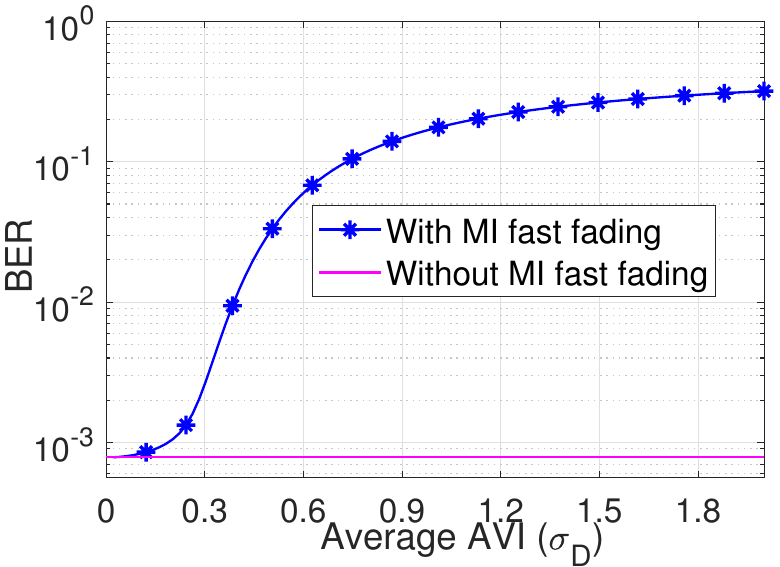}}   
\caption{Impact of MI fast fading on MIC performance: (a) Outage probability v.s. Rx average AVI ($\sigma_{\mathrm{D}}$); (b) Ergodic capacity v.s.  Rx average AVI ($\sigma_{\mathrm{D}}$) under the SNR of 10; and (c) BER v.s.  Rx average AVI ($\sigma_{\mathrm{D}}$) with $\frac{E_b}{N_o}=10$ \cite{Chen2025Statistical}. The solid lines in Figs. \ref{fig_chp2performance:pout}, \ref{fig_chp2performance:er} and \ref{fig_chp2performance:ber} are calculated from~\eqref{eqn_chp2pout},~\eqref{eqn_chp2ec0} and~\eqref{eqn_chp2ber0}, respectively. The dotted lines in Fig. \ref{fig_chp2performance:pout} are obtained through Monte Carlo simulations.}
\label{fig_chp2performance} %% label for entire figure
\vspace{-1.2em}
\end{figure}

Recently, Chen et al. \cite{Chen2025Statistical} established a fundamental performance limit  for the BER of the MI fast fading channel  under FSK modulation, as given by
\begin{equation}\label{eqn_chp2ber0}
%\mathrm{BER} \geq Q\left(\sqrt{\tfrac{E_b}{N_0}\mathbb{E}(J_{\mathrm{SD}})}\right),
\overline{\text{BER}} \!=\! \int_{0}^{\infty} Q \left( \sqrt{ \tfrac{E_b x}{N_o} } \right) f_{J_{\mathrm{SD}}}(x) \, dx \!\geq\! Q\left(\sqrt{\tfrac{E_b}{N_0}\mathbb{E}(J_{\mathrm{SD}})}\right),
\end{equation}
where $Q(\cdot)$ is the  Q-function, and $E_b$ is the energy per bit. As shown in Fig. \ref{fig_chp2performance:er},  there is a 100-fold increase in BER  at an average AVI angle of $37^\circ$. Notably, \eqref{eqn_chp2ber0} indicates that the BER supremum increases with the average MI fast fading gain $\mathbb{E}(J_{\mathrm{SD}})$, indicating that a higher $\mathbb{E}(J_{\mathrm{SD}})$ does not guarantee better  MIC  performance.

Table \ref{tbl_fastfadingperformance} summarizes the effects of MI fast fading on typical MIC performance metrics. Despite the few positive effects, MI fast fading exerts more pronounced negative impacts on MIC performance through the PDF $f_{J_{\mathrm{SD}}}(x | (\sigma_{\mathrm{S}}, \sigma_{\mathrm{D}}))$, which  depends heavily on the AVI inputs $\sigma_{\mathrm{S}}$ and  $\sigma_{\mathrm{D}}$.  In practical MIC systems, such AVI inputs frequently demonstrate substantial temporal variability, especially in the case of congested traffic conditions, leading to significant fluctuations in performance metrics over time. This characteristic markedly distinguishes MI fast fading from EMWC fast fading, as time-dependent performance variability  is  less pronounced in EMWC.

\subsection{Summary and Lessons Learned}\label{sectsub3_sll}

This section discusses the four decomposed gains influencing channel power gain (Table \ref{tbl_gain}) and MI fast fading (Table \ref{tbl_fastfadingdiff}). Table \ref{tbl_gain} suggests optimization directions for each factor, such as improving circuit gain in antenna designs.

{}\label{resppage_discoveriesincludingrandom} %for response letter page reffering

Recent discoveries, including random coil misalignment \cite{Dumphart2016Stochastic} and MI fast fading \cite{Ma2020Channel, Ma2024Fast}, challenge the widely accepted assumption in \cite{Li2019Survey} that MIC channels are quasi-static. However, these findings address only limited aspects of non-static MI channels. In contrast to EMWCs, research on non-static MI channels is still in its early stages. Key issues include: 1) No universal statistical model exists for MI fast fading, analogous to the Rayleigh model for EMWCs, since the existing distribution models of  AVI (e.g., BCS distribution)  are  scenario-specific; 2) The unexplored effects of MI fast fading on existing MIC theorems, such as channel estimation, CMICs, MI MIMO, and MI protocols; and 3) MI fast fading brings negative impacts on the performance of the MIC systems.

\begin{figure}[t]
\centering
\includegraphics[width=2.7in,height=1.3in]{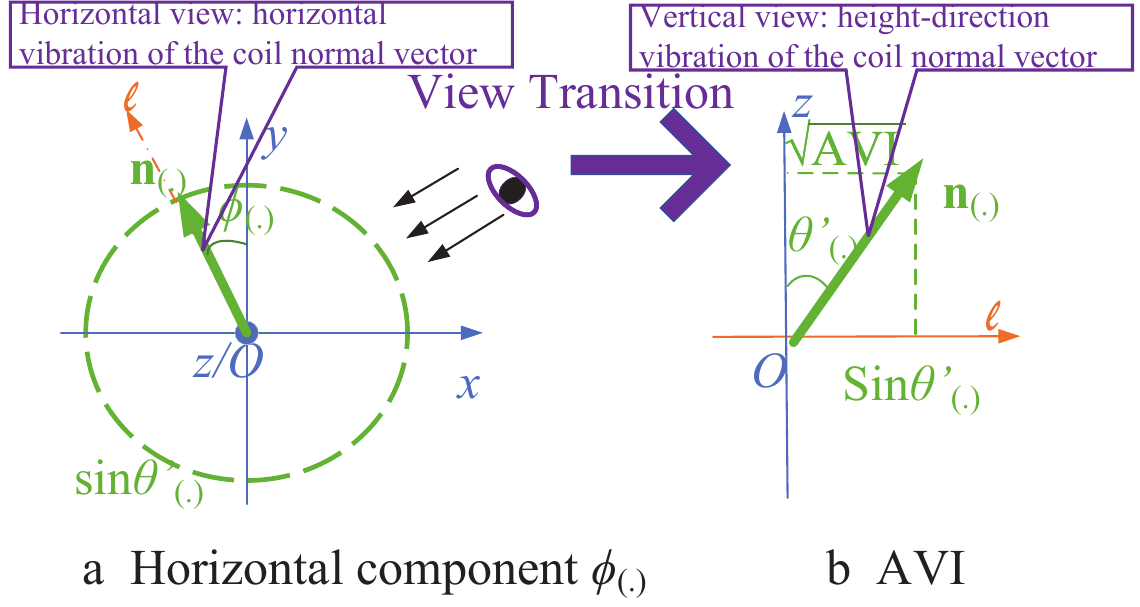}\\
\vspace{-0.5em}
\caption{Advised antenna vibration modeling in a 3-D space with independent random variables;  $\mathbf{n}_{(\cdot)}$, $\phi_{(\cdot)}$ and $\theta'_{(\cdot)}$ denote the normal vector, horizontal and vertical components  of the antenna vibration (angle) of node $(\cdot)$,  respectively.%SU - SVU, PU->PVU P1->P0 %SU - SVU, PU->PVU P1->P0
}\label{fig_fastfadingmodel}
\vspace{-0.01em}
\end{figure}

\begin{figure}[t]
\centering
\includegraphics[width=3.5in,height=2.1in]{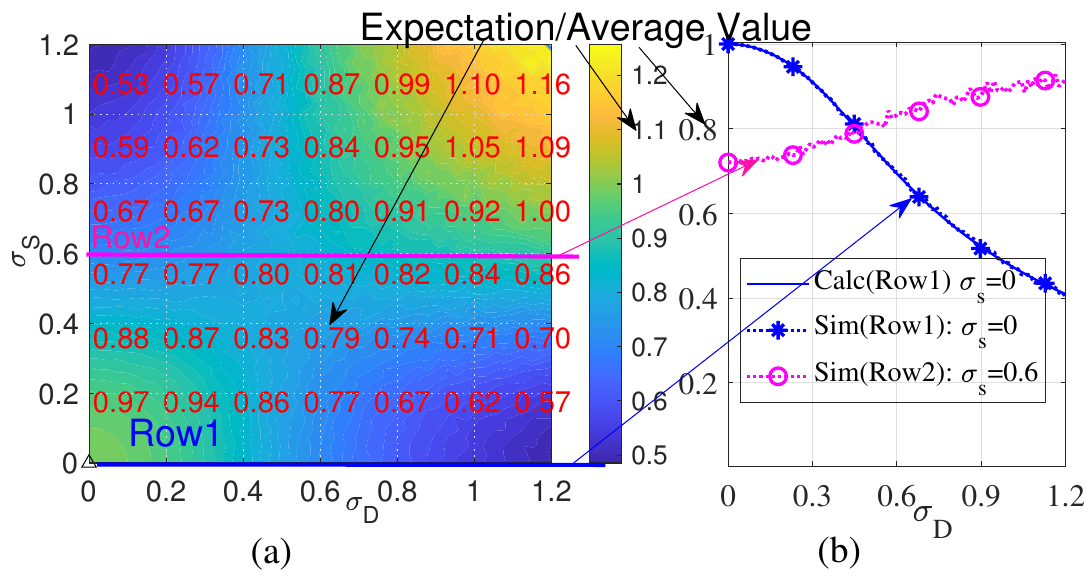}\\
\vspace{-0.5em}
\caption{Expectation of MI fast fading $J_{\mathrm{SD}}$ for the model in Fig. \ref{fig_fastfadingmodel}: (a) Ad-hoc link in 3D space; (b) Two specific links; Here, the AVIs of $\mathrm{S}$ and $\mathrm{D}$ follow  BCS distributions, with a boundary $\varsigma=0.8$ and their respective average AVIs $\sigma^2_{\mathrm{S}}$ and $\sigma^2_{\mathrm{D}}$. Their horizontal components $\phi_{\mathrm{S}}$ and $\phi_{\mathrm{D}}$ follow the uniform distribution within $[0, 2\pi)$. The dotted lines of Row 1 and Row 2 in  Fig. \ref{fig_sec2Ejsd}(b) are obtained through Monte Carlo simulations (sim), and the solid line of Row 1 in  Fig. \ref{fig_sec2Ejsd}(b)  is the calculation (calc) from~\eqref{eqn_chp2JExpect}. Here, the overlap between the calculated curve of Row 1 and its simulated counterpart validates~\eqref{eqn_chp2JExpect}.  The calculation curve for Row 2 cannot be presented  due to the lack of a universal expression.  %SU - SVU, PU->PVU P1->P0 %SU - SVU, PU->PVU P1->P0
}\label{fig_sec2Ejsd}
\vspace{-1.21em}
\end{figure}

The challenges involved include: 1) The antenna vibration’s limited dependent components make it difficult to apply the CLT, a critical method for obtaining fast fading models in EMWCs; and 2) The expectation/variance of MI fast fading remains random due to its strong dependence on antenna vibration velocity. To address these challenges, we introduce a universal geometric model for antenna vibration (Fig. \ref{fig_fastfadingmodel}), which transforms all dependent components into independent ones. This model enables the estimation of MI fast fading gain via Monte Carlo simulation, as shown in Fig.~\ref{fig_sec2Ejsd}.

The practical takeaways or common pitfalls include: 1) No existing  MI fast fading models may  hold when the near-field and weak coupling conditions  are not satisfied; 2) the eddy gain  requires correction in  TTE practice, as TTE media are rarely homogeneous; 3) for VMICs, the ferromagnetic iron in the vehicle body can distort the magnetic field, potentially causing significant deviations in existing MI channel models; and 4) improving  or uniformizing the polarization gain $J_{\mathrm{SD}}$  benefits MICs, whereas doing this for the average MI fast fading gain $\mathbb{E}[J_{\mathrm{SD}}]$  may be detrimental due to increased  outage probability and BER.

%\vspace{6 mm}

\section{Design of Point-to-point TTE MIC }\label{sect_p2p}

%Especially, the channel power gain of an MI link serves as the foundation for analyzing the MIC and its network optimization, including channel capacity, MIC range, 3-dB bandwidth, power control, and localization. It even affects the upper-layer optimization solutions, such as modulation, channel coding, media access control (MAC) protocol,  network connectivity, and routing decisions. 
This section introduces the MI antenna designs, and compares the advantages and disadvantages of four MI antenna types, including the  RPMA design as a recent research hotspot in MI antennas.    Subsequently, we discuss the key performance metrics of  point-to-point (P2P) MI links, with a particular focus on the challenging derivations of the 3-dB MI bandwidth and MIC range for a TTE MI link. Moreover, we  discuss the  upper-layer P2P MI techniques, such as channel estimation, modulation, and FEC.  Recent challenges and approaches about these techniques are briefly summarized in Table \ref{tbl_chp2P2P}.

\begin{table*}[htp] \scriptsize%\footnotesize
\centering
\label{tbl_chp2P2P}
{
\caption{Overview of related research on P2P MIC techniques \\The marker`\hloitem' describe the methods; The markers `\hllitem' and `\hlxitem'  represent pros and cons, respectively.}
\vspace{-0.8em}
\scalebox{0.95}{
\begin{threeparttable}
\begin{tabular}{m{0.08\textwidth}<{\centering}|m{0.065\textwidth}<{\centering}|m{0.37\textwidth}<{\centering}|m{0.37\textwidth}<{\centering}|m{0.04\textwidth}<{\centering}}
\hline
\textbf{Aspects} & \textbf{\emph{Refs.}} & \textbf{Addressed issues \& methods} & \textbf{Remaining issues (\& Proposed approaches)} &\textbf{Priority}$^\dagger$ \\
\hline \hline
%%%%%%%%%%%%%%%%%%%%%%%%%%%%%%%%%%%%%%%%%%%%%%%%%%%%%%%%%%%%%%%%%%%%%%%%%%%%%%%%%%%%%%%%%%%%%%%%%%%%%%%	
Channel modeling & \cite{Sun2010Magnetic, Ma2024Fast, Liu2021Mechanical, Guo2015M2I} &\begin{itemize}[leftmargin=15pt]
\item[\hllitem  \hloitem]  Two typical MI fast fading models
\item[\hllitem  \hloitem]  See Table \ref{tbl_gain} in Section \ref{sect_sub2channel}
\vspace{-1.0em}
\end{itemize} &\begin{itemize}[leftmargin=15pt]
\item[\hlxitem  \hloitem]   More universal MI fast fading modeling
\item[\hlxitem  \hloitem]  See Table \ref{tbl_gain} in Section \ref{sect_sub2channel}
\vspace{-1.0em}
\end{itemize} & \hlprioritem\hlprioritem\hlprioritem \\ \hline
%%%%%%%%%%%%%%%%%%%%%%%%%%%%%%%%%%%%%%%%%%%%%%%%%%%%%%%%%%%%%%%%%%%%%%%%%%%%%%%%%%%%%%%%%%%%%%%%%%%%%%%
Antenna design & \cite{Ma2024Fast, Liu2021Mechanical, Guo2017Practical} & \begin{itemize}[leftmargin=15pt] 
\item[\hllitem  \hloitem]  Coil, orthogonal MIMO, RPMA, M$^2$I
\item[\hllitem  \hloitem] See Table \ref{tbl_antennatype}
\vspace{-1.0em}
\end{itemize}  & \begin{itemize}[leftmargin=15pt]
\item[\hlxitem  \hloitem] Orientation sensitivity; Practical deployment issues
\item[\hlxitem  \hloitem] See Table \ref{tbl_antennatype} \vspace{-1.0em}\end{itemize} & \hlprioritem\hlprioritem \\ \hline		
%%%%%%%%%%%%%%%%%%%%%%%%%%%%%%%%%%%%%%%%%%%%%%%%%%%%%%%%%%%%%%%%%%%%%%%%%%%%%%%%%%%%%%%%%%%%%%%%%%%%%%%	
Capacity & \cite{sun2012capacity, Guo2015M2I} & \begin{itemize}[leftmargin=*] 
\item[\hllitem ]  Expressions of capacity for coil based and M$^2$I MICs
\vspace{-1.0em}
\end{itemize}  & \begin{itemize}[leftmargin=*]
\item[\hlxitem] Expressions of capacity for RPMA-based MICs \vspace{-1.0em}
\end{itemize} & \hlprioritem\hlprioritem \\ \hline
%%%%%%%%%%%%%%%%%%%%%%%%%%%%%%%%%%%%%%%%%%%%%%%%%%%%%%%%%%%%%%%%%%%%%%%%%%%%%%%%%%%%%%%%%%%%%%%%%%%%%%%	
Bandwidth & \cite{Ma2019Effect, Jiang2015Capacity} & \begin{itemize}[leftmargin=*] 
\item[\hllitem ]  Expressions~\eqref{eqn_chp2bwIneq},~\eqref{eqn_chp2BwDipole}, and~\eqref{eqn_chp2JBwCouple} (see Table \ref{tbl_bandexpress})
\vspace{-1.0em}
\end{itemize}  & \begin{itemize}[leftmargin=*]
\item[\hlxitem] Bandwidth of RPMA-based MICs \vspace{-1.0em}
\end{itemize} & \hlprioritem\hlprioritem\hlprioritem \\ \hline	
%%%%%%%%%%%%%%%%%%%%%%%%%%%%%%%%%%%%%%%%%%%%%%%%%%%%%%%%%%%%%%%%%%%%%%%%%%%%%%%%%%%%%%%%%%%%%%%%%%%%%%%
MIC Range & \cite{Zhou2017Maximum} & \begin{itemize}[leftmargin=*] 
\item[\hllitem ]  Expression for short MIC range (see~\eqref{eqn_chp2eqrsdzhou})
\vspace{-1.0em}
\end{itemize}  & \begin{itemize}[leftmargin=*]
\item[\hlxitem]  Expressions for TTE MIC with significant eddy gain \vspace{-1.0em}
\end{itemize} & \hlprioritem\hlprioritem\hlprioritem \\ \hline		
%%%%%%%%%%%%%%%%%%%%%%%%%%%%%%%%%%%%%%%%%%%%%%%%%%%%%%%%%%%%%%%%%%%%%%%%%%%%%%%%%%%%%%%%%%%%%%%%%%%%%%%
\multirow{3}{0.06\textwidth}[-16pt]{\centering Channel estimation} & \cite{Kisseleff2014Transmitter}& \begin{itemize}[leftmargin=*]
\item[\hllitem] Lacking in sufficient CSI
\item[\hloitem] \textbf{Transmitter-side}  estimation  without explicit training sequence
\vspace{-1.0em}
\end{itemize}&\begin{itemize}[leftmargin=*]
\item[\hlxitem] \textbf{Insufficient feedback for the dyadic backscatter MI channel estimation}
\vspace{-1.0em}
\end{itemize} &\hlprioritem\\
\cline{2-5}  
%%%%%%%%%%%%%%%%%%%%%%%%%%%%%%%%%%%%%%%%%%%%%%%%%%%%%%%%%%%%%%%%%%%%%%%%%%%%%%%%%%%%%%%%%%%%%%%%%%%%%%%   
&\cite{Tan2017environment} &\begin{itemize}[leftmargin=*]
\item[\hllitem] Unknown propagation environment (e.g., medium conductivity)
\item[\hloitem] \textbf{Environment-aware method} using Kernel function to learn the positive/negative factors in the environment
\vspace{-1.0em}
\end{itemize}&\begin{itemize}[leftmargin=*]
\item[\hlxitem] Unclear in the case of  VLF TTE MICs and  MIC with fast fading
\vspace{-1.0em}
\end{itemize}&\hlprioritem\\
\cline{2-5}
%%%%%%%%%%%%%%%%%%%%%%%%%%%%%%%%%%%%%%%%%%%%%%%%%%%%%%%%%%%%%%%%%%%%%%%%%%%%%%%%%%%%%%%%%%%%%%%%%%%%%%%
&\cite{Guo2019Inter} & \begin{itemize}[leftmargin=*]
\item[\hllitem] For inter-media MI backscatter channel
\item[\hloitem] \textbf{Joint channel estimation and data transmissions}, and stratified medium model for high penetration
\vspace{-1.0em}
\end{itemize}& \begin{itemize}[leftmargin=*]
\item[\hlxitem]Lacking the analysis  for \textbf{VLF and long-distance} MIC
\vspace{-1.0em}
\end{itemize}&\hlprioritem \\
\cline{1-5}
%%%%%%%%%%%%%%%%%%%%%%%%%%%%%%%%%%%%%%%%%%%%%%%%%%%%%%%%%%%%%%%%%%%%%%%%%%%%%%%%%%%%%%%%%%%%%%%%%%%%%%%
\multirow{2}{0.06\textwidth}[-6pt]{\centering Modulation} & \cite{Kisseleff2014Modulation}&   \begin{itemize}[leftmargin=*]
\item[\hllitem] Basic MI Modulation schemes and filter design %for uncoded transmission of MIC and MI waveguide
\item[\hloitem] Frequency-division multiplexing (FDM) with multiple sub-bands
\vspace{-2.0em}
\end{itemize}&\begin{itemize}[leftmargin=*]
\item[\hlxitem] Incompatibility with RPMA systems
\item[\hlxitem] Shortage similar to FDM in EMWCs
\vspace{-1.0em}
\end{itemize}&\hlprioritem\hlprioritem\\
\cline{2-5}
%%%%%%%%%%%%%%%%%%%%%%%%%%%%%%%%%%%%%%%%%%%%%%%%%%%%%%%%%%%%%%%%%%%%%%%%%%%%%%%%%%%%%%%%%%%%%%%%%%%%%%%
&\cite{Rezaei2020Mechanical, Zhang2023Rotation, Glickstein2020Power, Liu2023Chirp} & \begin{itemize}[leftmargin=*]
\item[\hllitem] Basic modulation  for RPMA-based MICs
\item[\hloitem] Amplitude shift keying, on-off keying, frequency-shift keying, Chirp-rate shift keying
\vspace{-1.0em}
\end{itemize}& \begin{itemize}[leftmargin=*]
\item[\hlxitem] Requiring additional GTSs and energy for overcoming inertia
\vspace{-1.0em}
\end{itemize}&\hlprioritem\hlprioritem\\
\cline{1-5}
%%%%%%%%%%%%%%%%%%%%%%%%%%%%%%%%%%%%%%%%%%%%%%%%%%%%%%%%%%%%%%%%%%%%%%%%%%%%%%%%%%%%%%%%%%%%%%%%%%%%%%%
\multirow{2}{0.06\textwidth}[-6pt]{\centering Channel coding} &\cite{lin2015distributed} &   \begin{itemize}[leftmargin=*]
\item[\hllitem] FEC with higher channel estimation requirement
\item[\hloitem] EMW-based multilevel cyclic BCH coding
\vspace{-1.0em}
\end{itemize}&\begin{itemize}[leftmargin=*]
\item[\hlxitem] Ignoring non-static MI channel cases
\item[\hlxitem] FEC's resistance characteristics to MI fast fading
\vspace{-1.0em}
\end{itemize}&\hlprioritem\\
\cline{2-5}
%%%%%%%%%%%%%%%%%%%%%%%%%%%%%%%%%%%%%%%%%%%%%%%%%%%%%%%%%%%%%%%%%%%%%%%%%%%%%%%%%%%%%%%%%%%%%%%%%%%%%%%
&\cite{Chen2023Novel} & \begin{itemize}[leftmargin=*]
\item[\hllitem] Polar code for variable attenuating MI channel
\item [\hloitem] Bhattacharyya parameter estimation algorithm 
\vspace{-1.0em}
\end{itemize}& \begin{itemize}[leftmargin=*]
\item[\hlxitem] Higher channel estimation requirements with low  capacity
\item[\hlxitem] Ignoring $\mathcal{E}_{\mathrm{SD}}$ and $J_{\mathrm{SD}}$ in their algorithm.
\item[\hlxitem] FEC's resistance characteristics to MI fast fading
\vspace{-1.0em}
\end{itemize}&\hlprioritem\hlprioritem\hlprioritem\\
\hline      
\end{tabular}
\begin{tablenotes}  
\footnotesize  
\item[$\dagger$] Priority: Priority level of remaining issues and  proposed approaches for exploration. Here, low priority (\hlprioritem) indicates that:  1) the remaining issues have been explored in subsequent MIC literature; 2) the existing EMWC schemes  are compatible with MIC for these issues; or 3) exploring  these issues is optional.
\end{tablenotes}  
\end{threeparttable}
} 
}  
\vspace{-1.5em}
\end{table*}

\subsection{Antenna Design}\label{sect_sub2antenna}

In this subsection, we introduce four MI antenna designs, with the RPMA design. Table \ref{tbl_antennatype} summarizes their key features. Detailed studies are presented below.

\begin{table*}[t!]  
\caption{Comparison among different antenna types}
%\begin{minipage}{\textwidth} 
\vspace{-0.8em}
\scalebox{0.93}{
\centering
\begin{tabular}{m{1.9cm}<{\centering}|m{1.0cm}<{\centering}|m{2.1cm}<{\centering}|m{4.5cm}<{\centering}|m{5.2cm}<{\centering}|m{2.0cm}<{\centering}}
\hline
\textbf{Antenna types} & \textbf{Direction} & \textbf{Focuses on} $G_{\mathrm{SD}}$& \textbf{Advantages} & \textbf{Disadvantages}  & \textbf{\emph{Refs.}}\\
% Antenna types &Direction & Focuses on $G_{\mathrm{SD}}$& Advantages  & Disadvantages  & Refs.\\		
\hline \hline
%\raggedleft
\centering
%%%%%%%%%%%%%%%%%%%%%%%%%%%%%%%%%%%%%%%%%%%%%%%%%%%%%%%%%%%%%%%%%%%%%%%%%%%%%%%%%%%%%%%%%%%%%%%%%%%%%%%		
Coil& Rx/Tx & $J_{\mathrm{SD}}$, $\mathcal{C}_{\mathrm{SD}}$ & \begin{enumerate} [leftmargin=3mm]
\item  Low system complexity 
\item  Easy deployment 
\item  Lower costs 
\vspace{-1.0em}
\end{enumerate}
& \begin{enumerate}  [leftmargin=3mm]
\item Strong orientation sensitivity
\item Low bandwidth and energy   efficiency
\item Remarkable  fast fading in mobile MICs
\vspace{-1.0em}
\end{enumerate}
&\cite{Sun2010Magnetic, zhang2014cooperative, Ma2024Fast} \\ %\cline{2-5}
\hline
%%%%%%%%%%%%%%%%%%%%%%%%%%%%%%%%%%%%%%%%%%%%%%%%%%%%%%%%%%%%%%%%%%%%%%%%%%%%%%%%%%%%%%%%%%%%%%%%%%%%%%%		
Orthogonal MIMO coils&Rx/Tx  &$\mathcal{J}_{\mathrm{SD}}$ &\begin{enumerate} [leftmargin=3mm]
\item  Low orientation sensitivity
\item  High receive sensitivity
\item  Higher capacity and bandwidth
\vspace{-1.0em}
\end{enumerate}
&\begin{enumerate} [leftmargin=3mm]
\item  Challenging deployment in vehicle and TTE environments
\item  Complex circuit and protocol designs
\vspace{-1.0em}
\end{enumerate}
&\cite{Zhang2024Cooperative,Zhang2017Connectivity} \\\hline
%%%%%%%%%%%%%%%%%%%%%%%%%%%%%%%%%%%%%%%%%%%%%%%%%%%%%%%%%%%%%%%%%%%%%%%%%%%%%%%%%%%%%%%%%%%%%%%%%%%%%%%	    
RPMA&Tx & $\mathcal{C}_{\mathrm{SD}}$ &\begin{enumerate} [leftmargin=3mm]
\item  Low energy consumption under VLF
\item  Smaller antenna size for TTE and mobile MICs
\item  Lower cross-talk effects
\vspace{-1.0em}
\end{enumerate}
&\begin{enumerate} [leftmargin=3mm]
\item   Only for  Tx antennas
\item   High maintenance costs 
\item Limited $\mathbf{m}_{\mathrm{S}}$  by permanent magnets
\item  Longer GTS requirement due to inertia
\vspace{-1.0em}
\end{enumerate} & \cite{Liu2021Mechanical,Golkowski2018Novel,Rezaei2020Mechanical,Sun2022Research,Glickstein2020Power,Liu2023Chirp, Slevin2022Broadband,Zhang2023Rotation, Li2024Rotating,Zhu2024Piezo, Cheng2024Bionic, Wang2024Radiation, Cui2024Optimization}\\
\hline
%%%%%%%%%%%%%%%%%%%%%%%%%%%%%%%%%%%%%%%%%%%%%%%%%%%%%%%%%%%%%%%%%%%%%%%%%%%%%%%%%%%%%%%%%%%%%%%%%%%%%%%		
M$^2$I & Rx/Tx & $\mathcal{S}_{\mathrm{SD}}$ & \begin{enumerate} [leftmargin=3mm]
\item  Extremely high receive sensitivity
\item  Higher capacity and bandwidth
\vspace{-1.0em}
\end{enumerate}
& \begin{enumerate} [leftmargin=3mm]
\item High costs
\item Larger radius (e.g., $\frac{50{\rm mm}}{15{\rm mm}}$ times) than coil
\vspace{-2.0em}
\end{enumerate}
& \cite{Guo2015M2I,guo2016m2i,Guo2017Practical, Sharma2017Metamaterial, Li2019Antenna, Li2022Optimal}\\
\hline		
\end{tabular}
}
%\end{minipage}
\label{tbl_antennatype}
\vspace{-1.3em}
\end{table*}

\subsubsection{Coil Tx/Rx Antenna} \label{sectsubsub4_cta}
\begin{figure}[t]
\centering
\includegraphics[width=1.8in]{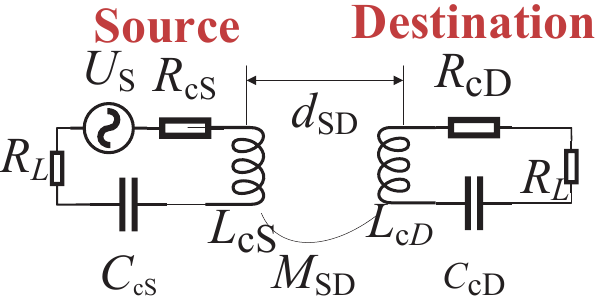}\\
\vspace{-0.5em}
\caption{ Equivalent circuit model for the coils-based link S$\rightarrow$D, where $U_{\mathrm{S}}$ denotes the instantaneous voltage in the Tx antenna, and $d_{\mathrm{SD}}$ denotes the distance between source $\mathrm{S}$ and destination $\mathrm{D}$. Additionally, $R_{\mathrm{cS}}$,  $R_{\mathrm{cD}}$, $L_{\mathrm{cS}}$,  $L_{\mathrm{cD}}$, $R_L$, $R_{\mathrm{cS}}$ and $R_{\mathrm{cD}}$ are described in Table \ref{tbl_symbol}, respectively.   %SU - SVU, PU->PVU P1->P0 %SU - SVU, PU->PVU P1->P0
}\label{fig_circuit}
\vspace{-0.01em}
\end{figure}

The coil is the most widely used antenna in the field of MICs \cite{Li2019Survey, Kisseleff2018Survey} owing to its high receive sensitivity, ease of implementation, and deployment. It is very suitable for TTE applications with limited free space. The utilization of coils dates back to research on NFCs, a short-range MIC that commonly employs a 13.56 MHz carrier frequency according to ISO/IEC 18092 standards.   The MIC link is established through mutually inductive coupling between the Tx coil and the Rx coil (see Fig.~\ref{fig_circuit}). However, unlike standard NFC,  MICs for UG-WSNs and UW-WSNs exhibit weak coupling due to longer communication distances. The Tx coil is typically modeled as a magnetic dipole, and the bandwidth of such MIC is much smaller than the standard NFC, especially for the TTE MIC applications. Due to this weak coupling, the Tx coil with current $I_{\mathrm{S}}$ generates the magnetic moment with a norm $|\mathbf{m}_{\mathrm{S}}| =   \pi^2 a_{\mathrm{cS}}^2 a_{\mathrm{cD}}^2 N_{\mathrm{S}} N_{\mathrm{D}} I_{\mathrm{S}}$ and current loss compensating coefficient $\aleph_{\mathrm{SD}} = \left| \tfrac{(2\pi f)^2 R_L}{Z_{\mathrm{S}}Z^2_{\mathrm{D}}} \right|$. Such $\aleph_{\mathrm{SD}}$ determines predominatly the frequency bandwidth of an MI link.
\begin{comment}
\begin{subequations}
\begin{align}
|\mathbf{m}_{\mathrm{S}}| =   \pi^2 a_{\mathrm{S}}^2 a_{\mathrm{D}}^2 N_{\mathrm{S}} N_{\mathrm{D}} I_{\mathrm{S}}, \ \ \ \ \
\aleph_{\mathrm{SD}} = \left| \tfrac{(2\pi f)^2 R_L}{Z_{\mathrm{S}}Z^2_{\mathrm{D}}} \right|, \notag
\end{align}
\end{subequations}
\end{comment}
The overall circuit impedances of the Tx and Rx coils are denoted by   $Z_{\mathrm{S}}$$=$$j 2\pi f {L_{\mathrm{cS}}}$$+$$\frac{1}{j2\pi f C_{\mathrm{cS}}}$$+$$R_{\mathrm{cS}}$$+$$R_L$ and  $Z_{\mathrm{D}}$$=$$j 2\pi f {L_{\mathrm{cD}}}$$+$$\frac{1}{j2\pi f C_{\mathrm{cD}}}$$+$$R_{\mathrm{cD}}$$+$$R_L$, respectively. The channel power gain $G_{\mathrm{SD}}$ of coils-based link S$\rightarrow$D has the specific circuit gain $\mathcal{C}_{\mathrm{SD}}$ which can be expressed by
\begin{equation}\label{eqn_chp2FrakCsd}
\begin{aligned}
\mathcal{C}_{\mathrm{SD}} &=  \frac{|\mathbf{m}_{\mathrm{S}}|^2\aleph_{\mathrm{SD}}}{|I_{\mathrm{S}}|^2}
&= (\pi a_{\mathrm{cS}}^2 a_{\mathrm{cD}}^2 N_{\mathrm{S}} N_{\mathrm{D}})^2 \left| \tfrac{(2\pi f)^2 R_L}{Z_{\mathrm{S}}Z^2_{\mathrm{D}}} \right|.
\end{aligned}
\end{equation}
From~\eqref{eqn_chp2GSDNear} and~\eqref{eqn_chp2FrakCsd}, the circuit gain  $\mathcal{C}_{\mathrm{SD}}$ in $G_{\mathrm{SD}}$ increases with frequency $f$. Meanwhile, the eddy gain $\mathcal{E}_{\mathrm{SD}}$  in $G_{\mathrm{SD}}$ also increases with $f$. When $f$ is sufficiently large,  the condition $\sigma_\mathrm{u} \gg 2\pi f \epsilon_\mathrm{u}$ no longer holds. The effect of circuit gain $\mathcal{C}_{\mathrm{SD}}(f)$ becomes  significant. Consequently, coils-based MIC systems with a higher frequency achieve higher performance compared to RPMA. Also, the expressions for $Z_{\mathrm{S}}$ and $Z_{\mathrm{D}}$ indicate that the coil resonance makes the MI channel  frequency-selective. This results in a narrow 3-dB bandwidth and a negative impact on the design of upper-layer protocols.

\begin{figure}[t]
\centering
\subfigure[]{
\label{fig_sec2antenna:mimo} %% label for second subfigure
\includegraphics[width=0.7in]{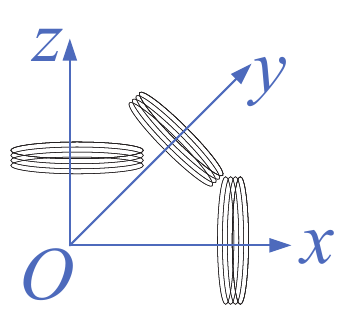}}  
\subfigure[]{
\label{fig_sec2antenna:rpma} %% label for second subfigure
\includegraphics[width=1.70in]{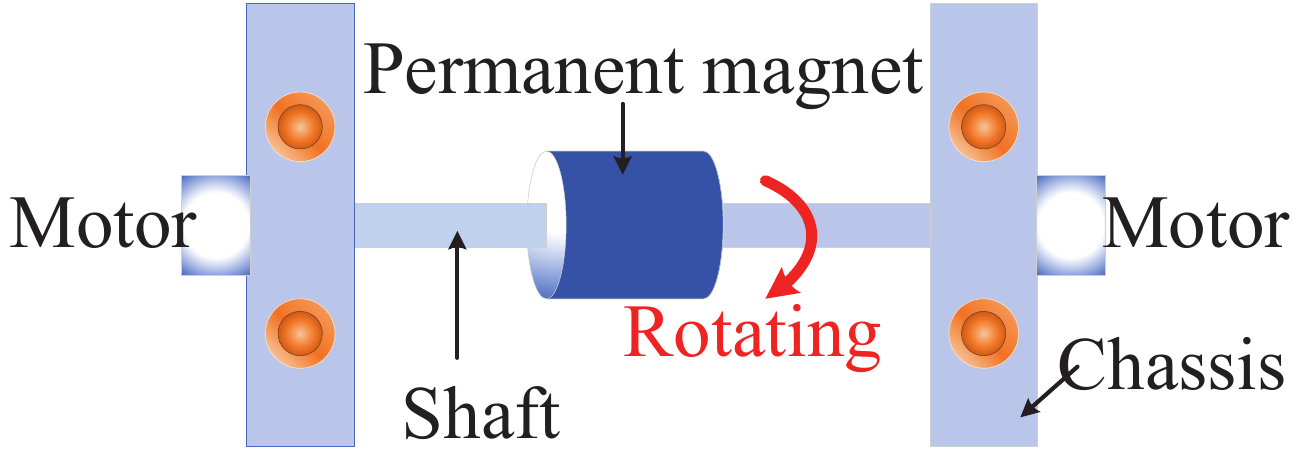}}  
\subfigure[]{
\label{fig_sec2antenna:m2i} %% label for second subfigure
\includegraphics[width=0.70in]{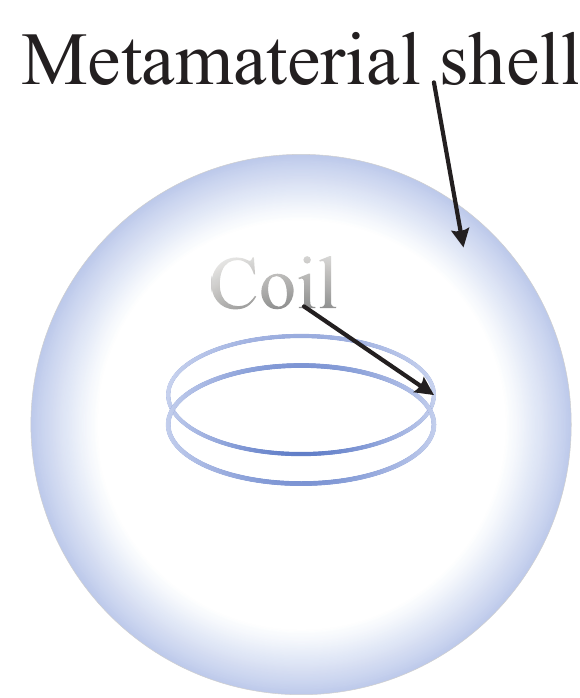}} 
\caption{Non-conventional  MI antenna types: (a) Orthogonal MIMO coils~\cite{Zhang2017Connectivity}; (b) A typical RPMA~\cite{Rezaei2020Mechanical}; (c) M$^2$I antenna~\cite{Guo2015M2I}.  }
\label{fig_sec2antenna} %% label for entire figure
\vspace{-0.9em}
\end{figure}

\subsubsection{Orthogonal MIMO coils} \label{sectsubsub4_omc}

SISO coil-based MIC systems exhibit antenna orientation sensitivity due to polarization gain $J_{\mathrm{SD}}$.  This can cause remarkable fast fading in mobile MIC applications. To address this, Lin \emph{et al.}  developed the orthogonal MIMO coil antennas \cite{lin2015distributed}, consisting of three orthogonal coils, as shown in Fig. \ref{fig_sec2antenna:mimo}. This antenna reduces the orientation sensitivity and enhances the MIC channel capacity through spatial diversity. Their directional pattern showed that the minimal mutual inductance increased from 0 to $\frac{1}{3}$ of maximal mutual inductance. However, for the TTE applications using VLF-LA~\cite{zhang2014cooperative, Ma2019Effect},  deployment of such an antenna faces challenges due to limited free space and crosstalk among coils.

\subsubsection {RPMA} \label{sectsubsub4_rpma}

The RPMA is a mechanical antenna  (see Fig. \ref{fig_sec2antenna:rpma}) that can generate a magnetic field of 960 Hz \cite{Golkowski2018Novel} and achieves a transmission bit rate of over 12.5 bits/s\cite{Sun2022Research, Zhang2023Rotation}. Bickford \emph{et al.} have been designing RPMA since 2017\cite{Bickford2017Low}.  Recently,  the RPMA was used for cross-medium communication in  \cite{Zhu2024Piezo, Cheng2024Bionic}, where the research \cite{Zhu2024Piezo} focused on omnidirectional communication and the research \cite{Cheng2024Bionic} applied the bionic methodology.  

Apart from the channel modeling, researchers also studied modulation for RPMA-based systems such as amplitude shift keying (ASK)~\cite{Rezaei2020Mechanical}, on-off keying (OOK)~\cite{Zhang2023Rotation}, frequency-shift keying (FSK)~\cite{Glickstein2020Power} and Chirp-rate shift keying (CSK)~\cite{Liu2023Chirp}. When using a single magnet as an RPMA,  circuit gain $\mathcal{C}_{\mathrm{SD}}$ changes, while the other gains in~\eqref{eqn_chp2GSDNear} remain unchanged. According to~\cite{Liu2021Mechanical}, the moment generated by a single magnet can be calculated by $|\mathbf{m}_{\mathrm{S}}|=\frac{ V_\mathrm{m}}{I_{\mathrm{S}} \mu_{\rm 0}}$, where $V_\mathrm{m}$ denotes the volume of the RPMA. The factor $\frac{1}{I_{\mathrm{S}}}$  represents the remanence (denoted by $\mathcal{B}_{\rm rm}$) of RPMA. Thus, the circuit gain can be calculated by
\begin{equation}\label{eqn_chp2frakCRPMA}
\begin{aligned}
\mathcal{C}_{\mathrm{SD}} &=  \tfrac{|\mathbf{m}_{\mathrm{S}}|^2\aleph_{\mathrm{SD}}}{16\pi^2|I_{\mathrm{S}}|^2}
&= \left(\tfrac{\mathcal{B}_{\rm rm} V_\mathrm{m}}{4\pi \mu_{\rm 0}}\right)^2 \aleph_{\mathrm{S}}(f)\aleph_{\mathrm{D}}(f),
\end{aligned}
\end{equation}
where $\mu_{\rm 0}$ is the vacuum permeability; $\frac{1}{\aleph_{\mathrm{S}}(f)}$ and  $\frac{1}{\aleph_{\mathrm{D}}(f)}$ are the friction loss of RPMA and the circuit loss of the receiving device, respectively.  Here, $\frac{1}{\aleph_{\mathrm{S}}(f)}$ increases as the frequency $f$ increases. In contrast,   $\frac{1}{\aleph_{\mathrm{D}}(f)}$ decreases as the frequency increases since the Rx antenna is a general coil-based MIC device. According to \cite{Chen2025Statistical}, for a link with a Tx RPMA (as shown in Fig. \ref{fig_sec2antenna:rpma}) and an Rx coil, the circuit loss can be expressed as  $\frac{1}{\aleph_{\mathrm{D}}(f)}=\frac{2Z_{\mathrm{D}}}{\pi^2 f a_{\mathrm{cD}}}$. The experiment in the sea near Sanya city~\cite{Liu2021Mechanical} indicated that the RPMA Rx node received a magnetic field of 25 nT at a distance of 10 m.

Mechanical constraints, such as inertia and energy loss, may limit frequency agility or energy efficiency. For the RPMA  shown in Fig. \ref{fig_sec2antenna:rpma}, the instantaneous input power is $P^{\rm RPMA}_{\mathrm{S}}$$=$$(\tau_{\rm fr}$$+$$\tau_{\rm nr})2\pi f$$/$$\eta$\cite{Liu2021Mechanical},  where $\eta$ denotes the energy conversion efficiency, and $\tau_{\rm fr}$ denotes the friction loss.  The inertia constraint  $\tau_{\rm nr}$$=$$\frac{I_{\rm nr}2\pi f}{dt_{\mathrm{S}}}$ depends linearly on the frequency $f$, indicating a higher energy proportion of $\tau_{\rm nr}$  in high-frequency systems. The moment of inertia $I_{\rm nr}$$\simeq$ $3.75 \times 10^{-4}\ \mathrm{kg{\cdot}m^2}$ leads to a rated angular acceleration of approximately $539\ \mathrm{Hz/s}$\cite{Liu2021Mechanical}.

Some issues were not considered in previous studies. By comparing~\eqref{eqn_chp2frakCRPMA} with~\eqref{eqn_chp2FrakCsd}, it can be found that RPMA-based system can achieve higher performance under VLF conditions and lower performance under high-frequency conditions. This phenomenon can be attributed to the increase in friction and inertia associated with higher rotating speeds.   Additionally, the lifespan of an RPMA-based MIC device may be shorter than that of a coil-based one due to friction.  Particularly,  the RPMA-based MIC requires an additional guaranteed time slot (GTS) to overcome inertia, resulting in additional data rate loss and a variable data rate under the same packet length.  Most importantly, permanent magnets on Earth are unable to generate ultra-strong magnetic moments sufficient to support MIC over ultra-long distances (e.g., 1,500 m). Thus, the massive RPMA MIMO  can be promising due to negligible crosstalk among RPMAs.

\subsubsection{M$^2$I  antenna} \label{sectsubsub4_m2i}
\begin{comment}
\begin{figure}[t]
\centering
\includegraphics[width=0.7in]{sec2m2i.eps}\\
\vspace{-0.5em}
\caption{ An M$^2$I antenna.   %SU - SVU, PU->PVU P1->P0 %SU - SVU, PU->PVU P1->P0
}\label{fig_sec2m2i}
\vspace{-0.01em}
\end{figure}
\end{comment}

From~\eqref{eqn_chp2GSDNear}, we observe that space gain $\mathcal{S}_{\mathrm{SD}}\propto \frac{1}{d^6_{\mathrm{SD}}}$, indicating that the magnetic field attenuates very fast along the signal path. To improve the magnetic fields around the MI receivers, Guo \emph{et al.} \cite{Guo2015M2I} proposed the M$^2$I  enhanced technique by enclosing MI antennas with metamaterial shells (see Fig. \ref{fig_sec2antenna:m2i}).  Unlike natural inductors, metamaterial is an artificial material with arbitrary permeability and permittivity.  It compensates antenna-generated reactive power and converts the reactive power into real power that extends the MIC range.  They also proposed an easily implementable M$^2$I antenna with a shell  made of many small coils for practical use in \cite{ Guo2017Practical,guo2016m2i}. Subsequently, Li \emph{et al.} \cite{Li2022Optimal} designed an active and reconfigurable M$^2$I antenna under given meta-sphere size and power constraints.  Although the simulations in~\cite{Guo2015M2I} indicated that the MIC distance increased from 20 m to 60 m under a capacity of 1 bit/s,   deploying M$^2$I  may be challenging due to the large space occupied by the metamaterial shell. For example, the meta-sphere shell in \cite{Guo2015M2I} has a radius six times that of a coil, i.e., a volume about 8,649 times larger. This creates a critical trade-off between performance and deployability. In the scenarios similar to deep mining, this volume penalty is justifiable. However, in narrow boreholes or mobile applications, conventional coils or RPMAs remain preferable.

\subsection{Channel Capacity and Bandwidth}\label{subsect_capa}

As early as 2013, Sun \emph{et al.} \cite{Sun2013Increasing} suggested that the capacity $C_{\mathrm{SD}}$  for plat MIC channel can be derived from Shannon's equation $\mathfrak{C}_{\mathrm{SD}} = B_{\rm w}\log_2(1 + \frac{P_{\mathrm{S}f}}{N_{of}}G_{\mathrm{SD}})$. For P2P MIC, the MI bandwidth is important to enhance channel capacity, especially when the Tx  and noise  power spectral densities (PSDs) are predetermined and cannot be altered, respectively. 

The bandwidth widely refers to the 3-dB bandwidth, which is defined as the frequency range where signal power halves.
There are primarily two methods (see Table \ref{tbl_bandexpress}) that can evaluate the 3-dB bandwidth of the MI links.  
When the noise PSD is fixed, the bandwidth can be obtained by  solving the inequality with respect to $f$, i.e., 
\begin{equation}\label{eqn_chp2bwIneq}
\left|\tfrac{G_{\mathrm{SD}}(f) P_{\mathrm{S}f}(f)}{N_{{\rm o}f}}\right| \geq \tfrac{1}{2}\left|\tfrac{G_{\mathrm{SD}}(f_0) P_{\mathrm{S}f}(f_0)}{N_{{\rm o}f}}\right|,
\end{equation} 
which is referred to as the inequality calculation.
\begin{comment}
\begin{equation}\label{chp2BwIneq}
\begin{aligned}
\left|\tfrac{G_{\mathrm{SD}}(f) P_{\mathrm{S}}(f)}{N_{\rm o}}\right| \geq \tfrac{1}{2}\left|\tfrac{G_{\mathrm{SD}}(f_0) P_{\mathrm{S}}(f_0)}{N_{\rm o}}\right|
\end{aligned}
\end{equation}\end{comment}
For the TTE MICs, the Tx coil is treated as a magnetic dipole.  Assuming identical Tx and Rx coils, the study\cite{Ma2019Effect} obtained the solution for the bandwidth  of a P2P coil-based MIC link $B^{\rm dipole}_{ \mathrm{w,SD}}=B_{\rm w}\left(\frac{1}{8}(R_{\mathrm{cD}}+R_L)^3\right)$, where $B_{\rm w} (\cdot)$ is
\begin{equation}\label{eqn_chp2BwDipole}
\begin{aligned}
B_{\rm w}(\mathcal{Z}_C) &\!\simeq\! \sqrt{\varpi_{\mathrm{w}}(\mathcal{Z}_C)\!+\!\varrho_{\mathrm{w}}(\mathcal{Z}_C) }-\sqrt{\varpi_{\mathrm{w}}(\mathcal{Z}_C)\!-\!\varrho_{\mathrm{w}}(\mathcal{Z}_C)},  \\
\varpi_{\rm w} (\mathcal{Z}_C) &= f_0^2 \!+\! 2 \pi ^2 C_{\mathrm{cD}}^2 f_0^4 \left[\mathcal{Z}_C^{-\frac{2}{3}}\!-\!(R_{\mathrm{cD}} + R_L)^2\right] , \\
\varrho_{\rm w}(\mathcal{Z}_C) &\!=\!  \sqrt{ \left\{ f_0^2\!+\!2 \pi ^2 C_{\mathrm{cD}}^2 f_0^4 \left[\mathcal{Z}_C^{-\frac{2}{3}}-(R_{\mathrm{cD}} \!+\! R_L)^2\right]\right\}^2\!-\! f_0^4}  .
\end{aligned}
\end{equation}
However,~\eqref{eqn_chp2BwDipole} involved  the assumptions of  VLF-LA ($\frac{1}{2\pi f_0^2 C_{\mathrm{cD}}} \gg 2\pi M_{\mathrm{SD}}$) and $a_{\mathrm{cS}}  =  a_{\mathrm{cD}}$. If these assumptions are not met,  the bandwidth $B^{\rm dipole}_{ \mathrm{w,SD}}$ can be obtained through numerical methods. This inequality calculation can be extended to the non-coil-based MIC,  but it may not yield closed-form analytical expressions.

Also,  studies have considered MICs with shorter communication ranges, particularly in BAN applications.  In these scenarios, the Tx coil is not treated as a magnetic dipole~\cite{Azad2012Link,Jiang2015Capacity}.  The bandwidth and channel capacity become more complex. To put it simply, since the loosely coupled coils generate the boundary conditions based on the coupling coefficient $k_{c}=\frac{M_{\mathrm{SD}}}{\sqrt{L_{\mathrm{cS}}L_{\mathrm{cD}}}}$,  both the bandwidth and channel capacity are piecewise functions of $Q_{\mathrm{S}}$ and $Q_{\mathrm{D}}$. Here, $Q_{\mathrm{S}}=\frac{2\pi f_0 L_{\mathrm{cS}}}{R_{\mathrm{cS}}+R_L}$ and $Q_{\mathrm{D}}=\frac{2\pi f_0 L_{\mathrm{cD}}}{R_{\mathrm{cD}}+R_L}$ are the loop quality factors of Tx and Rx devices, respectively. In \cite{Jiang2015Capacity}, the bandwidth of this model  is estimated as
\begin{equation}\label{eqn_chp2JBwCouple}
\begin{aligned}
&B^{\rm coupling}_{\mathrm{w, SD}}=\begin{cases}
\begin{aligned}
& \tfrac{f_0}{Q_{\mathrm{S}}},   &\text{if }Q_{\mathrm{S}} > Q_{\mathrm{D}} ;          \\
&\tfrac{f_0}{Q_{\mathrm{D}}},    &\text{if }Q_{\mathrm{S}} < Q_{\mathrm{D}}.  
\end{aligned}
\end{cases}
\end{aligned} 
\end{equation}
Obviously, ~\eqref{eqn_chp2JBwCouple} cannot be used to obtain the non-coils-based MIC, such as RPMA-based MI links.

\begin{table}[t]  
\caption{Comparison of two bandwidth calculations  }
%\begin{minipage}{\textwidth} 
%\tiny
\scalebox{0.93}{
\centering
\begin{threeparttable} 
%\begin{tabular}{ p{2.3cm} p{4.5cm}}
\begin{tabular}{m{0.7cm}<{\centering} |m{2.3cm}<{\centering} |m{3.4cm}<{\centering}| m{1.1cm}<{\centering}}
\hline 
\textbf{\emph{Refs.}}  & \textbf{Advantage}   &
\textbf{Limitations}  &
\textbf{Typical Eqs.}   \\
\hline \hline
%\raggedleft
\centering
%%%%%%%%%%%%%%%%%%%%%%%%%%%%%%%%%%%%%%%%%%%%%%%%%%%%%%%%%%%%%%%%%%%%%%%%%%%%%%%%%%%%%%%%%%%%%%%%%%%%%%%			
\cite{Ma2019Effect}	&Possibility for non-coils MIC &Closed-form solution with homogeneous VLF-LA limit& \eqref{eqn_chp2bwIneq},\eqref{eqn_chp2BwDipole} \\\hline
%%%%%%%%%%%%%%%%%%%%%%%%%%%%%%%%%%%%%%%%%%%%%%%%%%%%%%%%%%%%%%%%%%%%%%%%%%%%%%%%%%%%%%%%%%%%%%%%%%%%%%%	
~\cite{Azad2012Link,Jiang2015Capacity} &Not limited by homogeneous coils &Only for coil-based MIC&\eqref{eqn_chp2JBwCouple}\\ \hline	
%%%%%%%%%%%%%%%%%%%%%%%%%%%%%%%%%%%%%%%%%%%%%%%%%%%%%%%%%%%%%%%%%%%%%%%%%%%%%%%%%%%%%%%%%%%%%%%%%%%%%%%	 
\end{tabular}
\end{threeparttable}
}
\label{tbl_bandexpress}
\vspace{-1.0em}
\end{table}

As $f_0$$=$$\frac{1}{2\pi\sqrt{L_{\mathrm{cS}}C_{\mathrm{cS}}}}$ holds, two calculation methods~\eqref{eqn_chp2BwDipole} and~\eqref{eqn_chp2JBwCouple} imply that the P2P MIC bandwidth under VLF is roughly independent of the resonance frequency $f_0$. Under the simulation parameters of Table \ref{tbl_sim}, the bandwidth is around 450 Hz when $ 1 \ \mathrm{ kHz} \leq f_0 \leq  1 \ \mathrm{ MHz}$, which is evaluated based on~\eqref{eqn_chp2bwIneq}. 

{}\label{resppage_wefocuson} %for response letter refering

\begin{figure}[t!]
\centering{\includegraphics[width=65mm]{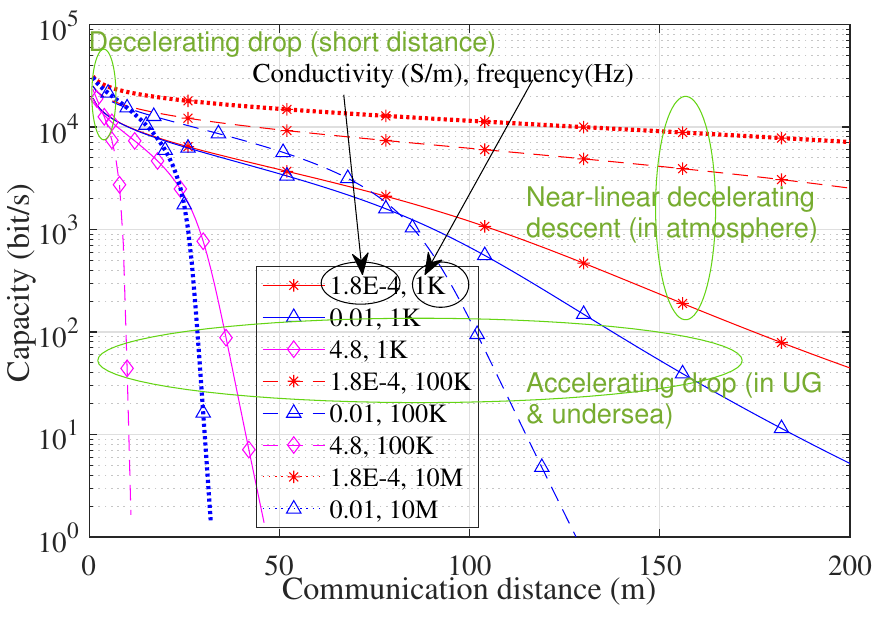}}
\caption{MATLAB simulation of MIC channel capacity v.s.  distance under different environments.  All simulation parameters  are listed in Table \ref{tbl_sim},  except for the frequency $f$ (shown in the legend) and the MIC distance $d_{\mathrm{SD}}$ (as the simulated independent variable). This simulation is based on $\mathfrak{C}_{\mathrm{SD}} = B_{\rm w}\log_2\left(1 + \frac{P_{\mathrm{S}f}}{N_{of}}G_{\mathrm{SD}}\right)$; see \eqref{eqn_chp2GSDNear} and~\eqref{eqn_chp2BwDipole}. 
}\label{fig_chp2capadsd}
\vspace{-0.0em}
\end{figure}

Subsequently, we focus on the MI channel capacity. Fig. \ref{fig_chp2capadsd} shows that higher frequencies boost MIC capacity in shorter ranges, and also enhance capacity over long distances in air media. In these cases, the trend of capacity decreasing with frequency is determined by the space gain $\mathcal{S}_{\mathrm{SD}}$, while the eddy gain  $\mathcal{E}_{\mathrm{SD}}$ is negligible. However, for TTE and undersea MICs, the capacity drops sharply with increasing frequency and distance. This is due to the fact that the eddy gain $\mathcal{E}_{\mathrm{SD}}$ becomes the key determinant of the channel power gain $G_{\mathrm{SD}}$.
The frequency characteristics of MIC bandwidth and capacity motivate researchers and engineers to adopt VLF-LA methods in TTE MIC networks, evident from~\cite{zhang2014cooperative, Zhang2017Connectivity, Ma2019Antenna, Ma2019Effect, Ma2024Fast, Ma2020Channel, VitalCanaryComm}. Even at the VLF of 1 kHz, the MIC capacity in low-conductivity  media (0.01 S/m, e.g., soil) is over 320-fold greater than  in high-conductivity  media (4.8 S/m, e.g., seawater) at 45 m MIC distance. Fig. \ref{fig_chp2capadsd} also shows that MIC channel capacity drops below 10 kbits/s beyond a distance of 60 m.

\subsection{Communication Range }\label{sect_submicrange}
%Here, we derive the closed-form expression of the MIC communication range for coil-based MICs and the optimal frequency to maximize the point-to-point MI link communication range in TTE scenarios.

The communication range refers to the distance between the transmitter and receiver for correct data reception. It is the key performance metric for TTE communications. 
For the MICs, many studies focus on this metric. For example, Sun \emph{et al.} conducted simulations and experiments for communication range and proposed the MI waveguide to extend the communication range \cite{Sun2010Magnetic}. Guo \emph{et al.} \cite{Guo2015M2I} proposed the M$^2$I enhanced communication to increase the communication range. Zhang \emph{et al.} used a relay to extend the MIC range\cite{zhang2014cooperative}. 

For  MICs, most studies (e.g., \cite{Zhang2015Effective, Sun2010Magnetic, Sheinker2019Localization, Wang2019Efficient}) have simplified the channel model by omitting polarization and eddy gains to address their respective concerns, and evaluated their concern performance metrics (e.g., path loss\cite{Sun2010Magnetic} and capacity\cite{kisseleff2013channel}) w.r.t. distance via simulations or experiments.  Obtaining closed-form expression of the MIC range may initially seem straightforward. For example, Zhou \emph{et al.} \cite{Zhou2017Maximum}  derived the MIC range $d^*_{\mathrm{SD}}$ as 
\begin{equation}\label{eqn_chp2eqrsdzhou}
(d^*_{\mathrm{SD}})^{3}=\frac{\mu |\mathbf{m}_{\mathrm{S}}|}{4\pi S_{\min}}\vert 3(\mathbf{m}_{\mathrm{S}}\cdot \mathbf{r}_{0})|\mathbf{r}_{0}|-\mathbf{m}_{\mathrm{S}}|,
\end{equation} 
where denotes $\mathbf{r}_{0}$ unit vector of S$\rightarrow$D, $S_{\min}$ denotes the minimum detectable signal strength. Obviously, \eqref{eqn_chp2eqrsdzhou} does not consider the eddy gain $\mathcal{E}_{\mathrm{SD}}$.
In TTE applications, the significant eddy current produced by large-scale underground materials cannot be ignored, as illustrated by comparing  ``Decelerating descent" with ``Accelerating descent" in Fig. \ref{fig_chp2capadsd}. 
%Thus,  deriving the closed-form expression of the MIC range becomes much challenging. 
%Here, we derive the closed-form expressions for the  MIC  range and the optimal frequency to maximize the P2P MI link communication range in TTE scenarios.

%  \begin{figure}[t]
%	\vspace{0em}
%	\centering{\includegraphics[width=63mm]{snrdsd.eps}}
%	\caption{Description of the  MIC range. The values of the horizontal coordinates of the solid rectangles represent the  MIC ranges in which the SNRs are sufficiently strong to receive correct data. All MATLAB simulation parameters are listed in Table \ref{tbl_sim}, and the SNRs are obtained from the left side of~\eqref{eqn_chp2eqrsd}.
%		\vspace{-0.0em}
%	}\label{fig_chp2snrdsd}
%	\vspace{-0em}
%\end{figure}

\begin{figure}[t]
\centering
\subfigure[ ]{
\label{fig_chp2snrdsd:desc} %% label for second subfigure
\includegraphics[width=1.60in, height=1.23in]{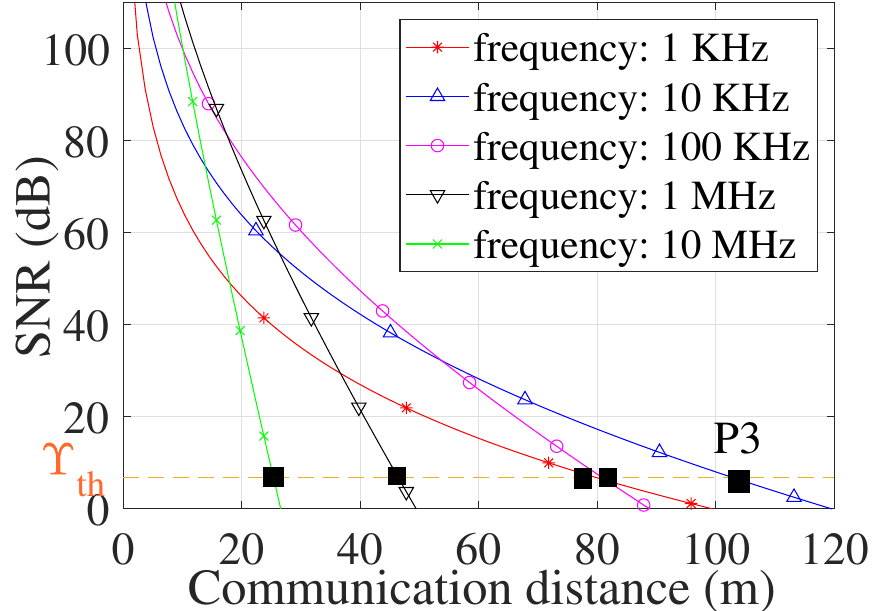}}    
\subfigure[]{
\label{fig_chp2snrdsd:freq} %% label for second subfigure
\includegraphics[width=1.70in]{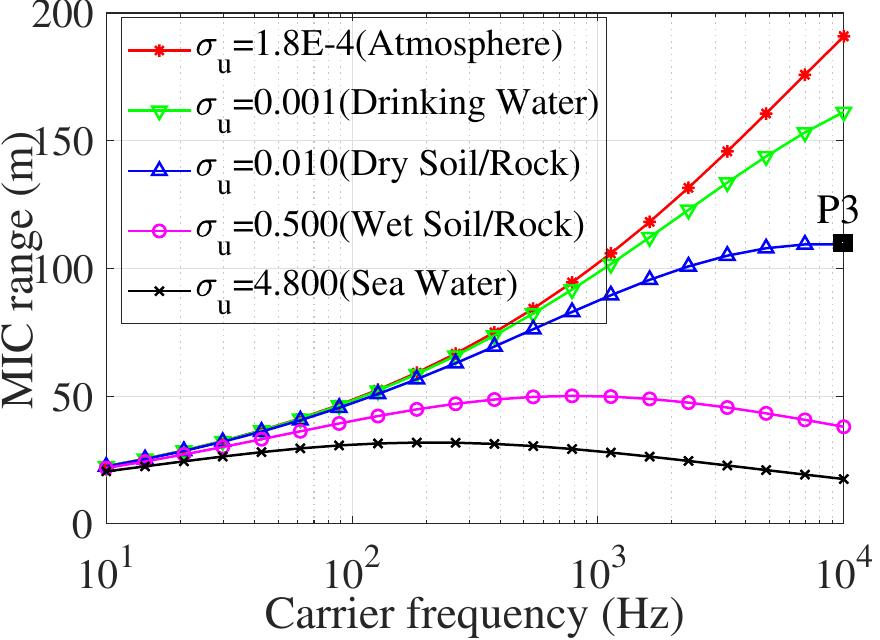}}    
\caption{MIC range simulations: (a) Description of the  MIC range; (b) MIC range v.s. carrier frequency under different UG materials. The values of the horizontal coordinates of the solid rectangles represent the  MIC ranges in which the SNRs are sufficiently strong to receive correct data. All MATLAB simulation parameters are listed in Table \ref{tbl_sim}, and the SNRs are obtained from the left side of~\eqref{eqn_chp2eqrsd}.
%		\vspace{-0.0em}
}
\label{fig_chp2snrdsd} %% label for entire figure
\vspace{-0.0em}
\end{figure}

Assuming a required SNR threshold  $\Upsilon_{\rm th}$ for correct data reception (see Fig. \ref{fig_chp2snrdsd}), the  MIC range $d^*_{\mathrm{SD}}$ of MI link S$\rightarrow$D can be obtained by solving $\frac{P_{\mathrm{S}} G_{\mathrm{SD}}(d^*_{\mathrm{SD}})}{N_{\rm o}} = \Upsilon_{\rm th}$ w.r.t. $d^*_{\mathrm{SD}}$, i.e.
\begin{equation}\label{eqn_chp2eqrsd}
\begin{aligned}
\tfrac{P_{\mathrm{S}}\mathcal{S}_{\mathrm{SD}}\mu^2_{\rm u}}{N_{\rm o}}\tfrac{1}{{d}^*_{\mathrm{SD}} }e^\frac{-d^*_{\mathrm{SD}}}{\delta_{\rm u}} J_{\mathrm{SD}} = \Upsilon_{\rm th}.
\end{aligned}
\end{equation}
Fig. \ref{fig_chp2snrdsd:desc} describes the MIC range for TTE scenarios via MATLAB simulation. The simulation  shows that the MIC system has the largest range at 10 kHz, while this frequency is neither the highest nor the lowest in this simulation. Further, the maximal MIC range exceeds 100 m but remains below 50 m in wet soil, as shown in Fig. \ref{fig_chp2snrdsd:freq}. Fig. \ref{fig_chp2snrdsd:freq} illustrates the relationship between carrier frequency and the MIC range across UG media with varying conductivities: for high-conductivity media, the range shows a non-monotonic trend with increasing frequency, while for low-conductivity media, it exhibits a more consistent increase.  On the other hand, in the lower-conductivity materials, further optimization of the MIC range is possible once the solution to \eqref{eqn_chp2eqrsd} is obtained.

\subsection{Upper-Layer P2P Techniques} \label{sect_upp2p}
\begin{figure}[t!]
\centering{\includegraphics[width=49mm]{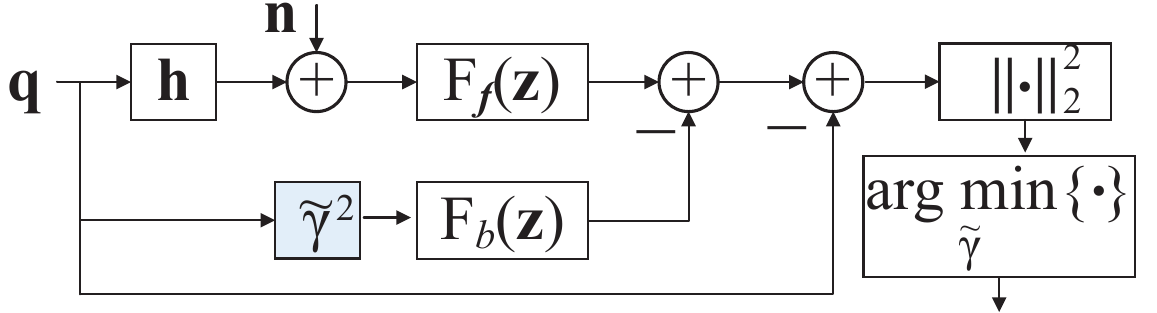}}
\caption{Detection of changes in mutual inductance $M$ for MI channel estimation \cite{Kisseleff2014Transmitter}. $\widetilde{\gamma}$ is the estimated positive coefficient and $M=\widetilde{\gamma} M_0$, where $M_0$ denotes the initial value of the mutual inductance. $\mathbf{q}$ denotes a training sequence. $\mathbf{h}$ stands for the discrete-time channel impulse response. $\mathbf{n}$ denotes the noise. $\mathbf{z}$ is the equalized received signal. The transfer functions F$_f(\cdot)$ and F$_b(\cdot)$ are discrete-time feedforward and feedback filters, respectively. 
}\label{fig_sec4estimation} 
% \vspace{-1.0em}
\end{figure}

This subsection discusses the  upper-layer P2P techniques,  including channel estimation, modulation, and FEC schemes. Compared to EMWC, applying these schemes to TTE MIC faces issues of inefficient bandwidth and channel capacity.  Chen \emph{et al.} \cite{Chen2023Novel} proposed a Polar coding scheme with code lengths of 256 and 1024, the capacity of the TTE MIC channel may not accommodate such a large frame.

\subsubsection{Channel estimation}\label{sectsubsub4_channelesti}
Compared to EMWC, while the MI channel estimation faces inefficient bandwidth for CSI exchanging, the assumption of larger MI channel coherence time helps improve channel estimation performance. For this feature, the transmitter-side channel estimation method (see Fig.~\ref{fig_sec4estimation}) was proposed in \cite{Kisseleff2014Transmitter}.  Such a method does not require an explicit training sequence. However, insufficient feedback challenges the dyadic backscatter channel (DBC) estimation.  Guo \emph{et al.}  \cite{Guo2019Inter} designed a system for joint channel estimation and data transmissions. They considered the effects of  UG material conductivity and sensor depth, which may not be the focus in the EMWC field.

\subsubsection{Modulation}\label{sectsubsub4_modulation} 
The modulation scheme encounters the issues caused by low bandwidth and frequency-selectivity, which allows a higher-order modulation scheme. Kisseleff \emph{et al.}  \cite{Kisseleff2014Modulation} proposed a modulation approach like frequency-division multiplexing (FDM) for MI links. Like EMWCs, the FDM scheme faces inter-subcarrier interference. It may not be suitable for RPMA systems. Hence, researchers applied amplitude shift keying\cite{Rezaei2020Mechanical}, frequency shift keying\cite{Zhang2023Rotation, Glickstein2020Power}, chirp-rate shift keying\cite{Liu2023Chirp}, and on-off keying\cite{Zhang2023Rotation} to MIC systems. Due to the inertia of RPMA, additional delays should be considered when converting the mechanical states.

%channel coding
\subsubsection{Channel coding}\label{sectsubsub_channelcoding}
Channel coding is important in the physical layer. Lin \emph{et al.} \cite{lin2015distributed} used the BCH code for the FEC enhancing the MI link transmission reliability. However, the BCH code-based design did not consider the underground MI channel which is not quasi-static. To tackle this, Chen \emph{et al.} \cite{Chen2023Novel} proposed a Polar code construction scheme with Bhattacharyya parameters specially optimized for MI UG-WSNs. Here,  one of the inputs is SNR w.r.t circuit gain $\mathcal{C}_{\mathrm{SD}}$ and the space gain $\mathcal{S}_{\mathrm{SD}}$; in other words, the authors did not consider the eddy gain $\mathcal{E}_{\mathrm{SD}}$ and polorization gain $J_{\mathrm{SD}}$. Compared to traditional coding, this work considered multiple distributions within a codeword from underground conductors. However, this method requires frequent CSI exchange, which may pose a challenge for the VLF-LA MIC. %Additionally, deep JSCC techniques can be considered for some specific data transmissions such as image transmission. 

\begin{figure}[t!]
\centering{\includegraphics[width=65mm]{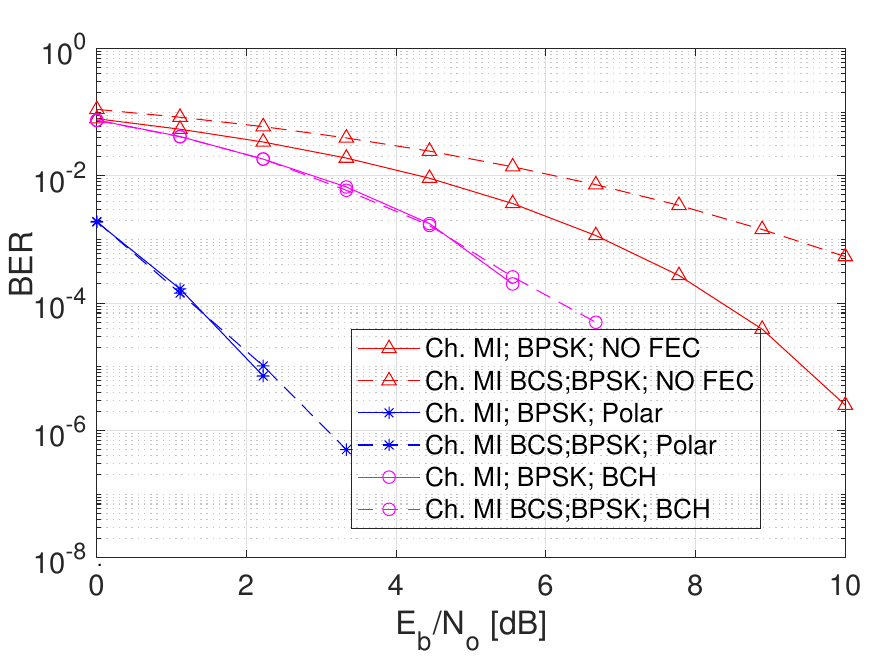}}
\caption{BER simulations of BPSK Modulation \& FECs for MI channel and MI fast fading channel, where the AVI follows the BCS distribution (MI BCS): $(\sigma^2_{\mathrm{D}}$$=$$0.5^2, \varsigma$$=$$0.8)$; FECs: BCH (n=15, k=11), Polar coding (n$=$128, k$=$66). The FECs exhibit an anti-fast fading characteristic.
}\label{fig_chp2bersnr}
\vspace{-0.0em}
\end{figure}

By reviewing the literature, e.g.,  \cite{lin2015distributed, Chen2023Novel}, on modulation and FEC in a P2P MI channel and considering  MI fast fading effects, we conduct  comparative  simulations in Fig. \ref{fig_chp2bersnr}. It is noticed that an appropriate FEC strategy can significantly reduce BER, with Polar coding offering far greater advantages than BCH. On the other hand, the FEC  mitigates the adverse impact of MI fast fading on BER  at low SNRs. As shown in Fig. \ref{fig_chp2bersnr}, with FEC employed, the dashed and solid lines nearly coincide. Although we minimize the physical layer frame length for simulating current modulation and FEC schemes, it remains excessive for TTE MIC channels, indicating room for improvement in these schemes provided by the existing MIC literature, as listed in Table \ref{tbl_chp2P2P}.

\subsection{Summary and Lessons Learned} \label{sectsub4_sll}

This section discusses the addressed and remaining challenges in MI antenna designs and MIC performance metrics (such as 3-dB bandwidth \cite{Ma2019Effect, Jiang2015Capacity}, channel capacity \cite{kisseleff2013channel}, and MIC range) in TTE scenarios. We introduce four antenna designs, each with its respective advantages and disadvantages (see Table \ref{tbl_antennatype}).

Notably, RPMA has emerged as a key area of research due to its minimal crosstalk among Txs. Future work could explore the application of massive MIMO and beamforming techniques, though these are difficult to implement in coil-based systems because of crosstalk. RPMA, however, faces challenges related to additional delay and energy consumption required to overcome inertia. For other antenna types, the design focus should be on optimizing $\mathcal{C}_{\mathrm{SD}}$ and $J_{\mathrm{SD}}$. Moreover, advancements in sensor and quantum technologies, such as high-sensitivity TMR sensors, could significantly reduce antenna size and crosstalk, making it easier to apply massive MIMO and beamforming techniques in MICs. Additionally, we  discuss channel estimation, modulation, and FEC strategies for MIC, validate their effectiveness through simulations, and demonstrate their role in mitigating  MI fast fading-induced BER degradation.

Regarding MIC performance metrics, two key aspects remain underexplored: the closed-form expressions for the 3-dB bandwidth and the MIC range in TTE scenarios (as detailed in Table \ref{tbl_bandexpress}). While numerical methods can approximate both metrics, they are time-consuming. Furthermore, significant potential exists to improve the MIC range once its expression is fully developed, as suggested by \eqref{eqn_chp2eqrsd}.

The practical takeaways or common pitfalls include: 1) The GTSs in physical layer protocol designs for RPMA inertia have been substantially overlooked, and so has the lifespan of the RPMA; 2) the spatial inhomogeneity of TTE materials  may act as a dominant source of systematic errors, causing inconsistencies between theoretical P2P MIC predictions and experimental observations; and 3) the P2P upper-layer solution tends to overlook the eddy gain and polarization gain.

%\vspace{6 mm} 

%\section {MI Networks}

\section{MI Relay and Cooperative MIC} \label{sect_cmi}

\begin{table*}[htp] \scriptsize%\footnotesize
\centering
\caption{Overview of related research on MI relay techniques \\The marker`\hloitem' describe the methods; The markers `\hllitem' and `\hlxitem'  represent pros and cons, respectively.}
\label{tbl_relay}
\vspace{-0.8em}
\scalebox{0.98}{
\begin{threeparttable}
\begin{tabular}{m{0.05\textwidth}<{\centering}|m{0.06\textwidth}<{\centering}|m{0.03\textwidth}<{\centering}|m{0.34\textwidth}<{\centering}|m{0.34\textwidth}<{\centering}|m{0.05\textwidth}<{\centering}}
\hline
\textbf{Types}  &  \textbf{Aspects} & \textbf{Refs.} & \textbf{Addressed issues \& methods} & \textbf{Remaining issues (\& Proposed approaches)} &\textbf{Priority}$^\dagger$  \\
\hline \hline
%% %% \multirow{NumberOfRows}{CellWidth\textwidth}[-Fromtop]{\centering Passive Relay}
%%%%%%%%%%%%%%%%%%%%%%%%%%%%%%%%%%%%%%%%%%%%%%%%%%%%%%%%%%%%%%%%%%%%%%%%%%%%%%%%%%%%%%%%%%%%%%%%%%%%%%%
\multirow{5}{0.06\textwidth}[-20pt]{\centering Passive relay}
& \multirow{3}{0.06\textwidth}[-3pt]{\centering MI waveguide} & \cite{Sun2010Magnetic, kisseleff2013channel} & \begin{itemize}[leftmargin=*]
\item[\hllitem] Increase the channel power gain \cite{Sun2010Magnetic} and  capacity\cite{kisseleff2013channel} \textbf{by several orders of magnitude}% compared to the P2P  link
\item[\hloitem] Channel modeling \cite{Sun2010Magnetic}; waveguide capacity analysis\cite{kisseleff2013channel} %of MI waveguide (with linearly equidistant aligned coils)
\vspace{-1.0em}
\end{itemize}& \begin{itemize}[leftmargin=*]
\item[\hlxitem] Challenge in \textbf{aligning and deploying coils} to ensure the same mutual inductance between adjacent relays
\vspace{-1.0em}
\end{itemize}&\hlprioritem\hlprioritem\\
\cline{3-6}
%%%%%%%%%%%%%%%%%%%%%%%%%%%%%%%%%%%%%%%%%%%%%%%%%%%%%%%%%%%%%%%%%%%%%%%%%%%%%%%%%%%%%%%%%%%%%%%%%%%%%%%		
&    & \cite{Sun2013Optimal} & \begin{itemize}[leftmargin=*]
\item[\hllitem]  Allowing slight \textbf{misalignment of  coils} and being  robust to \textbf{node failure} 
\item[\hloitem] \textbf{MST} and \textbf{TC algorithms} to reduce the number of relays
\vspace{-1.0em}
\end{itemize} & \begin{itemize}[leftmargin=*]
\item[\hlxitem]  Crosstalk effect in misaligned coils
\vspace{-1.0em}
\end{itemize}&\hlprioritem\\
\cline{2-6}
%%%%%%%%%%%%%%%%%%%%%%%%%%%%%%%%%%%%%%%%%%%%%%%%%%%%%%%%%%%%%%%%%%%%%%%%%%%%%%%%%%%%%%%%%%%%%%%%%%%%%%%
& MPRlA & \cite{Ma2015Topology} & \begin{itemize}[leftmargin=*]
\item[\hllitem] \textbf{Bandwidth increased} by over 15\% compared to a quadrilateral array
\item[\hloitem] %Channel modeling for a regular 
Hexagonal MPRlA using KVL equations
\vspace{-1.0em}
\end{itemize}&\begin{itemize}[leftmargin=*]
\item[\hlxitem] Challenge of coil alignment
\vspace{-1.0em}
\end{itemize} &\hlprioritem\\
\cline{2-6}
%%%%%%%%%%%%%%%%%%%%%%%%%%%%%%%%%%%%%%%%%%%%%%%%%%%%%%%%%%%%%%%%%%%%%%%%%%%%%%%%%%%%%%%%%%%%%%%%%%%%%%%
& Crosstalk effect &\cite{Li2019Survey} & \begin{itemize}[leftmargin=*]
\item[\hllitem] Existence of crosstalk effect
\vspace{-1.0em}
\end{itemize}& \begin{itemize}[leftmargin=*]
\item[\hlxitem] Lack theoretical analysis; overlook \textbf{positive} crosstalk effect
\item[\hlxitem] Crosstalk among local \textbf{high-density nodes} in a network
\item[\hloitem] Proposed Transformer based framework (Fig. \ref{fig_sec5crossmig})
\vspace{-1.0em}
\end{itemize}&\hlprioritem\hlprioritem\hlprioritem\\
\hline 
%%%%%%%%%%%%%%%%%%%%%%%%%%%%%%%%%%%%%%%%%%%%%%%%%%%%%%%%%%%%%%%%%%%%%%%%%%%%%%%%%%%%%%%%%%%%%%%%%%%%%%%
\multirow{5}{0.06\textwidth}[-20pt]{\centering Active relay}
& \multirow{3}{0.06\textwidth}[-3pt]{\centering Stationary CMIC-$n$AR } & \cite{Kisseleff2015On} & \begin{itemize}[leftmargin=*]
\item[\hllitem] Capacity increasing by an order of magnitude
\item[\hloitem] \textbf{CMIC-$n$AR}  modeling and capacity analysis using \textbf{AF}, \textbf{DF} and \textbf{FF} schemes %(with a single hop model)
\vspace{-1.0em}
\end{itemize}& \begin{itemize}[leftmargin=*]
\item[\hlxitem] Approximate \textbf{coil alignment required;  higher energy consumption and \textbf{protocol complexity} than the MI waveguide}
\vspace{-1.0em}
\end{itemize}&\hlprioritem\hlprioritem\\
\cline{3-6}
%%%%%%%%%%%%%%%%%%%%%%%%%%%%%%%%%%%%%%%%%%%%%%%%%%%%%%%%%%%%%%%%%%%%%%%%%%%%%%%%%%%%%%%%%%%%%%%%%%%%%%%
&  & \cite{Khalil2021Optimal} & \begin{itemize}[leftmargin=*]
\item[\hllitem] The number of relay reduced by about $\frac{2}{8}$
\item[\hloitem] Formulating  optimum relay placement problems 
%for both throughput and the number of relays 
using \textbf{RCP based approach}
\vspace{-1.0em}
\end{itemize} & \begin{itemize}[leftmargin=*]
\item[\hlxitem] Strict coil alignment required
\vspace{-1.0em}
\end{itemize}&\hlprioritem\\
\cline{2-6}
%%%%%%%%%%%%%%%%%%%%%%%%%%%%%%%%%%%%%%%%%%%%%%%%%%%%%%%%%%%%%%%%%%%%%%%%%%%%%%%%%%%%%%%%%%%%%%%%%%%%%%%
& \multirow{3}{0.06\textwidth}[-16pt]{\centering Stationary CMIC-1NR} & \cite{zhang2014cooperative} &  \begin{itemize}[leftmargin=*]
%\item[\hllitem] OCMI yielding over 20\% increase in MIC range than CMI
\item[\hllitem] OCMI yielding over 20\% increase in CMIC range.
% \item[\hloitem] Fix CMIC-1NR model with a fixed-position relay and an optimal angle CMI (OCMI) approach to increase the MIC range
\item[\hloitem] Fixed-coil-based CMIC-1NR; optimal angle CMI (\textbf{OCMI}) for larger ranges
\vspace{-1.0em}
\end{itemize}&\begin{itemize}[leftmargin=*]
\item[\hlxitem] Limited application scenarios due to fixed relay position
\vspace{-1.0em}
\end{itemize}&\hlprioritem\\
\cline{3-6}
%%%%%%%%%%%%%%%%%%%%%%%%%%%%%%%%%%%%%%%%%%%%%%%%%%%%%%%%%%%%%%%%%%%%%%%%%%%%%%%%%%%%%%%%%%%%%%%%%%%%%%%
&  & \cite{Ma2019Effect} & \begin{itemize}[leftmargin=*]
\item[\hllitem] Potentially capacity increasing over 50\%
%\item[\hloitem] Modeling for arbitrarily placed coils using AF schemes with the derivation of the CMI bandwidth 
\item[\hloitem] Non-fixed coil-based CMIC-1NR; AF schemes; CMI bandwidth derivation
\vspace{-1.0em}
\end{itemize}& \begin{itemize}[leftmargin=*]
\item[\hlxitem]  Space constraint of tunnels
\item[\hlxitem] Bandwidth expression restricted to isomorphic coils
\vspace{-1.0em}
\end{itemize}&\hlprioritem\\
\cline{3-6}
%%%%%%%%%%%%%%%%%%%%%%%%%%%%%%%%%%%%%%%%%%%%%%%%%%%%%%%%%%%%%%%%%%%%%%%%%%%%%%%%%%%%%%%%%%%%%%%%%%%%%%%
&  & \cite{Ma2019Antenna} & \begin{itemize}[leftmargin=*]
%\item[\hloitem] APO algorithm  using geometric approximation modeling  and random-search   approaches			
\item[\hllitem]  Excellent \textbf{global} optimal search ability and \textbf{convergence}
\item[\hloitem] Geometric approximation  and random-search   approaches
\vspace{-1.0em}
\end{itemize} & \begin{itemize}[leftmargin=*]
\item[\hlxitem] Validation of multi-relay CMIC systems with arbitrary APOs
\vspace{-1.0em}
\end{itemize}&\hlrevisioneq{\hlprioritem\hlprioritem}\\
\cline{2-6}
%%%%%%%%%%%%%%%%%%%%%%%%%%%%%%%%%%%%%%%%%%%%%%%%%%%%%%%%%%%%%%%%%%%%%%%%%%%%%%%%%%%%%%%%%%%%%%%%%%%%%%%
&Mobile underwater CMIC  & \cite{Zhang2024Cooperative} & \begin{itemize}[leftmargin=*]	
\item[\hllitem]  Obtain the \textbf{PDF} of CMI channel for \textbf{underwater} scenarios
%\item[\hloitem] Statistical characteristics of CMI channel using uniform distribution antenna angle input
\item[\hloitem] Derivation based on \textbf{uniform distribution} antenna angle input	
\vspace{-1.0em}
\end{itemize}&  \begin{itemize}[leftmargin=*]
\item[\hlxitem]  Uniform-based PDF  not applicable to mobile underground CMI channels
\vspace{-1.0em}
\end{itemize}&\hlprioritem\hlprioritem\\
\hline
%%%%%%%%%%%%%%%%%%%%%%%%%%%%%%%%%%%%%%%%%%%%%%%%%%%%%%%%%%%%%%%%%%%%%%%%%%%%%%%%%%%%%%%%%%%%%%%%%%%%%%%
Hybrid  relay	&Stationary CMIC & \cite{Li2019Survey} & \begin{itemize}[leftmargin=*]
\item[\hloitem] Place passive relays between two adjacent active relays	
\item[\hllitem]  Energy saving
\vspace{-1.0em}
\end{itemize} & \begin{itemize}[leftmargin=*]
\item[\hlxitem] Challenging antenna deployment for TTE and mobile applications
\vspace{-1.0em}
\end{itemize}&\hlprioritem\hlprioritem\\
\hline
%%%%%%%%%%%%%%%%%%%%%%%%%%%%%%%%%%%%%%%%%%%%%%%%%%%%%%%%%%%%%%%%%%%%%%%%%%%%%%%%%%%%%%%%%%%%%%%%%%%%%%%
\end{tabular}
\begin{tablenotes}  
\footnotesize  
\item[$\dagger$] Priority: Priority level of remaining issues and  proposed approaches for exploration. Here, low priority (\hlprioritem) indicates that:  1) the remaining issues have been explored in subsequent MIC literature; 2) the existing EMWC schemes  are compatible with MIC for these issues; or 3) exploring  these issues is optional.
\end{tablenotes}  
\end{threeparttable}
}   
\vspace{-1.3em}
\end{table*}

\begin{figure}[t]
\centering
\subfigure[ ]{
\label{fig_sec3relayoil:n} %% label for second subfigure
\includegraphics[width=1.60in]{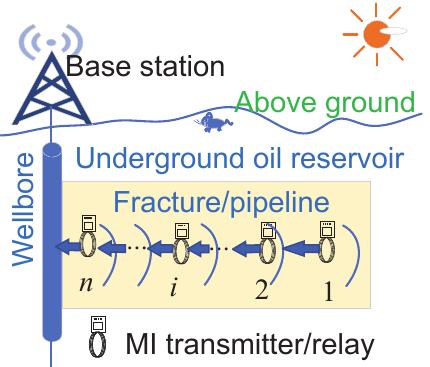}}    
\subfigure[]{
\label{fig_sec3relayoil:1} %% label for second subfigure
\includegraphics[width=1.60in]{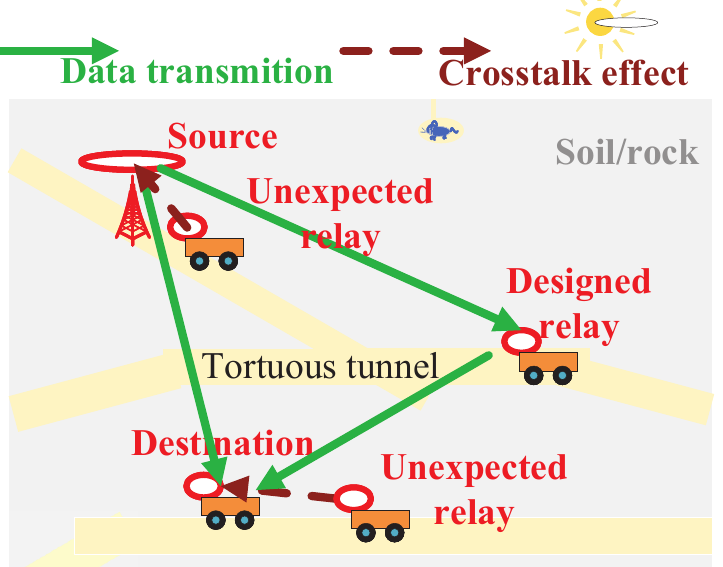}}    
\caption{ Use cases of relay and CMIC: (a) Pipeline case \cite{Alshehri2017Optimal,guo2014channel}; (b) Mobile MIC case. In Fig. \ref{fig_sec3relayoil:n}, MI waveguide, CMIC-$n$AR and hybrid relay techniques can be used; in Fig. \ref{fig_sec3relayoil:1}, the CMIC-1NR technique can be used.
}
\label{fig_sec3relayoil} %% label for entire figure
\vspace{-0.9em}
\end{figure}

Improving the communication range and achievable rate are the two key goals of MI studies for TTE applications. Existing research has demonstrated that these two performance metrics can be increased by one or several orders of magnitude through the use of MI relays. Most MI relay techniques require sufficient space for relay deployment.  This section categorizes the MI relay into the passive MI relay (including MI waveguide) and active MI relay (called CMIC). Furthermore, we consider the advances in these two relay techniques, particularly highlighting their respective limitations.  Additionally, we theoretically analyze the crosstalk effect phenomenon among MI relays through our derivations and simulations, which are overlooked in the existing studies. Table \ref{tbl_relay} outlines the challenges and methodologies of the MI relay techniques.

\subsection{Use Cases and Scenarios}\label{sec_chp4ucs}

 For the TTE scenarios, there are two  typical use cases, as shown in Figs. \ref{fig_sec3relayoil:n} and \ref{fig_sec3relayoil:1}.

In the first use case, such as  oil reservoirs, straight tunnels and pipelines provide  sufficient space and a clear path for signal propagation. In these scenarios, MI coils can be aligned and deployed as a stable linear topology along these pathways, ensuring reliable communication between MI Txs/relays and the base station above ground. For example,  Guo and Abdallah  proposed  the framework of a linear pipeline/oil sensor network topology with dense nodes in \cite{Sun2011MISE, Alshehri2017Optimal}. In this framework, most MI nodes do not require an independent power supply. As shown in Fig. \ref{fig_sec3relayoil:n}, the energy of MI modes is induced by the EMW generated by a large dipole. The required power to transmit data from a sensor to its neighbor sensor 3 m apart is about -50 dBm. Thus, the passive relay techniques can be used to achieve high data rate and energy performance.

In the second use case, the CMIC with one non-aligned relay  (CMIC-1NR) topology\cite{Ma2019Antenna} is designed for more dynamic and less predictable environments. For example, in underground Internet of Vehicles (IoV) systems, the mobile nodes cannot form a stable  linear topology.  In such scenarios, nodes are typically sparsely and randomly distributed, with each requiring a power supply to generate signals. Thus, the CMIC topology is suited for these conditions.  However, randomly distributed nodes might form locally dense  clusters, as shown in Fig. \ref{fig_sec3relayoil:1}. Within these clusters, certain nodes inadvertently act as unexpected passive relays,  inducing a crosstalk effect.  In what follows, we discuss the MI passive relay, CMIC techniques, and MI crosstalk effect in detail.

\subsection{MI Waveguide and Passive Relay Techniques} \label{sectsub5_mwprt}

The MI passive relay passes signals through without requiring an active power source for amplification or processing. In this subsection, we introduce two MI passive relay techniques i.e., MI waveguide and MPRlA. We highlight their advantages and disadvantages for TTE applications. We also discuss the previously unstudied MI crosstalk effect, with a simulation to validate this effect. 

%We also analyze the crosstalk effect of an unexpected MI relay. This effect indicates that passive relays may negatively impact the channel power gain of shorter MIC links, except in the cases of MI waveguides\cite{Sun2010Magnetic, kisseleff2013channel}, and  MPRlA\cite{Ma2015Topology}. 

\subsubsection{MI Waveguide} \label{sectsubsub5_miwaveguid}

\begin{figure}[t]
\centering{\includegraphics[width=70mm, height=20mm]{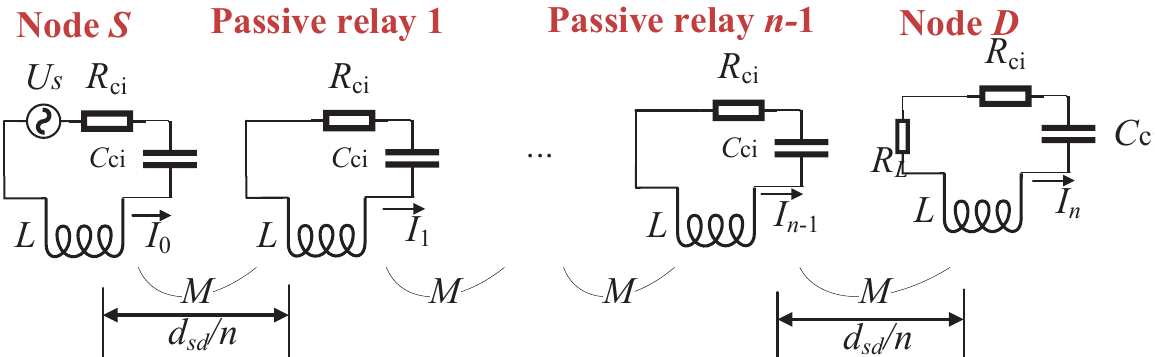}}
\caption{MI waveguide channel model. Here, $M$,  $L$, and $C_{\mathrm{ci}}$  denote the mutual inductance between two adjacent relays, inductivity, and matching capacitor for  $f_0=\frac{1}{2\pi \sqrt{LC_{\mathrm{ci}}}}$, respectively.  
}\label{fig_sec3waveguide}
\vspace{-0.0em}
\end{figure}

Research on passive relays began with the introduction of the MI waveguide approach in \cite{Shamonina2002Magneto, Ishtiaq2021Performance}. In TTE environments, the overall cost of the relay deployment, including price, time, and risks, is enormous. However, for the non-emergency and non-mobile TTE MIC applications, the MI waveguide can be considered due to its excellent performance. 

In \cite{Sun2010Magnetic, kisseleff2013channel}, Sun's and Kisseleff's teams increased the MIC range and channel capacity several times through the waveguide techniques. The typical structure of the MI waveguide is shown in Fig. \ref{fig_sec3waveguide} where identical passive relays are placed at equal intervals between $\mathrm{S}$ and $\mathrm{D}$. The channel power gain of the MI waveguide link was derived  as~\cite{Sun2010Magnetic, kisseleff2013channel}:
\begin{equation}\label{eqn_chp3gwg}
\begin{aligned}
&	G_{\mathrm{wg}} = \tfrac{R_L}{\mathrm{Sn}(Z_M, Z_L, n) \mathrm{Sn}(Z_M, Z_L, n+1) },\\
&\mathrm{Sn}(Z_M, Z_L, k) =  \mathrm{Fn}(Z_M, k) + Z_L \cdot \mathrm{Fn}(Z_M, k-1),   \\
&\mathrm{Fn}(Z_M, k)=\tfrac{\left(\tfrac{(Z_M+\sqrt{Z_M^{2}-4})}{2}\right)^{k+1}-\left(\tfrac{(Z_M-\sqrt{Z_M^{2}-4})}{2}\right)^{k+1}}{\sqrt{Z_M^{2}-4}},
\end{aligned}
\end{equation}
through Kirchhoff's voltage law (KVL), where $Z_M = \frac{Z_{\mathrm{LC}} + R_\mathrm{ci}}{j2\pi f M}$, and $Z_L = \frac{R_L}{j2\pi f M}$. The experiment in \cite{Sun2010Magnetic} indicated that the channel power gain increased by several orders of magnitude when using MI waveguide. Later, the MI waveguide networks were investigated. However, the number of MI passive relays determines the complexity of the MI waveguide system.  To reduce the number of passive relays, Sun \emph{et al.} proposed the minimal spanning tree (MST) algorithm and TC algorithm  coils\cite{Sun2010Deployment,Sun2013Optimal}. Their simulations indicated that the number of relays decreased from about 2,500 to 1,300 when using the MST algorithm at a sensor density of 10$^{-3}$ nodes/m$^2$. These studies indicate that the MI waveguide techniques are constrained by the challenge of coil alignment in the TTE environment, even in the UW-WSNs.

\subsubsection{ MPRlA} \label{sectsubsub5_mprla}

\begin{figure}[t!]
\centering{\includegraphics[width=35mm]{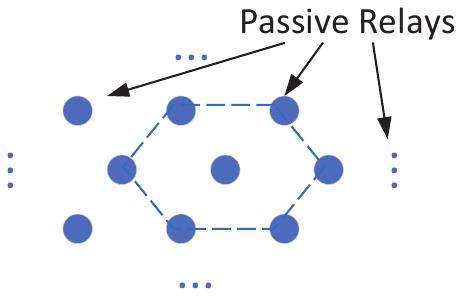}}
\caption{Regular hexagonal MPRlA. Here, identical passive relays are uniformly placed in a 2D plane, and each relay serves as the center of a hexagon whose vertices are its nearest neighboring relays. 
}\label{fig_sec3hexwg}
\vspace{-0.0em}
\end{figure}

The MPRlA is  categorized as a 2D/3D MI waveguide\cite{Wang2019Energy} or  non-linear MI waveguide\cite{Li2019Survey},  as exemplified in \cite{Masihpour2010Cooperative, Ma2015Topology}. Ma et al.~\cite{Ma2015Topology} proposed a topology for a regular hexagonal MPRlA (see  Fig. \ref{fig_sec3hexwg}). The passive relays transmit propagation energy to adjacent coils and return reflection energy.  They modeled the channel of this regular hexagonal MPRlA using the KVL and found that the bandwidth of the regular hexagonal MPRlA increased by over 15\% compared to a quadrilateral array. However,  the MPRlA techniques also face the challenge of coil alignment.

\subsubsection{Crosstalk effect of short-range passive relays}\label{sectsubsub_cespr}
\begin{figure}[t!]
\centering{\includegraphics[width=45mm]{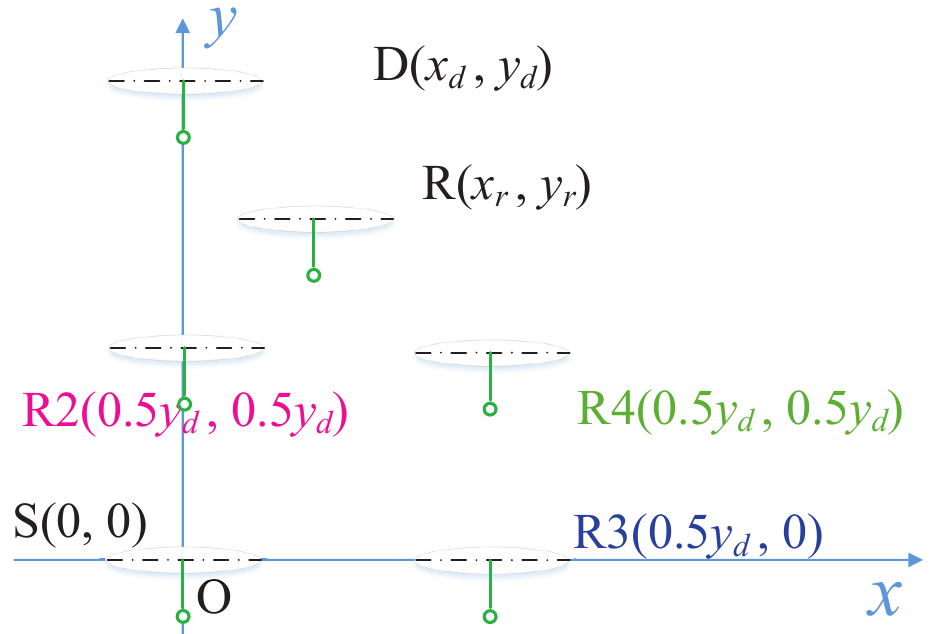}}
\caption{Example of Crosstalk effect from an unexpected passive relay. Suppose that a passive relay $\mathrm{R}$ moves (sufficiently slowly to avoid MI fast fading) along the path  R$\rightarrow$ R2 $\rightarrow$ R3 $\rightarrow$ R4.    
}\label{fig_sec3passivenetwork}
\vspace{-0.0em}
\end{figure}

In \cite{Li2019Survey}, Li \emph{et al.}  briefly mentioned the existence of crosstalk among the aligned relays.  Consider the crosstalk effect caused by an arbitrarily placed unexpected passive relays, including the randomly moving Rx or idle coils belonging to other networks.  Consider an MI link $\mathrm{S}$$\rightarrow$$\mathrm{D}$ with an unexpected passive relay $\mathrm{R}$ within the MIC range of $\mathrm{S}$. The locations of $\mathrm{S}$, $\mathrm{R}$, and $\mathrm{D}$ are shown in Fig. \ref{fig_sec3passivenetwork}.
% The parameters of $\mathrm{S}$, $\mathrm{R}$ and $\mathrm{D}$ are as listed in Table \ref{tbl_sim}, with specific adjustments $a_{\mathrm{cS}} = a_{\mathrm{c}r} = a_{\mathrm{cD}} = 0.6$ m, $N_{\mathrm{S}} = N_r = N_{\mathrm{D}} = 15$,  
% Let $\mathcal {K} = \{S, R, D\} $,
According to KVL, we have
\begin{equation}\label{eqn_chp3Kf1}
\begin{aligned}
&I_{\mathrm{S}}Z_\mathrm{LC}+  I_{\mathrm{D}} \cdot j2\pi f M_{\mathrm{SD}} +I_{\mathrm{R}} \cdot j2\pi f M_{\mathrm{SR}} = U_{\mathrm{S}}  \\
& I_{\mathrm{S}} \cdot j2\pi f M_{\mathrm{SD}} +  I_{\mathrm{R}} \cdot j2\pi f M_{\mathrm{RD}} =- \!I_{\mathrm{D}}Z_\mathrm{LC} \\
& I_{\mathrm{S}} \cdot j2\pi f M_{\mathrm{SR}} +  I_{\mathrm{D}} \cdot j2\pi f M_{\mathrm{RD}} =- \!I_{\mathrm{R}}Z_\mathrm{LC},   \\
\end{aligned}
\end{equation}
\begin{comment}
\begin{equation}\label{eqn_chp3Kf1}
\begin{aligned}
&I_{\mathrm{S}}Z_{Z_\mathrm{LC}}+  \sum\limits_{l\in \mathcal{K} \verb|\| \{S\}} I_{l}\cdot j2\pi f M_{sl} = U_{\mathrm{S}},  \\
& I_{\mathrm{S}} j2\pi f M_{sk} + \sum\limits_{l\in \mathcal{K} \verb|\| \{S,k\}} I_{l} j2\pi f M_{kl} =- \!I_{k}Z_{Z_\mathrm{LC}}.   \\
\end{aligned}
\end{equation}
\end{comment}
where $Z_\mathrm{LC}$$=$$j2\pi f L_{\mathrm{cS}}$$+$$\tfrac{1}{j2\pi f C_{\mathrm{cS}}}$$+$$R_{\mathrm{cS}}$$+$$R_L$. Let $Z_{\mathrm{SR}}$$=$$j2\pi f M_{\mathrm{SR}}$,  $Z_{\mathrm{SR}}$$=$$j2\pi f M_{\mathrm{SR}}$ and  $Z_{\mathrm{RD}}$$=$$j2\pi f M_{\mathrm{RD}}$.
Solving these equations, we obtain the current in coil $\mathrm{D}$: 
\begin{subequations}\label{eqn_chp3IdPassive}
\begin{align}
I_{\mathrm{D}} &= -\frac{U_{\mathrm{S}} (Z_{\mathrm{SD}} Z_{\mathrm{LC}}^2 - Z_{\mathrm{pa1}}(\mathrm{S},\mathrm{D},\mathrm{R}))}{Z_{\mathrm{LC}}^4 + Z_{\mathrm{pa2}}(\mathrm{S},\mathrm{D},\mathrm{R})}, \label{eqn_chp3IdPassive:0}  \\  
Z_{\mathrm{pa1}} &(\mathrm{S},\mathrm{D},\mathrm{R}) =Z_{\mathrm{RD}} Z_{\mathrm{SR}} Z_\mathrm{LC},  \label{eqn_chp3IdPassive:1}\\
Z_{\mathrm{pa2}} &(\mathrm{S},\mathrm{D},\mathrm{R}) = 2 Z_{\mathrm{RD}} Z_{\mathrm{SD}} Z_{\mathrm{SR}} Z_\mathrm{LC}  - Z_{\mathrm{RD}}^2 Z_\mathrm{LC}^2 \notag \\ 
& \ \ \ \    - Z_{\mathrm{SD}}^2 Z_\mathrm{LC}^2  -  Z_{\mathrm{SR}}^2 Z_\mathrm{LC}^2,   \label{eqn_chp3IdPassive:2}   
\end{align}
\end{subequations} 
where we call $Z_{\mathrm{pa1}}(\mathrm{S},\mathrm{D},\mathrm{R})$ and $Z_{\mathrm{pa2}}(\mathrm{S},\mathrm{D},\mathrm{R})$ the \emph{crosstalk impedances} of link S$\rightarrow$D. Thus, we can define the\emph{ crosstalk effect}  as the phenomenon occurring when a link has at least one non-zero crosstalk impedance. 
Also, using KVL equations as~\eqref{eqn_chp3Kf1}, we can derive the more complex expressions of crosstalk impedances $Z_{\mathrm{pa1}}(\mathrm{S},\mathrm{D},\mathrm{R}, ...)$ and $Z_{\mathrm{pa2}}(\mathrm{S},\mathrm{D},\mathrm{R}, ...)$  when a link has more than two passive relays. 

It is noticed in~\eqref{eqn_chp3IdPassive} that the MI crosstalk effect directly impacts the performance of the MIC network through the crosstalk impedances  $Z_{\mathrm{pa1}}(\mathrm{S},\mathrm{D},\mathrm{R}, ...)$ and $Z_{\mathrm{pa2}}(\mathrm{S},\mathrm{D},\mathrm{R}, ...)$. For example, when $Z_{\mathrm{pa1}}(\mathrm{S},\mathrm{D},\mathrm{R}, ...) > 0$ and $Z_{\mathrm{pa2}}(\mathrm{S},\mathrm{D}, \mathrm{R}, ...) > 0$, the received current $I_{\mathrm{D}}$ decreases, thereby inducing the negative crosstalk effect. This effect  may reduce the MIC range and increase the BER.  Conversely, the negative impedances  $Z_{\mathrm{pa1}}(\mathrm{S},\mathrm{D},\mathrm{R}, ...)$  and  $Z_{\mathrm{pa2}}(\mathrm{S},\mathrm{D},\mathrm{R}, ...)$ can result in positive crosstalk effects, which may improve the MIC range and capacity. The MI waveguide conforms to this case. For long-range MIC using VLF-LA with low $f$, where   $M_{\mathrm{SD}}$,   $M_{\mathrm{SR}}$ and   $M_{\mathrm{RD}}$  are sufficiently small  to satisfy $Z_\mathrm{LC}\!\gg\! j2\pi f M_{\mathrm{SD}}$, $Z_\mathrm{LC}\!\gg\!j2\pi f M_{\mathrm{SR}}$ and $Z_\mathrm{LC}\!\gg j2\pi\!f M_{\mathrm{RD}}$,  the crosstalk impedances can be ignored.

%These impedances can either increase or decrease the channel power gain, similar to how conductive material generates eddy currents. 

\begin{figure}[t]
\centering
\subfigure[VLF ($f_0$$=$$10$  kHz).]{
\label{fig_Ejyz:3} %% label for second subfigure
\includegraphics[width=1.60in]{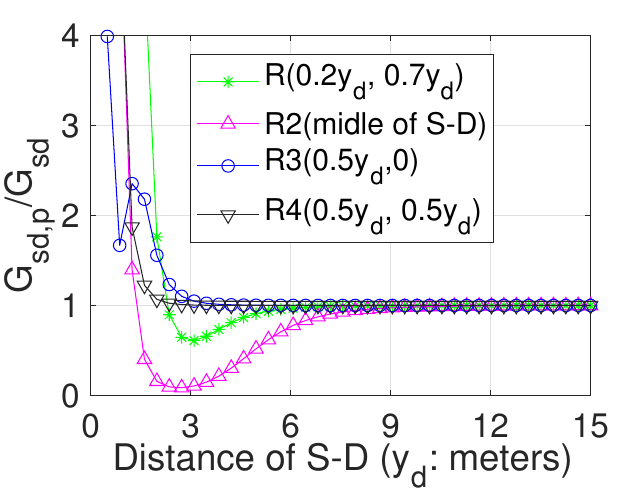}}    
\subfigure[MF ($f_0$$=$$1$ MHz).]{
\label{fig_Ejyz:4} %% label for second subfigure
\includegraphics[width=1.60in]{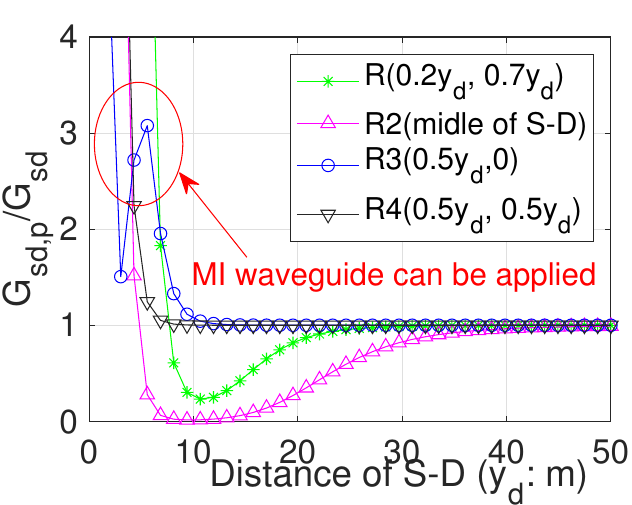}}    
\caption{Ratio $\frac{G_{SD, \mathrm{p} } }{G_{\mathrm{SD}} }$ for crosstalk effect as shown in Fig.\ref{fig_sec3passivenetwork}, where $G_{SD, \mathrm{p} }$ represents the power gain of the MI link S$\rightarrow$D when a passive relay is used, i.e., $G_{SD, \mathrm{p}}=\frac{I^2_{\mathrm{D}} R_L}{I_{\mathrm{S}}U_{\mathrm{S}}}$, where $I_{\mathrm{S}}$ and $I_{\mathrm{D}}$ are obtained from~\eqref{eqn_chp3Kf1} and~\eqref{eqn_chp3IdPassive}, respectively. This ratio quantifies how much the power gain improves (or deteriorates, if $<$ 1) with the introduction of a passive relay.  The simulation parameters are listed in Table \ref{tbl_sim}, except for $a_{\mathrm{cD}}= a_{\mathrm{c}r} = 0.6$ m, and $N_{\mathrm{cD}} = N_{\mathrm{c}r} = 15$. 
}
\label{fig_sec3crosstalk} %% label for entire figure
\vspace{-0.9em}
\end{figure}

This is verified by our simulation results as shown in Fig. \ref{fig_sec3crosstalk}. This simulation also shows that the short-range or higher-frequency MI links encounter a significantly more pronounced crosstalk effect. When  the ratio $\frac{G_{\mathrm{SD}, \mathrm{p} } }{G_{\mathrm{SD}} }$ exceeds 1, the passive relay serves as part of the MI waveguide and has a positive effect on MIC. For long-range MIC, all curves in Fig. \ref{fig_sec3crosstalk} converge to 1, indicating the elimination of crosstalk.   It is worth noting that the CMIC  and large-scale MI network also have the crosstalk effect issue according to KVL equations.  

From~\eqref{eqn_chp3IdPassive} and $M_{ij}$$=$$\frac{\pi a_{\mathrm{c}i}^2 a_{\mathrm{c}j}^2 N_i N_j \mu_{\mathrm{u}}}{2 d_{ij}^3} \mathcal{J}_{ij}\sqrt{\mathcal{E}_{ij}}$ with  $i, j$$\in$$\{\mathrm{S}, \mathrm{D}, ...\}$, it can be concluded that both $Z_{\mathrm{pa1}}(\mathrm{S},\mathrm{D},...)$ and $Z_{\mathrm{pa2}}(\mathrm{S},\mathrm{D},...)$ are multimodal functions. Determining their signs is challenging. To address this, future studies are needed on the spatial distribution of positive crosstalk effects. We also propose a deep learning framework to predict the signs of $Z_{\mathrm{pa1}}(\mathrm{S},\mathrm{D}, ...)$ and $Z_{\mathrm{pa2}}(\mathrm{S},\mathrm{D}, ...)$, as will be described in Section \ref{sect_chp6crosstalk}.

\subsection{Cooperative MIC (CMIC)} \label{sectsub5_cmic}

In this subsection, we delineate two active relay techniques, i.e., CMIC with multiple aligned relays (CMIC-$n$AR)  and CMIC-1NR. We also summarize  distinct challenges associated with cooperative communication between  MIC and  EMWC.

\begin{figure}[t]
\centering
\subfigure[CMIC-$n$AR ]{
\label{fig_sec3cminet:cmar} %% label for second subfigure
\includegraphics[width=1.30in]{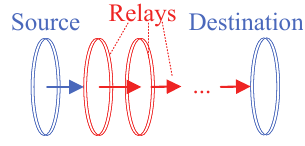}}    
\subfigure[CMIC-1NR]{
\label{fig_sec3cminet:conar} %% label for second subfigure
\includegraphics[width=1.30in]{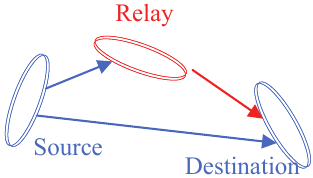}}    
\caption{Two types of the CMIC. The topology of CMIC-$n$AR  is similar to the MI waveguide, and signals are transmitted one by one. The coils exhibit a near-perfect alignment.  In CMIC-1NR, the coil of the relay does not need to be aligned, and a diversity combining method should be used to combine the relay and source signals.} 
\label{fig_sec3cminet} %% label for entire figure
\vspace{-0.9em}
\end{figure}

\begin{table*}[t!]  
\caption{Comparison of cooperative EMW and CMI communications}
%\begin{minipage}{\textwidth} 
%\tiny
\vspace{-0.9em}
\scalebox{0.93}{
\begin{threeparttable}
\centering

\begin{tabular}{ p{2.6cm}  p{5.0cm} p{2.0cm} p{5.6cm} p{2.0cm}}
\hline
\textbf{Comparison}   & \hspace{1.9em}\textbf{EMW} & \textbf{\emph{Refs.}} & \textbf{CMI} & \textbf{\emph{Refs.}}  \\
\hline 
%\raggedleft

Channel characteristic  &  Small-scale fading   & \cite{Hong2010Cooperative}  & Quasi-static fading \tnote{$\dagger$}&\cite{zhang2014cooperative, Kisseleff2015On, Ma2019Antenna}  \\

Key issue   &Reducing  outage probability  & \cite{Hong2010Cooperative}  & Enhancing Rx signal strength   & \cite{zhang2014cooperative, Kisseleff2015On, Ma2019Antenna}  \\

Methods   & AF, DF, Compress-and-farword  & \cite{Hong2010Cooperative}  & AF, DF, FF  & \cite{Kisseleff2015On, Ma2019Antenna, Li2019Survey} \\			

Benefits  &  Universal & \cite{Hong2010Cooperative} & Limited by coil locations and orientations  & \cite{zhang2014cooperative, Ma2019Antenna, Ma2019Effect} \\			

\hline
\centering
\end{tabular}
\vspace{-1.0em}
\begin{tablenotes}  
\footnotesize  
\item[$\dagger$] With Advances in MI fast fading research, limited literature on the  CMI channel with non-quasi-static fading (\emph{e.g.}, \cite{Zhang2024Cooperative}) has been released.
\end{tablenotes}
\end{threeparttable}
}
\vspace{-1.1em}
%\end{minipage}
\label{tbl_cmi}
\end{table*}

The MI waveguide enhances channel capacity and range but requires numerous underground relays.  %Its crosstalk effect also poses deployment challenges (\emph{e.g.}, alignment)  in TTE scenarios, especially for the mobile MIC.
For several decades, cooperative communication has been a focus in the EMWC, FSOC, and acoustic communication \cite{Hong2010Cooperative, Li2018Cooperative, Li2021Performance, Liu2015Adaptive, Ye2024Relay}. These communication channels experience significant small-scale fading. Cooperative relays use spatial and time diversity to mitigate these fading effects, which significantly reduces outage probability and enhances achievable rates. Various cooperative communication schemes were proposed, such as amplified-and-forward (AF),  decode-and-forward (DF), and filter-and-forward (FF) schemes \cite{Li2019Survey, Hong2010Cooperative}. 

However, the traditional MIC channel model is quasi-static without small-scale fading. Such quasi-static property renders outage probability physically meaningless.  Spatial and time diversity offer limited benefits for mitigating small-scale fading in traditional MICs. Fortunately,   researchers discovered that active relays can enhance signal strength at the Rx coil under certain conditions. The CMIC can be categorized into two types: CMIC-$n$AR \cite{Kisseleff2015On, Li2019Survey, Ishtiaq2021Performance, Khalil2021Optimal} and  CMIC-1NR~\cite{zhang2014cooperative, Ma2019Antenna, Ma2019Effect}. The typologies of the CMIC-$n$AR  and CMIC-1NR are as shown in Figs. \ref{fig_sec3cminet:cmar} and \ref{fig_sec3cminet:conar}, respectively. 

%Specifically, in 2014, Zhang \emph{et al.} proposed the concept of CMIC using an active relay at a fixed position\cite{zhang2014cooperative}. They propose an antenna orientation optimization method to improve the MIC range. One year later, Kisseleff \emph{et al.} began investigating the achievable rate of CMIC. 

The topology of CMIC-$n$AR   is similar to that of  MI waveguide~\cite{Kisseleff2015On}. The simulations indicated that the data rate increased from 10$^4$ to over 10$^5$ at the distance of 70 m when using a DF or FF relay.
%Using the AF, DF, and FF schemes, their CMI system made the overall data rate up to an order of magnitude (achieving a 1000\% data rate increase). 
To reduce energy consumption, Li \emph{et al.}~\cite{Li2019Survey} proposed a hybrid relay structure that combines the waveguide and active relay. Their simulation showed that the energy consumption decreased from 1 J to 0.3 J at the distance of 96 m. In 2021, Ishtiaq  \emph{et al.} \cite{Ishtiaq2021Performance} developed a mathematical model for evaluating the performance of the multi-hop MI link, including the hop state. Also, the multi-relay optimization has caught researchers' attention. In \cite{Khalil2021Optimal},  Khalil \emph{et al.} proposed a CMI  system similar to an MI waveguide without $\mathrm{S}$,  achieving the maximum throughput and reducing the number of relays. The transmission rate must be positive, leading to the constraint inequalities shaped as convex functions with negative values\cite{Khalil2021Optimal}. Therefore, reverse convex programming (RCP) was introduced as a solution.
% In this system, there are several active relays, a sink node, and several sensor nodes. The active relays and sink nodes constitute a topology similar to the MI waveguide without $\mathrm{S}$. Each sensor node connects to its closest relay or sink node.
However, similar to the MI waveguide, the CMIC-$n$AR  is suitable for underwater and limited underground scenarios. 

\begin{figure}[t!]
\centering{\includegraphics[width=65mm]{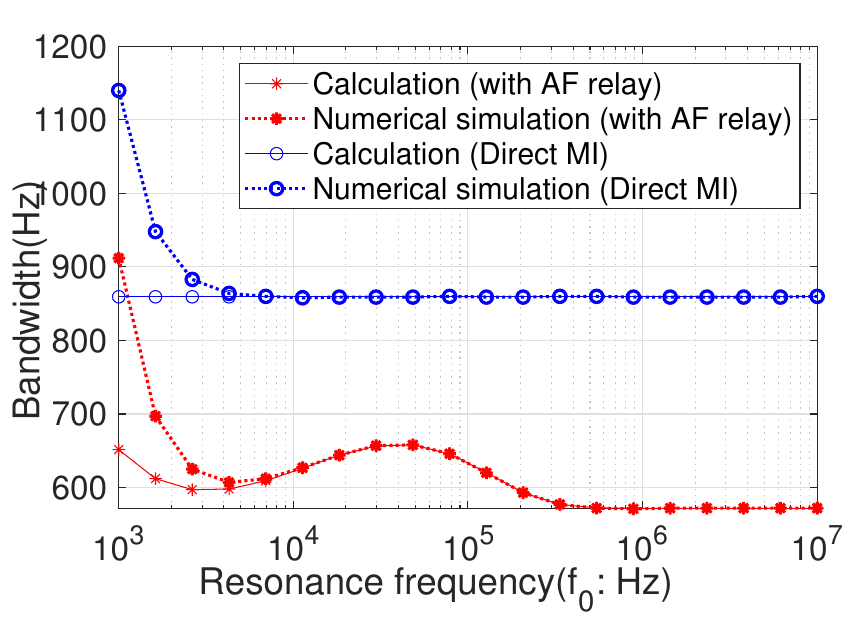}}
\caption{Comparison of the bandwidth between AF-based CMI and DMI links. The simulation parameters are listed in Table \ref{tbl_sim}, except for $\theta_{\mathrm{D}}=30^\circ$. Here, solid lines are calculated from   $B^{\rm dipole}_{ \mathrm{w},\mathrm{SD}}=B_{\rm w}\left(\frac{1}{8}(R_{\mathrm{cD}}+R_L)^3\right)$ and   $B_{\mathrm{w}}^{\mathrm{AF}}=	B_{\mathrm{w}}(\mathcal{Z}_{C,\mathrm{AF}})$. respectively, The function $B_{\mathrm{w}}^{\mathrm{AF}}(\cdot)$ is given by~\eqref{eqn_chp2BwDipole} in this paper. The impedance $\mathcal{Z}_{C,\mathrm{AF}}$ is given by Equation (14) in \cite{Ma2019Effect}.
}\label{fig_sec3bafb1} 
\vspace{-0.0em}
\end{figure}

\begin{figure}[t!]
\centering{\includegraphics[width=65mm]{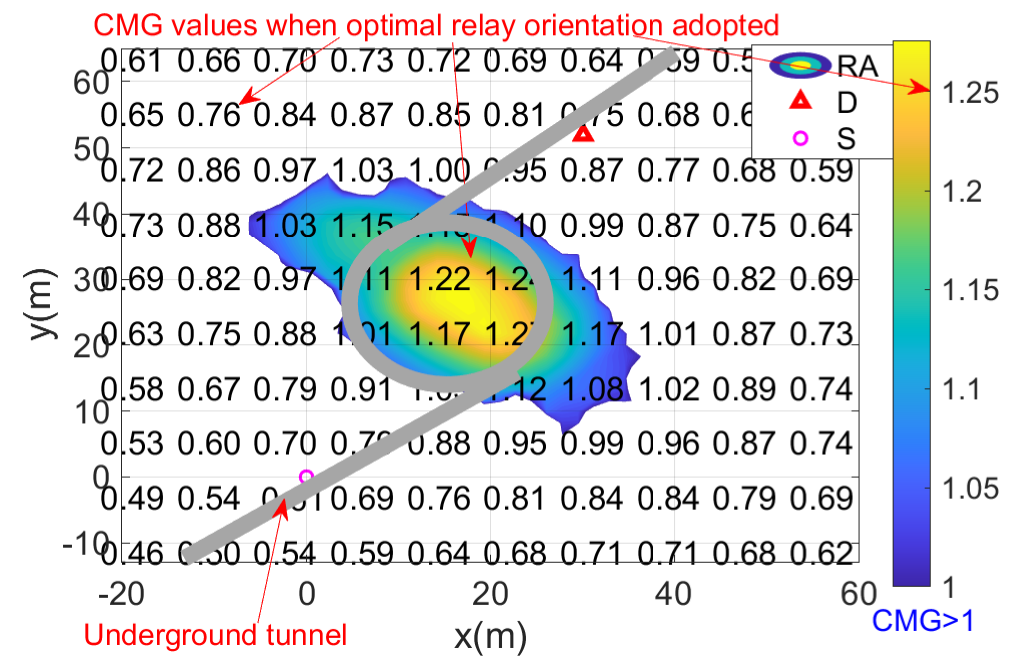}}
\caption{Effects of relay positions on the CMI performance, and an example of tunnel constraints.  The simulation parameters are listed in Table \ref{tbl_sim}, except for $\theta_{\mathrm{D}}=30^\circ$. Here, CMG  ( defined in \cite{Ma2019Effect}) is the ratio of capacities of the CMI link to the DMI link, and the area where CMG$>$1 is called the relay area (RA). The thick gray curve represents an arbitrary underground tunnel in which  $\mathrm{S}$, $\mathrm{R}$, and $\mathrm{D}$  are constrained to reside.
}\label{fig_sec3cmgxyth} 
\vspace{-0.0em}
\end{figure}

Due to the disadvantage of CMIC-$n$AR  for TTE MICs, some researchers explored the CMIC-1NR  by optimizing $J_{\mathrm{SD}}$.  Zhang \emph{et al.} \cite{zhang2014cooperative} proposed the CMIC-1NR. Later, Ma \emph{et al.}  \cite{Ma2019Effect, Ma2019Antenna} further investigated CMIC-1NR systems to boost MIC achievable rates.  In \cite{Ma2019Effect}, they studied the achievable rate of AF-based CMIC with an arbitrary relay APO. Specifically, they derived the closed-form expressions for CMIC bandwidth. The bandwidth of the  CMI link varies with the APO and is smaller than that of the Direct magnetic induction (DMI) link (see Fig. \ref{fig_sec3bafb1}). Even with the smaller bandwidth, an active relay may enhance the achievable rate of the MI link (see Fig. \ref{fig_sec3cmgxyth}), especially for weak signals. Unlike the DMI link,  this improvement is APO-dependent. This complicates TTE applications due to underground space constraints.  In Fig. \ref{fig_sec3cmgxyth}, despite maximal  CMIC achievable rate gain (CMG) at the RA center, the relay cannot be placed there due to tunnel constraints. Ma \emph{et al.} \cite{Ma2019Antenna} addressed this with a geometric modeling approach and a random-search algorithm to find the optimal APO for relay deployment in tunnels. Their simulations showed that compared to the Gradient Descent Algorithm, the number of iterations for their algorithm decreased from 600 to 100, and the number of local optima decreased from 11 to 1. However, despite the contributions of CMIC-1NR, it can be summarized that all the CMIC-1NR techniques mentioned in this survey are significantly subjected to the APOs. Also, the CMG of CMIC-1NR is much smaller than that of CMIC-$n$AR.

With advances in MI fast fading research, researchers have focused on the CMI channel with random polarization gain
$J_{\mathrm{SD}}$. In 2024, Zhang \emph{et al.} \cite{Zhang2024Cooperative} investigated the CMIC systems with an unidirectional coil and a tri-directional active relay, respectively. Assuming that norm directions of coils $\mathrm{S}$,  $\mathrm{R}$, and $\mathrm{D}$ follow the uniform distributions, they derived closed-form expressions for the PDFs of received SNRs. Their simulations indicated that the ergodic rate increased from 6 bps/Hz to 16  bps/Hz at a distance of 20 m and a Tx power of 20 dBw, even using an AF relay.  However, for the terrestrial mobile MIC, it is obvious that the probability of a weak antenna vibration  $\theta'_{\mathrm{D}}\simeq0$ is much greater than that of a  strong one $\theta'_{\mathrm{D}}\simeq90^\circ$. Thus, the uniform distribution-based model deviates significantly from the feature of non-underwater mobile MICs.

\subsection{Summary and Lessons Learned} \label{sectsub5_sll}

Table \ref{tbl_relay} summarizes the key issues, methods, and remaining challenges in MI relay research, covering MI passive relay \cite{Sun2013Optimal}, MI active relay (CMIC) \cite{Zhang2024Cooperative}, and hybrid relay techniques \cite{Li2019Survey}. These studies show that both passive relay and CMIC techniques can significantly enhance channel capacity and extend the MIC range, sometimes by orders of magnitude.

The passive relay technique is energy-efficient with simple protocol complexity, but it faces deployment challenges, particularly in antenna alignment due to the negative crosstalk effect. A potential solution involves the use of an iron core coil to collect MI signals, although careful attention is needed to mitigate the effects of the eddy current. Additionally, implementing passive relays in RPMA-based systems is challenging due to minimal crosstalk effects. Moreover, passive relay and CMIC-$n$AR  techniques are unsuitable for mobile MICs due to their dynamic topology, which prevents the formation of stable systems with positive crosstalk effects (see~\eqref{eqn_chp3IdPassive} and Fig.~\ref{fig_sec3crosstalk}).
In contrast, CMIC significantly reduces the number of required relays. The CMIC-1NR solution, which eliminates the need for coil alignment, can be applied to RPMA-based systems and mobile MICs. We introduce the unexpected passive relay phenomenon, specifically the crosstalk effect in relays, and highlight its role in MI waveguides as a special case.

The practical takeaways or common pitfalls include: 1) A primary practical challenge in MI relay is antenna alignment. Thus, the MI waveguide for TTE suits only scenarios like straight tunnels, which is infeasible for mobile MICs; 2) increasing active relays does not guarantee throughput improvement; and 3) a high-density multi-node MI network exhibits the crosstalk. However, determining the spatial distribution of positive crosstalk remains challenging, complicating the passive relay  deployment.

\begin{table*}[htp] \scriptsize%\footnotesize
\centering
%	\tiny
\caption{Overview of related research on MIC in line with the  OSI framework  \\The marker `\hloitem'  describes the methods; The markers `\hllitem' and `\hlxitem' describe addressed and remaining issues,  respectively}
\label{tbl_osi}
\vspace{-0.9em}
\scalebox{0.89}{
\begin{threeparttable}
%	\adjustbox{max width=\textwidth, scale=0.85}{
\begin{tabular}{m{0.05\textwidth}<{\centering}|m{0.06\textwidth}<{\centering}|m{0.08\textwidth}<{\centering}|m{0.38\textwidth}<{\centering}|m{0.38\textwidth}<{\centering}|m{0.03\textwidth}<{\centering}}
\hline
\textbf{OSI Lys}  &  \textbf{Aspects} & \textbf{Refs.} & \textbf{Methods and addressed issues }&\textbf{Remaining issues}& \textbf{Priority} \\
\hline \hline
%% %% \multirow{NumberOfRows}{CellWidth\textwidth}[-Fromtop]{\centering Passive Relay}
%%%%%%%%%%%%%%%%%%%%%%%%%%%%%%%%%%%%%%%%%%%%%%%%%%%%%%%%%%%%%%%%%%%%%%%%%%%%%%%%%%%%%%%%%%%%%%%%%
\multirow{3}{0.03\textwidth}[-25pt]{\centering Ly1}
& {\hspace{-0.95em}P2P \& relay-based MICs} & \cite{Sun2010Magnetic, Ma2019Effect,Kisseleff2014Transmitter,Kisseleff2014Modulation,Chen2023Novel} & \begin{itemize}[leftmargin=*]
\item[\hllitem] {Channel modeling, channel capacity, MIC range, channel estimation, channel coding and modulation (see Sections  \ref{sect_sub2channel}, \ref{sect_p2p}, and \ref{sect_cmi})}
\item[\hloitem] {Detailed in Tables \ref{tbl_gain}, \ref{tbl_chp2P2P}, and \ref{tbl_relay}} 
\vspace{-1.0em}
\end{itemize}& \begin{itemize}[leftmargin=*]
\item[\hlxitem] {A universal  MI fast fading model; TTE MIC range; RPMA channel capacity; MI crosstalk effects; CMIC with MI fast fading}
\item[\hlxitem] {Detailed in Tables \ref{tbl_gain}, \ref{tbl_chp2P2P}, and \ref{tbl_relay} }
\vspace{-1.0em}
\end{itemize}& {-}\\
\cline{2-6}
%%%%%%%%%%%%%%%%%%%%%%%%%%%%%%%%%%%%%%%%%%%%%%%%%%%%%%%%%%%%%%%%%%%%%%%%%%%%%%%%%%%%%%%%%%%%%%%%%%%%%%%
&\multirow{2}{0.06\textwidth}[-8pt]{\centering{ Power allocation}}  & \cite{lin2015distributed,Singh2021Optimal} & \begin{itemize}[leftmargin=*]
\item[\hllitem] {Power allocation issue for a \textbf{stationary} MI channel} % for both direct and waveguide MICs
\item[\hloitem] {Nash game}
\vspace{-1.0em}
\end{itemize}& \begin{itemize}[leftmargin=*]
\item[\hlxitem] { Not suitable for  \textbf{time-varying channel} over a longer time span}
\item[\hlxitem] {Quantization-induced precision loss}
\vspace{-1.0em}
\end{itemize}&{\hlprioritem}\\	
\cline{3-6}
%%%%%%%%%%%%%%%%%%%%%%%%%%%%%%%%%%%%%%%%%%%%%%%%%%%%%%%%%%%%%%%%%%%%%%%%%%%%%%%%%%%%%%%%%%%%%%%%%%%%%%%
& & \cite{Ma2024Fast} & \begin{itemize}[leftmargin=*]
%\item[\hllitem] Giving the power control solution for the cellular MI network with MI fast fading channels due to the dramatically fluctuating average AVI
\item[\hllitem]{Power allocation issue for an \textbf{MI fast fading} channel} %due to the dramatically fluctuating average AVI
\item[\hloitem] {\textbf{Nash game-based multiagent RL}  with Bellman iteration}
\vspace{-1.0em}
\end{itemize} & \begin{itemize}[leftmargin=*]
\item[\hlxitem] {\textbf{Slow convergence} due to lack of information exchange;}    \item[\hlxitem] {\textbf{Sacrifice precision} for faster convergence}
\vspace{-1.0em}
\end{itemize}&{\hlprioritem\hlprioritem}\\   
\cline{2-6}
%%%%%%%%%%%%%%%%%%%%%%%%%%%%%%%%%%%%%%%%%%%%%%%%%%%%%%%%%%%%%%%%%%%%%%%%%%%%%%%%%%%%%%%%%%%%%%%%%%%%%%%
& {Frequency allocation} &\cite{Li2024Resource}&\begin{itemize}[leftmargin=*]
\item[\hllitem] {For \textbf{lower system delays} ($>40$\% decrease) for cluster-based UW-WSN}
\item[\hloitem] {Multi-variable alternating iterative resource allocation algorithm}
\vspace{-1.0em}
\end{itemize}
&\begin{itemize}[leftmargin=*]
\item[\hlxitem] {No considerations of network throughput and energy consumption}
\vspace{-1.0em}
\end{itemize}&{\hlprioritem\hlprioritem}\\
\cline{2-6}
\hline 
%%%%%%%%%%%%%%%%%%%%%%%%%%%%%%%%%%%%%%%%%%%%%%%%%%%%%%%%%%%%%%%%%%%%%%%%%%%%%%%%%%%%%%%%%%%%%%%%%%%%%%%
\multirow{3}{0.03\textwidth}[-15pt]{\centering Ly2}
& \multirow{2}{0.06\textwidth}[-3pt]{\centering MAC} & \cite{Ahmed2016Multi, Ahmed2019Design} & \begin{itemize}[leftmargin=*]
\item[\hllitem] {Low-cost (\$100); low energy (Currents: Rx/Tx=0.49/253 mA)}
\item[\hloitem] {Designing three packet types (reservation, acknowledge, data)}%and a state transition machine for MAC packets exchanging
\vspace{-1.0em}
\end{itemize} & \begin{itemize}[leftmargin=*]
\item[\hlxitem]{Lacking analysis for SISO case for VLF-LA}
\vspace{-1.0em}
\end{itemize}&\hlrevisioneq{\hlprioritem}\\
\cline{3-6}
%%%%%%%%%%%%%%%%%%%%%%%%%%%%%%%%%%%%%%%%%%%%%%%%%%%%%%%%%%%%%%%%%%%%%%%%%%%%%%%%%%%%%%%%%%%%%%%%%%%%%%%
&  & \cite{Ahmed2024Design} & \begin{itemize}[leftmargin=*]
\item[\hllitem] {Considering both  energy (Rx/Tx: 0.49mA/0.74mA) and throughput (46-144 bytes/cycle) }
\item[\hloitem] Contention-based protocol  using hybridizing three configurations of the orthogonal MIMO coils
\vspace{-1.0em}
\end{itemize}&\begin{itemize}[leftmargin=*]
\item[\hlxitem] {Low energy efficiency for  SISO-coil; low EPR; high collision probability} %for a higher EPR in a VLF-LA system 
\item[\hlxitem] {Too large MAC headers for  a  VLF-LA system} %requiring bit-level compression by jointing upper-layer headers (e.g., TCP/IP headers) 
\vspace{-1.0em}
\end{itemize}&\hlrevisioneq{\hlprioritem\hlprioritem\hlprioritem}\\
\cline{2-6}
%%%%%%%%%%%%%%%%%%%%%%%%%%%%%%%%%%%%%%%%%%%%%%%%%%%%%%%%%%%%%%%%%%%%%%%%%%%%%%%%%%%%%%%%%%%%%%%%%%%%%%%
& LLC & No \emph{refs.} & - & \begin{itemize}[leftmargin=*]
\item [\hlxitem] {Adaptive retransmission strategy for unnecessary packets}	
\item[\hlxitem] Compatibility   of existing solutions for MIC
\vspace{-1.0em}
\end{itemize}&{\hlprioritem}\\
\cline{2-4}
\hline
%%%%%%%%%%%%%%%%%%%%%%%%%%%%%%%%%%%%%%%%%%%%%%%%%%%%%%%%%%%%%%%%%%%%%%%%%%%%%%%%%%%%%%%%%%%%%%%%%%%%%%%
\multirow{5}{0.03\textwidth}[-90pt]{\centering Ly3}
& \multirow{3}{0.06\textwidth}[-16pt]{\centering Connectivity} & \cite{Sun2011Dynamic} & \begin{itemize}[leftmargin=*]
\item[\hllitem] {Dynamic connectivity for a \textbf{2D model}} %for AG-UG, UG-AG, AG-UG channels}
\item[\hloitem] {Designing framework of connectivity probability bounds}
\vspace{-1.0em}
\end{itemize}& \begin{itemize}[leftmargin=*]
\item[\hlxitem] {Only applicable to the 2D model}
\item[\hlxitem] {Disregarding attenuation differences with directions}
\vspace{-1.0em}
\end{itemize}&{\hlprioritem}\\
\cline{3-6}
%%%%%%%%%%%%%%%%%%%%%%%%%%%%%%%%%%%%%%%%%%%%%%%%%%%%%%%%%%%%%%%%%%%%%%%%%%%%%%%%%%%%%%%%%%%%%%%%%%%%%%%
&  & \cite{Gulbahar2012Communication} & \begin{itemize}[leftmargin=*]
\item[\hllitem]  $k$-connectivity for an underwater grid network 
\item[\hloitem] Derivation based on power and BER
\vspace{-1.0em}
\end{itemize} & \begin{itemize}[leftmargin=*]
\item[\hlxitem] {Disregarding attenuation differences with directions}
\vspace{-1.0em}
\end{itemize}&{\hlprioritem}\\
\cline{3-6}
%%%%%%%%%%%%%%%%%%%%%%%%%%%%%%%%%%%%%%%%%%%%%%%%%%%%%%%%%%%%%%%%%%%%%%%%%%%%%%%%%%%%%%%%%%%%%%%%%%%%%%%
&  & \cite{Zhang2015Effective, Zhang2017Connectivity} & \begin{itemize}[leftmargin=*]
\item[\hllitem] {\textbf{Randomly} deployed in a \textbf{3D} space, and  considering\textbf{ attenuation differences with directions.}}
\item[\hloitem] {Probability theorem; gradient descent method; homogeneous Poisson Point Process} 
\vspace{-1.0em}
\end{itemize}& \begin{itemize}[leftmargin=*]
\item[\hlxitem] {\textbf{Violating homogeneous condition} in  heterogeneous networks or mobile MIC.}          \vspace{-1.0em}
\end{itemize}&{\hlprioritem\hlprioritem}\\
\cline{2-6}
%%%%%%%%%%%%%%%%%%%%%%%%%%%%%%%%%%%%%%%%%%%%%%%%%%%%%%%%%%%%%%%%%%%%%%%%%%%%%%%%%%%%%%%%%%%%%%%%%%%%%%%
& \multirow{2}{0.06\textwidth}[-3pt]{\centering Data collection   and node  deployment} & \cite{Wang2018Data} &                     \begin{itemize}[leftmargin=*]
\item[\hllitem] Optimal data collection and network lifetime (21.2\%$\sim$38.3\% higher) with low energy consumption (0.2$\sim$0.34 J)
\item[\hloitem]{HENPC algorithm and ant colony optimization}
\vspace{-1.0em}
\end{itemize}&\begin{itemize}[leftmargin=*]
\item[\hlxitem] {Large delay caused by  ant colony algorithm}
\vspace{-1.0em}
\end{itemize}&{\hlprioritem\hlprioritem}\\
\cline{3-6}
%%%%%%%%%%%%%%%%%%%%%%%%%%%%%%%%%%%%%%%%%%%%%%%%%%%%%%%%%%%%%%%%%%%%%%%%%%%%%%%%%%%%%%%%%%%%%%%%%%%%%%%
&  & \cite{Wei2022Power} & \begin{itemize}[leftmargin=*]
\item[\hllitem] {May increasing the network lifespan (by $\frac{1}{31}$$\sim$$\frac{5}{18}$ ).} % with the same  energy.}
\item[\hloitem]  AANSFR algorithm %which alternately optimizes the AUV path planning and network data flow routing
\vspace{-1.0em}
\end{itemize} &\begin{itemize}[leftmargin=*]
\item[\hlxitem] {Non-guaranteed optimality due to the constraint of  \cite[Eq. 10(a)]{Wei2022Power}}
\vspace{-1.0em}
\end{itemize} &{\hlprioritem}\\
\cline{2-6}    
%%%%%%%%%%%%%%%%%%%%%%%%%%%%%%%%%%%%%%%%%%%%%%%%%%%%%%%%%%%%%%%%%%%%%%%%%%%%%%%%%%%%%%%%%%%%%%%%%%%%%%%
& \multirow{3}{0.06\textwidth}[-10pt]{\centering Routing} & \cite{Wang2019Efficient, Alsalman2021Balanced} & \begin{itemize}[leftmargin=*]
\item[\hllitem] Balancing the network  latency  (reduced by 18\% in \cite{Alsalman2021Balanced}), and  energy efficiency (increasing by 16\% in \cite{Alsalman2021Balanced})
\item[\hloitem]  Q-learning based energy-delay routing (QL-EDR) \cite{Wang2019Efficient} and  Balanced routing protocol based on Q-earning \cite{Alsalman2021Balanced} algorithms
\vspace{-1.0em}
\end{itemize}& \begin{itemize}[leftmargin=*]
\item[\hlxitem]{Overlooking the frequency features}
% of the MI channel
\item[\hlxitem]{Slow convergence of Q-learning}
\vspace{-1.0em}
\end{itemize}&{\hlprioritem\hlprioritem}\\
\cline{3-6}
%%%%%%%%%%%%%%%%%%%%%%%%%%%%%%%%%%%%%%%%%%%%%%%%%%%%%%%%%%%%%%%%%%%%%%%%%%%%%%%%%%%%%%%%%%%%%%%%%%%%%%%
&  & \cite{Liu2022Qlearning,Liu2024Frequency} & \begin{itemize}[leftmargin=*]
\item[\hllitem] {\textbf{Frequency-selective property} in a dynamic multilayer MI UG-WSN} %considering network throughput and energy consumption
\item[\hloitem] {Distributed Q-learning-based algorithm;
% to provide routing description; 
formulating frequency-switchable routing decision problem} 
\vspace{-1.0em}
\end{itemize} &\begin{itemize}[leftmargin=*]
\item[\hlxitem] {Slow convergence for Q-learning} 
\item[\hlxitem] {\textbf{Poor} efficiency of \textbf{routing tables exchange} in narrow-band MI channels} %TTE MIC with VLF-LA
\vspace{-1.0em}
\end{itemize}&{\hlprioritem\hlprioritem\hlprioritem}\\
\cline{2-6}      
%%%%%%%%%%%%%%%%%%%%%%%%%%%%%%%%%%%%%%%%%%%%%%%%%%%%%%%%%%%%%%%%%%%%%%%%%%%%%%%%%%%%%%%%%%%%%%%%%%%%%%%
& \multirow{3}{0.06\textwidth}[-6pt]{\centering Topology} & \cite{Sun2011Dynamic} & \begin{itemize}[leftmargin=*]
\item[\hloitem] Ad-hoc (high  self-organizing capability; limited scalability) 
\vspace{-1.0em}
\end{itemize} &-
& -\\
\cline{3-6}
%%%%%%%%%%%%%%%%%%%%%%%%%%%%%%%%%%%%%%%%%%%%%%%%%%%%%%%%%%%%%%%%%%%%%%%%%%%%%%%%%%%%%%%%%%%%%%%%%%%%%%%
&  & \cite{Alshehri2017Optimal} & \begin{itemize}[leftmargin=*]
\item[\hloitem] Linear topology (simple protocol; single point of failure) 
\vspace{-1.0em}
\end{itemize}&-&-\\
\cline{3-6}
%%%%%%%%%%%%%%%%%%%%%%%%%%%%%%%%%%%%%%%%%%%%%%%%%%%%%%%%%%%%%%%%%%%%%%%%%%%%%%%%%%%%%%%%%%%%%%%%%%%%%%%
&  & \cite{Ma2024Fast} & \begin{itemize}[leftmargin=*]
\item[\hloitem] Cellular topology (frequency reuse;  local network congestion)
\vspace{-1.0em}
\end{itemize}&\begin{itemize}[leftmargin=*]
\item[\hlxitem]{Basic resource allocation/reuse schemes}
\item[\hlxitem]{A basic MI cellular network protocol stack}
\vspace{-1.0em}
\end{itemize}&{\hlprioritem\hlprioritem\hlprioritem}\\
\cline{2-6}
%%%%%%%%%%%%%%%%%%%%%%%%%%%%%%%%%%%%%%%%%%%%%%%%%%%%%%%%%%%%%%%%%%%%%%%%%%%%%%%%%%%%%%%%%%%%%%%%%%%%%%%
& IP & No \emph{refs.}&- & \begin{itemize}[leftmargin=*]
\item[\hlxitem] {Low efficiency in TTE MIC due to the \textbf{large IP header} (over 20 bytes) and limited channel capacity (\textbf{IP-HC scheme required})}
\item[\hlxitem] {Compatibility   of existing solutions for MIC}
\vspace{-1.0em}
\end{itemize}&\hlrevisioneq{\hlprioritem\hlprioritem\hlprioritem}\\
\hline
%%%%%%%%%%%%%%%%%%%%%%%%%%%%%%%%%%%%%%%%%%%%%%%%%%%%%%%%%%%%%%%%%%%%%%%%%%%%%%%%%%%%%%%%%%%%%%%%%%%%%%%
Ly4  & TCP & No \emph{refs.} &- & \begin{itemize}[leftmargin=*]
\item[\hlxitem]{\textbf{Fairness} issue; \textbf{RTT suppression} issue; \textbf{congestion control} scheme for TCP connections w.r.t. the SNR in Rx MI nodes}
\item[\hlxitem] {Compatibility   of existing solutions for MIC}
\vspace{-1.0em}
\end{itemize}&{\hlprioritem\hlprioritem\hlprioritem}\\
\cline{2-4}
\hline
%%%%%%%%%%%%%%%%%%%%%%%%%%%%%%%%%%%%%%%%%%%%%%%%%%%%%%%%%%%%%%%%%%%%%%%%%%%%%%%%%%%%%%%%%%%%%%%%%%%%%%%
Ly5-7 & Applications & cf. Table  \ref{tbl_app} & - &\begin{itemize}[leftmargin=*]
\item[\hlxitem] {MCNSI, TTE infrared image transmission;  TTE IoV applications;  DARPA Subterranean Challenge}
\vspace{-1.0em}
\end{itemize} &{\hlprioritem\hlprioritem}\\
\hline
%%%%%%%%%%%%%%%%%%%%%%%%%%%%%%%%%%%%%%%%%%%%%%%%%%%%%%%%%%%%%%%%%%%%%%%%%%%%%%%%%%%%%%%%%%%%%%%%%%%%%%%
\multirow{2}{0.03\textwidth}[-10pt]{CLO}  
& \multirow{2}{0.06\textwidth}[-10pt]{\centering -} & \cite{lin2015distributed} & \begin{itemize}[leftmargin=*]
\item[\hllitem]  {\textbf{Statistical QoS guarantee} and obtaining both optimal energy savings and throughput gain concurrently}
\item[\hloitem]   {DEAP framework jointing Ly1, 
%(Modulation, FEC, and power control),
Ly2,
%(DS-CDMA)
and Ly3 }
%(geographical routing algorithm)
\vspace{-1.0em}
\end{itemize}&\begin{itemize}[leftmargin=*]
\item[\hlxitem]{ \textbf{Not suitable for the mobile MIC} in an MI fast fading channel due to its drastically fluctuating average AVI}
\vspace{-1.0em}
\end{itemize}&{\hlprioritem\hlprioritem}\\
\cline{3-6}
%%%%%%%%%%%%%%%%%%%%%%%%%%%%%%%%%%%%%%%%%%%%%%%%%%%%%%%%%%%%%%%%%%%%%%%%%%%%%%%%%%%%%%%%%%%%%%%%%%%%%%%
&  & \cite{Singh2021Optimal} & \begin{itemize}[leftmargin=*]
\item[\hllitem] Optimal energy efficiency and low computational complexity, with higher throughput than EDAP; % for both direct and waveguide MICs
\item[\hloitem] {Distributed Energy-throughput efficient Cross-Layer solution using Naked Mole
Rat algorithm (\textbf{DECN})}
\vspace{-1.0em}
\end{itemize}& \begin{itemize}[leftmargin=*]
\item[\hlxitem] { \textbf{Not suitable for the mobile MIC} in an MI fast fading channel due to its dramatically fluctuating average AVI}
\vspace{-1.0em}
\end{itemize}&{\hlprioritem\hlprioritem}\\
\cline{3-6}
%%%%%%%%%%%%%%%%%%%%%%%%%%%%%%%%%%%%%%%%%%%%%%%%%%%%%%%%%%%%%%%%%%%%%%%%%%%%%%%%%%%%%%%%%%%%%%%%%%%%%%%
\hline
\end{tabular}
\begin{tablenotes}  
\footnotesize  
\item[$\dagger$] Priority: Priority level of remaining issues and  proposed approaches for exploration. Here, low priority (\hlprioritem) indicates that:  1) the remaining issues have been explored in subsequent MIC literature; 2) the existing EMWC schemes  are compatible with MIC for these issues; or 3) exploring  these issues is optional.
\end{tablenotes}  
\end{threeparttable}
\vspace{-1.5em}
}
\end{table*}

\vspace{6 mm}
\section{MI Network and Its Architecture} \label{sect_network}

\begin{figure}[t!]
\centering{\includegraphics[width=69mm]{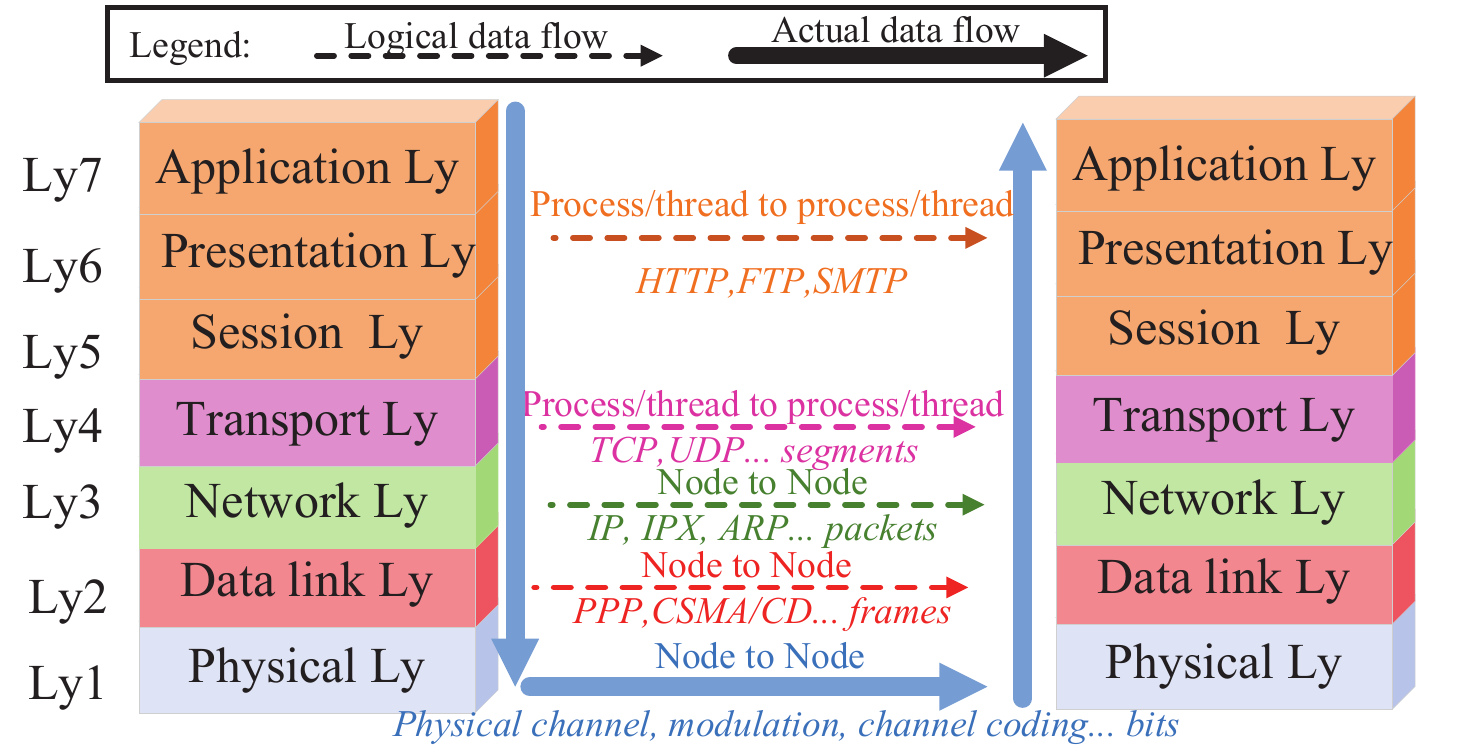}}
\caption{The OSI framework, which has seven layers.
}\label{fig_sec5osi} 
% \vspace{-1.0em}
\end{figure}

In wireless communications, key focal points include improving network throughput, increasing access to users, and reducing energy consumption. However, recent techniques in EMWC using signal reflection and refraction \cite{Xing2021Achievable, Xing2023Joint, Xing2024Efficient} are challenging to apply to MIC signals due to their lack of reflection and refraction.  Consequently, the role of MI network architecture becomes more important than that of the EMWC network. In this section, we categorize recent literature on multi-node MIC with reference to  the widely used OSI framework, 
%including the physical layer, data link layer, network layer, and application layer. 
The OSI framework (see Fig. \ref{fig_sec5osi}) originated from the International Organization for Standardization (ISO) in 1984. It serves as a conceptual framework for designing network communication protocols and facilitating communication between different systems.  %In this section, the issues and methodologies on the multi-node MIC network are arranged based on the OSI framework, as listed in Table \ref{tbl_osi}. 
The research on the MI network and differences from other communication networks are summarized in  Tables \ref{tbl_osi} and~\ref{tbl_osidiff}, respectively,  with more details provided below.

\subsection{Physical Layer (Ly1)} \label{sectsub6_ly1}

\begin{table*}[htp] \scriptsize%\footnotesize
\centering
%	\tiny
\caption{Comparison of  wave-based and MI communications based on the OSI framework\\ The markers `\hlwiitem', `\hlacitem' and `\hlmiitem'  describe EMWC,   acoustic communication and MIC,  respectively}
\label{tbl_osidiff}
\vspace{-1.0em}
%	\scalebox{0.9999}{
\begin{threeparttable} 
%	\adjustbox{max width=\textwidth, scale=0.85}{
\begin{tabular}{m{0.03\textwidth}<{\centering}|m{0.12\textwidth}<{\centering}|m{0.05\textwidth}<{\centering}|m{0.33\textwidth}<{\centering}| m{0.35\textwidth}<{\centering} }
\hline
\textbf{OSI Lys}  &  \textbf{Aspects} & \textbf{Typical \emph{refs.}} & \textbf{Other communications compared to MIC} & \textbf{MICs}  \\
\hline \hline
%% %% \multirow{NumberOfRows}{CellWidth\textwidth}[-Fromtop]{\centering Passive Relay}
%%%%%%%%%%%%%%%%%%%%%%%%%%%%%%%%%%%%%%%%%%%%%%%%%%%%%%%%%%%%%%%%%%%%%%%%%%%%%%%%%%%%%%%%%%%%%%%%%%%%%%%
\multirow{3}{0.03\textwidth}[-15pt]{\centering Ly1}
& Channel estimation & \cite{Kisseleff2014Transmitter} & \begin{itemize}[leftmargin=*]
\item[\hlwiitem] Estimation not based mutual inductance
\item[\hlwiitem] Overlooking the medium conductivity 
\vspace{-1.0em}
\end{itemize} 
& 
\begin{itemize}[leftmargin=*]
\item[\hlmiitem] Estimation based mutual inductance
\item[\hlmiitem] Significantly influenced by medium conductivity 
\vspace{-1.0em}
\end{itemize} 
\\
\cline{2-5}
%%%%%%%%%%%%%%%%%%%%%%%%%%%%%%%%%%%%%%%%%%%%%%%%%%%%%%%%%%%%%%%%%%%%%%%%%%%%%%%%%%%%%%%%%%%%%%%%%%%%%%%
& Modulation & \cite{Kisseleff2014Modulation} & \begin{itemize}[leftmargin=*]
\item[\hlwiitem]  Negligible frequency dependent eddy currents effect
\item[\hlwiitem]  Larger frequency  bandwidth
\vspace{-1.0em}
\end{itemize}

&  \begin{itemize}[leftmargin=*]
\item[\hlmiitem]  Non-Negligible frequency dependent eddy currents effect
\item[\hlmiitem]  Much smaller frequency bandwidth
\vspace{-1.0em}
\end{itemize}
\\
\cline{2-5}
%%%%%%%%%%%%%%%%%%%%%%%%%%%%%%%%%%%%%%%%%%%%%%%%%%%%%%%%%%%%%%%%%%%%%%%%%%%%%%%%%%%%%%%%%%%%%%%%%%%%%%%
& Channel coding & \cite{Chen2023Novel} & \begin{itemize}[leftmargin=*]
\item [\hlwiitem] Negligible conductor on the propagation path
\vspace{-1.0em}
\end{itemize} & \begin{itemize}[leftmargin=*]
\item [\hlmiitem] Variable  codeword duration from underground conductors 
\vspace{-1.0em}
\end{itemize} 
\\
\cline{2-5}

%%%%%%%%%%%%%%%%%%%%%%%%%%%%%%%%%%%%%%%%%%%%%%%%%%%%%%%%%%%%%%%%%%%%%%%%%%%%%%%%%%%%%%%%%%%%%%%%%%%%%%%
& {Resource allocation} & \cite{Ma2024Fast} & \begin{itemize}[leftmargin=*]
\item [\hlwiitem] {Sufficient bandwidth for  allocation table exchange}
\vspace{-1.0em}
\end{itemize} & \begin{itemize}[leftmargin=*]
\item [\hlmiitem] {Insufficient bandwidth  for  allocation table exchange}
\vspace{-1.0em}
\end{itemize} 
\\
\cline{2-4}
\hline
%%%%%%%%%%%%%%%%%%%%%%%%%%%%%%%%%%%%%%%%%%%%%%%%%%%%%%%%%%%%%%%%%%%%%%%%%%%%%%%%%%%%%%%%%%%%%%%%%%%%%%%
\multirow{2}{0.03\textwidth}[-3pt]{\centering Ly2}
& MAC & \cite{Ahmed2024Design} & \begin{itemize}[leftmargin=*]
\item[\hlwiitem] Often ignoring the directional nature of EMW
\vspace{-1.0em}
\end{itemize} 
& 
\begin{itemize}[leftmargin=*]
\item[\hlmiitem] Requiring further bit-level compression for MAC headers 
\item[\hlmiitem] With directional nature of magnetic fields

\vspace{-1.0em}
\end{itemize} 
\\
\cline{2-5}
%%%%%%%%%%%%%%%%%%%%%%%%%%%%%%%%%%%%%%%%%%%%%%%%%%%%%%%%%%%%%%%%%%%%%%%%%%%%%%%%%%%%%%%%%%%%%%%%%%%%%%%
& LLC & \cite{Daladier2009Adaptive}$^\dagger$ & \begin{itemize}[leftmargin=*]
\item [\hlacitem] Acoustic channel quality based retransmission scheme
\vspace{-1.0em}
\end{itemize} &
\begin{itemize}[leftmargin=*]
\item [\hlmiitem]  MI channel quality based retransmission scheme$^\dagger$
\vspace{-1.0em}
\end{itemize}
\\
\cline{2-5}
\hline
%%%%%%%%%%%%%%%%%%%%%%%%%%%%%%%%%%%%%%%%%%%%%%%%%%%%%%%%%%%%%%%%%%%%%%%%%%%%%%%%%%%%%%%%%%%%%%%%%%%%%%%
\multirow{4}{0.03\textwidth}[-6pt]{\centering Ly3} & Connectivity&\cite{Zhang2017Connectivity}&
\begin{itemize}[leftmargin=*]
\item [\hlwiitem]  Negligible attenuation differences with directions 
\vspace{-1.0em}
\end{itemize} & \begin{itemize}[leftmargin=*]
\item [\hlmiitem]  Attenuation differences with directions 
\vspace{-1.0em}
\end{itemize} \\
\cline{2-5}
%%%%%%%%%%%%%%%%%%%%%%%%%%%%%%%%%%%%%%%%%%%%%%%%%%%%%%%%%%%%%%%%%%%%%%%%%%%%%%%%%%%%%%%%%%%%%%%%%%%%%%%
& Deployment&\cite{Sun2013Optimal,Kisseleff2013Interference}&
\begin{itemize}[leftmargin=*]
\item [\hlwiitem] Higher protocol complexity
\item [\hlwiitem] Less APO dependent
\vspace{-1.0em}
\end{itemize} & \begin{itemize}[leftmargin=*]
\item [\hlmiitem]  Full utilization of MI waveguide advantages
\item [\hlmiitem]  Higher APO dependent
\vspace{-1.0em}
\end{itemize} \\
\cline{2-5}
%%%%%%%%%%%%%%%%%%%%%%%%%%%%%%%%%%%%%%%%%%%%%%%%%%%%%%%%%%%%%%%%%%%%%%%%%%%%%%%%%%%%%%%%%%%%%%%%%%%%%%%
& Routing&\cite{Liu2024Frequency}&
\begin{itemize}[leftmargin=*]
\item [\hlwiitem] Less frequency-selective and APO-selective
\vspace{-1.0em}
\end{itemize} & \begin{itemize}[leftmargin=*]
\item [\hlmiitem]  Significant frequency-selective and APO-selective
\vspace{-1.0em}
\end{itemize} \\
\cline{2-5}
%%%%%%%%%%%%%%%%%%%%%%%%%%%%%%%%%%%%%%%%%%%%%%%%%%%%%%%%%%%%%%%%%%%%%%%%%%%%%%%%%%%%%%%%%%%%%%%%%%%%%%%
& IP&\cite{Parrein2023Internet}$^\dagger$&
\begin{itemize}[leftmargin=*]
\item [\hlacitem] IP-HC scheme using acoustic network information
\vspace{-1.0em}
\end{itemize} & \begin{itemize}[leftmargin=*]
\item [\hlmiitem] IP-HC scheme depending on MI channel$^\dagger$
\vspace{-1.0em}
\end{itemize} \\
\hline
%%%%%%%%%%%%%%%%%%%%%%%%%%%%%%%%%%%%%%%%%%%%%%%%%%%%%%%%%%%%%%%%%%%%%%%%%%%%%%%%%%%%%%%%%%%%%%%%%%%%%%%
Ly4 & TCP &\cite{Matsuda2002Performance}$^\dagger$&
\begin{itemize}[leftmargin=*]
\item [\hlwiitem] Less RTT suppression
\vspace{-1.0em}
\end{itemize} & \begin{itemize}[leftmargin=*]
\item [\hlmiitem] Significant RTT suppression$^\dagger$
\vspace{-1.0em}
\end{itemize} \\
\hline
%%%%%%%%%%%%%%%%%%%%%%%%%%%%%%%%%%%%%%%%%%%%%%%%%%%%%%%%%%%%%%%%%%%%%%%%%%%%%%%%%%%%%%%%%%%%%%%%%%%%%%%
CLO & - &\cite{lin2015distributed, Ma2024Fast}&
\begin{itemize}[leftmargin=*]
\item [\hlwiitem]  $G_{\mathrm{SD}}\propto d^{-2}_{\mathrm{SD}}$ 
\item [\hlwiitem] Stable and determined moments of fast fading gain
\vspace{-1.0em}
\end{itemize} & \begin{itemize}[leftmargin=*]
\item [\hlmiitem] $G_{\mathrm{SD}}\propto d^{-6}_{\mathrm{SD}}e^{-d_{\mathrm{SD}}/\delta_{\rm u}}$
\item [\hlmiitem]  Random  moments of MI fast fading gain
\vspace{-1.0em}
\end{itemize} \\
\hline
%%%%%%%%%%%%%%%%%%%%%%%%%%%%%%%%%%%%%%%%%%%%%%%%%%%%%%%%%%%%%%%%%%%%%%%%%%%%%%%%%%%%%%%%%%%%%%%%%%%%%%%
\end{tabular}
\begin{tablenotes}  
\footnotesize  
\item[] The superscript $^\dagger$ indicates the issue lacking relevant literature on MIC-related research and MIC solutions, to the best of our knowledge, to date.
\end{tablenotes}  
\end{threeparttable}
\vspace{-1.31em}
%	}
\end{table*}

Most research on MIC  focused on the physical layer issues, including channel modeling,  performance,  estimation, modulation, coding, and resources allocations. We have discussed the issues of  physical layer schemes on P2P and MI relay-based system, i.e., channel modeling, key performance metrics, {channel estimation, modulation, and channel coding  in Sections~\ref{sect_p2p} and \ref{sect_cmi}.
For the MI physical layer with multiple nodes, the focus has been primarily on the resource allocations, including power allocation\cite{Ma2024Fast,lin2015distributed,Li2024Resource}, as well as frequency and bandwidth allocations\cite{Li2024Resource}.
% The capacity of MIC  is much lower than that of EMWC,  requiring additional bandwidth for network control frames (\emph{e.g.}, CSI). This is a common issue that is not limited to the MI physical layer. 

\subsubsection{Power control} \label{sectsubsub6_pcc}
From the network's perspective,  power control aims to optimize signals, reduce interference, and enhance efficiency.  Lin el al.~\cite{lin2015distributed} pioneered the study of power optimization formulation from the entire multi-hop network. To formulate the power control problem, they used the Nash game with   the utility function as shown in Table \ref{tbl_sec5resallocation}.
%	$\mathfrak{G}(\{1, ..., i, ..., n\}, \mathbf{P}_i, u_i)$ with the utility function}
The study presented in \cite{lin2015distributed} did not consider mobile MI networks with the  unpredictable AVI inputs. To address this, Ma \emph{et al.} \cite{Ma2024Fast} proposed a power allocation algorithm for cellular VMI networks. They used the Nash game-based RL (Q-learning) with   the utility function, as shown in Table \ref{tbl_sec5resallocation}. 
%\begin{equation}\label{eqn_chp5Qkt1}
%	\begin{aligned}
%		\hspace{-0.6em}Q_i^{(t\!+\!1)}(s_i, p_i) &\!:=\!Q_i^{(t)}(s_i, p_i)\!+\!\alpha^{(t)}\big[u_i(s_i,  p_i, \mathbf{p}_{-i})   \\
%		&  +\!\gamma_i \max\limits_{p'_i\in\mathbf{P}_i}Q^{(t)}_i(s'_i, p'_i)\!-\!Q_i^{(t)}(s_i, p_i )\big],
%	\end{aligned}
%	\vspace{-0.0em}
%\end{equation}
%where \hlrevision{$\alpha^{(t)}$ denotes the learning rate, $t$ represents the time slot, $s_i$ denotes the current state, $(\cdot)'$ represents the next state, the utility function $u_i$ follows a similar form to ~\eqref{eqn_chp5ui}, where the  data rate $\mathcal{C}_i(p_i, p_{-i})$ is replaced by the ergodic  rate $\mathbb{E}[\mathcal{C}_i(p_i, p_{-i})|\sigma^{(t)}_i]$ under AVI  $(\sigma^{(t)}_i)^2$ during time slot $t$.}
The RL solution from \cite{Ma2024Fast} addresses the problem of the unpredictable AVI inputs in MIC systems. 

Due to the  convergence limit, it can be difficult for the RL algorithm to obtain a high-precision power allocation policy. 
%Thus, the deep RL can be considered to address this precision issue. 
Moreover, since this power allocation algorithm requires no information exchange among the nodes, it is feasible for low-capacity channels. By contrast, the expectation of EMW fast fading is predictable in most cases; in this sense, RL may not be needed for an EMW cellular network.

\subsubsection{Frequency and bandwidth allocation} \label{sectsubsub6_fbas}
Effective frequency/bandwidth allocation  can  enhance capacity, QoS, and cut the network energy consumption.  Due to the narrow MI bandwidth and limited research on MI cellular topology, frequency/bandwidth allocation has garnered limited attention. Nonetheless, Li \emph{et al.}\cite{Li2024Resource} studied   bandwidth allocation involved schemes, applied to the AUVs and cluster-based UW-WSNs.  A joint optimization was proposed to obtain allocations of Tx power, bandwidth, and computational resources. The key goal (see Table \ref{tbl_sec5resallocation}) was to minimize total system delay $T$. 
The authors developed a centralized multi-variable alternating iterative resource allocation algorithm.  The core of the algorithm lies in its two-stage iterative process within a single loop, including obtaining power allocations with fixed bandwidth, and  obtaining bandwidth  allocations with fixed power. Although this study greatly reduces the system delay  by up to 40\%,   it does not account for frequency allocation for the performance improvement of the capacity, QoS, and network energy consumption.

Table \ref{tbl_sec5resallocation} compares the  three resource allocation schemes discussed above. The first two methods suit low-speed channels, minimize inter-user information exchange but converge slowly, where the second, though better for dynamic channels, converges extremely slowly. The third, with fast convergence, fits higher-speed networks (e.g., hybrid MIC).

\begin{table*}[htp] \scriptsize%\footnotesize
\centering
%	\tiny
\caption{Comparison of three resources allocation methods on MIC research}
\label{tbl_sec5resallocation}
\vspace{-0.9em}
\scalebox{0.96}{
\begin{threeparttable} 
%	\adjustbox{max width=\textwidth, scale=0.85}{
\begin{tabular}{m{0.03\textwidth}<{\centering}|m{0.28\textwidth}<{\centering}|m{0.51\textwidth}<{\centering}|m{0.12\textwidth}<{\centering} }
\hline
\textbf{\emph{Refs.}} & \textbf{Goal function / utility function} & \textbf{ Explanations} & \textbf{Methods}   \\
\hline \hline
%% %% \multirow{NumberOfRows}{CellWidth\textwidth}[-Fromtop]{\centering Passive Relay}
\hline
\cite{lin2015distributed}&$u_i(p_i, p_{-i})$$=$$\! \mathcal{C}_i(p_i, p_{-i}) - w\mathfrak{E}_i(p_i, p_{-i}) $ & \textbf{Power allocation for stationary MICs:}  $\mathcal{C}$$=$  data rate, $\mathfrak{E}$$=$ total power, $i$$=$this user, $-i$$=$competing  users, $p$$=$power allocations, $w$$=$weight& Nash game\\ \hline
%%%%%%%%%%%%%%%%%%%%%%%%%%%%%%%%%%%%%%%%%%%%%%%%%%%%%%%%%%%%%%%%%%%%%%%
\cite{Ma2024Fast}&$u_i(p_i, p_{-i})$$=$$\mathbb{E}[\mathcal{C}_i(p_i, p_{-i})|\sigma^{(t)}_i]$ & \textbf{Power allocation for mobile MICs:}  $\mathbb{E}[\mathcal{C}]$$=$ergodic data rate,  $i$$=$this user, $-i$$=$competing  users, $p$$=$power allocations, $\sigma^{(t)}$$=$average AVI  during time slot $t$ &Joint Nash game and RL algorithm\\ \hline
%%%%%%%%%%%%%%%%%%%%%%%%%%%%%%%%%%%%%%%%%%%%%%%%%%%%%%%%%%%%%%%%%%%%%%%%%%%%%%
\cite{Li2024Resource}&	$\min\limits_{\mathbf{P}_{\mathrm{UA}}, \mathbf{P}_{\mathrm{CH}}, \mathbf{B}, \mathbf{F}} T$ & \textbf{Bandwidth allocation for  MICs:}  $T$$=$system delay, $\mathbf{P}$$=$power allocations, $\mathbf{B}$$=$bandwidth allocations, $\mathbf{F}$$=$computational resources, UA$=$acoustic node, CH$=$MI cluster header & Centralized optimization \\ \hline					
\end{tabular}
\end{threeparttable}
}
\vspace{-0.0em}
\end{table*}

\subsection{Data Link Layer (Ly2)} \label{sectsub6_ly2}

The data link layer is responsible for ensuring reliable data transfer between adjacent network nodes through functions such as error detection, error correction, and flow control. It is divided into the MAC and logical link control (LLC) sub-layers. While only the MAC sub-layer is channel-dependent. 

\subsubsection{MAC} \label{sectsubsub6_mac}

The low bandwidth of an MIC link presents challenges for real-time MAC protocol implementation. Also, many EMWC MAC solutions cannot be directly applied to MI UG-WSN  due to the directional nature of magnetic fields~\cite{Ahmed2016Multi}.  Ahmed \emph{et al}. \cite{Ahmed2016Multi} provided valuable insights into energy-efficient MI-based MAC protocols.  They designed three packet types, i.e., reservation, acknowledge, and data packets, and a general state transition machine for the MAC packets exchanging. They also designed an MI device with this MAC protocol that achieves less than \$100 in cost, 60 $\mu$A in the sleep mode, 0.49 mA in the Rx mode and 253 mA in the Tx mode. However, this work did not consider the throughput.
Recently, in \cite{Ahmed2024Design}, they proposed an improved MAC protocol and algorithm that considers detailed MI channel parameters.  This work balanced the energy and throughput performance metrics by hybridizing three configurations of the orthogonal MIMO coils, each corresponding to a different packet type (as detailed in Fig.~\ref{fig_sec5mac}). They applied the CSMA-based scheme, and pointed out that their MI MAC protocol exhibits low energy efficiency under the SISO-coil-based MIC. That is, it may encounter problems with the MIC that employs VLF-LA.  Furthermore, the large MAC header (8 bytes) can be further compressed at the bit-level for the VLF-LA channel.

\subsubsection{LLC} \label{sectsubsub6_llc}
The LLC sub-layer is often only a thin adaptation sub-layer that provides the reliability of communication, e.g., through data transfer, flow control, and error control. Since the LLC sub-layer is designed to be independent of the physical layer and media in the OSI model, it has received limited attention in MIC research. Only a few studies have explored LLC protocols for other communication channels (e.g., acoustic communication channel~\cite{Daladier2009Data,Daladier2009Adaptive} and quantum communication~\cite{Dahlberg2019Link}). Specifically, Daladier \emph{et al.} \cite{Daladier2009Data,Daladier2009Adaptive} proposed the SW-MER protocol to enhance throughput and reliability by employing an adaptive retransmission strategy that takes advantage of acoustic channel quality to minimize redundant packet transmissions. This approach can serve as a reference LLC for MIC, as both acoustic and MIC channels share low data rate characteristics.

\begin{figure}[t]
\centering
\subfigure[Wakeup (W) packet, acknowledgment (ACK) packet, and data packet ]{
\label{fig_sec5mac:pkt} %% label for second subfigure
\includegraphics[width=3.3in]{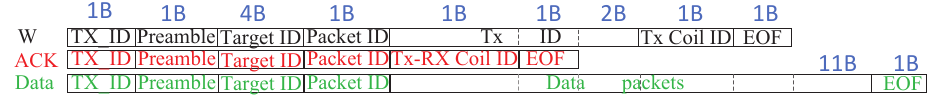}}    
\subfigure[Time slots]{
\label{fig_sec5mac:time} %% label for second subfigure
\includegraphics[width=3.30in]{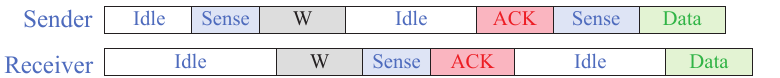}}    
\caption{A typical MI MAC solutions in \cite{Ahmed2024Design}. (a) The packets are designed into three types. (b) Complete communication cycle between a Tx and Rx, where the Tx node starts with a W packet; the Rx node, after successfully receiving the W packet, acknowledges with an ACK packet; upon receiving the	ACK packet, the Tx node then sends out the whole data packet~\cite{Ahmed2024Design}. However, for TTE MICs, the packet size should be further compressed.}
\label{fig_sec5mac} %% label for entire figure
\vspace{-0.9em}
\end{figure}

\subsection{Network Layer (Ly3)} \label{sectsub6_ly3}

The network layer routes data packets efficiently to their destinations. In this subsection, we introduce the studies on the functionality of the MI network, including connectivity, data collection, node deployment, and MIC routing. Here,  routing in MIC has recently represented a new research aspect.

%In the past five years, researchers have focused on designing the MI network protocol and addressing issues of connectivity, routing, and node deployment.

\subsubsection{Connectivity} \label{sectsubsub6_connectivity}
Connectivity is a cornerstone of networks and is crucial for designing network layer protocols, including routing algorithms, traffic control, and quality of service (QoS). It has been widely studied in EMWC research. For $n$ network nodes uniformly distributed within a circular area of $\pi \mathcal{R}^2$,  Gupta \emph{et al.} \cite{Gupta1998Critical} derived  $\mathcal{R} (n)$$=$$\sqrt{\frac{\log(n)+c(n)}{\pi n} }$, where $c(n)$ is a correction function. They proved that these nodes are connected with probability one if $\lim\limits_{n\rightarrow\infty}c(n)$$\rightarrow$$\infty$.  The connectivity analysis differs fundamentally between MIC and EMWC. As the MIC is immune to shadowing, the geometric disk/sphere models used in  EMWC network studies \cite{Gupta1998Critical, Wattenhofer2001Distributed, Bettstetter2002On, Wang2013Mobility} are inadequate for the MIC channel. 

Since 2011, there has been little literature on the connectivity in the MIC field.  Sun \emph{et al.} \cite{Sun2011Dynamic} conducted a 2-D connectivity analysis of UG-WSN using probability derivation, yet overlooked attenuation differences. Gulbahar and Akan \cite{Gulbahar2012Communication} performed a $k$-connectivity analysis on an underwater MI grid network with Tx coils at the fixed positions and directional angles. Zhang \emph{et al.} \cite{Zhang2017Connectivity} performed a connectivity analysis of TTE MI UG-WSN and proposed an optimization algorithm to address the connectivity adaptability of frequent frequency-switching w.r.t. APO-switching. For the mobile MIC, the MI fast fading results in an irregular shape of the MI coverage space.In heterogeneous MI UG-WSNs, the nodes are not homogeneous. These features violate their assumptions in the connectivity model (e.g., Poisson point process) and represent open issues.

\subsubsection{Deployment strategic and data collection}  \label{sectsubsub6_dsdc}

The strategic deployment and data collection of MI sensors are also essential for effective local network design. Fig. \ref{fig_chp2HField} and Eq.~\eqref{eqn_chp2Gsd} show that the APOs significantly impact the polarization gain $J_{\mathrm{SD}}$. Optimizing APO is a distinct issue from EMWC~\cite{Kisseleff2013Interference, zhang2014cooperative, Ma2019Effect, Ma2019Antenna}. Specifically, Kisseleff \emph{et al.} \cite{Kisseleff2013Interference} optimized the antenna deployment to avoid channel interference from other nodes. Recently, focus has been placed on node deployment for efficient data collection and routing protocols. For a 3D-UW-WSN, Wang \emph{et al.} designed an optimal deployment strategy and a clustering algorithm to prolong the network lifetime by 38.3\% and 21.2\% compared to the EELEACH-C algorithm~\cite{Wang2018Data}. This algorithm is also compatible with the 3D-UG-WSN. Wei \emph{et al.} \cite{Wei2022Power} studied the power-efficient AUV data collection schemes in an MI and acoustic hybrid sensor. They proposed the alternating anchor nodes selection and flow routing (AANSFR) for AUV data collection, where the 3dB MI bandwidth is considered. Using AANSFR, the lifespan could increase from 14 hours to 18 hours.

\subsubsection{Routing}  \label{sectsubsub6_routing}

\begin{figure}[t!]
\centering{\includegraphics[width=70mm, height=25mm]{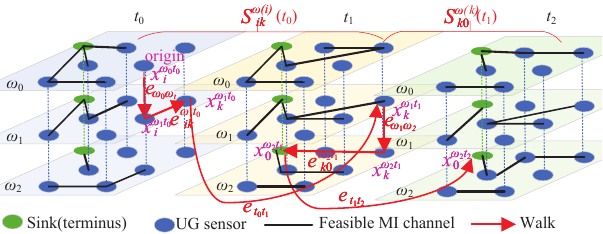}}
\caption{Schematic of a walk (arrow trajectories) for the frequency-selective  MI links (from sensor $i$, through sensor $k$, to the sink 0)\cite{Liu2024Frequency}. In this figure, $\omega$ denotes the operating frequency; $t$ denotes the time; $x$ denotes the node state;  $S(\cdot)$ denotes a step.  The objective of the optimization problem is to maximize the total capacity of related walks.
}\label{fig_sec4routing} 
\vspace{-0.0em}
\end{figure}

Although routing design is a kernel function model in the network layer,  studies on routing in MI UW-WSNs and UG-WSNs are limited. From~\eqref{eqn_chp2Gsd}, it is observed that  circuit gain $\mathcal{C}_{\mathrm{SD}}(f)$ and eddy gain $\mathcal{E}_{\mathrm{SD}}(f)$ are the functions of carrier frequency $f$. These gains represent the frequency-selective property. Compared to traditional routing strategies, this property makes the MI UG-WSN appear to have entirely different topological structures at different operating frequencies~\cite{Liu2024Frequency}. Likewise, as the MIC based on VLF-LA is orientation sensitivity (called APO-selective property), the mobile MI-UG-WSN can also encounter different topological structures due to the unpredictable moment of MI fast fading gain $J_{\mathrm{SD}}$.  

For the frequency-selective property,   Liu \textit{et al.} \cite{Liu2022Qlearning, Liu2024Frequency} proposed the frequency-switchable strategies for routing decisions, using a distributed Q-learning-based algorithm. Specifically, in \cite{Liu2022Qlearning}, they mapped the frequency-switchable MI UG-WSN to a multilayer network and proposed a distributed Q-learning-based algorithm to provide a description on its routing decision. They also evaluated the convergence of the algorithm and network lifetime. As Q-learning is time-consuming, in \cite{Liu2024Frequency}, they redefine the routing decision problem with frequency switchability in dynamic MI-WUSNs (see Fig.~\ref{fig_sec4routing}). Their simulations showed that the throughput increased from 40 to over 45 when the connectivity is 1.  

Improving network lifetime and reducing the transmission delay are two conflicting goals in routing studies. To balance these goals, Wang \emph{et al.} \cite{Wang2019Efficient} proposed two algorithms based on RL, \emph{i.e.}, the QL-EDR algorithm, and the path selection strategy algorithm.  Alsalman \emph{et al.} \cite{Alsalman2021Balanced} proposed a balance routing protocol based on machine learning (BRP-ML)  to reduce network latency and energy consumption. In these studies \cite{Liu2024Frequency, Wang2019Efficient, Alsalman2021Balanced, Liu2022Qlearning},  the eddy gain $\mathcal{E}_{\mathrm{SD}}$ and polarization gain $J_{\mathrm{SD}}$ are ignored. Moreover, their centralized-based multiagent RL algorithms require the exchange of band tables among nodes. This poses a practical challenge due to the low data rate in the MIC with VLF-LA.

\subsubsection{Internet Protocol (IP)}  \label{sectsubsub6_ip}

In the TCP/IP framework,  the IP, including IPv4\cite{rfc791ip} and IPv6\cite{rfc8200ip}, was originally designed to be channel-independent. To the best of our knowledge,  there has been no literature specifically addressing the issues related to IP in the field of the MIC. However,  the most significant challenge is IP's low efficiency in TTE MIC due to the large IP header (over 20 bytes) and the limited channel capacity  of TTE MI links.   In the fields of other communication techniques, such as bandwidth-constrained EMWC\cite{Jia2022End} and acoustic communication\cite{Parrein2023Internet}, this challenge was mitigated or addressed via IP header compression schemes. Specifically, Parrein \emph{et al.}~\cite{Parrein2023Internet}  proposed an acoustic protocol based on the static context header compression (SCHC) protocol  to reach the header compression ratio of 99.74\% using lower-layers' information. This also can be applied to the MIC.

\subsection{Transport Layer (Ly4) and TCP} \label{sectsub6_ly4}
The transport layer, akin to the LLC sub-layer and including TCP \cite{rfc793tcp}, ensures end-to-end communication with reliable data transfer, flow control, and error recovery. It was originally designed to be channel-agnostic. There is limited literature on MI transport layer research. 
%\subsubsection{\hlrevision{TCP optimization schemes}}
Particularly for TCP, fairness and congestion control issues pose a challenge and can be potentially optimized utilizing the characteristics of the specific channel. For example, TCP connections with shorter round-trip times (RTTs) often hinder the throughput of those with longer RTTs (i.e., RTT suppression issue). In the EMWC field, these issues have received some  attention, e.g., \cite{Matsuda2002Performance, Xu2020When}. In \cite{Matsuda2002Performance}, the congestion window was modeled as a function of the SNR of the EMW link, suggesting that one can formulate the corresponding optimization problem of the RTT for TCP w.r.t. the MI channel power gain $G_{\mathrm{SD}}$.

Besides the TCP schemes, in the Internet of Things (IoT), constrained devices proliferate under strict power/bandwidth limits, motivating researchers to develop IoT-specific protocols with transport layer adaptations, %--ones also applicable to TTE MIC systems, 
such as the Constrained Application Protocol (CoAP)\cite{rfc7252}, Delay-Tolerant Networking (DTN)\cite{rfc4838}, and ROBust Header Compression\cite{Jonsson2004RObust}. The lightweight design of CoAP  and its support for unreliable transport (established on UDP) align with the low-power requirements of TTE MIC devices.  The bundle protocol of DTN can  mitigate challenges posed by intermittent connectivity arising from  low bandwidth and data rate. These protocols and mechanisms, validated through established RFC standards,  offer potential directions for adapting transport-layer or cross-layer functionalities in TTE MIC networks.

\subsection{Cross-layer Optimization (CLO)} \label{sectsub6_clo}

The OSI framework promotes modularity and standardization with high cohesion and low coupling. However, it faces efficiency challenges in MI networks due to strict bandwidth, energy, and latency constraints. In this scenario, the lower layer may call upper-layer parameters or algorithms to improve the network performance.  Notably, for UG-WSNs, the solution for optimizing energy consumption in Ly1 may require the routing decision results in Ly3.  

Lin \emph{et al.} \cite{lin2015distributed} developed a distributed environment-aware protocol (DEAP) based on the Nash game to satisfy the statistical quality of service (QoS) requirement. This protocol achieved both optimal energy savings and throughput gain concurrently in 2015. The DEAP considers the interactions of different layer functionalities, such as power control, modulation and FEC in Ly1, the distributed MAC schemes in Ly2 and geographical routing algorithm in Ly3. 

In 2021, Singh \emph{et al.} \cite{Singh2021Optimal} developed a Distributed Energy-Throughput Efficient Cross-layer solution using the Naked Module Rat algorithm (NMRA), also called the DECN approach.  While this approach is similar to DEAP one, the DECN approach can apply to direct and waveguide MI links.  
Their simulations showed that when using the DECN approach with 50 nodes, the energy consumption decreased from 160$\sim$390 J/packet to 100 J/packet, whereas the normalized throughput increased from 2$\sim$6 packets to 11 packets/s.
Compared to wave-based communication (e.g., EMWCs), the DEAP and DECN approaches fully consider the fact that the received power diminishes linearly to the distance $d^6_{\mathrm{SD}}$ in an MIC channel, which diminishes much faster than in the EMW channel (in the order of $d^2_{\mathrm{SD}}$).

Compared to EMWC, it can be concluded that all MI CLO solutions use distributed algorithms. This is due to the fact that the capacity and bandwidth of an MI channel are much lower than those of an EMW channel. This lack of information exchange can cause slow convergence of algorithms.

\subsection{Summary and Lessons Learned} \label{sectsub6_sll}
Table \ref{tbl_osi} summarizes the issues addressed, their corresponding methods, and the remaining issues of research  on the multi-node MI network under the OSI-originated framework, including the physical, data link, network, transport, and application layers. It can be summarized that the issues of the MI network primarily stem from the coil resonance feature, i.e., low MI bandwidth and significant frequency-selectivity.  

Apart from MI fast fading, many remaining issues need to be addressed: 1) The physical layer support for RPMA-based MICs is inadequate, even lacking a universal expression for channel capacity, with key challenges stemming from mechanical inertia and friction. Despite the packets having the same length, the RPMA MI channel may lead to variable data rates;
2) the high collision probability of existing MI MAC solutions\cite{Ahmed2024Design,Ahmed2016Multi,Ahmed2019Design}  has not been addressed due to the coil resonance feature;
3) most solutions in Ly3, especially the routing solutions\cite{Wang2019Efficient, Alsalman2021Balanced, Liu2024Frequency}, have overlooked the gain factors $\mathcal{E}_{\mathrm{SD}}$ and $J_{\mathrm{SD}}$. These solutions in Ly3 have limited support for MI UG-WSNs and UW-WSNs with longer ranges;
4) the channel-independent OSI-based solutions, particularly TCP and IP, need to be validated due to issues, such as excessively large frame headers and RTT suppression.  These are crucial for the  SAGUMI;  and
5) all solutions using  RL \cite{Wang2019Efficient, Alsalman2021Balanced, Ma2024Fast}, especially those using distributed RL with the Nash game,  may face low convergence and precision. %For this challenge, our proposed MI framework as depicted in Fig.~\ref{fig_sec5tcp} can help improve the precision result via the online deep learning techniques (e.g., deep RL). 

Regarding these five challenges 1) -- 5), our proposed framework, the status quo, potential solutions, and research gaps are elaborated on in the subsequent sections. Table \ref{tbl_osi_map} summarizes the reviewed techniques mapped to each OSI layer and their respective technical readiness levels (TRLs),  revealing that most existing MI techniques lack experimental evidence, particularly for multi-node solutions.

The practical takeaways or common pitfalls include: 1) Most multi-node MI protocol studies have assumed near-field and weak coupling conditions, which are more prone to breakdown in TTE environments than  in general MIC environments;  2) existing MI upper-layer protocols tend to overlook the eddy gain and polarization gain; 3) TCP/IP stack application to TTE MICs may suffer from the congestion mechanism failures and retransmission storms; 4) existing MI upper-layer protocols overlook that the MI channel bandwidth may be insufficient to support the exchange of their large control data, such as  routing and Q tables; and 5) the frequent variation in average AVIs may render  many existing MI solutions from Ly1 to Ly7 for both P2P and multi-node networks inapplicable.

\begin{table*}[htp] \scriptsize%\footnotesize
\centering
%	\tiny
\caption{Technical Readiness  of MI techniques mapped to OSI layers}
\label{tbl_osi_map}
\vspace{-0.8em}
\scalebox{0.96}{
\begin{threeparttable}
%	\adjustbox{max width=\textwidth, scale=0.85}{}
\begin{tabular}{m{0.05\textwidth}<{\centering}|m{0.22\textwidth}<{\centering}|m{0.18\textwidth}<{\raggedright\arraybackslash}|m{0.43\textwidth}<{\centering}|m{0.03\textwidth}<{\centering}}
\hline
\textbf{OSI Lys}  &  \textbf{Technology} & \textbf{TRL $^\dagger$} & \textbf{Future research directions } & \textbf{Priority} \\
\hline \hline
%% %% \multirow{NumberOfRows}{CellWidth\textwidth}[-Fromtop]{\centering Passive Relay}
\multirow{4}{0.03\textwidth}[-10pt]{\centering Ly1}&MI fast fading& \ding{72}  & A universal statistical modeling &\hlprioritem\hlprioritem\hlprioritem\\ \cline{2-5}
&Channel modeling &\ding{72}\ding{72}\ding{72}\ding{72} & Mixed-field \& multilayer  medium cases   & \hlprioritem\hlprioritem\\ \cline{2-5}
&Antenna design &\ding{72}\ding{72}\ding{72}\ding{72} & TMR Rx antenna; deployment   & \hlprioritem\hlprioritem\\ \cline{2-5}
&Channel estimation &\ding{72} & For  VLF-LA and MI fast fading cases    & \hlprioritem\hlprioritem\\ \cline{2-5}
& Modulation &\ding{72}\ding{72}\ding{72}\ding{72} $^\ddagger$ & For RPMA-based links   & \hlprioritem\\ \cline{2-5}
&Channel coding &\ding{72}\ding{72} & Deep JSCC   & \hlprioritem\hlprioritem\\ \cline{2-5}
&Passive relays & \ding{72}\ding{72}\ding{72} & Positive MI crosstalk& \hlprioritem\hlprioritem\\ \cline{2-5}
&Active relays / CMIC & \ding{72} &Arbitrarily deployed multi-relays& \hlprioritem\hlprioritem\\ \cline{2-5}
&Resource allocation &\ding{72} & Bandwidth and frequency allocations; balancing precision and convergence    & \hlprioritem\hlprioritem\\ \hline
\multirow{2}{0.03\textwidth}[-3pt]{\centering Ly2}&MI MAC& \ding{72}\ding{72}  & Header compression;  conflict reduction  &\hlprioritem\hlprioritem\hlprioritem\\ \cline{2-5}
&LLC & \ding{73} (0) &Optimization for VLF-LA case& \hlprioritem\\ \hline
\multirow{2}{0.03\textwidth}[-4pt]{\centering Ly3}&Connectivity& \ding{72} & Heterogeneous or mobile MI networks  &\hlprioritem\hlprioritem\\ \cline{2-5}
&Data collection and node deployment & \ding{72} & Algorithm convergence &\hlprioritem  \\ \cline{2-5}
&Routing & \ding{72} & Minimizing information exchange among nodes &\hlprioritem\hlprioritem  \\ \cline{2-5}
&IP & \ding{73} (0) & IP-HC &\hlprioritem\hlprioritem\hlprioritem  \\ \hline
Ly4 &TCP& \ding{73} (0) & Fairness, congestion control, connection issues due to extremely low bandwidth &\hlprioritem\hlprioritem\hlprioritem\\
\hline
\end{tabular}
\begin{tablenotes}  
\footnotesize  
\item[$\dagger$] \ding{72}: Technology Readiness Level (TRL) (cf. \cite{Mankins1995Technology}) 1--3  (basic research); \ \ding{72}\ding{72}:TRL 4--5 (lab validation); \  \ding{72}\ding{72}\ding{72}: TRL 6--7 \ (system-level testing); \ \ding{72}\ding{72}\ding{72}\ding{72}: TRL 8--9 (industrial application); \ding{73}: No MI-specific references available for this TRL level.	
\item[$\ddagger$] The TTE MIC products have entered the market \cite{VitalCanaryComm}, and modulation is an essential integrated technical module.
\end{tablenotes}
\end{threeparttable}
}
\vspace{-1.38em}
\end{table*}

\section{Promising MI Network Framework with TCP/IP and Linux Support}\label{sect_linux}

This section proposes a Linux-and-TCP/IP-supported MI network framework to  implement most existing MIC protocols and algorithms, including those  in Sections \ref{sect_p2p}, \ref{sect_cmi}, and \ref{sect_network}. %{including P2P MI solutions, relay techniques, and multi-hop, multi-node MI network designs. }

\subsection{Significance and Architecture Overview}\label{sectsub7_sao}

\subsubsection{Significance}\label{sectsubsub7_significance}
While MIC techniques have been applied in various underground applications, few studies explore their compatibility with standard protocols, particularly the TCP/IP framework, due to the limited bandwidth of MIC systems. With ongoing advancements in MIC performance and the expansion of SAGUMI applications, integrating MIC with the TCP/IP framework is becoming an inevitable and essential trend. While deep learning has proven effective in addressing EMWC challenges \cite{Zhao2025Generative, Ferrag2023Edge}, its application to MIC remains largely unexplored due to the difficulties of using robust neural network platforms in embedded systems. %such as MIC systems.

On the other hand, Linux excels in large-scale wireless applications, such as mobile ad-hoc networks (MANETs) and Android (built on the Linux kernel) smartphones, enabling dynamic routing and offering scalability via its modular kernel and TCP/IP stack. Moreover,  Linux offers a wealth of open-source resources across diverse research domains, particularly in wireless network protocol stacks (e.g., IEEE 802.15) and neural network platforms (e.g., TensorFlow, PyTorch, and RKNN). These resources facilitate rapid development by providing robust tools and platforms for research.

As illustrated in Fig. \ref{fig_sec5tcp}, we propose a Linux-based MIC framework. This framework tackles multi-hop MIC challenges by integrating TCP/IP, Linux kernel modules, and the MI solutions discussed in the preceding section, balancing protocol compatibility with UG-WSN requirements.

\begin{figure}[t!]
\centering{\includegraphics[width=3.1in]{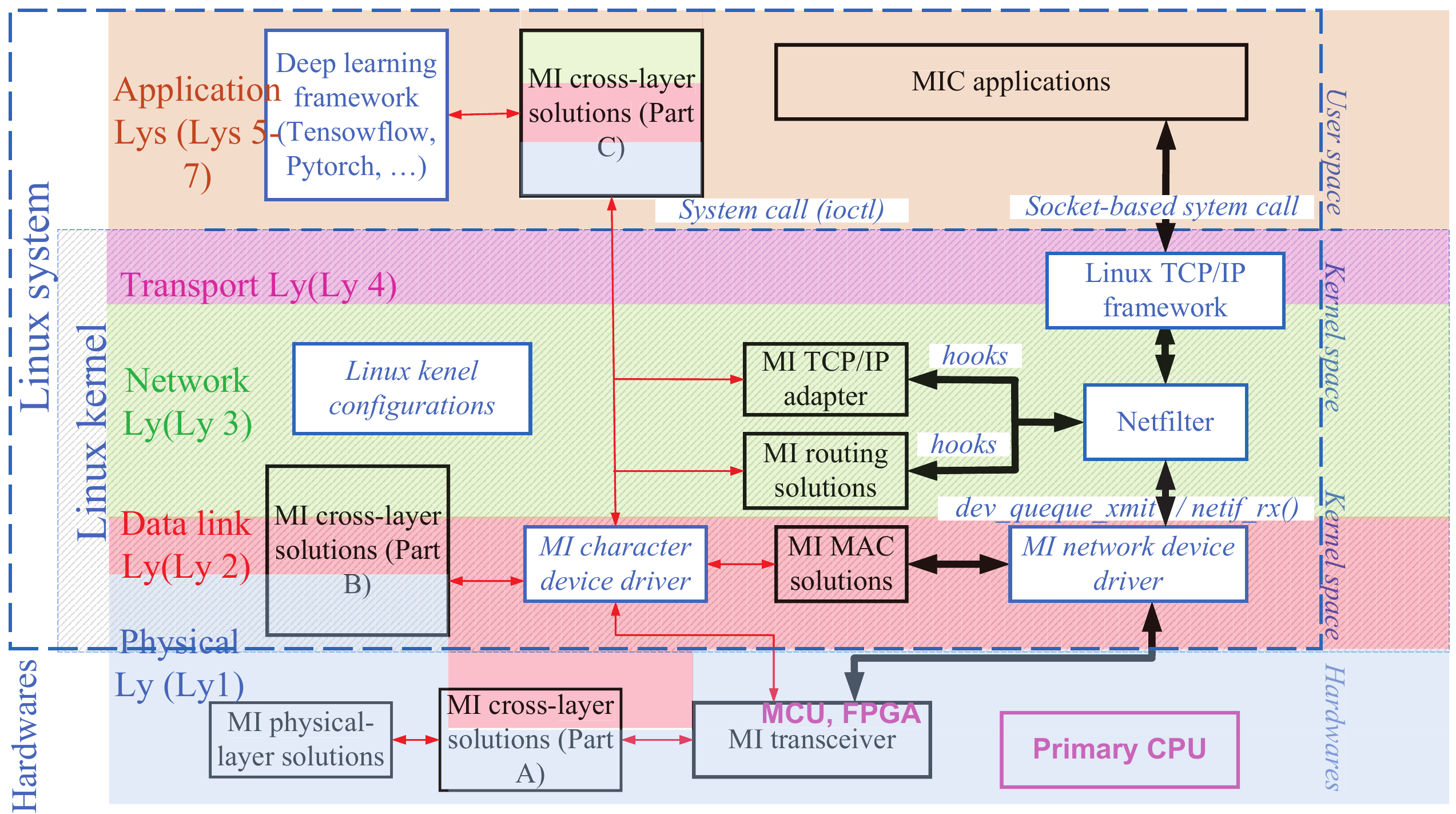}}
\caption{Proposed  framework for TCP/IP \& Linux support.  The thick arrow line  represents effective Tx/Rx data, and the thin red arrow line represents control data for the MI network protocol stack. The box with a white background and italic text represents the interface between  Linux system and  MI protocol. The box with a white background represents the  Linux model.
}\label{fig_sec5tcp} 
\vspace{-0.0em}
\end{figure}

\subsubsection{Linux wireless network device driver-inspired architecture}\label{sectsubsub7_lwndda}
Building on the support schemes for TCP/IP and Linux in EMWCs, such as the Linux driver designs \cite{Jonathan2015Linux} for WiFi and Bluetooth, we come up with a Linux-based MIC framework. This framework, as illustrated in Fig. \ref{fig_sec5tcp}, draws parallels with EMWC solutions. It incorporates MI cross-layer solutions (Part A) and MI transceivers, which are hardware and firmware-based models designed for protocols and algorithms that rely on analog signals, high parallel computation, or high-accuracy real-time processing, such as channel modeling and estimation. The MI character device (MCD) driver  and MI network device (MND)  driver   are designed to handle the retrieval and storage of MIC data. The MCD driver forwards control data, such as CSI, while the MND driver handles effective data streams, enabling MIC applications to directly access the full TCP/IP protocol through the \texttt{socket()} interface.

\subsubsection{MI-specific models}\label{sectsubsub7_msm}
In this framework, several  models differ from conventional Linux wireless drivers:
%traditional EMWC-based solutions:
\begin{itemize} 

\item \emph{MI TCP/IP adapter}: This model adapts the native Linux TCP/IP framework with MI-specific TCP/IP optimization schemes proposed by researchers, such as TCP/IP header compression and RTT optimization for the MIC. 

\item \emph{MI routing solutions model}: This model can implement and help evaluate specific MI routing algorithms proposed by researchers. These algorithms can be invoked similar to the MI TCP/IP adapter.

\item \emph{Cross-layer solutions model (Part C)}: This model supports deep learning-based solutions, such as deep JSCC. However, as it resides in the user space of the operating system, it may not be suitable for algorithms with stringent real-time requirements. 
\item \emph{Cross-layer solutions model (Part B)}: To reduce device size and energy consumption, Some  MI modulation and channel coding can be incorporated into the MI cross-layer solutions (Part B) model rather than in FPGA or MI cross-layer solutions (Part A) model. The TTE MIC system uses the VLF-LA technique, which allows the primary CPU sufficient time to process, enabling the removal of FPGA and MI cross-layer (Part A) models in some simpler MI nodes. 
\end{itemize}

\subsection{System  Architecture and Implementation}\label{sect_chp5sia}

Fig. \ref{fig_sec5tcp} describes the system implementation of the Linux-based MIC framework, focusing on the interface between Linux TCP/IP stack and MI solutions models.

\subsubsection{Linux MCD  and MND drivers} \label{sectsubsub7_lmmd}

%The MCDD and MND models are  virtually identical to the most architectures of the EMWC drivers.  They both facilitate data exchange between hardware and the Linux system while providing an architectural foundation for MI-specific models. The MCDD performs the following operations.

The MCD and MND drivers models mirror EMWC driver architectures, facilitating hardware-Linux data exchange and laying a foundation for MI-specific models. The MCD driver operates as follows.

\begin{itemize}
\item  Register a character device associated with  several Linux  file operations,  whose members include \texttt{open}, \texttt{close}, \texttt{read}, \texttt{write}, and \texttt{ioctl} callbacks, to the  kernel;
\item  Invoke MI-specific models (MI cross-layer solutions Parts A and  B) via hardware interrupts and Linux/user APIs;
\item The \texttt{write} and \texttt{ioctl} callbacks can manage downlink control streams (from  application layer to  physical layer);
\item The \texttt{read} and \texttt{ioctl} callbacks can manage uplink control streams (from  physical layer to  application layer).
\end{itemize}

The MND driver operates as follows.
\begin{itemize}
\item  Register a Linux MND with integrating its netfilter hooks with the  MI TCP/IP adapter and MI routing models. Linux can invoke  models via these hooks automatically; 
\item   Invoke MI MAC solutions via the  callbacks of the MND;
\item Manage routing and downlink data streams (from  application layer to  physical layer) via the  MND callbacks (e.g., \texttt{ndo\_start\_xmit});
\item  Manage routing and uplink data streams (from  physical layer to  application layer) via the  MND callbacks (e.g., \texttt{netif\_rx}).
\end{itemize}

\subsubsection{MI physical-layer solutions and  MI cross-layer solutions (Part A)} \label{sectsubsub7_parta}

In line with the most EMWC solutions that employ   hardware-centric implementation in PHY/MAC in dedicated chips, the models of MI physical-layer solutions and  MI cross-layer solutions (Part A) reside in MCUs or FPGAs (not Linux).  Given that a few MI solutions exhibit strong dependency on analog signal processing and impose stringent timing constraints, these requirements exceed the performance capabilities of the Linux kernel. For this reason, these models handle modulation/coding and  cross-layer logic (e.g., interference mitigation)  using FPGA/MCU cores. These models interface with Linux system via custom bus protocols, e.g., serial peripheral interface (SPI), or memory-mapped I/O.

\begin{algorithm}[h]\label{alg_tcp}
\footnotesize %\small
%	\KwIn{The BS index $k$}
\tcc{\sf Conventions: The \emph{italic text} represents a Linux API, and Model\_name.Function\_name(Input\_data) represents a function of a  model  with parameter Input\_data}
\tcc{\sf   \textbf{Boot/initialization stages  of Linux}}
Register  MCD and  MND; \\
\tcc{\sf IP-HC in TCP/IP adapter  called after routing} 
Register a netfilter hook named MI\_adapter with \emph{hooknum}=\emph{NF\_IP\_POST\_ROUTING}, \emph{hook}=MI\_IPHC, priority$\geq$100; \\
Register a netfilter hook named MI\_routing with \emph{hooknum}=\emph{NF\_IP\_LOCAL\_OUT},  \emph{hook}=frequency\_switchable\_routing (cf. \cite{Liu2024Frequency}), priority$\leq$-400; 	\\
\tcc{\sf \textbf{Runtime stage of Linux}}
\For{\rm Loop}
{
\textbf{Call}  MI\_Channel\_Estimation in Algorithm \ref{alg_esti} which generates MI\_CSI; \\
User calls \emph{sendto}(Destination\_IP, Destiation\_port, MI\_packet) which generates MI\_IP\_packet; \\
Linux TCP/IP  calls the hook MI\_routing.frequency\_switchable\_routing(MI\_IP\_packet) to complete routing  based on \cite{Liu2024Frequency}, see Fig. \ref{fig_sec4routing}; \\
Linux TCP/IP calls the hook MI\_adapter.MI\_IPHC(MI\_IP\_packet) and generates MI\_IPHC\_packet;\\
Obtain MI\_IPHC\_packet via  callback MND.\emph{ndo\_start\_xmit()}; \\
MND.\emph{ndo\_start\_xmit}() calls MI\_MAC.Mac\_based\_on\_\cite{Ahmed2016Multi}(MI\_IPHC\_packet,  MI\_CSI) to complete MAC  and generates MI\_MAC\_packet;\\
MND.\emph{ndo\_start\_xmit}() calls MI\_cross\_layer\_Part\_B.BPSK\_Polar\_based\_on\_\cite{Chen2023Novel}(MI\_MAC\_packet, MI\_CSI) and generates MI\_BPSK\_packet and control bytes;\\
This BPSK and Polar encoder function downloads the MI\_BPSK\_packet and control bytes to the  MI transceiver's corresponding buffer and hardware registers, respectively; 
}
Unregister all devices and netfilter hooks \\
\caption{ Example of a TCP/IP transmit process. }
\end{algorithm}
\vspace{-0.0em}

\begin{algorithm}[h]\label{alg_esti}
\footnotesize %\small
%	\KwIn{The BS index $k$}
%\tcc{\sf Convention: The \emph{italic text} represents a Linux API, and Variable\_name represents a variable name}
MCD calls MI\_cross\-layer\_Part\_B.MI\_estimation\_based\_on\_\cite{Kisseleff2014Transmitter} (MI\_CSI); \\
MCD calls MI\_cross\-layer\_Part\_B.MI\_fast\_fading() to require  MI\_cross\-layer\_Part\_C.MI\_fast\_fading() via  Linux callback \emph{read}(); \\
MI\_cross\-layer\_Part\_C.MI\_fast\_fading() completes the MI fast fading estimation based on Fig. \ref{fig_sec5avipre} and Pytorch framework; \\
MI\_cross\-layer\_Part\_C.MI\_fast\_fading() responds the updated MI\_CSI to the  MI\_cross\-layer\_Part\_B via Linux callback \emph{write}();\\
\Return  updated MI\_CSI
\caption{  MI\_Channel\_Estimation() based on deep learning framework. }
\end{algorithm}
\vspace{-0.0em}

\subsubsection{MI cross-Layer solutions (Part B)}\label{sectsubsub7_partb}

For the VLF within the Linux kernel's capabilities, this model can directly implement most physical and MAC layer solutions.
\begin{itemize}
\item  Perform MI channel estimations without deep learning (e.g., \cite{Kisseleff2014Transmitter}) in the Linux Interrupt Service Routine registered by the MCD driver upon receiving the control stream from the MI transceiver;
\item Perform modulations (e.g., BPSK) and FECs (e.g., Polar codec \cite{Chen2023Novel}) upon MAC frame/packet arrival at the MND driver, as shown in Algorithm \ref{alg_tcp} (Lines 11 and 12);
\item  Forward the control stream to MI cross-Layer solutions (Part C) via \texttt{read}, \texttt{write} and \texttt{ioctl}.
\end{itemize}

\subsubsection{MI cross-layer solutions (Part C)}\label{sectsubsub7_partc}

This model facilitates access to the interfaces of deep learning platforms (e.g., PyTorch), as these interfaces and platforms can only run in user space. Following are the details (cf. Algorithm \ref{alg_esti}):

\begin{itemize}
\item  MI cross-layer solutions (Part B) forward required data to Part C via \texttt{read} callback;
\item Part C executes algorithms using the interfaces of the deep learning platform in the user space; 
\item  Part C responds the results to Part B via  \texttt{write} callback.
\end{itemize}

\subsubsection{MI TCP/IP adapter and MI routing solutions}\label{sectsubsub7_mtmrs}

The MI TCP/IP adapter and routing solutions integrate with the Linux TCP/IP framework primarily through netfilter hook points \cite{Benvenuti2005Understanding}, which correspond to distinct stages in the local machine’s internal routing process, i.e.,  \texttt{NF\_xx\_PRE\_ROUTING} (prior to routing decisions), \texttt{NF\_xx\_LOCAL\_IN} (before packets destined for the local system enter the protocol stack), \texttt{NF\_xx\_FORWARD} (for packets being forwarded), \texttt{NF\_xx\_LOCAL\_OUT} (before packets originated from the local system exit the protocol stack), and \texttt{NF\_xx\_POST\_ROUTING} (after the routing decision has been made but before the packets are transmitted on the network interface). Here, \texttt{xx} denotes the  protocol type (e.g., IPv4, IPv6, or bridge protocol).

Upon Linux startup, the NMD driver registers netfilters for MI TCP/IP adapter and routing solutions.  The  MI routing netfilter should be set to the highest priority to prevent other routing schemes. MI-specific routing functions attach to hook points \texttt{NF\_xx\_LOCAL\_OUT} and \texttt{NF\_xx\_PRE\_ROUTING}, as shown in Algorithm \ref{alg_tcp} (Line 2).   MI TCP/IP adapter (especially IP-HC) is set to low priority to avoid kernel packet modification. Then, the  functions of the MI TCP/IP adapter  are associated to \texttt{NF\_xx\_POST\_ROUTING}, as shown in Algorithm \ref{alg_tcp} (Line 3). These netfilters are automatically invoked by the Linux TCP/IP framework.

\subsubsection{MI MAC solutions}\label{sectsubsub7_mimac}

This model packages the MAC header (see Fig. \ref{fig_sec5mac:pkt}) and maintains the state machine as the one in \cite{Ahmed2016Multi}. MI MAC functions for uplink streams are triggered by MI transceiver interrupts; those for downlink streams are invoked via  the \texttt{ndo\_start\_xmit} callback of the MND  (see Algorithm \ref{alg_tcp}, Line 11).

\subsubsection{Example} \label{sectsubsub7_example}

Consider an MI protocol stack integrated into the Linux TCP/IP framework. This stack includes physical-layer solutions, such as channel estimation, MI fast fading prediction (see Fig. \ref{fig_sec5avipre}), BPSK modulation, and Polar coding (as in \cite{Chen2023Novel}). It also incorporates the MAC solution from \cite{Ahmed2016Multi} and the routing solution from\cite{Liu2024Frequency}. The TCP/IP transmit process is described in Algorithm \ref{alg_tcp}. When an application sends a packet via the Linux Socket API \texttt{sendto}, the packet passes through the Linux TCP/IP framework, including netfilters associated with MI routing and IPHC schemes, and arrives at the MND. The MND  invokes the MI MAC algorithm, BPSK modulation, and Polar encoding to generate a physical-layer frame, which is then downloaded to the MI transceiver.

\subsection{Summary} \label{sectsub7_sll}
The proposed framework enables researchers to leverage the abundant Linux resources available for communication networks and deep learning, such as OpenZigbee\cite{zboss2023}, TensorFlow, and RKNN\cite{rknn2025}. As a result, it can accelerate MIC research and development.

The proposed framework manages control flow across OSI layers using Linux interrupt service routines and MCD callbacks \texttt{read}/\texttt{write}. It handles data flow and protocol encapsulation across OSI layers via interrupt service routines, MND callbacks  \texttt{ndo\_start\_xmit}/\texttt{netif\_rx}, and socket APIs. MI-specific network layer protocols are managed through Linux netfilter hooks.

The following sections provide further insights into related MI techniques, which can also be supported by this framework.

\begin{table*}[t!]  \scriptsize
\caption{Future issues, promising work and advised methodologies  for TTE MICs }
%\begin{minipage}{\textwidth} 
%\tiny
\centering
\vspace{-0.8em}
\scalebox{0.99}{
\begin{threeparttable} 
\begin{tabular}{m{1cm}<{\centering}|m{1.4cm}<{\centering}|m{1.0cm}<{\centering}|m{6.6cm}<{\centering}|m{6.4cm}<{\centering}}
\hline
\textbf{Types} &  \textbf{Research  aspects} & \textbf{OSI layer} & \textbf{Future issues \& works}  & \textbf{Advised methodologies} \\
\hline \hline
%\raggedleft
\centering
%%%%%%%%%%%%%%%%%%%%%%%%%%%%%%%%%%%%%%%%%%%%%%%%%%%%%%%%%%%%%%%%%%%%%%%%%%%%%%%%%%%%%%%%%%%%%%%%%%%%%%%
\multirow{6}{0.03\textwidth}[-45pt]{ P2P MICs } &MI fast fading  &  Ly1 &\begin{enumerate}[leftmargin=3mm]
\item  A universal statistical model
\item  A velocity-dependent expectation/variance
\item See Table \ref{tbl_futurefastfading}
\vspace{-1.0em}
\end{enumerate}  & \begin{enumerate}[leftmargin=3mm]
\item  Maxwell-equations-based derivation and FEMs
\item  Probability-theorem-based derivation
\item   Deep JSCC
\item  Transformers model or LLM to predict average AVI 
\vspace{-1.0em}
\end{enumerate}\\ \cline{2-5}
%%%%%%%%%%%%%%%%%%%%%%%%%%%%%%%%%%%%%%%%%%%%%%%%%%%%%%%%%%%%%%%%%%%%%%%%%%%%%%%%%%%%%%%%%%%%%%%%%%%%%%%
&Antenna design  &  Ly1 &\begin{enumerate}[leftmargin=3mm]
\item  Inertia issue of a mechanical antenna
\item  Using high-sensitivity and small-size magnetic sensor
\vspace{-1.0em}
\end{enumerate}  &-\\ \cline{2-5}			
%%%%%%%%%%%%%%%%%%%%%%%%%%%%%%%%%%%%%%%%%%%%%%%%%%%%%%%%%%%%%%%%%%%%%%%%%%%%%%%%%%%%%%%%%%%%%%%%%%%%%%%		
&MCNSI  & \hspace{-0.6em} Ly1--Ly3, Ly7 & \begin{enumerate}[leftmargin=3mm]
\item  Balancing between  MI, navigation and sensing
\item Chaotic RSSIs 
\vspace{-1.0em}
\end{enumerate} & \begin{enumerate}[leftmargin=3mm]
\item   GAN techniques and DWE algorithm
\item  Formulating a  joint optimization problem
\vspace{-1.0em}
\end{enumerate}\\ \cline{2-5}					
%%%%%%%%%%%%%%%%%%%%%%%%%%%%%%%%%%%%%%%%%%%%%%%%%%%%%%%%%%%%%%%%%%%%%%%%%%%%%%%%%%%%%%%%%%%%%%%%%%%%%%%	
& \hspace{-0.7em}Mixed-field channel model  & Ly1& \begin{enumerate}[leftmargin=3mm]
\item  Difficulties in  sub-modeling of channel power gain $G_{\mathrm{SD}}$
\vspace{-1.0em}
\end{enumerate}   & \begin{enumerate}[leftmargin=3mm]
\item  Maxwell-equations-based derivation
\vspace{-1.0em}
\end{enumerate}\\ \cline{2-5}	
%%%%%%%%%%%%%%%%%%%%%%%%%%%%%%%%%%%%%%%%%%%%%%%%%%%%%%%%%%%%%%%%%%%%%%%%%%%%%%%%%%%%%%%%%%%%%%%%%%%%%%%		
&Inhomoge- neous  medium & Ly1 &  \begin{enumerate}[leftmargin=3mm]
\item  Boundary conditions for multi-layer materials
\item  Boundary conditions between  the near-field region  and radiation field region
\item  Statistical characteristic of dynamic medium
\vspace{-1.0em}
\end{enumerate}   & \begin{enumerate}[leftmargin=3mm]
\item  Maxwell-equations-based derivation
\item  Geometrical approximation by jointing regular shapes
\item  Data fusion and attention mechanism
\vspace{-1.0em}
\end{enumerate}\\ \cline{2-5}			
%%%%%%%%%%%%%%%%%%%%%%%%%%%%%%%%%%%%%%%%%%%%%%%%%%%%%%%%%%%%%%%%%%%%%%%%%%%%%%%%%%%%%%%%%%%%%%%%%%%%%%%		
& JSCC  &Ly1, Ly7& \begin{enumerate}[leftmargin=3mm]
\item Image, audio \& video data transmissions
\item Content-based communication
\item High-dimensional data communication
\vspace{-1.0em}
\end{enumerate} & \begin{enumerate}[leftmargin=3mm]
\item Deep JSCC
\item Multimodal  semantic communication
\item Offline distilled LLM
\vspace{-1.0em}
\end{enumerate} \\ \cline{2-5}	

\hline
%%%%%%%%%%%%%%%%%%%%%%%%%%%%%%%%%%%%%%%%%%%%%%%%%%%%%%%%%%%%%%%%%%%%%%%%%%%%%%%%%%%%%%%%%%%%%%%%%%%%%%%
\hspace{-0.9em}MI Relay &CMIC & Ly1  & \begin{enumerate}[leftmargin=3mm]
\item  Spatial distribution of Cross-talk effects
\item  Multiple active relays with misaligned antenna
\vspace{-1.0em}
\end{enumerate} & \begin{enumerate}[leftmargin=3mm]
\item  KVL equations
\vspace{-1.0em}
\end{enumerate}\\ \cline{2-5}		

\hline	
%%%%%%%%%%%%%%%%%%%%%%%%%%%%%%%%%%%%%%%%%%%%%%%%%%%%%%%%%%%%%%%%%%%%%%%%%%%%%%%%%%%%%%%%%%%%%%%%%%%%%%%		
\multirow{2}{0.04\textwidth}[-25pt]{\hspace{-0.9em}MI network } & Heteroge- neous MI Network &  Ly1--Ly7 &\begin{enumerate}[leftmargin=3mm]
\item  Channel licensing and spectrum sensing
\item   Connectivity issues
\item  Power and throughput optimization 
\vspace{-1.0em}
\end{enumerate} & \begin{enumerate}[leftmargin=3mm]
\item  Percolation theory
\item   Multiagent RL
\vspace{-1.0em}
\end{enumerate} \\ \cline{2-5}	
%%%%%%%%%%%%%%%%%%%%%%%%%%%%%%%%%%%%%%%%%%%%%%%%%%%%%%%%%%%%%%%%%%%%%%%%%%%%%%%%%%%%%%%%%%%%%%%%%%%%%%%		
&MI MAC   & Ly2 &\begin{enumerate}[leftmargin=3mm]
\item  The SISO Antenna case
\item  Balancing the frame error ratio and EPR
\item  Power and throughput optimization 
\item  Communication security  issues
\vspace{-1.0em}
\end{enumerate} & \begin{enumerate}[leftmargin=3mm]
\item   Machine learning for CSMA-CA issue
\item   Using antenna orientation information for frame error ratio \& EPR issue
\item  Bit-level compression for the MAC header
\vspace{-1.0em}
\end{enumerate}\\ \cline{2-5}	
%%%%%%%%%%%%%%%%%%%%%%%%%%%%%%%%%%%%%%%%%%%%%%%%%%%%%%%%%%%%%%%%%%%%%%%%%%%%%%%%%%%%%%%%%%%%%%%%%%%%%%%		
&MI routing  & Ly3 & \begin{enumerate}[leftmargin=3mm]
\item Effects of $\mathcal{E}_{\mathrm{SD}}$ and $J_{\mathrm{SD}}$ 
\vspace{-1.0em}
\end{enumerate}   & - \\ \cline{2-5}
%%%%%%%%%%%%%%%%%%%%%%%%%%%%%%%%%%%%%%%%%%%%%%%%%%%%%%%%%%%%%%%%%%%%%%%%%%%%%%%%%%%%%%%%%%%%%%%%%%%%%%%
&MI TCP/IP   & Ly1-Ly4 &\begin{enumerate}[leftmargin=3mm]
\item  Large  TCP/IP packet header unsuitable for ultra-narrow-band MI channel
\item  Significant RTT suppression for TCP connections
\item  Excessive duration for connection establishment
\vspace{-1.0em}
\end{enumerate} & \begin{enumerate}[leftmargin=3mm]
\item   Header compression techniques
\item  RTT optimization based on MI channel conditions
\item  Optimizing data chunking and aggregation
\item  Intelligent retransmission strategy
\item  Machine learning  solution under MI fast fading
\vspace{-1.0em}
\end{enumerate}\\ \cline{2-5}	

\hline
%%%%%%%%%%%%%%%%%%%%%%%%%%%%%%%%%%%%%%%%%%%%%%%%%%%%%%%%%%%%%%%%%%%%%%%%%%%%%%%%%%%%%%%%%%%%%%%%%%%%%%%
Implemen- tations &\hspace{-0.9em}MI network \hspace{-0.9em}framework &  \makecell[l]{\vspace{-0.6em}\\\hspace{-0.0em}Ly1-Ly7} & \begin{enumerate}[leftmargin=3mm]
\item  {Experiments and testing for TTE MIC systems}
\item  TCP/IP support
\vspace{-1.0em}
\end{enumerate}  &  
Fig. \ref{fig_sec5tcp} \\ \cline{2-5}	
\hline			
\end{tabular}
%\end{minipage}
%			\begin{tablenotes}  
%				\footnotesize  
%				\item[] 
%			\end{tablenotes}	

\end{threeparttable}
}
\vspace{-1.31em}
\label{tbl_future}
\end{table*}

\section{Research Challenges and Future Directions}\label{sect_future}

Sections \ref{sect_sub2channel}--\ref{sect_network} have provided a comprehensive overview, encompassing a wide range of MIC topics of state-of-the-art methodologies and theoretical frameworks, including the MI channel modeling, P2P MIC, MI relay, and MI network architecture. Many algorithms and solutions discussed within this comprehensive review can be implemented and/or evaluated through the MI framework proposed in Section \ref{sect_linux}; see Table \ref{tbl_linux}.  For instance, the channel modeling and estimation solutions in \cite{Sun2010Magnetic, Kisseleff2014Transmitter, Ma2024Fast} require processing analog signals and can be implemented in the MI transceiver model.

In this section, we summarize the potential challenging issues that were not addressed in the literature. We also introduce  new promising techniques (e.g.,  deep JSCC, MCNSI, etc.) for future research. 
Table \ref{tbl_future} outlines the remaining issues, their potential solutions, and novel techniques for MIC. 

\begin{table*}[t!]  \scriptsize
\caption{Our proposed prototype for MI network implementation to support existing and Future Studies }
%\begin{minipage}{\textwidth} 
%\tiny
\centering
\vspace{-0.8em}
\begin{threeparttable} 
\begin{tabular}{p{4.2cm}|p{7.4cm}|p{5cm}}
\hline
\textbf{Models in Fig. \ref{fig_sec5tcp}}  & \textbf{Recommended  solutions$^\dagger$}   & \textbf{Typical \emph{refs.}}\\
\hline \hline
%\raggedleft
\centering

{MI transceiver} & Channel modelings and estimations&\cite{Sun2010Magnetic, Ma2024Fast, Ma2019Effect, Guo2015M2I, Li2015Capacity, Kisseleff2014Transmitter}   \\ 
\hline
%%%%%%%%%%%%%%%%%%%%%%%%%%%%%%%%%%%%%%%%%%%%%%%%%%%%%%%%%%%%%%%%%%%%%%%%%%%%%%%%%%%%%%%%%%%%%%%%%%%%%%%		
MI cross-layer solutions (Part A) & Modulations; Polar coding; JSCC; high real-time CSMA-CA &\cite{Kisseleff2014Modulation, Chen2023Novel}\\ \cline{1-3}	
MI MAC solutions  &  Contention based MI MAC protocols&\cite{Ahmed2024Design,Ahmed2016Multi, Ahmed2019Design}  \\ \cline{1-3}		
MI cross-layer solutions (Part B)  &Modulations, FEC; Polar coding;  data collection  &\cite{Kisseleff2014Modulation,Ma2024Fast,lin2015distributed, Chen2023Novel, Wei2022Power} \\ \cline{1-3}					
MI routing solutions & MI connectivity and routing solutions  &\cite{Wang2019Efficient, Alsalman2021Balanced, Liu2022Qlearning, Liu2024Frequency, Zhang2017Connectivity}\\ \cline{1-3}	

MI TCP/IP adapter &  IP-HC, RTT optimization, intelligence retransmission, etc. &Future research  \\ \cline{1-3}	

MI cross-layer solutions (Part C) & Deep JSCC; algorithms with deep RL &Future research  \\ \cline{1-3}

MI character device driver & Reading/writing protocol control data  from/to the MI transceiver &\cite{Jonathan2015Linux}\\ \cline{1-3}							

MI network  driver  & Reading/writing effective data from/to the MI transceiver &\cite{Jonathan2015Linux}\\ \cline{1-3}

\hline

\end{tabular}
%\end{minipage}
\begin{tablenotes}  
\footnotesize  
\item[$\dagger$] In this table, we prioritize the solutions of the MIC under the VLF-LA case
\end{tablenotes}	
\end{threeparttable}
\vspace{-1.31em}
\label{tbl_linux}
\end{table*}

\subsection{MI Fast Fading in Mobile MIC Systems}\label{sectsub8_mffmms}

Until 2020, no concept of an MI fast fading channel was formally introduced. The research on MI fast fading is still in its early stages (see Table \ref{tbl_fastfadingdiff}). In this subsection, we discuss the  challenges and remaining tasks for channel statistical characteristics, and elucidate the potential ramifications on established MIC (from Ly1 to Ly3), including outage probability, channel estimation, MI MIMO, and CMIC. Table \ref{tbl_futurefastfading} summarizes the typical issues arising from the introduction of MI fast fading into MI channels.

\begin{table*}[t!]  \scriptsize
\caption{Potential further  issues if  MI fast fading channels are introduced  }
%\begin{minipage}{\textwidth} 
%\tiny
\centering
\vspace{-0.8em}
\begin{tabular}{p{2cm}|p{3.0cm}|p{2.6cm}|p{8.5cm}}
\hline
\textbf{OSI  layers} & \textbf{Research  aspects} &  \textbf{Involved \emph{refs.}} & \textbf{Typical issues with  introduction of MI fast fading } \\
\hline \hline
%\raggedleft
\centering
%%%%%%%%%%%%%%%%%%%%%%%%%%%%%%%%%%%%%%%%%%%%%%%%%%%%%%%%%%%%%%%%%%%%%%%%%%%%%%%%%%%%%%%%%%%%%%%%%%%%%%%
\multirow{5}{*}{ Physical layer }&Channel modeling &  \cite{Dumphart2016Stochastic},\cite{Ma2020Channel},\cite{Ma2024Fast} &  For more universal scenarios  \\ \cline{2-4}
%%%%%%%%%%%%%%%%%%%%%%%%%%%%%%%%%%%%%%%%%%%%%%%%%%%%%%%%%%%%%%%%%%%%%%%%%%%%%%%%%%%%%%%%%%%%%%%%%%%%%%%		
& Achievable   rate &\cite{Sun2013Increasing, Ma2019Antenna, Ma2019Effect, Zhang2024Cooperative} &
Random outage probability in most cases    \\    \cline{2-4}

& Channel estimation&\cite{Kisseleff2014Transmitter} &
Unpredictable CSI; Spectrum obtaining   \\  \cline{2-4}

%	& \multirow{1}{*}{\hspace{-0.4em}Modulation} &\makecell{ \vspace{-1.em}\\\cite{Kisseleff2014Modulation} }& \makecell{\vspace{-0.9em}\\\hspace{-0.9em} Outage probability } \\ \cline{2-4}

& Channel coding&\cite{Chen2023Novel} &Unpredictable average AVI \\ \cline{2-4}

&{Power control} &\cite{Ma2024Fast} &Frequent disruptions of Nash equilibrium due to  unstable  average AVIs  \\ \cline{2-4}

& MIMO \& CMI &\cite{Li2015Capacity, Ma2019Effect, Ma2019Antenna, Zhang2024Cooperative} &  Spatial diversity; Outage probability reducing \\ \cline{2-4}

\hline

Data link layer & {MAC}  & \cite{Ahmed2016Multi, Ahmed2024Design} &   CSMA-related time / time-slot  \\ \cline{3-4}

\hline

\multirow{2}{*}{  Network layer }& Connectivity &  \cite{Sun2011Dynamic,Zhang2017Connectivity} &  Irregular shape of MI coverage space  \\  \cline{2-4}

& Routing&\cite{Chen2023Novel} &Variable latency and energy consumption \\ \cline{2-4}

\hline

\multirow{2}{*}{Cross-layer}& Cross-layer protocols &  \cite{lin2015distributed, Singh2021Optimal} &  Unpredictable average AVI; QoS requirement  \\ \cline{2-4}

& Networking &- &Variable network topology \\ \cline{2-4}	
\hline

\end{tabular}
\vspace{-1.31em}
%\end{minipage}
\label{tbl_futurefastfading}
\end{table*}

\subsubsection{Statistical modeling when the CLT does not apply} \label{sectsubsub8_smcna}

 Previous literature has derived closed-form expressions of the CDF/PDF and expectation of the MI fast fading. These expressions are only applicable in typical scenarios, such as underwater\cite{Dumphart2016Stochastic,Zhang2024Cooperative}, 2D TTE \cite{Ma2020Channel} and 3D TTE  using MI cellular network\cite{Ma2024Fast}. 
Firstly, unlike EMWCs, MI fast fading is not caused by signal propagation. As illustrated in Fig. \ref{fig_fastfadingmodel}, we modeled MI fast fading using four independent random variables  ($\phi_{\mathrm{S}}, \theta’_{\mathrm{S}}, \phi_{\mathrm{D}}, \theta’_{\mathrm{D}}$) with distinct distributions. These four random variables are not suitable for applying  CLT since  CLT requires a large number of independent random variables. Thus, deriving a universally used CDF/PDF, such as the Rayleigh model, becomes challenging. We utilize Monte Carlo simulations to obtain  universal models for MIC links (e.g., Fig. \ref{fig_sec2Ejsd}). However,  these models have high time complexity for network algorithms and protocols.
Secondly,  antenna design,  antenna carrier, and its mechanical degrees of freedom affect the MI fast fading. For example, the antennas composed of orthogonal MIMO coils, RPMA, M$^2$I, and magnetoresistance exhibit different CDFs of MI fast fading. The vibration model of the backpack antenna may not follow the boundary $p(x)$ distribution.  The mechanical degrees of freedom of the vehicle influence the distributions of horizontal components of antenna vibration $\phi_{\mathrm{S}}$ and $\phi_{\mathrm{D}}$. These interdisciplinary issues also pose the challenge to derivation of a universal statistical model.

It is noticed that the practical relevance of existing MI fast fading models remains untested. Monte Carlo simulations have validated three such models, each with AVIs distributed as uniform, BCS, and boundary $p(x)$. Despite the validation, the practical applicability of these models is yet to be confirmed. Future research can give priority to experimental validation using measured disturbance profiles across diverse environments to enhance their practical applicability.

\subsubsection{Outage probability with velocity-dependent AVIs} \label{sectsubsub8_opva}

MI fast fading influences the achievable rate through the expectation  $\mathbb{E}(J_{\mathrm{SD}})$ and  outage probability. For  mobile MIC, the expectation $\mathbb{E}(J_{\mathrm{SD}})$ is determined by the average AVIs $\sigma_{\mathrm{S}}$ and $\sigma_{\mathrm{D}}$. These AVIs are velocity-dependent, making the AVIs statistically unpredictable due to their dependence on the driver's mindset. Additionally, unlike the traditional MIC, the outage probability regains its physical meaning due to fast fading. Both the unpredictability of $\mathbb{E}(J_{\mathrm{SD}})$ and the outage probability of a mobile MIC link have a significant impact on the existing solutions for MIC networks. Unlike the EMWC, such an outage probability is still random in most cases. However, as the vehicle velocity exhibits a human activity, the attention-based deep learning (e.g., Transformer and BERT models) and large language models (LLMs) can be considered to address the challenges caused by unpredictable $\mathbb{E}(J_{\mathrm{SD}})$. 

\subsubsection{Lack of frequency spectrum for channel estimation}\label{sectsubsub8_lfsce}

For the study of MIC channel estimations \cite{Kisseleff2014Transmitter}, an MIC channel is  assumed to be quasi-static.  For  mobile MIC with a fast fading channel and a given  time $t$, the mutual inductance $M_{\mathrm{SD}}= M_{\mathrm{SD}}(\phi_{\mathrm{S}}(t), \theta’_{\mathrm{S}}(t), \phi_{\mathrm{D}}(t), \theta’_{\mathrm{D}}(t)) = M_{\mathrm{SD}}(t)$    changes faster during the coherence time $T_c$.  An mobile MIC channel exhibits time-selective fading. 
Let $T_c = \mathcal{N}_t T_t$, where $\mathcal{N}_t$  is the number of symbols between two pilot signals and $T_t$ is the duration of a symbol. The receive signal is  $\tilde{y}$ with the CDF $F_{J(\phi_{\mathrm{S}}(t), \theta’_{\mathrm{S}}(t), \phi_{\mathrm{D}}(t), \theta’_{\mathrm{D}}(t))}(y)$. For time-selective fading, one of the widely used methods is interpolation between symbols 1 and $\mathcal{N}_t$. Hence, the frequency spectrum of $J(\phi_{\mathrm{S}}(t), \theta’_{\mathrm{S}}(t), \phi_{\mathrm{D}}(t), \theta’_{\mathrm{D}}(t))$ is crucial. As the CDF of $J_{\mathrm{SD}} $ helps  obtain this frequency spectrum,  it is an open issue for MIC channel estimation. Additionally, the statistically unpredictable expectation $\mathbb{E}(G_{\mathrm{SD}})$ makes it hard to obtain accurate CSI, which is also an issue for channel estimation.

\subsubsection{Unpredictable bandwidth for  Spatial diversity for MIMO and CMICs} \label{sectsubsub8_ubsdmc}

For traditional MIC links with quasi-static channels, MIMO and CMI techniques enhance signal strength at the Rx node. The received SNR determines the feasibility of MIMO and CMI in these scenarios. For MIC with fast fading MI, outage probability must also be considered. For example, in Fig.\ref{fig_sec3cmgxyth}, CMIC is feasible in RA with $\Upsilon_{\rm AF} > \Upsilon_{\mathrm{SD}}$, where $\Upsilon_{\rm AF}$ and $\Upsilon_{\mathrm{SD}}$ are the received SNRs of CMI and DMI links  at $\mathrm{D}$, respectively. For mobile MIC, not all points in RA are suitable for a CMI relay due to the presence of outage probability conditions. Similarly, some points outside RA may be suitable for a CMI relay.

In addition, the achievable rate of an AF-relay CMI system can be expressed as  \cite{Ma2019Effect}
\begin{equation}
\begin{aligned}
\mathfrak{C}_{\rm AF}  &= \frac{1}{2}\int_{f_0-\frac{1}{2}B_{\rm AF}(\mathbf{v})}^{f_0+\frac{1}{2}B_{AF}(\mathbf{v})} \log_2(1 + \Upsilon_{\rm AF}(f))df,	
\end{aligned}
\end{equation}
where the bandwidth  $B_{\rm AF}$  is a function of  the APO  $\mathbf{v}$ and, in turn, a function of MI fast fading gains $J_{\mathrm{SD}}$, $J_{\mathrm{SR}}$ and $J_{\mathrm{RD}}$.  This poses a  challenge to the ergodic rate calculation of mobile CMI links.

\subsubsection{Non-uniform and irregular MI coverage}\label{sectsubsub8_nimc}
For a stationary MI SISO link, the polarization gain satisfies $J_{\mathrm{SD}}= 1 + 3\cos^2\theta_{\mathrm{SD}}$,  where $\theta_{\mathrm{SD}}$ is the angle between the norm vector of the coil and line $\mathrm{SD}$, implying that MI coverage space has petal shape. However, it is observed in Fig. \ref{fig_sec2Ejsd}  that the shape of  MI coverage space may become irregular in the presence of MI fast fading. This is a challenging issue for MI connectivity modeling.  On the other hand, the MI fast fading may increase the channel power gain, as shown in Fig. \ref{fig_sec2Ejsd}, presenting an exciting opportunity for network optimization.

\subsubsection{Proposed framework for average AVI prediction}\label{sect_chp6pfap}
The average AVI  relies significantly on a vehicle's velocity, which depends on driver intent and is hard to predict using conventional methods (e.g., Kalman Filter). Attention-based deep learning can be an alternative to obtaining average AVIs. We propose a Transformer-based supervised deep  learning framework to obtain a discretized/classified  average AVI, as shown in Fig. \ref{fig_sec5avipre}. In this framework, historical sequence vectors act as inputs, pass through a Transformer for prediction, and then go through dimensionality reduction via a Dense layer. Final outputs  are normalized by the softmax function. The result is obtained by taking  the index of the maximum in the outputs tensor. The boundaries  $\varepsilon_{1}$,  $\varepsilon_{2}$, ...,  $\varepsilon_{k-1}$ are determined by analyzing historical average AVI distributions (including mean, variance, and extreme values), and aligning with practical operational requirement, based on the pre-designed $k$ classes.

 Assume a standard Transformer configuration with the number of attention heads 
$h$$=$$8$, multi-head attention blocks $b$$=$$12$, and hidden dimension $d$$=$$64$. The sequence lengths are constrained to $n$$=$$128$ due to on-device memory limits.
The computational complexities of the Transformer block and subsequent Dense layer are $\mathcal{O}(bn^2d+8nd^2)$ and  $\mathcal{O}(dk)$, respectively, with a total of 16,818,176 FLOPs when $k$$=$$5$.   This workload is executable on a low-speed MCU  (100 MHz ARM Cortex-M4) with an inference latency of up to 75 ms.

The remaining issues are listed in Table~\ref{tbl_futurefastfading}.

\begin{figure}[t!]
\centering{\includegraphics[width=89mm, height=27mm]{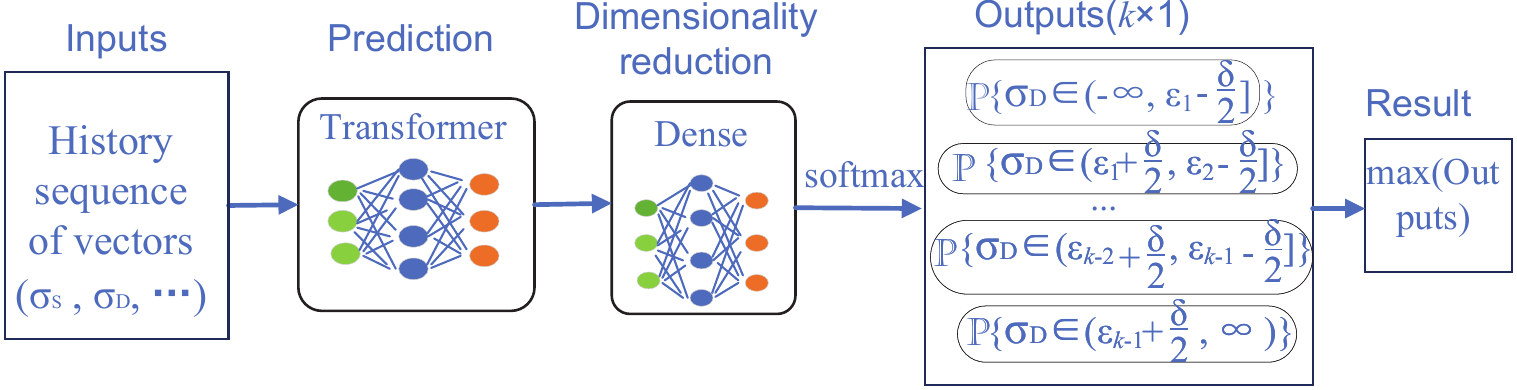}}
\caption{Proposed framework  of Rx average AVI $\sigma_{\mathrm{D}}$ prediction. This framework is a $k$-class classification supervised  deep learning framework where  predicted AVI  $\sigma_{\mathrm{D}}$ (output) is discretized into $k$ levels, and $\delta$ is the boundary margin.   Inputs include   average AVIs of Tx and Rx ($\sigma_{\mathrm{S}}, \sigma_{\mathrm{D}}$), road/traffic conditions, and vehicle velocity, and are assumed to be pre-denoised. 
}\label{fig_sec5avipre} 
\vspace{-1.0em}
\end{figure}

\begin{figure}[t!]
\centering{\includegraphics[width=88mm]{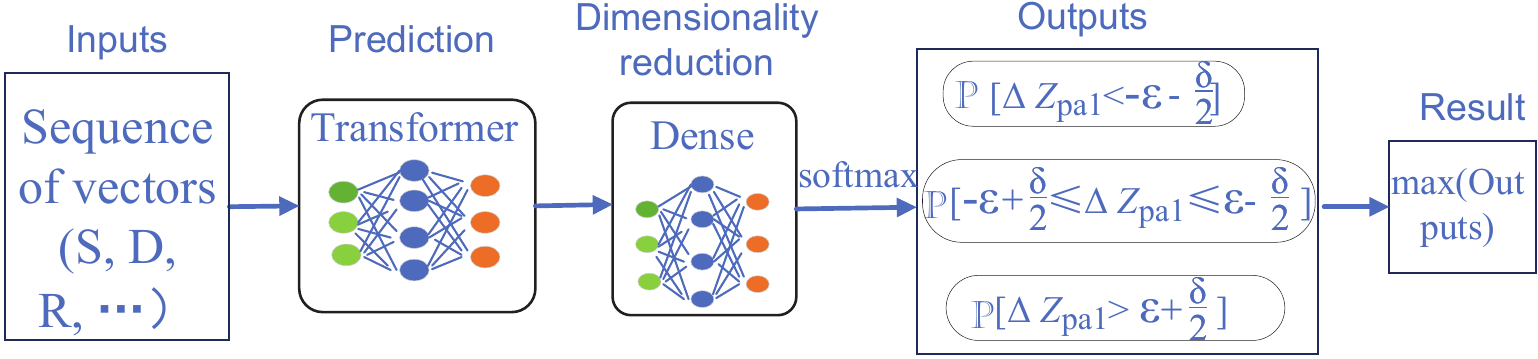}}
\caption{Proposed framework  of MI crosstalk impedances prediction. Here, the classifications $\mathbb{P}[\Delta Z_{\rm pa1}<-\varepsilon-\frac{\delta}{2}]$, $\mathbb{P}[-\varepsilon+\frac{\delta}{2}\leq\Delta Z_{\rm pa1}\leq-\varepsilon+\frac{\delta}{2}]$ and $\mathbb{P}[\Delta Z_{\rm pa1}>\varepsilon+\frac{\delta}{2}]$ represent a decrease, no change, and an increase in impedance $Z_{\rm pa1}$, respectively, where $\varepsilon$ is a boundary threshold and $\delta$ is the boundary margin.  Inputs are assumed to be pre-denoised.
}\label{fig_sec5crossmig} 
\vspace{-0.0em}
\end{figure}

\subsection{Antenna Design}\label{sectsub8_antennadesign}

Due to lower circuit gain $\mathcal{C}_{\mathrm{SD}}$, coil-based antennas are larger, especially in VLF-LA systems for TTE applications.  In  deep underground environments, the large size makes it difficult to use MIMO techniques. Several non-coil antenna types can be considered for future long-distance MICs:

\subsubsection{Mechanical antenna}\label{sectsubsub8_mba}
A mechanical antenna has a smaller size for  mid-range and mobile MICs~\cite{Liu2020Rotating, Tarek2018Power, Zhang2022Research}. Such an antenna  presents challenges stemming from  inertia.
\begin{itemize}
\item {Additional latency: The MIC protocols (e.g., modulations) need additional preset time slots between different symbols to overcome  antenna inertia.
}

\item{Additional energy consumption: Changing the mechanical state  requires additional energy to overcome inertia. Specifically, transmitting `010101' (5 state switches) requires more energy than `111000' (1 state switch).

\item{Variable data rate under the same packet length: For packets of the same length, transmitting `010101' takes 5 GTSs, while `111000' takes only 1 GTS, meaning `010101' transmits slower than `111000'.}

}
\end{itemize}

\subsubsection{Magnetic  sensor} \label{sectsubsub8_magneticsensor}
Mechanical antennas are often used in the Tx nodes, while magnetic sensors can be utilized at the Rx nodes. There are high-sensitivity magnetic sensors, such as TMR and quantum magnetic sensors. These sensors have potential applications in underground MICs. Due to the limitations of current magnetic sensor techniques, the receiver sensitivity of a magnetic sensor is much lower than that of a coil. Few studies investigate magnetic sensor-based MICs for underground applications. Nevertheless,  magnetic sensors have two advantages.
\begin{itemize} 
\item{Small antenna size: The size of a magnetic sensor is much smaller than that of a coil. For example,  a TMR sensor with a small outline package 8 (SOP8) has dimensions of 6 mm $\times$ 5 mm $\times$ 1.5 mm. These dimensions are several orders of magnitude smaller than  coils. }

\item{Low crosstalk effect:
The crosstalk effect among the magnetic sensors can be greatly reduced. Techniques such as massive MIMO and intelligent reflective surfaces can be introduced into MICs. }
\end{itemize}
Given these advantages and the advancement of magnetic sensor techniques, magnetic sensors emerge as a highly promising option for MIC antenna design.

\subsubsection{MI beamforming and massive MI MIMO} \label{sectsubsub8_mimo}

The massive MIMO technology plays a pivotal role in the fifth-generation  (5G) mobile communication systems and beyond. The MI MIMO and beamforming techniques are also widely employed for  wireless power transfer \cite{Kisseleff2015Beamforming, Kim2016Field, Li2020Enabling, Yang2017Magnetic, Xu2024Wireless} and short-range MIC\cite{Kisseleff2016Distributed, Wang2023Backscatter}. However, these techniques exhibit extremely low performance for  mid-range and long-range MICs due to the MI crosstalk effects among coils. There is no literature on massive MI MIMO.  With the advancements of  RPMA techniques and magnetic sensor technology, MI beamforming and massive MIMO techniques hold promise, as RPMA and magnetic sensors generate minimal MI crosstalk effects among coils. For a coil-based MI system, avoiding  crosstalk is crucial for the application of massive MIMO. Our simulation in Fig. \ref{fig_sec3crosstalk} may assist in addressing this challenge.
%Although Kim \emph{et al.}  developed the crosstalk line to mitigate crosstalk effects\cite{Kim2016Field}, this line has insufficient space for massive MI MIMO and beamforming deployments. Also, the  is impractical for mid-range and long-range MIC to 

\subsection{MI Crosstalk Effect Prediction Strategies}\label{sect_chp6crosstalk}
It is observed in~\eqref{eqn_chp3IdPassive} that the crosstalk impedances $Z_{\rm pa1}(\mathrm{S}, \mathrm{D}, \mathrm{R}, ...)$ and $Z_{\rm pa2}(\mathrm{S}, \mathrm{D}, \mathrm{R}, ...)$ are crucial for the crosstalk effect mitigation. Unfortunately, the expressions of crosstalk impedances are complex multimodal functions, posing a challenge for predicting  $Z_{\rm pa1}(\mathrm{S}, \mathrm{D}, \mathrm{R}, ...)$ and $Z_{\rm pa2}(\mathrm{S}, \mathrm{D}, \mathrm{R}, ...)$ using conventional methods. We propose a Transformer-based  framework (see Fig. \ref{fig_sec5crossmig}) for crosstalk impedance predictions for  crosstalk effect mitigation.  This framework is similar to that of average AVI prediction (see Fig. \ref{fig_sec5avipre}) with $k$$=$3.  Its computational complexity is   16,801,792  FLOPs with the number of attention heads $h$$=$$8$, multi-head attention blocks $b$$=$$12$, and hidden dimension $d$$=$$64$.   The sequence length is $n$$=$$128$. The latency remains under 75 ms on the 100 MHz Cortex-M4.

\subsection{MI Communication-Navigation-Sensing Integrated System} \label{sectsub8_mcis}

The MCNSI system aims to achieve high intelligence, supporting high-quality localization and sensing services while achieving high-quality communication. As summarized in Table \ref{tbl_mcnsi}, MIC studies have been focused on the stability of  signals among different positions in traditional research. By contrast, the MI localization/navigation focuses on  signal differences among different positions, while MI sensing emphasizes  sensitivity to the direction and strength of magnetic fields, both of which are highly susceptible to interference and noise.  

\begin{table*}[htp] \scriptsize%\footnotesize
\centering
%	\tiny
\caption{Comparison of  MIC, MI localization/navigation and MI sensing}
\label{tbl_mcnsi}
\scalebox{0.97}{
\begin{threeparttable} 
%	\adjustbox{max width=\textwidth, scale=0.85}{
\begin{tabular}{m{0.06\textwidth}<{\centering}|m{0.29\textwidth}<{\centering}|m{0.29\textwidth}<{\centering}|m{0.29\textwidth}<{\centering} }
\hline
\textbf{Aspects}  &  \textbf{MIC} & \textbf{MI localization / navigation (cf. \cite{Pasku2017Magnetic, Saeed2019Toward})} & \textbf{MI sensing (cf. \cite{Khan2021Magnetic})}  \\
\hline\hline
%% %% \multirow{NumberOfRows}{CellWidth\textwidth}[-Fromtop]{\centering Passive Relay}
Key       & Alternating magnetic fields & Magnetic fields differences  & Sensitivity to magnetic fields' direction \& strength\\ \hline
Techniques.       & Modulation, channel coding, MAC, routing & Magnetic fingerprint, ToA, inertial navigation  &Interference \& noise suppression\\  \hline
Signals & Active sources  & Active sources    & Passive sources  \\ \hline
Components  & Coil, RPMA, M$^2$I  & Coil, RPMA, Inertial measurement unit & TMR, Giant Magnetoresistance, Hall sensor, Coil   \\ \hline
Applications  & UG-WSN, UW-WSN, UG robot communication &AUV, Indoor position, UG Robot navigation & Oxygen content detect,  current measurement   \\
\hline
\end{tabular}
%	\begin{tablenotes}  
%		\footnotesize  
%		\item[$\dagger$]p
%	\end{tablenotes}  
\end{threeparttable}
}
\vspace{-1.31em}
\end{table*}

Recently, there have been some studies on joint localization and communication in 5G/6G technologies using  intelligent reflecting surfaces \cite{Xing2023Location, Wang2023Joint}. These signal-reflection-based techniques may not be suitable for MICs. There are key issues listed below according to these challenges and Table \ref{tbl_mcnsi}.
\begin{itemize}
\item Small MI signal difference: Since the space gain $\mathcal{S}_{\mathrm{SD}}$ in MI channel decays with the 6th power of distance (see \eqref{eqn_chp2GSDNear}),  the difference in MI signal strength between two points is much smaller than that of an EMW signal in a long-distance communication environment. 
\item Chaotic received signal strength indicator (RSSI): For  mobile MICs,  velocity-dependent MI fast fading channels cause the MI RSSI around the MI base station to be chaotic.
\item Balance between communication and navigation: In the MCNSI system, since small differences in MI signals are beneficial for MICs but detrimental to MI localization and navigation, the localization accuracy and achievable communication rate often cannot  reach their optima. We need to balance these two performance metrics.
\item Interference and noise: As both MIC and MI navigation subsystems are expected to generate sufficiently strong signals, these signals can interfere with  MI sensing.  The carrier frequency of TTE MIC using VLF-LA is closer to that of MI sensing, making interference suppression techniques (e.g., frequency band isolation) more challenging for the MI sensing subsystems.
\end{itemize}
For the first issue, we can use  dynamic weighted evolution/learning (DWE)  from \cite{Su2016Design}. Furthermore, we consider  generative adversarial networks (GANs) to obtain a super-resolution model of MI signal fingerprints. For the second issue,   RL techniques can be used for stochastically changing communication environments. For the third issue,  the key method is to obtain  closed-form expressions for the ratio of time slot allocation between  communication and navigation, and  formulate a joint optimization problem of  communication and navigation. For the fourth issue, the joint time-space-frequency isolation method can be adopted for appropriate electromagnetic compatibility design.

\subsection{Mixed-Field MI Channel Model} \label{sectsub8_mmcm}

The MIC channel is modeled within the near-field range ($k_0 d_{\mathrm{SD}} \ll 1$) in the vast majority of MIC research. There are a few works, such as \cite{Guo2021Joint, Guo2021Jointchannel, Liu2021Mechanical},  that focus on MIC in the  non-near-field  range (called mixed-field MIC) using~\eqref{eqn_chp2HField},~\eqref{eqn_chp2vech}, and~\eqref{eqn_chp2Gsd}, where $k_0 d_{\mathrm{SD}} \ll 1$ does not hold. 

Under the mixed-field conditions, the channel power gain $G_{\mathrm{SD}}$ is difficult to be sub-modeled as  $\mathcal{C}_{\mathrm{SD}}$, $\mathcal{S}_{\mathrm{SD}}$, $\mathcal{E}_{\mathrm{SD}}$ and $J_{\mathrm{SD}}$.  This makes the MIC channel model difficult to analyze since more effects of device components and environmental factors must be considered. For example, for the near-field model~\eqref{eqn_chp2GSDNear}, the antenna-vibration-based MI fast fading is primarily related to MI polarization gain $J_{\mathrm{SD}}$, which represents the APOs. For the mixed-field, more device and environment parameters ($\mu_{\mathrm{u}}$, $\epsilon_{\rm u}$, $\sigma_{\rm u}$) also determine the antenna-vibration-based MI fast fading, according to~\eqref{eqn_chp2vech}  and~\eqref{eqn_chp2Gsd}.
\begin{figure}[t!]
\centering{\includegraphics[width=69mm]{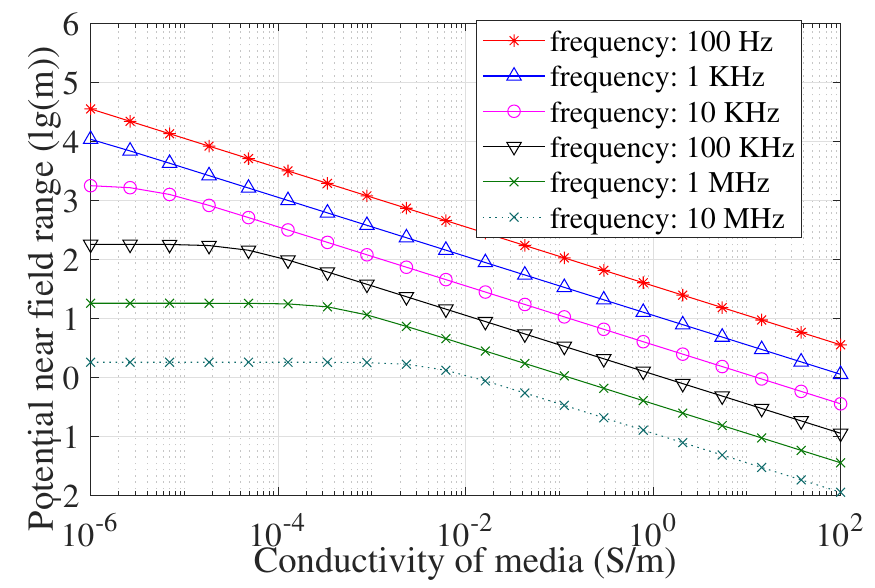}}
\caption{Near-field ranges.  The simulation parameters are listed in Table \ref{tbl_sim}, except for the media conductivity $\sigma_{\rm u}$ and frequency $f$. 
}\label{fig_sec5nearfieldrange} 
\vspace{-1.0em}
\end{figure}

In scenarios with a high conductivity medium, the near-field range is very small. As shown in Fig. \ref{fig_sec5nearfieldrange}, when the media conductivity is $\sigma_{\rm u}$$\simeq$0.01 S/m, the near-field range is less than 10 m under $f_0=100$  kHz. For TTE MIC channels with higher frequency and media conductivity, the radiation field cannot be disregarded. Studies on existing upper-layer MIC protocols relying on a near-field model may encounter issues.

\subsection {Multi-Layer and Inhomogeneous Media} \label{sectsub8_mim}
For  TTE scenarios, the underground medium is unlikely to be homogeneous and isotropic. There are three  scenarios:

\begin{itemize}
\item  Multi-layered medium:  The medium near mineral deposits is often multi-layered. In these layers, the mineral layer may have a high conductivity and even high permeability. This may invalidate  existing works. 

\item{Inhomogeneous medium with arbitrary geometric shapes: Such scenario appears more frequently than that with a multi-layered medium, such as underground tunnels like the gray thick line in Fig. \ref{fig_sec3cmgxyth}, urban subway systems, and subterranean business districts. }

\item{Dynamic medium: This scenario typically appears in  oil fields, underground rivers, and mobile MICs. In this scenario,  MIC channel exhibits  MI fast fading, despite its longer coherence time  than  symbol duration. }
\end{itemize}

For the first scenario, the boundary conditions between different layers and the boundary between the near-field region and  radiation field region need to be considered. For the second scenario, a geometrical approximation (see \cite{Ma2019Antenna}) can be used to transform geometric shapes. Methods based on Maxwell equations  (see \cite{guo2014channel, Guo2021Jointchannel}) and finite element methods (FEM) (see Fig. \ref{fig_sec2fem}) can  analyze the MIC channel model.
For the third scenario, the statistical characteristics of  dynamic medium are a key issue. While the derivation based on classical probability theorems can be used, the attention mechanism can be applied to obtain crucial channel information.

\subsection{Image Transmission and  Deep JSCC} \label{sectsub8_jscc}
\begin{figure}[t!]
\centering{\includegraphics[width=92mm]{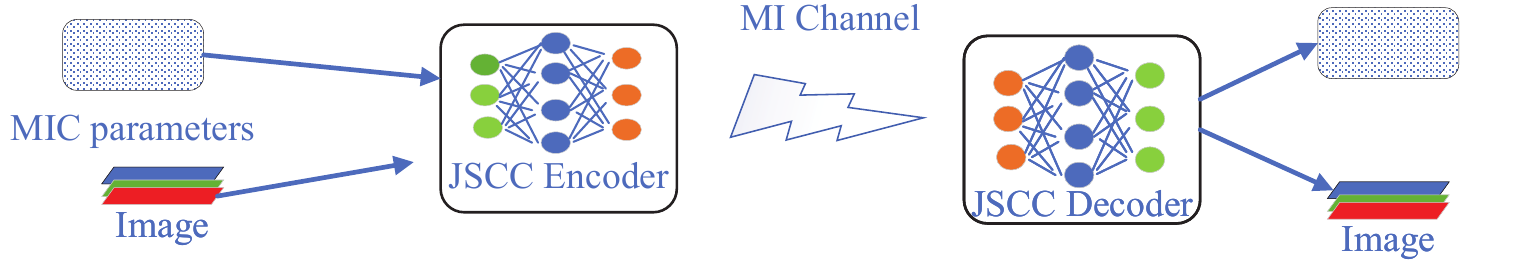}}
\caption{Block diagram of P2P MIC based on deep JSCC.
}\label{fig_sec5djscc} 
\vspace{-1.0em}
\end{figure}

Image and video transmission remains an intriguing yet unattainable function for TTE communications. It necessitates a higher achievable rate, potentially beyond Shannon's capacity for MICs utilizing   VLF-LA. Using separate source-channel coding  (SSCC), such as BCH and Polar coding, for further enhancing MI performance increases complexity with minimal gains,  hindering sustainable development. In the EMWC area, the JSCC  technique \cite{Zhai2005Joint} has been explored to tackle this issue. Meanwhile,  deep learning has been widely used to address issues on the physical layer of communication systems~\cite{Zhao2025Generative}. Deep JSCC \cite{Bourtsoulatze2019Deep,Xu2023Deep, Wu2024Deep} is a JSCC technique using deep learning.  The source code on deep JSCC is  available  on Github \cite{Zhang2025Semantics}.  It is a highly efficient approach for transmitting high-dimensional data.

Employing deep JSCC techniques, one can compress the high-dimensional data, including the underground environment information, through multimodal semantic or other goal-oriented analyses. If the deep JSCC technique is successfully applied to MICs (see Fig. \ref{fig_sec5djscc}),  the MIC achievable rate may exceed Shannon's capacity limit. Such a technique may make the transmission of images and even videos feasible.  

According to \cite{Xie2021Deep}, the deep JSCC framework can be formulated as
\begin{equation}
\mathbf{x}=\mathbb{C}_{\mathbf{\alpha}}(\mathbb{S}_{\mathbf{\beta}}(\mathbf{s})),
\end{equation}
where $\mathbf{x}$ is the encoded symbol; $\mathbf{s}$ is the input signal; $\mathbb{C}_{\mathbf{\alpha}}(\cdot)$ is the neutral network of the channel encoder with the parameter set $\mathbf{\alpha}$, including the MI device parameters and the underground environments; $\mathbb{S}_{\mathbf{\beta}}(\cdot)$ is the neutral network of the source encoder with the parameter set $\mathbf{\beta}$.

For text message transmission applications, deep semantic communication techniques\cite{Xie2021Deep,Huang2025D2}, which optimize text transmission rather than symbols transmission, can be employed to compress the overall effective data stream. Transfer learning (e.g., \cite{Xie2021Deep}) can be employed to reduce the number of pre-trained models required across different channel environments sharing the same statistical model. This is particularly relevant for  MI fast fading channels in ad-hoc links, as the PDF is difficult to derive.

\subsection{Cooperative MIC}  \label{sectsub8_cmic}

As depicted in Table \ref{tbl_cmi},  researchers have applied  either CMIC-1NR \cite{Ma2019Antenna,Ma2019Effect, zhang2014cooperative, Zhang2024Cooperative} or CMIC-$n$AR  \cite{Kisseleff2015On,Li2019Survey}. Consequently, some open issues arise, as follows.

\begin{itemize}
\item  Crosstalk effect of CMI system: The crosstalk effect in CMI systems is typically small and often ignored in long-distance CMIC investigations (see Fig. \ref{fig_sec3crosstalk}). For high node-density  networks, the key issue is the spatial distribution of negative crosstalk effects. Inevitably,   large-scale UG-WSNs with high-density nodes generate many sufficiently close node pairs. Obtaining this spatial distribution can help  upper-layer protocols, such as MAC and routing, avoid the crosstalk effect. 

\item{Multiple active relays with misaligned coils:  It has been witnessed that both CMIC-$n$AR  and CMIC-1NR can improve MIC performance. However, the improvement from multiple active relays with misaligned antennas (see Fig. \ref{fig_sec5cnar}) is less clear due to spatial diversity limitations.}
\end{itemize} 
For the first issue, we can use the KVL and fundamental matrix operations to obtain a closed-form expression of the MI crosstalk effect. However, it is still a challenge to find its spatial distribution.  

\begin{figure}[t!]
\centering{\includegraphics[width=41mm]{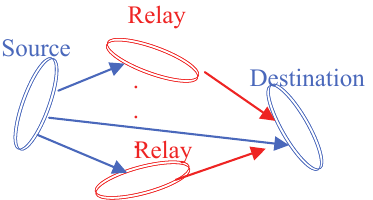}}
\caption{The topology of the CMIC with multiple active relays featuring misaligned coils.
}\label{fig_sec5cnar} 
\vspace{-1.0em}
\end{figure}

\subsection{MI Network and Architecture}  \label{sectsub8_mna}

Table \ref{tbl_osi} shows that  existing studies can form a basic protocol stack for a large-scale, runnable TTE network. Some open issues remain for further study.
%following research aspects across  OSI-originated layers.

\subsubsection{Heterogeneous MI Network} \label{sectsubsub8_hmn}

\begin{figure}[t!]
\centering{\includegraphics[width=60mm]{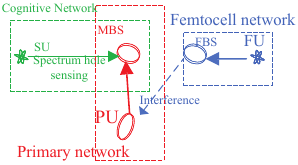}}
\caption{Examples of heterogeneous MI UG-WSN. Here,  primary users (PUs) and a macro base station (MBS) constitute the primary network. Secondary users (SUs) and an MBS can function as a part of the cognitive network. Femtocell users (FU) and a femtocell base station (FBS) constitute the femtocell network. 
}\label{fig_sec5htmi} 
\vspace{-1.0em}
\end{figure}

In TTE environments, free space for antenna deployment varies widely. It ranges from large spaces, such as subway stations and malls, to extremely small spaces, such as collapsed tunnels and mine shafts. We can deploy large antenna devices that constitute the backbone of a UG-WSN in large spaces and small antenna devices that constitute various levels of branch UG-WSNs in small spaces. Together, the backbone UG-WSN and the various levels of branch UG-WSNs form a heterogeneous MI UG-WSN (see Fig. \ref{fig_sec5htmi}),  including cognitive MI and femtocell networks.  The cognitive network addresses the issue of spectrum scarcity, while the femtocell network improves coverage and capacity in specific areas. Recent upper-layer protocol advancements highlight the significant potential of heterogeneous MI UG-WSNs on the following key aspects.
\begin{itemize}
\item { Channel licensing and spectrum sensing: The bandwidth of an MIC channel is extremely narrow (see Fig. \ref{fig_sec3bafb1}). Perceiving and utilizing spectrum holes is a fundamental task for spectrum reuse in a heterogeneous MI network. Parameters like the relationship between the resonance frequency $f_0$ and working frequency $f$ are crucial for channel licensing, spectrum sensing, and their optimization solution  (\emph{e.g.}, dynamic frequency selection).

\item{Connectivity: Existing investigations on MIC connectivity are under the assumption of uniform MI coverage. However,   heterogeneous MI UG-WSN  has different antennas, such as  SISO coil, RPMA, M$^2$C antenna, and Orthogonal MIMO coils. These MI antennas can generate MI coverage spaces with varying shapes and ranges.  This  significantly impacts the existing statistical model of MIC connectivity, such as the CDF of isolated MI nodes. }

\item{Power and throughput optimization:  Multiagent deep RL methods are widely used to address the joint power and throughput optimization problem under an unpredictable channel\cite{Chen2013stochastic, Shi2020Energy,Ortiz2022Cooperative}. The key issue is  insufficient bandwidth for exchanging cooperative packets among agents/players. This issue led to the abandonment of  cooperative multiagent RL method with high convergence performance.}% in studies like \cite{Ma2024Fast}. }
}
\end{itemize}

%\begin{comment}	
\subsubsection{MI MAC Solutions} \label{sectsubsub8_mimac}

The IEEE 802 standards divide the data link layer into the logical link control and MAC sub-layers. The logical link control sub-layer is responsible for non-media access-related functions. The functions seem to be compatible with the MIC network, but  not validated. For the MAC sub-layer, although the studies \cite{Ahmed2016Multi, Ahmed2019Design, Ahmed2024Design} have proposed MAC solutions, there are several open issues:
\begin{itemize}
\item { High orientation-sensitivity SISO: For the TTE scenarios, an SISO coil is a frequently used  MI antenna.  However, its orientation sensitivity may affect the existing MAC solutions, such as channel sensing and packet design.  }

\item{CSMA-CA: Reducing the probability of collision is an important indicator for low-bandwidth channels. The methods, such as supervised machine learning, send window optimization, and time slot optimization, can be introduced into  collision avoidance (CA) schemes to minimize the search range in the time domain.}

\item Frame error ratio (FER) and effective payload ratio (EPR): Lowering the FER  and increasing the EPR can improve the effective data rate. However, these two ratios often cannot be optimized simultaneously. Besides multi-objective and multiagent optimization methods,  antenna orientation information can  be utilized as the frame header, secret key, and even the beacon sent from the MI coordinator. Additionally, the MI MAC headers in \cite{Ahmed2024Design} require bit-level compression. 

\item Communication security: As  MIC has low bandwidth and a  frequency near resonance, it is more susceptible to interference than  EMWC. Existing security schemes in EMW-based MAC layers, e.g., those in the IEEE 802.15.4 standard, are insufficient to address this issue.  

%	\item{Energy saving: In the traditional MAC solutions, there are many MAC control headers and procedures for networking, such as CSMA-CA, timeslot competition,  frame synchronization, and connection creation. }

%\item{Frame synchronization for MIC with RPMA: The inertia of a rotating motor in MIC with RPMA can challenge MAC layer frame synchronization, influencing beacon and guarantee time slot (GTS) design. }
\end{itemize}  

%\end{comment}
\subsubsection{MI Routing Solutions}  \label{sectsubsub8_mirouting}
Routing is an important function in the network layer for a large-scale network. Since 2019, there have been many studies on MI routing issues, such as issues of lifetime, transmission delay, and routing decision algorithms. However, the channel models in studies on MI routing issues are air-based MIC channels, where the eddy gain $\mathcal{E}_{\mathrm{SD}}$ and polarization gain $J_{\mathrm{SD}}$ in TTE scenarios are ignored. Air-based MIC channels are similar to  EMWC. Whether  $\mathcal{E}_{\mathrm{SD}}(f)$ and   $J_{\mathrm{SD}}$ affect the MI routing solutions, e.g., the frequency switch scheme, remains an open issue.

\subsubsection{ MI Network Architecture}  \label{sectsubsub8_mna}
Most studies on MIC focus on the physical layer, such as MI channel modeling, channel estimation, and energy and capacity optimization. In recent years, researchers have increasingly dedicated efforts to upper-layer protocols like MAC and routing protocols (see Table \ref{tbl_osi}). The OSI-originated network framework adheres to the principle of high cohesion and low coupling in the field of software engineering, e.g., standard OSI, TCP/IP, and IEEE 802 frameworks,  This principle can  bring high communication performance, robustness, security, standardization, and compatibility to  MI networks. Although most solutions of OSI-originated network frameworks are compatible with the MI networks,   some issues can be considered for future studies and implementations.
\begin{itemize}
\item High cohesion and low coupling:  High cohesion and low coupling in a network framework benefit system stability and compatibility. To achieve higher MIC performance, cross-layer optimization methods have been proposed in the  literature (see Table \ref{tbl_osi}). These methods, to some extent, increase the coupling among different layers. Improving  MIC performance while maintaining the functionality of the mature network architecture is also a consideration in future research on MI optimization. In our  framework for future research and implementation (see Fig. \ref{fig_sec5tcp}), we  take this point into account. 
\item Standardization for MIC: As research on MIC network optimization temporarily boosts performance but disrupts the design principle of  ``low  coupling and high cohesion",  the standards for MIC are lacking.   While NFC has standards\cite{NFCStandard} (e.g., ISO 18092, NFCIP-1),  its short-range and fixed-frequency design does not align with the need of UG-WSN for wide coverage and heterogeneous network integration,  hindering access of large-scale TTE MIC  in  SAGUMI networks.

%\item{ZigBee support:}

\end{itemize}

\subsection{TCP/IP Support}\label{sect_future_tcpip}
 MIC device with TCP/IP support has broad application prospects. As most modules of TCP/IP were originally designed to be channel-agnostic, the research of  TCP/IP specific supporting for MI network has been largely overlooked. However, the general TCP/IP stack is not fully compatible with an ultra-narrow-band network due to its large packet header (i.e., low EPR).  As shown below, some techniques can be applied for  MIC-specific TCP/IP.

\begin{figure}[t]
\centering{\includegraphics[width=85mm]{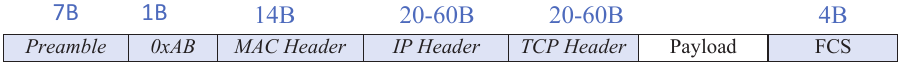}}
\caption{A typical TCP frame being propagated in a physical channel. In this figure, FCS denotes the Frame Check Sequence. Italicized text indicates the frame header, needing further compression for the VLF-LA channel.
}\label{fig_sec4tcppaket} 
\vspace{-0.0em}
\end{figure}

\subsubsection{Header Compression (HC)} \label{sectsubsub8_hc}
The TCP/IP packet header, including the MAC frame header, IP header, and TCP/UDP header, is quite large. For example, the traditional IPv4 header and  TCP header are both over 20 bytes in size (see Fig.~\ref{fig_sec4tcppaket}), causing 74\%  overhead\cite{Jia2022End}. Such large headers total over 320 bits, taking over 3 seconds to transmit on a 100 bits/s TTE MI channel. Using  header compression methods,  we can achieve a potential compression ratio of over 99\%. Besides  channel-agnostic solutions (e.g., \cite{Jonsson2004RObust}) proposed in  EMWC,  dynamic SNR-dependent header size w.r.t  frequency $f$ and  MI polarization gain $J_{\mathrm{SD}}(\theta_{\mathrm{S}},\theta_{\mathrm{D}})$  can be considered to achieve further compression.

\subsubsection{RTT Optimization (RTTO)} \label{sectsubsub8_rtto} 
In the TCP/IP framework, several schemes have been used to ensure multiple applications operate over a link through various connection schemes, such as TCP, HTTP, and LLC connections. Most of these connection schemes would face fairness challenges due to RTT  suppression in  MI channels with a low capacity. Since RTT suppression depends heavily on the transmission channel, we can potentially formulate the corresponding optimization problem w.r.t. the SNR $\Upsilon_{\mathrm{SD}}(f,J_{\mathrm{SD}}(\theta_{\mathrm{S}}, \theta_{\mathrm{D}}))$  of the MI links, which can be further transformed into the frequency and APO optimization problem.  Meanwhile,   the dynamic congestion window scheme based on  RTT and SNR can be adopted.

\subsubsection{Optimizing Data Chunking and Aggregation (ODCA)}  \label{sectsubsub8_odca}
Appropriate data chunking and aggregation help reduce the number of retransmissions. For example, the TTE MIC with a 100 bit/s data rate can use 15-byte data chunks much smaller than a typical maximum transmission unit size (1500 bytes). The receiver sends an aggregated ACK after accumulating multiple chunks. This strategy can be dynamic, adapting to the CSI of the MI channel determined by $f$ and $J_{\mathrm{SD}}(\theta_{\mathrm{S}},\theta_{\mathrm{D}})$.

\subsubsection{Intelligent Retransmission Strategy (IRS)}  \label{sectsubsub8_irs}
Priority retransmission involves immediate retransmissions of critical data (e.g., control commands) and delayed batch retransmissions of non-critical data (e.g., sensor readings). Context-aware packet loss detection dynamically adjusts retransmission timeouts based on MI channel quality predictions (e.g., SNR) to avoid unnecessary retransmissions.

\subsubsection{Machine-learning-based solution (ML-TCP/IP)}  \label{sectsubsub8_mltcpip}
For the specific MI channel, the aforementioned methods (i.e., HC, RTTO, ODCA, and IRS) can be further optimized by machine learning tools. Since the MI fast fading depends on driver's velocity, we can use  attention-based deep learning (e.g., Fig. \ref{fig_sec5avipre}) or  LLMs to predict the average AVIs $\sigma_{\mathrm{S}}$ and $\sigma_{\mathrm{D}}$, dynamically adjusting the parameters of the methods w.r.t. the expectation of channel power gain ($\mathbb{E}[G_{\mathrm{SD}}(\sigma_{\mathrm{S}} ,\sigma_{\mathrm{D}})]$) of the MI link. 

Table \ref{tbl_tcpsolutions} compares potential TCP/IP schemes (i.e., HC, RTTO, ODCA, IRS, and ML-TCP/IP) for TTE MIC systems, indicating that MI-specific optimizations can potentially establish relationships between their primary objectives and SNR-related parameters (e.g., APO and frequency).

\begin{table*}[htp] \scriptsize%\footnotesize
\centering
%	\tiny
\caption{Comparison of  potential TCP/IP schemes for TTE MICs}
\label{tbl_tcpsolutions}
\vspace{-0.8em}
\scalebox{0.99}{
\begin{threeparttable} 
%	\adjustbox{max width=\textwidth, scale=0.85}{
\begin{tabular}{m{0.08\textwidth}<{\centering}|m{0.26\textwidth}<{\centering}|m{0.27\textwidth}<{\centering}|m{0.06\textwidth}<{\centering}|m{0.20\textwidth}<{\centering}  }
\hline
\textbf{Schemes}  &  \textbf{Primary objective} & \textbf{Key mechanism} & \textbf{Complexity} & \textbf{MIC optimization directions}  \\
\hline\hline

%% %% \multirow{NumberOfRows}{CellWidth\textwidth}[-Fromtop]{\centering Passive Relay}
HC&Reduce payload overhead& Robust HC tunnel with CID negotiation\cite{Jia2022End} & Low & SNR-dependent header size\\ \hline 
RTTO&Improve fairness; minimize round-trip time& Optimize transmission/retransmission timing & Medium & RTT-SNR problem formulation\\ \hline
ODCA&Reduce retransmissions & Multiple small chunks with an aggregated ACK & Mediumn &CSI-aware optimization \\ \hline
IRS&Reduce retransmissions  &Priority retransmission scheme & Low & SNR-aware retransmission  \\\hline
ML-TCP/IP &Adapt protocol to  dynamics&AVI-driven scheme & High &Attention-based for AVI predictions \\

\hline
\end{tabular}
%	\begin{tablenotes}  
%		\footnotesize  
%		\item[$\dagger$]p
%	\end{tablenotes}  
\end{threeparttable}
}
\vspace{-1.31em}
\end{table*}

\subsection{Experiments and Testing in TTE MIC Systems}\label{sect_chp6ettms}

Some researchers have  developed  MIC testbeds for both  general UG scenario \cite{Tan2015Testbed} and  TTE scenario \cite{Zhang2017Connectivity}. Even though  TTE MIC products have  entered the market \cite{VitalCanaryComm}, a lack of robust experimental validation remains a key challenge in TTE MIC research. %This deficiency can be attributed to several factors.

\subsubsection{Unrepeatable and unrepresentative UG environments}  \label{sectsubsub8_uuue}
The  TTE environment is highly heterogeneous. Materials' conductivity, permittivity, and moisture content can vary significantly, even within a small area. Such unrepresentative environments may invalidate theoretical assumptions,  introducing measurement noise  that researchers struggle to eliminate. Moreover, VLF signals in TTE MIC are highly vulnerable to geomagnetic fluctuations and industrial interference. High-precision equipment, essential for capturing weak MI signals, is costly and scarce in academic labs. To address this challenge, a cross-scale channel model spanning centimeter-level  particles to kilometer-scale strata can be developed to match the validation requirements of target theories. A  data-driven method similar to the one depicted in Fig. \ref{fig_sec5avipre} can be considered to predict the environment.

\subsubsection{Antenna deployment challenge} \label{sectsubsub8_adc}
In deep subsurface environments, TTE MIC antennas (0.5$\sim$4 m) often exceed the space in narrow underground areas like mine tunnels. Flexible cables distort coil shapes from standard circles, especially on mobile vehicles, causing signal deviations from theoretical expectations.  For future MI fast fading validation,  high-permeability metallic components (e.g., vehicle bodies and tire bearings) act as unintended magnetic reflectors/absorbers, distorting field propagation and compromising the reliability of fading model (see Fig. \ref{fig_fastfadingmodel}) validation. Mitigation can focus on compact rigidizable antenna designs (e.g., shape-memory structures), magnetic shielding for vehicle-mounted systems, and calibration protocols accounting for metallic interference. In addition,  RPMA arrays are promising to replace  Tx coils for testing due to their smaller sizes.

\subsection{Summary of Challenges and Opportunities} \label{sectsub8_sll}

Table \ref{tbl_future} outlines the key future challenges, tasks, and recommended approaches for advancing MIC research, specifically in TTE applications. These future directions cover critical areas, such as P2P MIC communication, MI relay techniques, and overall MI network development.

One of the most pressing challenges is the development of a universal statistical model for MI fast fading. The lack of CLT support, combined with the complexity of antenna carrier configurations, has made this task particularly difficult. Furthermore, the impact of fast fading on existing MIC theorems is significant. Unlike traditional EMWC, the randomness and unpredictability of MI fast fading's expectation and variance, which are velocity-dependent, presents unique challenges that must be addressed.

There are several other  challenges related to performance metrics, antenna designs, MI MAC and routing protocols, channel modeling in inhomogeneous media,  CMIC techniques, TTE MI experiment and testing. A comprehensive understanding and resolution of these challenges is key to advancing MIC technology.
There are also  promising research directions and novel techniques that will significantly enhance MIC performance and broaden its applications. These include MCNSI, MI massive MIMO, deep JSCC for MIC, heterogeneous MI network techniques, and support for TCP/IP frameworks. Despite their potential, there is currently a scarcity of published research on these topics. In this paper, we propose an attention-based deep learning framework that is not limited to the prediction of the average AVI.
As described in Section \ref{sect_linux}, our MI network framework carefully considers the unresolved challenges and promising future research directions highlighted above.

\vspace{-0.0em}

\section{Conclusion}\label{sect conclusions}

Since 2020, research on MIC studies for TTE applications, including MI fast fading and CMIC, has increased significantly. Research in MIC continues to advance across all network layers. This survey provides a comprehensive overview of the latest developments in MIC technology within the TTE environment, covering aspects such as channel modeling in deep-penetration environments, MI fast fading channels, CMIC, and MI network architecture.
Specifically, we reviewed the MI channel modeling and P2P MIC techniques, proposed an fine-grained decomposition of MI channel power gain into four key factors, and outlined optimization directions based on these factors. We discussed the challenges and impacts of MI fast fading on MIC systems. We reviewed MI relay techniques, including the MI waveguide and CMIC, analyzed their performance in TTE environments, and explored theoretically MI cross-talk effects. Moreover, we summarized advancements in multi-node MIC and large-scale MI networks by using the OSI-originated framework as a guide, and identified outstanding issues, including channel estimation, modulation, and coding in physical layer functions, MAC protocols in link layer functions, and connectivity, data collection, and routing in network layer functions. 

Notably, we conceived a new and promising MI network framework with TCP/IP and Linux support, which enables researchers to leverage abundant research and development resources, accelerating MIC studies. We also summarized its challenges and open issues from the OSI-originated perspective, focusing on accessing SAGUMI in future network systems, and delineated the potential novel technical solutions, such as MCNSI, MI massive MIMO, deep JSCC for MIC, and heterogeneous MI network techniques.
\vspace{-0.0em}

%\appendices

%\section{Proof of Lemma \ref{prop_jbiconvex}}\label{sect_proofjbiconvex}

%The first derivative  of the  $J_{b,i+}$ and  $J_{b,i-}$ with respect to $X_i$ are

% \vfill

% by themselves when using endfloat and the captions off option.
\ifCLASSOPTIONcaptionsoff
\newpage
\fi

\bibliographystyle{IEEEtranet}
\bibliography{IEEEabrv, MIrefbrev}
% %% end hlma
%\end{comment}


% \begin{comment}

\begin{IEEEbiography}[{\includegraphics[width=1in,height=1.25in,clip,keepaspectratio]{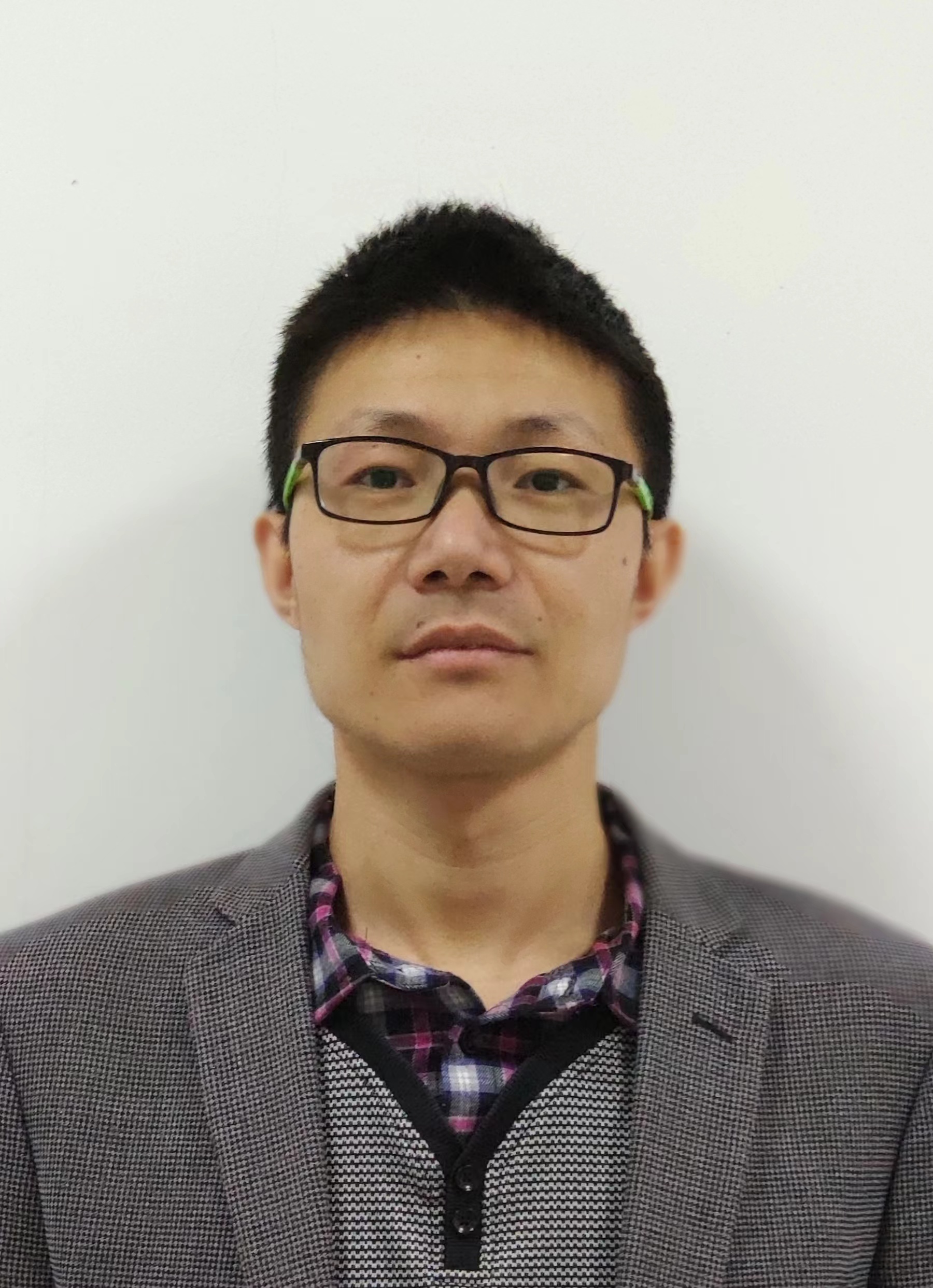}}]{Honglei Ma} received the Ph.D. degree from the  School of Electronics and Information Engineering, Tongji University, China, in 2020. He is currently with School of Electronic and Electrical Engineering, Shanghai University of Engineering Science. From 2009 to 2015, he served as a Senior Engineer  with the Chinese Academy of Sciences. His current research interests include magnetic induction communications, ad-hoc network, wireless sensor network, and object detection for  embedded systems.
\end{IEEEbiography}

\begin{IEEEbiography}[{\includegraphics[width=1in,height=1.25in,clip,keepaspectratio]{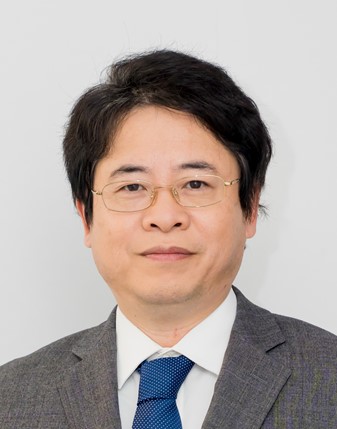}}]{Erwu Liu}
	(Senior Member, IEEE) received the Ph.D. degree from Huazhong University of Science and Technology, China, in 2001. He has been a Professor with Tongji University since 2011. Previously he was with Alcatel-Lucent (2001-2007) and Imperial College London (2007-2011). He studies localization \& sensing, AI \& blockchain, and wireless communications \& IoT, with 120+ papers published and 70+ patents granted/pending. Prof. Liu won the Microsoft Indoor Localization Competition (IPSN) in 2016 and 2018, and developed the indoor navigation system for China International Import Expo (CIIE). He is the Community Dev. Co-Chair of IEEE Blockchain Technical Community (BCTC), and leads the local group development of the IEEE BCTC in Asia/China. He leads the Shanghai Engineering Research Center for Blockchain Applications and Services (SERCBAAS). He is an IET Fellow, the Founding Editor-in-Chief of IET Blockchain, and the Founding Chair of the IEEE Global Blockchain Conference (GBC).
\end{IEEEbiography}
\vspace{1pt}

\begin{IEEEbiography}[{\includegraphics[width=1in,height=1.25in,clip,keepaspectratio]{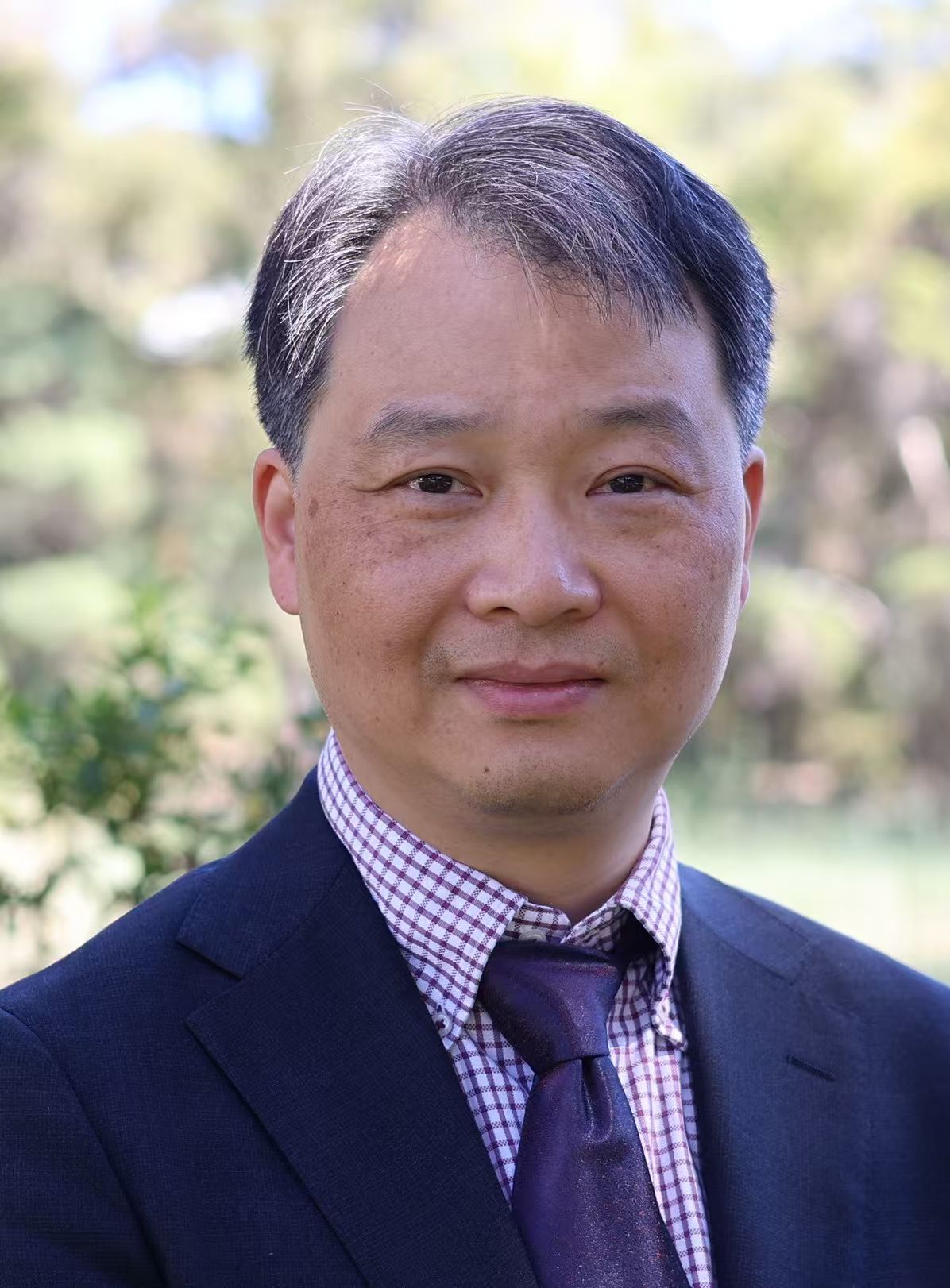}}]{Wei Ni}
	(Fellow, IEEE) 
	received the B.E. and Ph.D. degrees in Electronic Engineering from Fudan University, Shanghai, China, in 2000 and 2005, respectively. 
	He is the Associate Dean (Research) in the School of Engineering, Edith Cowan University, Perth, and a Conjoint Professor at the University of New South Wales, Sydney, Australia. 
	% He is also a Technical Expert at Standards Australia with focus on the international standardization of Big Data and AI.
	He was a Deputy Project Manager at the Bell Labs, Alcatel/Alcatel-Lucent from 2005 to 2008; a Senior Research Engineer at 
	% Devices R\&D, 
	Nokia from 2008 to 2009; and a Senior Principal Research Scientist and Group Leader at the Commonwealth Scientific and Industrial Research Organisation (CSIRO) from 2009 to 2025. 
	% He has co-authored four books, eleven book chapters, over 550 technical papers, 27 patents, and ten standard proposals accepted by IEEE. His research interests include machine learning, online learning, stochastic optimization, and their applications to system efficiency, integrity, and resilience.
	He has been an Editor for IEEE Transactions on Wireless Communications since 2018, IEEE Transactions on Vehicular Technology since 2022, IEEE Transactions on Information Forensics and Security and IEEE Communication Surveys and Tutorials since 2024, and IEEE Transactions on Network Science and Engineering since 2025. He served as Secretary, Vice-Chair, and Chair of the IEEE VTS NSW Chapter from  2015 to 2022, Track Chair for VTC-Spring 2017, Track Co-chair for IEEE VTC-Spring 2016, Publication Chair for BodyNet 2015, and Student Travel Grant Chair for WPMC 2014.
\end{IEEEbiography}

\begin{IEEEbiography}[{\includegraphics[width=1in,height=1.25in,clip,keepaspectratio]{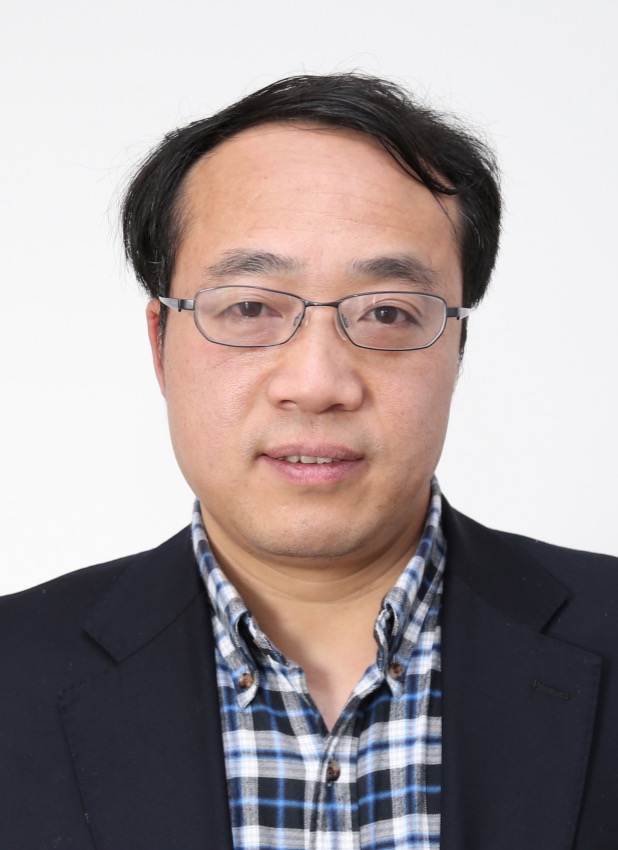}}]{Zhijun Fang}
	(Senior Member, IEEE) received the Ph.D. degree from Shanghai Jiao Tong University, Shanghai, China. He is currently a Professor and the Dean with the School of Electronic and Electrical Engineering, Shanghai University of Engineering Science. His current research interests include image processing, video coding, and pattern recognition. He was the General Chair of the Joint Conference on Harmonious Human Machine Environment (HHME) 2013 and the General Co-Chair of the International Symposium on Information Technology Convergence (ISITC) in 2014, 2015, 2016, and 2017. He received the ``Hundred, Thousand and Ten Thousand Talents Project" in China. He received several major program projects of the National Natural Science Foundation of China and the National Key Research and Development Project of the Ministry of Science and Technology of China.
\end{IEEEbiography}

%\vspace{4pt}
\begin{IEEEbiography}[{\includegraphics[width=1in,height=1.25in,clip,keepaspectratio]{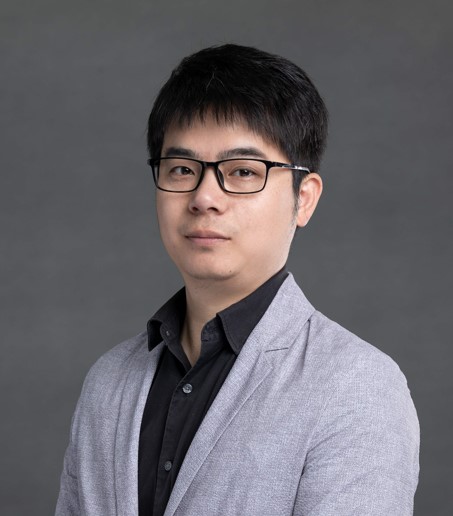}}]{Rui Wang}
	(Senior Member, IEEE) received the Ph.D. degree from Shanghai Jiao Tong University, China, in 2013. From August 2012 to February 2013, he was a Visiting Ph.D. Student with the Department of Electrical Engineering, University of California at Riverside. From October 2013 to October of 2014, he was a Post-Doctoral Research Associate with the Institute of Network Coding, The Chinese University of Hong Kong. He is currently a Professor with the College of Electronics and Information
	Engineering, Tongji University. He is also with the Shanghai Institute of Intelligent Science and Technology, Tongji University.  He has published over 60 articles. His research interests include wireless communications, artificial intelligence, and wireless positioning.  He is also an Associate Editor of IEEE ACCESS and an Editor of IEEE WIRELESS COMMUNICATIONS LETTERS.
\end{IEEEbiography}

\begin{IEEEbiography}[{\includegraphics[width=1in,height=1.25in,clip,keepaspectratio]{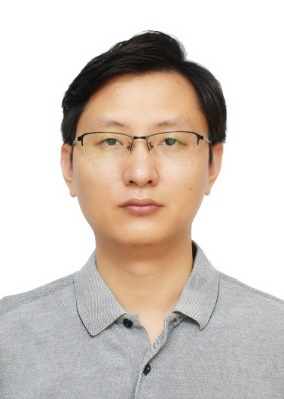}}]{Yongbin Gao}
	received the Ph.D. degree from Jeonbuk National University, South Korea. He is currently an Associate Professor with the School of Electronic and Electrical Engineering, Shanghai University of Engineering Science, Shanghai, China. He has published 30 SCI articles in prestigious journals.
\end{IEEEbiography}

\begin{IEEEbiography}[{\includegraphics[width=1in,height=1.25in, clip,keepaspectratio]{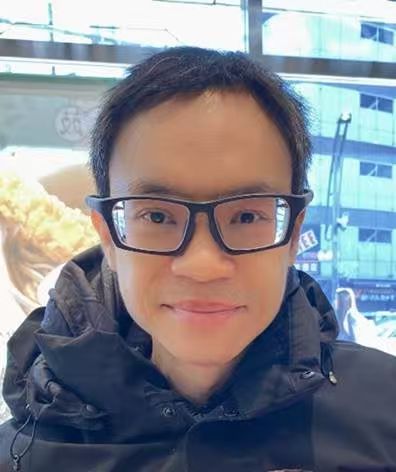}}]{Dusit Niyato}  (M'09-SM'15-F'17) is a professor in the College of Computing and Data Science, at Nanyang Technological University, Singapore. He received B.Eng. from King Mongkuts Institute of Technology Ladkrabang (KMITL), Thailand and Ph.D. in Electrical and Computer Engineering from the University of Manitoba, Canada. His research interests are in the areas of mobile generative AI, edge general intelligence, quantum computing and networking, and incentive mechanism design.
\end{IEEEbiography}

\begin{IEEEbiography}[{\includegraphics[width=1in,height=1.25in,clip,keepaspectratio]{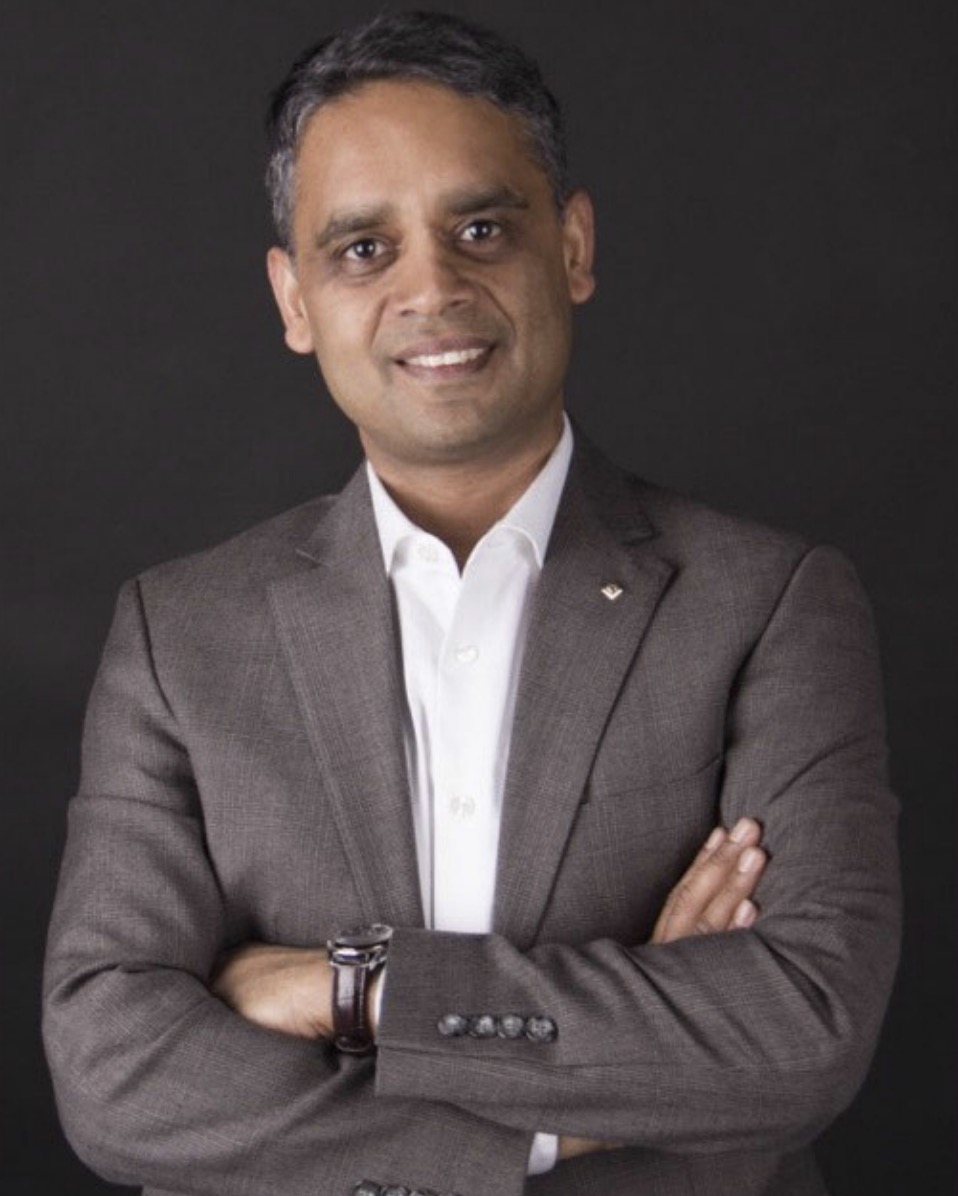}}]{Ekram Hossain}(Fellow, IEEE) is a Professor and the Associate Head (Graduate Studies) of the Department of Electrical and Computer Engineering, University of Manitoba, Canada. He is a Member (Class of 2016) of the College of the Royal Society of Canada. He is also a Fellow of the Canadian Academy of Engineering and the Engineering Institute of Canada. He has won several research awards, including the 2017 IEEE Communications Society Best Survey Paper Award and the 2011 IEEE Communications Society Fred Ellersick Prize Paper Award. He was listed as a Clarivate Analytics Highly Cited Researcher in Computer Science in 2017-2025. Previously, he served as the Editor-in-Chief (EiC) for the IEEE Press (2018–2021) and the IEEE Communications Surveys and Tutorials (2012–2016). He was a Distinguished Lecturer of the IEEE Communications Society and the IEEE Vehicular Technology Society. He served as the Director of Magazines (2020-2021) and the Director of Online Content (2022-2023) for the IEEE Communications Society.
\end{IEEEbiography}

% \vfill

\end{document}